\documentclass [12pt,oneside]{report}
\usepackage{setspace}

\input{amssymb.sty}
\usepackage[dvips]{graphicx}
\usepackage{rotate}
\usepackage{longtable}
\usepackage[dvips]{lscape}
\usepackage{supertabular}
\usepackage{rotate}
\begin{document}
\pagestyle{plain}
\baselineskip=18pt

\onehalfspacing
%\small\normalsize
\renewcommand{\thefootnote}{\fnsymbol{footnote}}
\bibliographystyle{unsrt}
%
% all the preliminary stuff
% all the preliminary stuff
\pagenumbering{roman}
\setcounter{page}{0}
\def\baselinestretch{1.5}
\begin{titlepage}         
\hoffset 0.in
\begin{center}
\large NORTHWESTERN UNIVERSITY\\
\vspace{0.45in}
{\Large\bf A Search for the Singlet-P State $h_c (1 ^1P_1)$ of 
Charmonium in Proton-Antiproton Annihilations
at Fermilab Experiment E835p}\\
\vspace{.35in}
\large A DISSERTATION\\
\vspace{.1in}
\large SUBMITTED TO THE GRADUATE SCHOOL\\
\mbox{IN~PARTIAL~FULFILMENT~OF~THE~REQUIREMENTS}\\
\vspace{.2in}
\large for the degree\\
\vspace{.05in}
\large DOCTOR OF PHILOSOPHY\\
\vspace{.1in}
\large Field of Physics and Astronomy\\
\vspace{.25in}
\large By\\
\vspace{.15in}
\large David N. Joffe\\
\vspace{.15in}
\large EVANSTON, ILLINOIS\\                        
\vspace{.1in}
\large December 2004\\   
\end{center}
\end{titlepage}
\def\baselinestretch{2.0}

\setcounter{page}{2}
%this page should contain the copyright info--I will format this later.
\newpage

\vspace*{\fill}
\begin{center}
\copyright  $\;$by David Noah Joffe 2004\\ 
\vspace{7 mm}
All Rights Reserved
\end{center}

% BEGIN ABSTRACT PAGE
\newpage
\chapter*{\centering \large ABSTRACT}
\addcontentsline{toc}{chapter}{ABSTRACT}
\vspace{-1cm}
	{\centering
		{\Large\bf
%%%% Put the title on the next line
A Search for the Singlet-P State $h_c (1 ^1P_1)$ of 
Charmonium Formed in Proton-Antiproton
Annihilations at Fermilab Experiment E835p}\\
}
%%%% If there is a subtitle, put it on the next line
\vspace{20pt}
%%%% Put your name on the next line
\begin{center}
David Noah Joffe
\end{center}	
%%%% Put in the text of your abstract, maximum 350 words.
\vspace{20pt}

We present the results of a search for the spin-singlet P-wave state $h_c (1 ^1P_1$) of charmonium formed through proton-antiproton annihilation at Fermilab experiment E835. The decay channels which were studied were $p\bar p \rightarrow J/\psi + X \rightarrow e^{+}e^{-} + X$, $p\bar p \rightarrow J/\psi + \pi^0 \rightarrow e^{+}e^{-} + \gamma\gamma$, $p\bar p \rightarrow J/\psi + \pi^0\pi^0 \rightarrow e^{+}e^{-} + 4\gamma$, and the neutral channel $p\bar p \rightarrow \eta_c\gamma \rightarrow (\gamma\gamma)\gamma$. The decay $p\bar p \rightarrow J/\psi \gamma \rightarrow e^{+}e^{-}\gamma$, into which $^1P_1$ decay is forbidden by C-parity conservation, was also examined for comparison. 

The $90\%$ confidence upper limits for the decay channels studied in the mass range 3525.1-3527.3 MeV for a $^1P_1$ resonance with a presumed width of 1.0 MeV were determined to be $B(p\bar p \rightarrow ^1P_1) \times B(^1P_1 \rightarrow J/\psi +X) \le 1.8 \times 10^{-7}$, $B(p\bar p \rightarrow \enskip ^1P_1) \times B(^1P_1 \rightarrow J/\psi + \pi^0) \le 1.2 \times 10^{-7}$, and $B(p\bar p \rightarrow ^1P_1) \times B(^1P_1 \rightarrow J/\psi \gamma) \le 1.0 \times 10^{-7}$. No evidence for a $^1P_1$ enhancement was observed in either of the two additional reactions studied; $p\bar p \rightarrow J/\psi + \pi^0\pi^0 \rightarrow e^{+}e^{-} + 4\gamma$ and $p\bar p \rightarrow \eta_c\gamma \rightarrow (\gamma\gamma)\gamma$.

\renewcommand{\baselinestretch}{2.0}\small\normalsize

\newpage

\vspace*{2.0 in}
\begin{center}
To my family and friends, but especially my parents, for all their support.
\end{center}

% BEGIN ACKNOWLEDGEMENTS PAGE
% This page is required and there is a lot which is required to be on it, see
% the standard.
\newpage
\chapter*{\Large Acknowledgements}
\addcontentsline{toc}{chapter}{Acknowledgements}
\vspace{-1.0cm}
\onehalfspacing
\setstretch{1.09}

First of all, I would like to thank my advisor, Kamal K. Seth for all his guidance and advice throughout the six years of our working relationship. I am very grateful to have had the opportunity to learn from such an experienced and knowledgable teacher, and his insight into experimental physics was invaluable to me.

I would also like to thank the members of our research group with whom I've been fortunate enough to have had the opportunity to work during my stay at Northwestern; Todd Pedlar, Xiaoling Fan, Pete Zweber, Ismail Uman, Willi Roethel, Sean Dobbs, Zaza Metreveli, and especially Amiran Tomaradze. Their assistance and friendship has helped make my Northwestern experience both productive and enjoyable, and I am very grateful to have had such a supportive research environment in which to work.

To all my collaborators in the E835 experiment, and especially our spokespersons, Stephen Pordes and Rosanna Cester, my heartfelt thanks and best wishes. I am also grateful to the many other students and postdocs without whom the year 2000 run of E835 would not have been successful. In particular I would like to thank Paolo Rumerio for all his help.

Finally, I would like to thank my friends and family, but especially my parents, for their constant and unwavering support of my education and of my career in physics. Thank you for everything.

\onehalfspacing

\tableofcontents
\listoffigures
\addcontentsline{toc}{chapter}{List of Figures}
\listoftables
\addcontentsline{toc}{chapter}{List of Tables}
\newpage
\pagestyle{myheadings}
\newpage

%Chapter 1--Introduction
\pagenumbering{arabic}
\baselineskip=24pt
\chapter{Introduction}

In the Standard Model of elementary particle physics, all matter is described as being composed of quarks and leptons, and the forces between them are described as being mediated by four gauge bosons; photons mediate the electromagnetic force, gluons mediate the strong hadronic or nuclear force, and $W^{\pm}$ and $Z^0$ bosons mediate the weak force. The quarks participate in all three forces, the charged leptons in the electromagnetic and weak forces, while the neutral leptons (neutrinos) participate only in the weak force. In this description of the natural world, these constituents are themselves structureless or pointlike; they are considered to be truly {\it fundamental}. Although the idea of explaining the world in terms of a series of ultimate constituents goes back at least as far as the ancient Greeks with the atomic theories of Democritus and Epicurus, the Standard Model is by far the most successful such description, effectively including all basic physical phenomena with the exception of gravity. The quarks and leptons which make up the basis of the Standard Model are shown in Table 1.1.

\renewcommand{\arraystretch}{0.5}
\begin{table}[t]
\centering
\caption[Fundamental constituents of the Standard Model]{Components of the Standard Model: the three generations of leptons and quarks, and the gauge bosons. The masses
for quarks are the so-called 'constituent quark' masses.~\cite{PDG2002}}
\begin{tabular}{|c||lll|lll|}
\multicolumn{7}{c}{\rule{0mm}{0mm}}\\
\hline
 & & & & & & \\
Generation & \multicolumn{3}{|c|}{Quarks} & \multicolumn{3}{|c|}{Leptons}\\
 & & & & & & \\
\cline{2-4}\cline{5-7}
 & & & & & & \\
 & & Charge & Mass & & Charge & Mass \\
 & & & & & & \\
\hline
\hline
 & & & & & & \\
I & $d$ & $-1/3$ & $\sim 300$ MeV & $e^{-}$ & $-$1 & 0.511 MeV\\
 & & & & & & \\
  & $u$& $+2/3$ & $\sim 300$ MeV & $\nu_{e}$ & 0 & $\leq$ 15 eV \\
 & & & & & & \\
\hline
 & & & & & & \\
II & $s$ & $-1/3$ & $\sim 500$ MeV & $\mu^{-}$ & $-$1 & 105.66 MeV \\
 & & & & & & \\
& $c$ & $+2/3$ & $\sim 1500$ MeV & $\nu_{\mu}$ & 0 & $\leq$  170 keV \\
 & & & & & & \\
\hline
 & & & & & & \\
III & $b$ & $-1/3$ & $\sim 4700$ MeV & $\tau^{-}$ & $-$1 & 1777.05 MeV \\
 & & & & & & \\
& $t$ & $+2/3$  & 173.8 GeV &$\nu_{\tau}$ & 0 & $\leq$ 18.2 MeV \\
 & & & & & & \\
\hline
\hline
 & & & & & & \\
Gauge & $\gamma$ & 0 & 0 & $W^{\pm}$ & $\pm 1$ & 80.42 GeV \\
 & & & & & & \\
Bosons & $g$ & 0 & 0 & $Z^0$ & 0 & 91.19 GeV \\
 & & & & & & \\
\hline
\end{tabular}
\end{table}

In the Standard Model, all strongly interacting, or hadronic, matter (and thus the vast majority of the matter of which our world is made) is composed of quarks, which interact via the exchange of gluons. The theory governing this quark-gluon interaction is known as Quantum Chromodynamics, or QCD. Many of the basic concepts of QCD date back to the 1960's, and the term 'quark' itself was coined by Gell-Mann in 1964~\cite{gellmann} in his discription of the 'eight-fold way' of understanding the symmetries among hadrons. Three quarks, 'up', 'down' and 'strange' were initially proposed to explain all hadronic states then observed. In this three quark model, the lowest-mass hadrons are explained as bound states of three quarks (baryons), three antiquarks (antibaryons), or a quark-antiquark pair (mesons). The lowest-mass baryons (and antibaryons) are arranged into an octet of $J^{P} = {1 \over 2}^{+}$ states and a decuplet of $J^{P} = {3 \over 2}^{+}$ states, where $J$ and $P$ are the spin and parity quantum numbers (see Figure 1.1). The lowest mass mesons are arranged into two nonets with $J^{P} = 0^{-}$ and $J^{P} = 1^{-}$. 

\begin{figure}
\begin{center}
\includegraphics[width=5.5in]{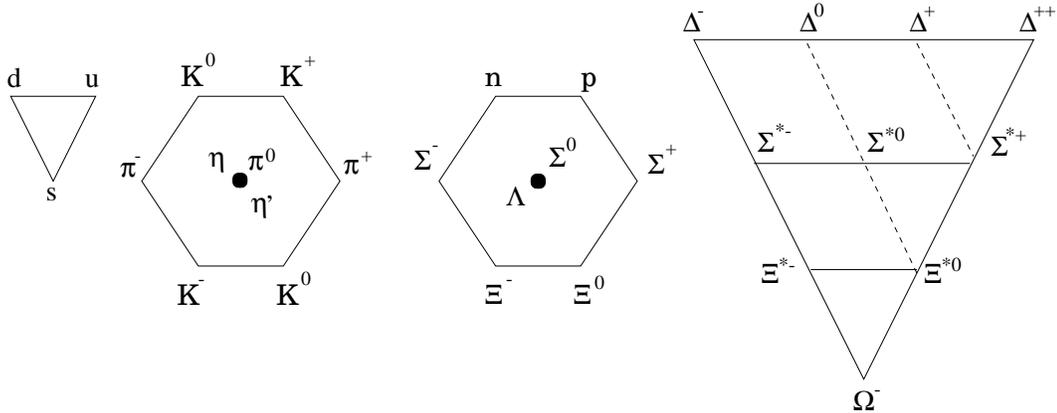}
\caption{SU(3) classification of light quarks,
and of hadrons into a $J^{P} = 0^{-}$ meson nonet,
a spin-1/2 baryon octet, 
and a spin-3/2 baryon decuplet.}
\end{center}
\end{figure}

In 1964, Greenberg proposed the term 'color' to represent the strong interaction charge`\cite{greenberg}. Three such 'colors' were necessary to explain the existance of the corner members of the baryon decuplets, each of which contains three identical quarks in relative s-states. The Pauli exclusion principle forbids this unless the three quarks are made distinguishable by assigning them a new quantum number, color, making them antisymmetric under color exchange. Although the quarks themselves were assigned three colors, termed {\it red}, {\it green}, and {\it blue}, only color neutral (or {\it white}) objects were allowed to exist freely in nature. The interaction between quarks was mediated by massless vector bosons called {\it gluons}, which carry both color and anti-color. This color interaction between quarks and gluons forms the basis of the theory of Quantum Chromodynamics (QCD) which will be discussed in detail in Chapter 2.

Although three quarks were sufficient to explain all hadrons which had been observed until the 1960's, a fourth quark was soon proposed. Bjorken and Drell first proposed its existance in order to allow the quarks to be grouped into two doublets in analogy to the two lepton doublets, $(e, \nu_e)$ and $(\mu, \nu_{\mu})$, then known to exist~\cite{bjorken}. They gave it the name 'charm' quark.

Further evidence for the existance of the fourth quark came when Cabibbo proposed \cite{cabibbo} that the quark states actually participating in the
weak interaction were the $u$ and an admixture, 
$~d_c=d\,\cos{\theta_c}+s\,\sin{\theta_c}$\,, of the physical 
quarks $d$ and $s$. This model successfully explained the experimentally-observed suppression of the 
strangeness-changing ($~\Delta S=1~$)
semileptonic weak decays with respect to the strangeness-conserving 
($~\Delta S=0~$) decays. The experimental value of the Cabibbo angle $\theta_c$ was found to be 0.25~rad. One consequence of the Cabibbo theory, however, was the existence of neutral currents with $~\Delta S=1~$, which were not observed in
nature. A famous paper by Glashow, Iliopoulos and Maiani\cite{Glashow} in 1970 solved this problem and provided strong theoretical support for the existence of a fourth quark, named {\it charm}.
The authors proposed that the charm quark $c$ would participate in the weak
interaction in a doublet with a state, 
$s_c=s\,\cos{\theta_c}-d\,\sin{\theta_c}$, orthogonal to the $d_c$ state.
Hence the two quark doublet for weak interaction would be:
\begin{displaymath}
{u \choose d_c=d\,\cos{\theta_c}+s\,\sin{\theta_c}} ~~~~ 
{c \choose s_c=s\,\cos{\theta_c}-d\,\sin{\theta_c}}~, 
\end{displaymath}
\begin{displaymath}
\mathrm{i.e.,} \quad
{u \choose 0.97d + 0.25s} ~~~~ 
{c \choose 0.97s - 0.25d}~. \quad\quad 
\end{displaymath}
Thus $s_c\bar s_c +d_c\bar d_c = s\bar s + d\bar d$ independently of the mixing angle $\theta$ leading to a vanishing term for the neutral currents with $~\Delta S=1~$, in agreement with the experimental observations.

Despite these theoretical successes, the quark model, and thus the core of the Standard Model itself, really only became universally accepted a decade later with the discovery of the $J/\psi$ in 1974.

This discovery was made essentially simultaneously by two separate groups working on opposite sides of the U.S. The east coast group, led by Ting at the Brookhaven National Laboratory (BNL), reported the observation of a particle which they called $J$, in the $e^{+}e^{-}$ invariant mass in the reaction $p + Be \rightarrow e^{+}e^{-} + X$ at 28 GeV~\cite{Aubert}. They reported a mass of 3.1 GeV, and a width ``consistent with zero''. The BNL results can be seen in Figure 1.2. The west coast group, led by Richter at the Stanford Linear Accelerator (SLAC), reported a resonance, which they called $\psi$, in the reactions $e^{+}e^{-} \rightarrow e^{+}e^{-}, \mu^{+}\mu^{-}, hadrons$~\cite{Augustin}. The SLAC results can be seen in Figure 1.3. They reported a mass of 3.105(3) GeV, and a width of $\le 1.3 MeV$. Both discoveries were published in the December 2, 1974 issue of the Physical Review Letters, and the resonance eventually came to be known as the $J/\psi$. The discovery was quickly confirmed by $e^{+}e^{-}$ annihilation experiments at Frascati~\cite{Bacci} and DESY~\cite{Braunschweig}.

The discovery of the $J/\psi$ prompted a spate of theoretical papers within weeks of the announcements, the most important of which were that of Appelquist and Politzer~\cite{Applequist} and De Rujula and Glashow~\cite{DeRujula} which proposed the interpretation of the $J/\psi$ as the bound state of a charm quark and an anticharm quark. 

\clearpage

With the discovery of the $J/\psi$, and its interpretation as a charm-anticharm $c\bar c$ bound state, the second family of the quark sector of the standard model became experimentally established. Other $c\bar c$ states were soon found, starting with the $\psi^{\prime}$, the first radial excitation of the $J/\psi$, which was discovered at SLAC only days after the $J/\psi$ was observed~\cite{Abrams}. 

\begin{figure}[htb]
\begin{center}
\includegraphics[width=12cm]
{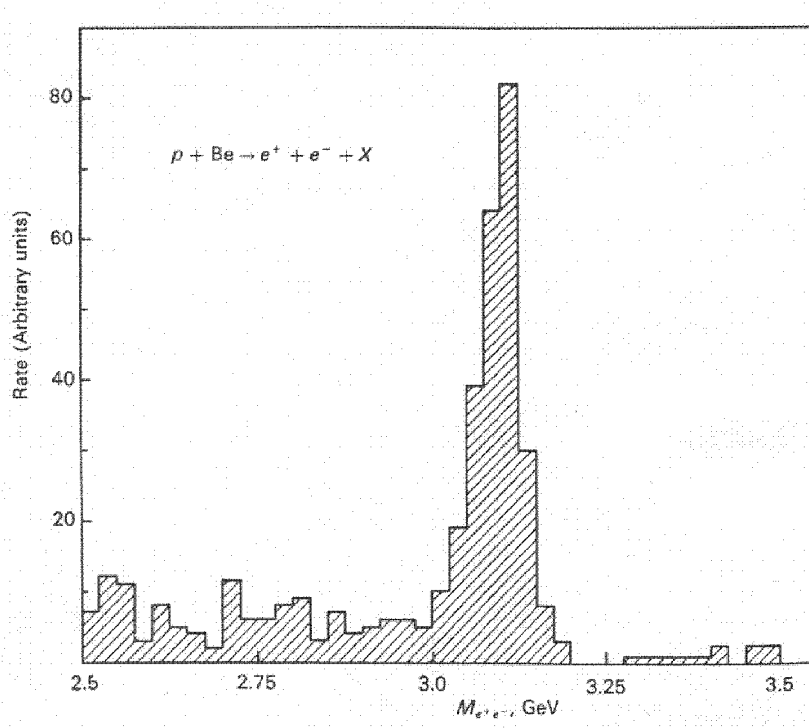}
\caption[Results of Aubert {\it et al.} (1974) indicating the narrow resonance  $J/\psi$ in the invariant mass distribution of $e^{+}e^{-}$ pairs produced in inclusive reactions of protons with a beryllium target.]{Results of Aubert {\it et al.} (1974) indicating the narrow resonance  $J/\psi$ in the invariant mass distribution of $e^{+}e^{-}$ pairs produced in inclusive reactions of protons with a beryllium target. The experiment was carried out with the 28-GeV AGS at the Brookhaven National Laboratory.~\cite{Aubert}~\cite{Perkins}}
\label{fig:1974a}
\end{center}
\end{figure}

\begin{figure}[htb]
\begin{center}
\includegraphics[width=10cm]
{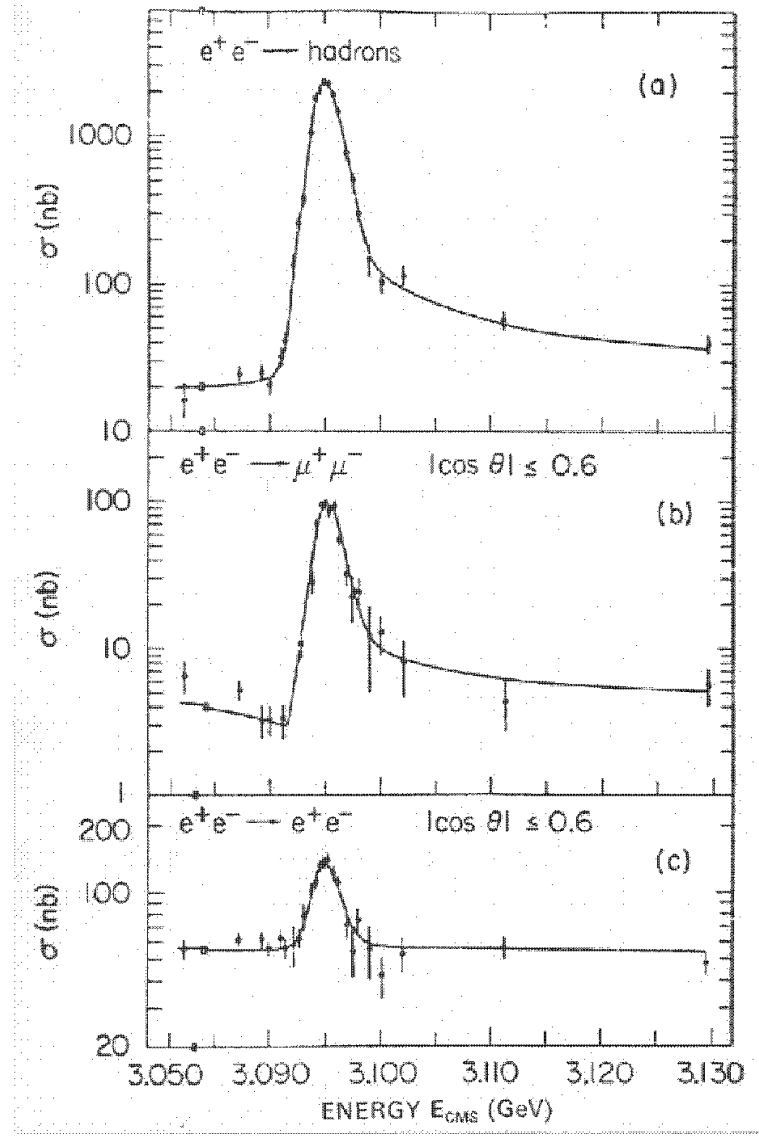}
\caption[Results of Augustin {\it et al.} (1974) showing the observation of the $J/\psi$ resonance of mass 3.1 GeV, produced in $e^{+}e^{-}$ annihilation at the SPEAR storage ring, SLAC.]{Results of Augustin {\it et al.} (1974) showing the observation of the $J/\psi$ resonance of mass 3.1 GeV, produced in $e^{+}e^{-}$ annihilation at the SPEAR storage ring, SLAC.~\cite{Augustin}~\cite{Perkins}}
\label{fig:1974b}
\end{center}
\end{figure}

\clearpage

Further evidence that the newly found $J/\psi$ and $\psi^{\prime}$ states were in fact bound states of $c\bar c$, was that their widths were soon determined to be very small, $<$~100 keV and $<$~300 keV respectively. Most strong interaction resonances with smaller masses were known to be much wider, as large as a few hundred MeV, i.e., three orders of magnitude larger than those of the newly observed states. This made it difficult to explain $J/\psi$ and $\psi^{\prime}$ in terms of the $u$, $d$, and $s$ quarks. By interpreting the $J/\psi$ and $\psi^{\prime}$ as $c\bar c$ states, and appealing to the Okubo-Zweig-Iizuka (OZI) rule~\cite{okubo}~\cite{zweig}~\cite{iizuka}, the narrowness of the states could be easily explained.

The OZI rule states that processes which can only be described by diagrams that contain disconnected lines (i.e. no quark flow) between the initial and final states should be strongly suppresed as compared to diagrams which contained connected lines (see Figure 1.4(a,b))

\begin{figure}[htb]
\begin{center}
\includegraphics[width=9cm]{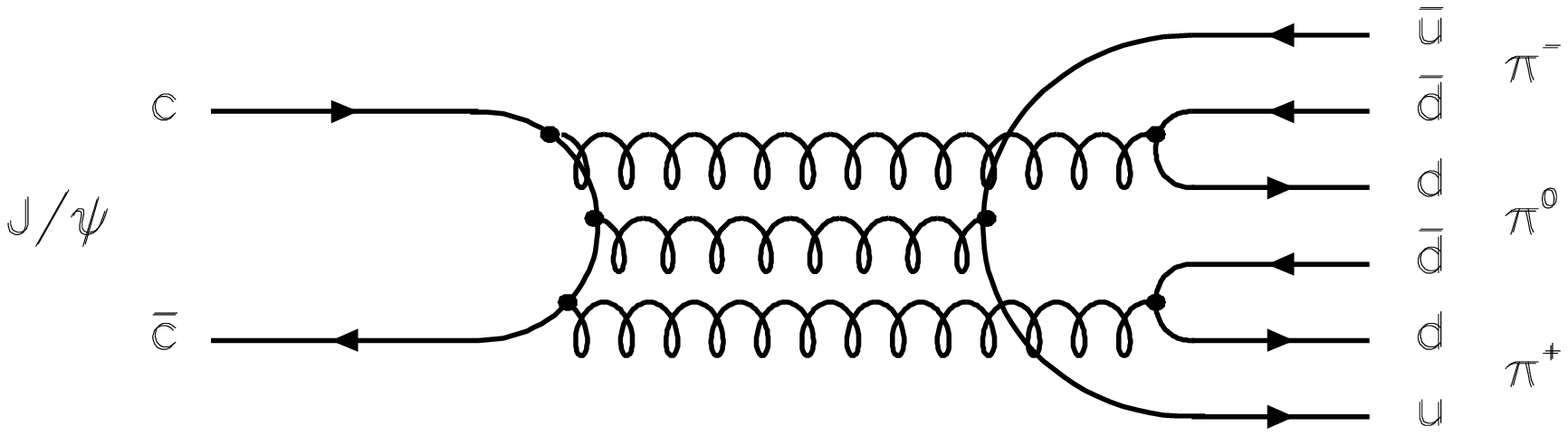}
\includegraphics[width=9cm]{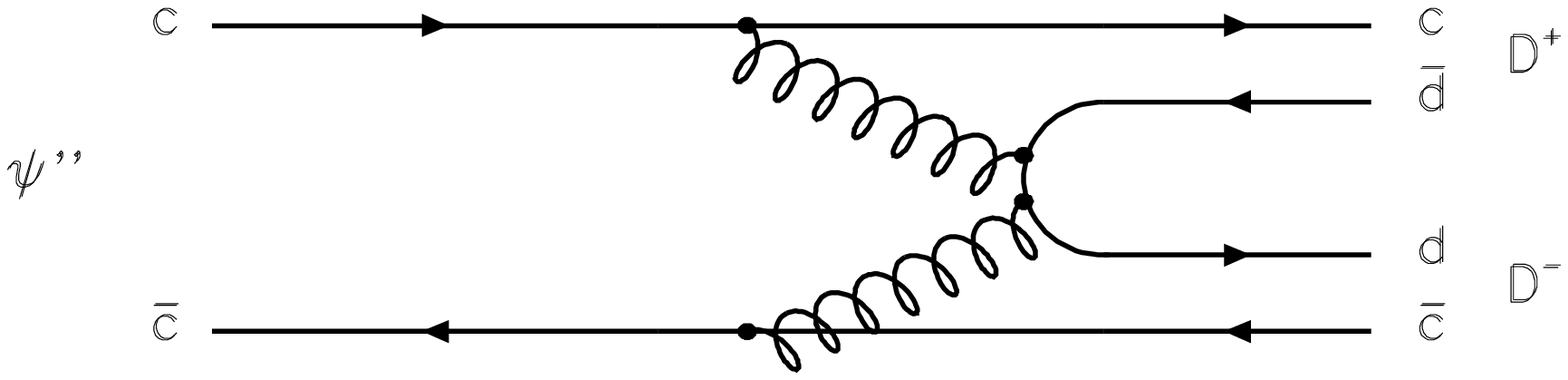}
\caption[Examples of disconnected and connected diagrams]
{Examples of disconnected and connected diagrams. Top (a): the disconnected diagram of the decay $J/\psi \rightarrow \pi^-\pi^0\pi^+$. Bottom (b): the connected diagram of the decay $\psi^{\prime\prime}(3770) \rightarrow D^+D^-$.}
\label{fig:jpsi_3pi}
\end{center}
\end{figure}

The OZI rule can be explained intuitively by noting that diagrams containing disconnected lines require the emission of ``hard'' or highly energetic gluons, which must carry the full four-momentum of the annihilating quark-antiquark pair (Fig. 1.4(a)). These gluons are much less likely to be produced than the ``soft'' gluons emmitted by a quark that continues to exist in the final state, as in the connected-line diagram (Fig. 1.4(b)). Thus, $c\bar c$ states which do not have enough mass to decay via the lowest-energy connected diagram ($D^{+}D^{-}$), must necessarily be narrow, thus expaining the small widths of the $J/\psi$ and $\psi^{\prime}$ resonances. After their spins had been determined by studying interference effects and angular distributions of decay products, the $J/\psi$ and $\psi^{\prime}$ were assigned the quantum numbers of the photon: $J^{PC} = 1^{--}$, where $J$ and $P$ are the spin and parity and $C$ is the charge conjugation parity.

After the discovery of the $J/\psi$ and the $\psi^{\prime}$, the SLAC group continued to make important new observations over the next several years with succesively improved detectors; Mark I, Mark II, Mark III, and finally Crystal Ball. After the $J/\psi$ and $\psi^{\prime}$, the SLAC group observed the $\chi_{cJ}$ resonances (the $0^{++}, 1^{++}$, and $2^{++}$ bound states of $c\bar c$), followed by the $0^{-+}$ $c\bar c$ ground state, the $\eta_c$, named in analogy with the light quark $\eta$. Figure 1.5 shows the Crystal Ball observation of the $\chi_{cJ}$ resonances, $\eta_c$, and $\eta_c^{\prime}$ (later shown not to be true)~\cite{crystalball}. The DESY, Orsay, and Frascati groups also made imporant contributions to the spectroscopy of $c\bar c$ states, which became known collectively as {\it charmonium}. The spectrum of charmonium states is shown in Figure 1.6. Charmonium states are labeled using the spectroscopic notation $n ^{2s+1}L_J$, where $n$ is the number of nodes in the radial excitation plus one, $s$ is the combined spin of the two quarks, $L$ is the orbital angular momentum, and $J$ is the total angular momentum. Parity and charge conjugation, as in any quark-antiquark state, are given by $P = (-1)^{L+1}$ and $C = (-1)^{L+S}$ respectively. A more abbreviated notation is to characterize the states just by their $J^{PC}$.

\begin{figure}[htb]
\begin{center}
\includegraphics[width=12cm]
{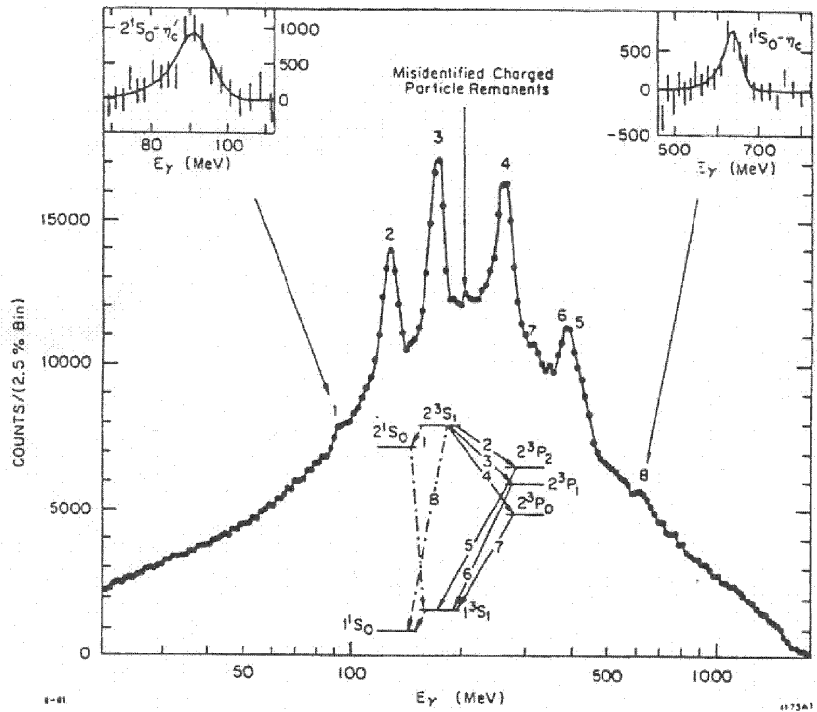}
\caption[Inclusive photon spectrum from $\psi^{\prime}$ decay, as measured by Crystal Ball (SLAC).]{Inclusive photon spectrum from $\psi^{\prime}$ decay, as measured by Crystal Ball (SLAC).~\cite{crystalball}}
\label{fig:crystalball}
\end{center}
\end{figure}

\begin{figure}[htbp]
\begin{center}
\includegraphics[width=14.5cm]
{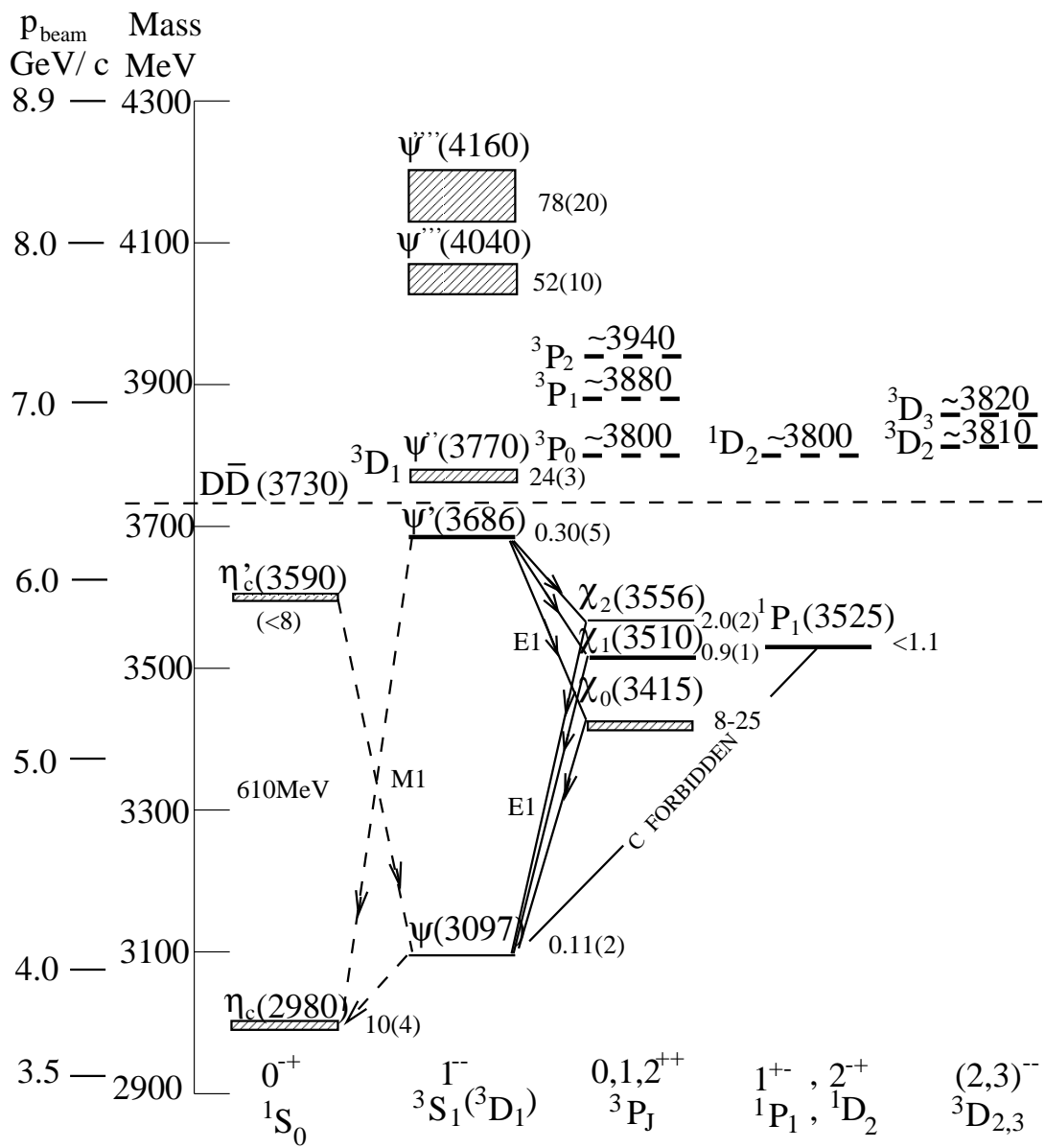}
\caption{Spectrum of charmonium resonances.}
\label{fig:ppprod}
\end{center}
\end{figure}

\clearpage

The last charmonium state to be discovered was the $\eta_c^{\prime}$, the radial excitation of the $\eta_c$ ground state, whose existance was firmly established only in 2003, by Belle~\cite{belle}, by our own group at CLEO\cite{cleo}, and BaBar~\cite{babar}. The CLEO results are shown in Figure 1.7.

\begin{figure}[htb]
\begin{center}
\includegraphics[width=11.5cm]
{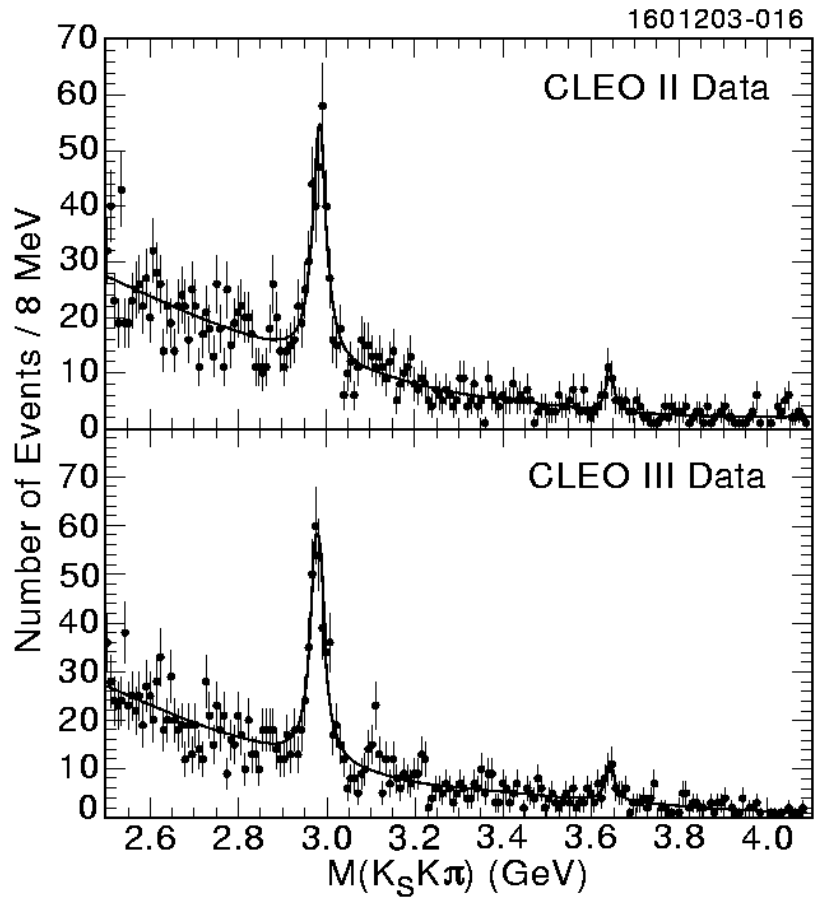}
\caption[$K_{S}^{0}K^{\pm}\pi^{\mp}$ invariant mass in the reaction 
$\gamma\gamma\rightarrow K_{S}^{0}K^{\pm}\pi^{\mp}$ 
from (a) CLEO II/II.V and (b) CLEO III data.]{$K_{S}^{0}K^{\pm}\pi^{\mp}$ invariant mass in the reaction 
$\gamma\gamma\rightarrow K_{S}^{0}K^{\pm}\pi^{\mp}$ 
from (a) CLEO II/II.V and (b) CLEO III data~\cite{cleo}. The $\eta_c^{\prime}$ is the enhancement at 3.643 GeV.}
% The curves in the
% figures are fit results using a second order polynomial background 
% and two Breit--Wigners for $\eta_{c}$ and $\eta_{c}^{\prime}$
% enhancements.
\label{fig:etacprime}
\end{center}
\end{figure}

\clearpage

\begin{figure}[htb]
\begin{center}
\includegraphics[width=3in]
{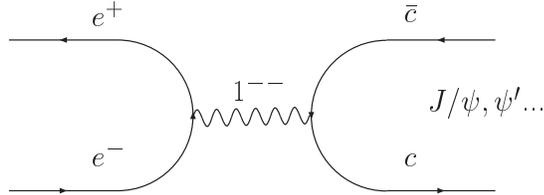}
%%%{chapter2/fig/pbar_chain.eps}
\caption{Formation of a vector charmonium resonance from $e^{+}e^{-}$ annihilation.}
\label{fig:eeprod}
\end{center}
\end{figure}

All studies of charmonium states described so far utilized $e^{+}e^{-}$ annihilation, in which only the $c\bar c$ vector states ($J^{PC} = 1^{--}$) are directly formed via the intermediate photon with $J^{PC} = 1^{--}$ (see Figure 1.8). All other states are only reached by decays, mostly radiative, from these vector states.

A major departure from this technique came with the R704 experiment at the ISR at CERN~\cite{Baglin}, which demonstrated that high resolution charmonium spectroscopy could be studied by charmonium formation in proton-antiproton annihilation (see Figure 1.9). This technique has the advantage of being able to directly populate $c\bar c$ mesons of {\it all} $J^{PC}$ via two and three gluon annihilations leading to C = +1 and C = $-1$ states respectively, and was used by Fermilab experiments E760 and E835 in their 1990, 1997, and 2000 runs.

\clearpage

\begin{figure}[htb]
\begin{center}
\includegraphics[width=12cm]
{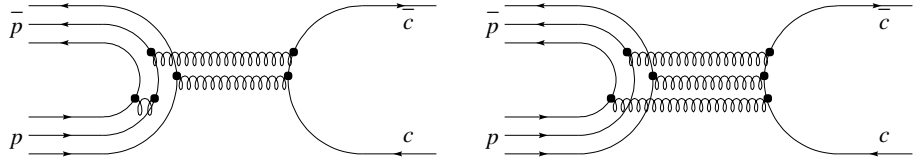}
%%%{chapter2/fig/pbar_chain.eps}
\caption{Formation of a charmonium resonance by two and three gluon $p\bar p$ annihilation.}
\label{fig:ppprod}
\end{center}
\end{figure}

Because it was a $p\bar p$ annihilation experiment which could directly populate the non-vector charmonium states, Fermilab E760 was in a unique position to study the $\chi_{cJ}$ states in greater detail, as well as to conduct a search for the $h_c$ or $1 ^1P_1 (1^{+-})$ state of charmonium. 

The $^1P_1$ resonance, which is the singlet-P partner of the $\chi_{cJ}$ resonances, is the final bound state of $c\bar c$ whose observation remains unconfirmed. The Crystal Ball experiment at SLAC failed to find it in the reaction $e^{+}e^{-} \rightarrow \psi^{\prime} \rightarrow ^1P_1 \pi^0$, $^1P_1 \rightarrow \gamma \eta_c$~\cite{porter}. During its 1990 run, E760 claimed observation of the $^1P_1$ in the reaction $p\bar p \rightarrow ^1P_1 \rightarrow J/\psi \pi^0 \rightarrow e^{+}e^{-}$ based on 15.9 pb$^{-1}$ of data taken. It claimed a $^1P_1$ mass of $3526.2 \pm 0.15 \pm 0.2$ MeV and a width of $\Gamma \le 1.1$ MeV (see Figure 1.10)~\cite{e760paper}. 

In 1997, an attempt was made to confirm this observation in the Fermilab experiment E835, the successor experiment to E760, using 38.9 pb$^{-1}$ of data. This attempt was inconclusive, however, due to instabilities in the antiproton beam energy. During the year 2000 run of E835 (E835p), 50.5 pb$^{-1}$ of data was taken dedicated to confirming the E760 observation of $^1P_1$. The analysis of these data in a variety of possible $^1P_1$ decay channels forms the topic of this dissertation. 

\clearpage

\begin{figure}[htb]
\begin{center}
\includegraphics[width=13cm]
{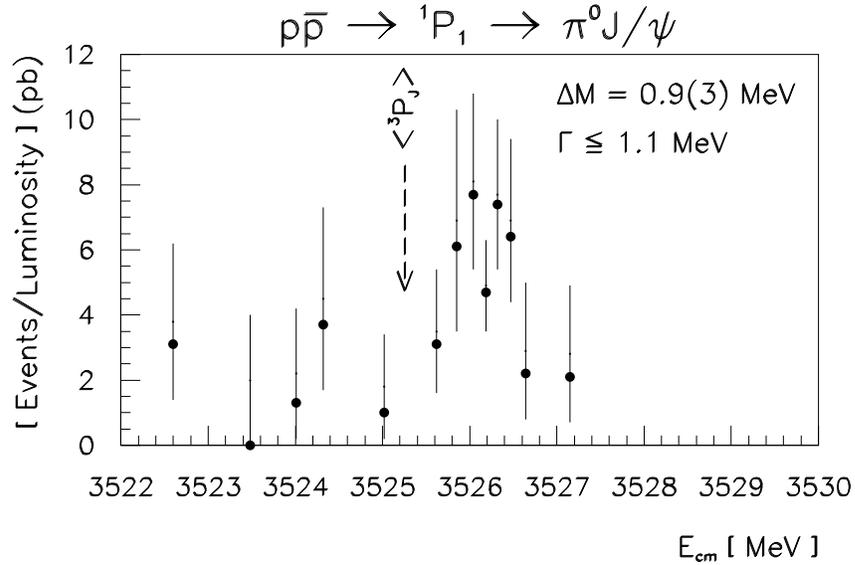}
\caption[The $^1P_1$ resonance as measured by Fermilab experiment E760.]{The $^1P_1$ resonance as measured by Fermilab experiment E760.~\cite{e760paper}}
\label{fig:E760}
\end{center}
\end{figure}

Chapter 2 of this thesis consists of a theoretical discussion of charmonium spectroscopy, in terms of the quark model and QCD, and an examination of theoretical predictions regarding the $^1P_1$. In Chapter 3, the experimental set-up of E835 is described in detail. In Chapter 4, the data selection and results of the $^1P_1$ search is discussed for reactions leading to final states containing $J/\psi$: $p\bar p \rightarrow ^1P_1 \rightarrow J/\psi + X \rightarrow e^{+}e^{-} + X$, $p\bar p \rightarrow ^1P_1 \rightarrow J/\psi + \pi^0 \rightarrow e^{+}e^{-} + \gamma\gamma$, and $p\bar p \rightarrow ^1P_1 \rightarrow J/\psi + \pi^0\pi^0 \rightarrow e^{+}e^{-} + 4\gamma$. Chapter 4 also describes the study of the reaction $p\bar p \rightarrow J/\psi + \gamma \rightarrow e^{+}e^{-} \gamma$, which is forbidden to occur via $^1P_1$ by C-parity conservation, and which was studied as a control. Chapter 5 describes our search for $h_c (^1P_1)$ in the reaction $p\bar p \rightarrow ^1P_1 \rightarrow \eta_c + \gamma \rightarrow (\gamma\gamma) + \gamma$ in the E835p data. Data tables for the final event selection in the  $p\bar p \rightarrow ^1P_1 \rightarrow J/\psi + \pi^0 \rightarrow e^{+}e^{-}+ \gamma\gamma$ reaction are given in Appendix A, and software used in determining luminosity during data taking is given in Appendix B.

%Chapter 2--QED, QCD and Charmonium
\baselineskip=24pt
\chapter{Theory}

\section{Quantum Chromodynamics}

In the Standard Model of particle physics, Quantum Chromodynamics, or QCD, is the established theory of strong interactions. QCD is the non-Abelian gauge theory of quarks and gluons, which carry a strong interaction charge called 'color'. The theory is non-Abelian because, unlike Quantum Electrodynamics (QED), the established theory of electromagnetic interactions, the gauge boson mediating the field itself carries the fundamental charge. Thus, while photons have no electric charge, and do not directly couple to each other, gluons, the QCD gauge bosons carry both a color and and anti-color, and do directly interact. The basic QCD interaction is invariant under the interchange of the different colors. The gluons are postulated to belong to an octet representation of the symmetry group SU(3), and the color states of the gluons in this octet can be expressed as:

\begin{equation}
r\bar b, r\bar g, b\bar g, b\bar r, g\bar r, g\bar b, {r\bar r - b\bar b \over \sqrt{2}}, {r\bar r + b\bar b - 2g\bar g \over \sqrt{6}}
\end{equation}

This non-Abelian field, combined with the much larger coupling constants for QCD, mean that although the formalism is pattered after the Abelian QED theory, in practice, QED and QCD are quite different. As can be seen in Figure 2.1, the color-charge of the gluons means that Feynmann diagrams with three and four gluon vertices are allowed, as well as the two gluon-quark vertices which would be expected in analogy to QED. Furthermore, the strong coupling constant $\alpha_s$, which governs the quark-gluon interaction, is much larger than its electromagnetic counterpart $\alpha_{em}$, and the value of $\alpha_s$ tends to infinity for small values of $q^2$ (large distances). Because of these complications, QCD, unlike QED, is a theory in which it is impossible, to solve problems analytically, and thus requires extremely cumbersome numerical calculations. QCD based computer models of varying degrees of sophistication are thus used to generate theoretical predictions with which the experimental data may be compared. A brief description of QCD and some of the more popular calculation techniques are given in this chapter, for further details a very good synopsis of QCD is given by Wilczek~\cite{wilczek}, and a summary of QCD calculational techniques as applied to charmonium is given by Seth~\cite{seth}. In the following we liberally borrow with permission from the review by Seth~\cite{seth}.

\begin{figure}[t]
\begin{center}
\includegraphics[height=1.5in]{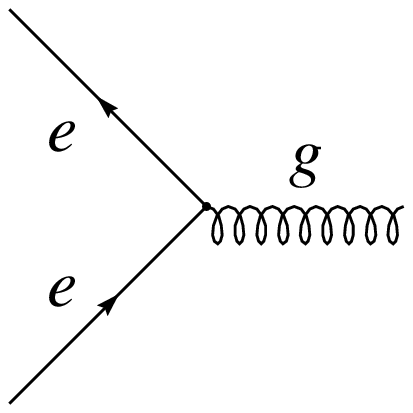}
\hfil
\includegraphics[height=1.5in]{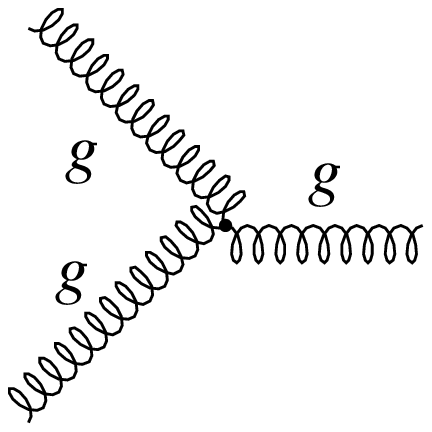}
\hfil
\includegraphics[height=1.5in]{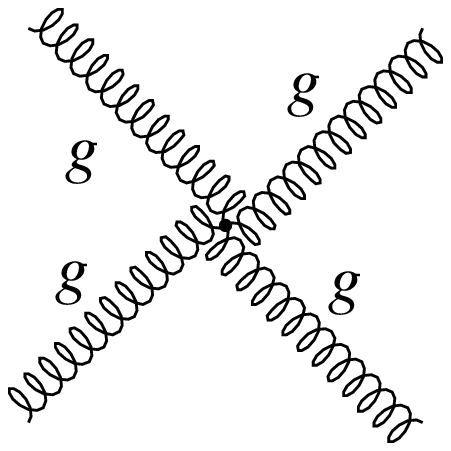}
\caption{The three basic QCD vertices.}
\end{center}
\end{figure}

Charmonium is a good system to test the assumptions of QCD. Unlike the case of light-quark hardons, for charmonium the value of $\alpha_s$ is sufficiently small $\sim 0.3$ to make perturburtive calculations possible. Furthermore, the relatively small binding energy compared to the rest mass of its constituents allow $c\bar c$ states to be described non-relativistically (with $v^2/c^2 \approx 0.25$). The fact that the charmonium resonances are eigenstates of $J^{PC}$ produces symmetry conserving simplifications. Finally, the bound $c\bar c$ states are well separated in energy and narrow in width, as opposed to the light-quark resonances which have large, often overlapping, widths, creating a complicated and difficult spectroscopy. 

The study of charmonium also has experimental advantages over the heavier quarks; the top quark decays too quickly to form any QCD bound states, and the production cross sections for $b\bar b$, or bottomonium, (particularly in $p\bar p$) are much smaller than that for charmonium, making direct production of anything other than the $J^{PC} = 1^{--}$ states all but impossible. 

By making precise measurements of the masses, widths, and branching ratios charmonium states, important information about the dynamics of the strong interaction may be extracted. For instance, by comparing the hadronic and electromagnetic branching ratios of appropriately chosen charmonium states, an estimate of the strong coupling constant $\alpha_s$ can be derived~\cite{kwong}. Unknown quantities, such as the squared absolute value of the wave function at the origin $\vert \psi(0) \vert^2$, or poorly measured quantities, such as branching ratios between the resonance and the initial state, may often cancel in the ratio, leaving $\alpha_s$ as the only unknown. For example, in the case of $\eta_c$, $\eta_c^{\prime}$, $\chi_{c0}$, and $\chi_{c2}$, this cancelation occurs when one compares the branching ratio into two photons and the branching ratio into two gluons. Different theoretical models may also provide predictions for the radiative partial widths of charmonium states which may be compared to the experimental results. Examples of this include the electric dipole transitions of the three $\chi_{cJ}$ states to $J/\psi$, and the magnetic dipole transition of $J/\psi$ to $\eta_c$, and $\psi^{\prime}$ to $\eta_c^{\prime}$. 

\subsection{The QCD Lagrangian}

The most fundamental expression of the theory of Quantum Chromodynamics begins with the QCD Lagrangian itself. This Lagrangian, which describes the interaction of quarks and gluons is:

\begin{equation}
{\cal L} = -{1 \over 4} G^{(a)}_{\mu\nu}G^{(a) \mu\nu} + \sum\limits_{q} \bar q (i {\cal D} - m_q)q
\end{equation}

Here, $G^{(a)}_{\mu\nu}$ is the gluon field strength tensor, with $A^{(a)}_{\mu\nu}$ as the eight gluon fields $(a=1,2,...,8)$, $q$ are the quark fields of six different flavors ($u,d,s,c,b,t$), each of three colors, $m_q$ are the quark masses, and ${\cal D}$ is the covariant derivative. In turn, the gluon field strength tensor and the covariant derivative are described by:

\begin{equation}
G^{(a)}_{\mu\nu} = \partial_{\mu}A^{(a)}_{\nu} - \partial_{\nu}A^{(a)}_{\mu} + g_s f^{abc} A_{\mu}^{(b)}A_{\nu}^{(c)}
\end{equation}

and

\begin{equation}
{\cal D} = \partial_{\mu} - igA_{\mu}^{(a)}\lambda^{(a)}/2
\end{equation}

where $\lambda$, the Gell-Mann matrices, and $f^{abc}$ are the generators and structure
constants, respectively, of SU(3).The gauge constant $g_s$, which determines the 
coupling between the quark and gluon fields, is related to the scale dependent 
(or "running") effective coupling constant of strong interaction by the relation:

\begin{equation}
\alpha_s = {g^2_s \over 4\pi}
\end{equation}

This coupling constant, $\alpha_s$ depends on the choice of the renormalization scale, At the energy scale $\mu$, in the lowest order,

\begin{equation}
\alpha_s(\mu) = {2\pi \over [11 - (2/3)n_f)] \times ln(\mu/\Lambda)}         
\end{equation}

where $n_f$ is the number of quark flavors with mass less than $\mu$, and 
$\Lambda$ is the QCD scale parameter, which depends on the number of quark flavors. At the mass of the $Z^0$ boson, $\alpha_s(Z^0)$ is measured to be 0.1172 $\pm$ 0.0020~\cite{PDG2002}.

As can be seen from the above expression for $\alpha_s$, QCD incorporates a unique property known as 'asymptotic freedom'; $\alpha_s$
decreases as one goes to higher energies or smaller distances, becoming zero at $\mu = \infty$. It is this feature of QCD which allows one to use perturbative methods at high energies (and small distances). At low energies $\alpha_s$ becomes large, and the use of perturbative QCD becomes highly questionable for the light ($u,d,s$) quarks. 

The non-linear terms in the field strength tensor $G_{\mu\nu}$ makes QCD
impossible to solve analytically, and difficult to work with even in lattice calculations. However, certain symmetries of the theory can be seen from the Lagrangian itself~\cite{wilczek}.

1.  The discrete symmetries P, C, and T are preserved by the strong interaction. 

2.  Quark flavor is conserved by the strong interaction, as are the quantities which can be derived from quark flavor such as isospin, electric charge, and baryon number.

3.  Approximate chiral and flavor symmetries are preserved for the light $u,d,s$ quarks, or  in the very high energy regime, when all quark masses are comparatively negligible. 

It is also worth noting that neither QCD nor anything else in the Standard Model predicts either the number of quark families, the scale parameter $\Lambda$, or the quark masses.  These must be added to the theory as input parameters.

Using the QCD Lagrangian, rules for calculating Feynman diagrams for strong interaction processes may be constructed.  Since this method of calculation treats interactions essentially as perturbations about the field-free ground state, they can be used to obtain predictions for reactions only at very
high momenta, where the strong coupling constant becomes small.  At small
momenta the strong coupling constant becomes too large, for these perturbative
methods (pQCD) to remain valid, and one is then forced to use numerical
solutions such as Lattice Gauge Calculations.

\subsection{Potential Models}

One technique used in calculations of hadronic resonances has been to replace the non-Abelian guage field theory of QCD by a non-relativistic potential model. Despite the fact that many have questioned the very existence of a static potential in the presence of an active gluon condensate~\cite{shifman}, potential model calculations have been very successful in predicting the masses of bound states of both charmonium and bottomonium. Non-relativistic models for charmonium are possible only because of the relatively large mass of the charm quark, which have $v^2/c^2 \approx 0.25$ in their bound states, as opposed to $v^2/c^2 \approx 0.8$ for the light quark mesons. Using corrections for relativistic, channel coupling, and radiative effects, the success of potential models for theoretical predictions even extends to some decays of charmonium states as well as their masses and widths.

\subsubsection{Spin-Independent Potentials}

As early as 1975, Appelquist and Politzer recognized that the single gluon exchange between a charm quark and antiquark should give rise to a Coulombic potential proportional to $1/r$ at small distances~\cite{Applequist}. They coined the name "orthocharmonium" for the $J/\psi$ in analogy with the $^3S_1$ orthopositronium, and went on to extend the analogy to predict the existence of $^1S_0$ "paracharmonium". With the announcement of the discovery of $\psi^{\prime}$ just a week after $J/\psi$ it became possible to flesh out the potential picture, and Appelquist and Politzer were able to predict the complete spectrum of bound charmonium states based on a charmonium potential which was expected to be intermediate between a Coulombic and a harmonic oscillator potential. The spectrum of charmonium states and a comparison with the positronium spectrum is shown in Figure 2.2.

\begin{figure}[htbp]
\begin{center}
\includegraphics[width=5in]{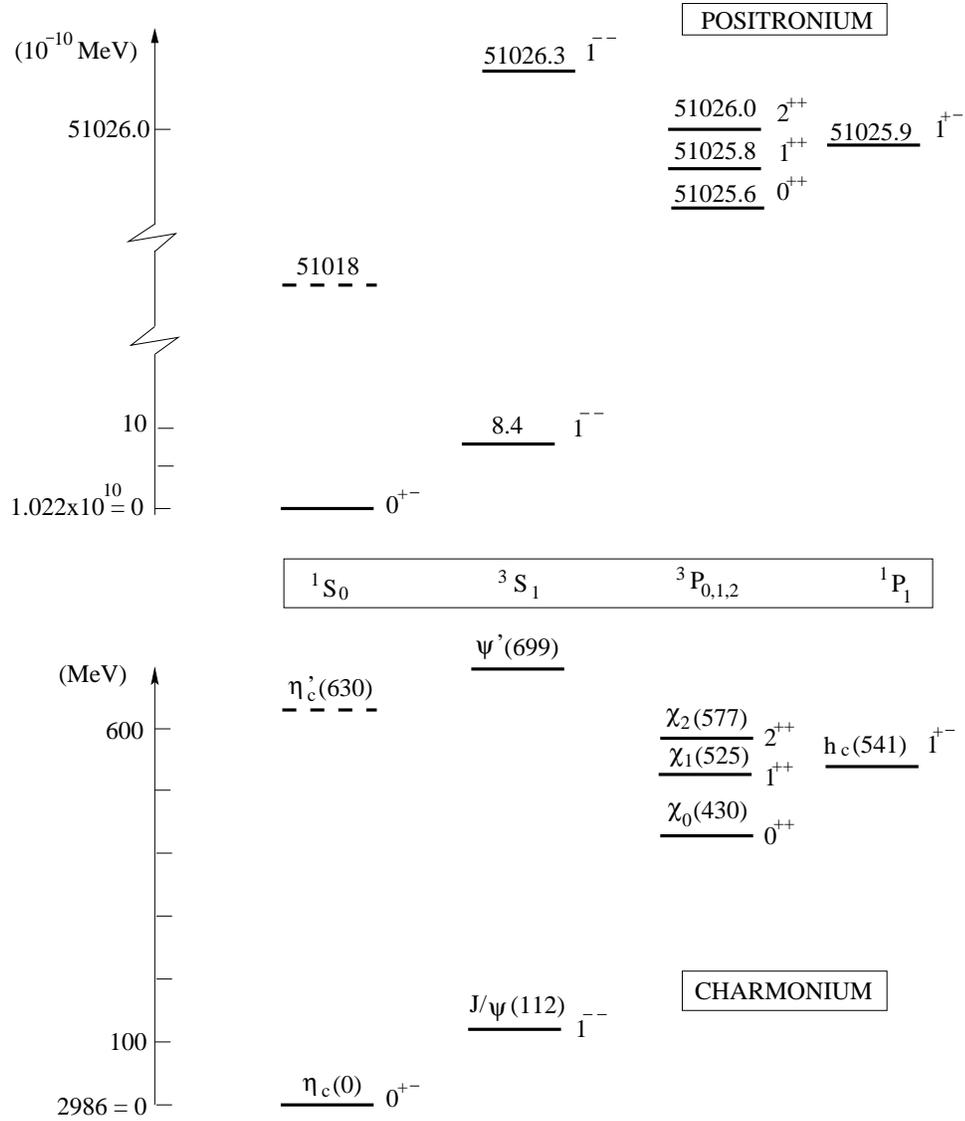}
\caption[A schematic showing the similarities between positronium (top)
and charmonium (bottom) spectra.]{A schematic showing the similarities between positronium (top)
and charmonium (bottom) spectra~\cite{todd}.}  
\end{center}
\end{figure}

\newpage

\clearpage

After the initial predictions of Appelquist and Politzer, another early potential model which was examined was a purely linear potential, used by Harrington, Park, and Yildez in 1975~\cite{harrington}. The most popular of these early potential models, however, is a combination of both the linear potential and Coulombic potentials. This model has become known as the Cornell potential. It was first proposed by Eichten {\it et al.} in 1975~\cite{eichten} and combines a one-gluon exchange Coulombic component, proportional to $1/r$, which is dominant at short distances, and an additional confinement term proportional to $r$, analogous to the "string tension" of a multigluon "flux tube", which is dominant at large distances. This linear confinement term was added in recognition of the fact that colored free quarks are never observed experimentally, but rather are confined to color singlet hadrons, either $q\bar q$ mesons, $qqq$ baryons, or $\bar q\bar q\bar q$ antibaryons. Other objects are possible as well, such as glueballs ($gg$, $ggg$), hybrids ($q\bar qg$), and multi-quark states $(q\bar qq\bar q)$. These objects are allowed in QCD as long as the total object is color neutral.

The Cornell Potential, with its linear confinement term, may be written as :  

\begin{equation}
V(r) = - k/r + r/a^2    
\end{equation}

where the constant $k$ is often identified with $4/3\alpha_s$, the $4/3$ factor being a numerical artifact of color symmetry, and the constant $1/a^2$ is of order $\sim 1$ GeV/fm. 

Subsequently, many alternative formulations of the spin-independent potential 
have been suggested. One important modification was motivated by the 
observation that the energy of the radial excitiation for $S$ states, $\Delta[M(2 ^3S_1) - M(1 ^3S_1)]$, is nearly the same for charmonium (589 MeV) and bottomonium (563 MeV). For this property to be truly independent of quark mass one must use a logarithmic potential. Furthermore, the $2S - 1P$ mass difference, which is zero in a purely Coulombic potential, is quite significant for charmonium, with $M(2 ^3S_1) - <M(1 ^3P_J)>$ = 161 MeV. These considerations lead Quigg and Rosner~\cite{quigg} in 1977 to propose the following potential:

\begin{equation}
V(r) = C ln(r/r_0)
\end{equation}

Incorporating this logaritmic feature, several different forms of potentials, 
interpolating between the Coulombic and confinement parts, have been proposed. One idea, first used by Richardson in 1979~\cite{richardson}, is to construct the potential in momentum space taking into account the momentum dependence of $\alpha_s$ as given by: 

\begin{equation}
V(q^2)= -(4/3)[12\pi/(33-2n_f)][q^2 \times ln(1+(q^2/\Lambda^2)]^{-1}
\end{equation}
                                                                        
This formulation has the advantage that it contains only one parameter, $\Lambda_{QCD}$, and in configuration space it varies as $1/r$ at small distances, and as $r$ at large distances. Improvements on this approach were later made by Buchm\"uller~\cite{buchmuller}\cite{buchmuller2}. In Table 2.1 we list the configuration space representations of some of the many empirical potentials used to fit the charmonium spectra. 

\begin{table}
\begin{tabular}{ll}
\hline
 & \\
Author & Potential \\
 & \\
\hline
 & \\
Applequist and Politzer~\cite{Applequist} & only Coulombic \\
 & \\
De Rujula and Glashow~\cite{DeRujula} & only Coulombic \\
 & \\
Kang and Schnitzer~\cite{kang} & only Linear \\
 & \\
Harrington~\cite{harrington} & only Linear \\
 & \\
Eichten~\cite{eichten} & $-k/r + r/a^2, \quad k = (4/3)\alpha_s$ \\
 & \\
Quigg and Rosner~\cite{quigg} & $A \cdot ln(r/r_0)$ \\
 & \\
Martin~\cite{martin} & $A + Br^{0.1}$ \\
 & \\
Celemaster and Georgi~\cite{celemaster} & $-(4/3)(1/8\pi b)(r ln(r/r_0))^{-1} - c r e^{-r/d} + a r$\\
 & \\
Celemaster and Henyey~\cite{celemaster} & $-(4/3)(1/(2\pi)^3)\int d^3k \cdot ex
p (i {\vec k}\cdot{\vec r}) (4\pi \alpha_s (k^2)/k^2)$ \\
 & \\
Richardson~\cite{richardson} & $-(4/3)(1/8\pi^3b)\int d^3k \cdot exp (i {\vec k}\cdot{\vec r})/k^2 ln(1 + k^2/\Lambda^2)$ \\
 & \\
\hline
\end{tabular}
\caption[Some QCD based spin-independent $q\bar q$ potentials.]{Some QCD based spin-independent $q\bar q$ potentials. (After Lucha {\it et al.}~\cite{lucha}). The constant $1/b = 48\pi^2/(33-2n_f)$. The potentials used by Buchm\"uller {\it et al.}~\cite{buchmuller} are essentially identical to that of Celemaster and Henyey, and are nearly equivalent to that of Richardson~\cite{richardson}, since $\alpha_s \approx 1/4\pi b ln(k^2/\Lambda^2).$}
\end{table}

All the different potential models obtain their parameters by fitting the spectra of the bound states of charmonium and bottomonium. The physical radius of these bound states range between $r$ = 0.2 and 1.0 fermi, and in this region all potentials, including the purely empirical ones due to Quigg and Rosner~\cite{quigg}, and Martin~\cite{martin} are essentially identical. These potential models share another important feature, which is that even at these very small distances, the quark-antiquark pair does not reside in a purely Coulombic potential, and the confinement contribution is significant even in the lowest lying states. A schematic showing the differences between the various potential models is shown in Figure 2.3.

\begin{figure}[t]
\centering
\includegraphics[width=4in]{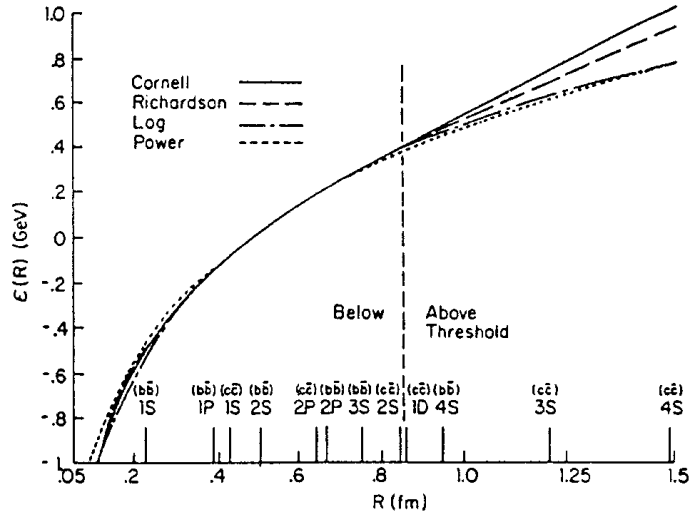}
\caption[Schematic showing various model potentials as a
function of quark separation distance $R$.]{Schematic showing various model potentials as a
function of quark separation distance $R$.  Various 
$c\bar c$ and $b\bar b$ levels are indicated by the
vertical lines~\cite{todd}.}   
\end{figure}

Combining Lattice Guage Calculation techniques with these potential models, static spin-independent interquark potentials have been calculated on the 
lattice in both quenched and unquenched approximations, and excellent fit to the lattice potential for $r > 0.3$ fm is obtained with the parametrization of the Cornell potential, with $k$ = 0.322, and $a$ = 2.56 GeV$^{-1}$~\cite{bali}. There is little difference between the quenched and unquenched lattice results for distance up to $r$ = 1.5 fm.  

\subsubsection{Spin-dependent Potentials}

Given the Coulombic nature of the short-range $q\bar q$ potential, and the spin dependence of the vector Coulomb potential, consisting of spin-orbit, spin-spin, and tensor components, the addition of spin-dependence to the emerging potential models was a natural step. The development of these spin-dependent potential models quickly followed the spin-independent versions, starting with Eichten et al. in 1975 ~\cite{eichten} and Henrique, Kellett and Moorhouse in 1976~\cite{henriques}. A full and systematic investigation of the spin dependent forces was done by Eichten and Feinberg~\cite{eichten2} in 1981 using a gauge-invariant formalism. Their representation of the spin dependent forces was done with a nonrelativistic potential, with a short-range vector exchange and a long-range scalar exchange. In this model, the spin dependent potential may be expressed as:

\vspace*{10pt}

$V_{SD} =$ $V^{\prime}(r)\vec{S}\cdot\vec{L} \enskip + \enskip 4\alpha_s/(3m^2r^3)\vec{S}\cdot\vec{L}$ 

$+ \enskip (32\pi/9mq^2)(\vec{S1}\cdot\vec{S2}) \delta(r)$ 

$+ \enskip (4\alpha_s/3mq^2r^3)\times(3(\vec{S1}\cdot\vec{R})(\vec{S2}\cdot\vec{R}) - \vec{S1}\cdot\vec{S2})$

\vspace*{10pt}

where $V^{\prime}(r)$ is the derivative of the spin independent central potential. Using the Cornell potential for $V(r)$, the first two terms combine to give the overall spin-orbit potential:

\begin{equation}
V_{SL} = (1/2mq^2r^2)(4\alpha_s/r + r/a^2)\vec{S}\cdot\vec{L}.
\end{equation}

The other two terms can be identified as the spin-spin potential, and the 
tensor potential. 

Precise measurements of the different resonance masses, or more particularly the differences between them, are a very effective way to test the spin-dependence of the different potential models. For instance, the tensor and spin-orbit interaction split the masses of the $\chi_{cJ} (1 ^3P_J), J = 0,1,2$ states (fine splitting). The spin-spin force splits the vector and psuedoscalar states, and this is repsonsible for the mass difference between $J/\psi$ and $\eta_c$, and between $\psi^{\prime}$ and $\eta_c^{\prime}$ (hyperfine splitting). A measurement of the deviation of the $^1P_1$ mass from the center of gravity of the $\chi_{cJ}$ states would indicate a departure from first order perturbation theory, since the spin-spin potential is a contact potential, which survives only with the finite wave function at the origin. Thus, this potential gives rise to hyperfine splitting between the triplet ($S = S1+S2 = 1$) and singlet ($S = S1+S2 = 0$) states only for $L$ = 0 S-wave states, and not for P-wave or any other higher L-states. This is the direct consequence of the long-range confinement potential having been assumed to be pure Lorentz scalar. There is no fundamental justification for this assumption, although support for it is observed in the results of quenched lattice calculations, for example, those of Bali, Schilling and Wachter~\cite{bali1997}. 

\subsection{Calculational Techniques}

In 1974, Wilson (1974)~\cite{wilson} showed how to quantize a gauge field theory on a discrete lattice in Euclidean space-time preserving exact gauge invariance, and applied this calculational tecnique to the strong coupling regime of QCD. In these Lattice Gauge Calculations, space-time is replaced by a four dimensional hypercubic lattice of size $L^3T$. The sites are separated by the lattice spacing $a$. In more recent calculations, asymmetric lattices have been used in which the lattice spacing for time is chosen to be much smaller than for space, with $a_s$ as small as 0.07 fm, and $a_s/a_t$ as large as 3.  The overall size of the lattice is generally 1-4 fermis.  

The quark and gluon fields are defined at discreet points on the lattice, and physical problems are solved numerically by Monte Carlo simulations using powerful computers, requiring only the quark masses as calculational input. Early 
lattice calculations were done in the "quenched" approximation, in which no 
quark-antiquark pairs are allowed to be excited from the QCD vacuum (or sea). 
Recently, the vast improvement in computers and computational techniques have made it  possible to do unquenched calculations in which $u\bar u$, $d\bar d$, and even $s\bar s$ quark-antiquark pairs may be excited from the vacuum. 

A useful variant of lattice calculations for heavy quarks is the so called Non 
Relativistic QCD (NRQCD), pioneered by Lepage and colleagues~\cite{lepage}. As heavy quarks (i.e. charm and bottom) are generally non-relativistic, renormalization group techniques may be used in these calculations to replace the relativistic Dirac action for the heavy quarks on the lattice by a non-relativistic Schrodinger action, which simplifies the calculations considerably. However, even with such simplifications, Lattice Gauge Calculations generally require immense computing resources. Yet despite the large amount of computation required lattice calculations are able to provide useful insight into static potentials for QCD, masses of charmonium states, and many decay characterstics. For an excellent review of lattice methods and their application to charmonium physics, see the Physical Report article by Bali~\cite{bali}. 
 
Another important technique used for QCD calculations is the QCD sum rule technique, which was introduced in 1979 by Shifman, Vainshtein and Zakharov~\cite{shifman}. The basic premise of the QCD sum rule technique is that the QCD vacuum is populated by large fluctuating fields whose strength is characterized by gluon and quark condensates

$<0\vert(\alpha_s/\pi)G_{\mu\nu}^{(a)}G^{(a) \mu\nu}\vert 0>,$ \quad and $<0\vert q\bar q \vert 0>$

When a pair of quarks is injected into this active vacuum, its dynamics is 
determined by the characterstics of the vacuum, and the subsequent formation of hadrons can be reliably calculated by dispersion relations. The heavy quark 
charmonium and bottomonium states are uneffected by light quark condensates, 
and are sensitive only to the gluon condensate. The sum rule technique is nearly saturated even by the lowest excitations of each $J^{PC}$, and is therefore next to impossible to apply to radial excitations or to resonances with higher orbital excitations than $J = 2$. 

One early spectacular success of the sum rule calculations was the correct 
prediction of the mass of the charmonium ground state, $\eta_c (1 ^1S_0)$~\cite{shifmanetac}. Highly successful QCD sum rule calculations of $1P$ charmonium states were made subsequently made by Reinders, Rubinstein and Yazaki~\cite{reinders}. 

Yet another QCD calculation technique is used to attempt to improve the relativistic problems caused by the singular nature of the Coulombic potential. The "smearing" technique is used to spread the singularity into a small region around the origin. This technique is used by Godfrey and Isgur~\cite{godfrey}.

In addition to lattice and sum-rule predictions, prediction may also be obtained using perturbative QCD (pQCD), in analogy to the perturbative QED (pQED) used for positronium. The technique is valid for large momenta and small values of $\alpha_s$, and has been used for charmonium annihilation calculations under the assumption that the $q\bar q$ wave function is purely color singlet, and that the annihilation is a short distance process. Strong radiative corrections for the electromagnetic, radiative, and hadronic decays of S and P wave quarkonia have also been made under this assumption by several authors~\cite{barbieri}~\cite{barbieri2}~\cite{barbieri3}. 

It has been argued, notably by Bodwin {\it et al.}~\cite{bodwin}, that the hadronic decays of charmonium, and particularly the P wave states, require taking acount of the possible $q\bar qg$ components, with $q\bar q$ in a color octet. The octet components may be determined only empirically from the existing data, but it is argued that the $^3P_1$ and $^1P_1$ hadronic decays take place dominantly through their octet components, and therefore these provide a good estimate. Using these suggestions, radiative corrections for octet decays have been calculated by several authors, and these have been summarized by Vairo~\cite{vairo}. A couple of problems should be noted about pQCD predictions. One problem is that unlike the case of positronium, in charmonium $m(q\bar q) \ne 2m(q)$, and thus there is an ambiguity about whether to evaluate pQCD expressions using $m(q\bar q)/2$ or $m(q)$, which may affect calculations of hadronic decay widths. The second problem relates to strong radiative corrections since, unlike the pQED case, $\alpha_s$ is large enough ($\sim 0.3$) for charm quarks that the lowest order gluon radiative corrections are often very large, up to 100$\%$. Despite this, radiative corrections calculated with pQCD have in some cases led to favorable agreement with experimental results.

\section{Theoretical Predictions}

A large number of theoretical predictions exist for the masses of charmonium states, including the $h_c (^1P_1)$. The majority of these predictions are based on potential model calculations, and differ only in the choice of the common parameters; the strong coupling constant, $\alpha_s$, and the charm quark mass, $m_c$. Some of these potential model and parameter choices are listed in Table 2.2. The Cornell potential, and the QCD based potentials modeled after that by Richardson are the most commonly used. Minor variations on these are used by several authors. 

The linear confinement potential is generally assumed to be scalar (S). Some authors have considered vector confinement (V) and mixtures of the two (V+S). Most calculations are non-relativistic. Some include relativistic corrections at the level of $v^2/c^2$. The only predictions based on lattice calculations are those by Bali~\cite{bali1997} and Okamoto~\cite{okamoto}.

A summary of predictions for the properties of the $^1P_1$ state are given in Tables 2.2 - 2.4. Mass and width predictions are shown in Tables 2.2 and 2.3. The $^1P_1$ mass is generally predicted to lie within a few MeV of the centroid of the $\chi_{cJ}$ states; $<M(\chi_J)>_{expt}$ = 3525.3 MeV. The predictions for the total width of $^1P_1$ lie in the range of 500-1000 keV. Predictions for the partial widths of various $^1P_1$ decays are shown in Table 2.4. The most prominent decay of the $^1P_1$ is expected to be the radiative transition to $\eta_c$, with predicted partial widths in the range of several hundred keV, based on the measured width of the E1 transition $^3P_1 \rightarrow \gamma J/\psi$.

\begin{table}
\begin{center}
\begin{tabular}{ccccccc}
\hline
 & & & & & & \\
Author & Year & $\Delta M_{hf}$ & Potential & $\alpha_s$ & $m_c$ & Conf. \\
 & & & & & & \\
 & & (MeV) & & & (GeV) & \\
 & & & & & & \\
\hline
 & & & & & & \\
 Eichten~\cite{eichten2} & 1979 & 0 & Cornell & 0.341 & 1.84 & S \\
 & & & & & & \\
 Ono~\cite{ono} & 1982 & +1.0 & & & & \\
 & & & & & & \\
 Gupta~\cite{gupta2} & 1982 & -1.4 & Cornell & 0.392 & 1.2 & S$^{*}$ \\
 & & & & & & \\
 McClary~\cite{mcclary} & 1983 & +5.3 & Cornell & 0.341 & 1.84 & S$^{*}$ \\
 & & & & & & \\
 Moxhay~\cite{moxhay} & 1983 & $\sim 0$ & Richardson & & 1.5 & Tensor$^{*}$ \\
 & & & & & & \\
 Godfrey~\cite{godfrey} & 1985 & +5 & Cornell$^{\dag}$ & 0.34 & 1.5 & S \\
 & & & & & & \\
 Pantaleone~\cite{pantaleone2} & 1986 & -3.6 & QCD & 0.24 & 1.48 & V + S \\
 & & & & & & \\
 Olsson~\cite{olsson} & 1987 & $+5.4 \pm 0.8$ & Cornell & & & V + S \\
 & & & & & & \\
 Igi~\cite{igi1}~\cite{igi2} & 1987 & $+24.1 \pm 2.5$ & QCD$^{\dag}$ & & 1.506 & \\
 & & & & & & \\
 Pantaleone~\cite{pantaleone} & 1988 & -1.4 & QCD & 0.33 & 1.48 & V + S \\
 & & & & & & \\
 Gupta~\cite{gupta4} & 1989 & -2.0 & Cornell & 0.392 & 1.2 & V + S$^{*}$ \\
 & & & & & & \\
 Badalian~\cite{badalyan} & 1990 & $-3.1 \pm 4.9$ & QCD & $\le 0.39$ & 1.35 & \\
 & & & & & & \\
 Dixit~\cite{dixit} & 1990 & $-8.0 \pm 8.0$ & Power Law & & & V \\
 & & & & & & \\
 Galkin~\cite{galkin} & 1990 & +8 & Cornell & 0.52 & 1.55 & V$^{*}$ \\
 & & & & & & \\
 Chakrabarty~\cite{chakrabarty} & 1991 & $0 \pm 70$ & Power Law & 0.25 & 1.5 & V \\ 
 & & & & & & \\
\hline
\end{tabular}
\end{center}
\caption[A summary of potential model predictions for the mass and width of the $^1P_1$ (part 1).]{A summary of potential model predictions for the mass and width of the $^1P_1$ (part 1). The hyperfine mass splitting $\Delta M_{hf} \equiv <M(\chi_J)> - M(^1P_1)$ where $<M(\chi_J)>_{expt}$ = 3525.3 MeV. Asterisks denote inclusion of some relativistic corrections. Daggers denote modified potentials.}
\end{table}

\clearpage

\begin{table}
\begin{center}
\begin{tabular}{ccccccc}
\hline
 & & & & & & \\
Author & Year & $\Delta M_{hf}$ & Potential & $\alpha_s$ & $m_c$ & Conf. \\
 & & & & & & \\
 & & (MeV) & & & (GeV) & \\
 & & & & & & \\
\hline
 & & & & & & \\
 Fulcher~\cite{fulcher} & 1991 & -3.0 & QCD & 1.30 & 0.54 & V + S \\
 & & & & & & \\
 Stubbe\cite{stubbe} & 1991 & $-23 \pm 13$ & QCD & & 1.5-1.8 & \\
 & & & & & & \\
 Lichtenberg~\cite{lichtenberg} & 1992 & +4 & QCD & & 1.82 & V \\
 & & & & & & \\
 Lichtenberg~\cite{lichtenberg2} & 1992 & $-1.3 \pm 4.2$ & QCD & $<0.44$ & 1.65 & S \\
 & & & & & & \\
 Beyer~\cite{beyer} & 1992 & +15 & Cornell & $\sim 0.5$ & 1.9-2.3 & S \\
 & & & & & & \\
 Halzen~\cite{halzen} & 1992 & $-0.7 \pm 0.2$ & Cornell & 0.28 & 1.2 & \\
 & & & & & & \\
 Chen~\cite{chen2} & 1992 & $-1.4 \pm 0.5$ & QCD & 0.22 & 1.48 & \\   
 & & & & & & \\
 Eichten~\cite{eichten3} & 1994 & +1 & QCD & 0.31 & 1.48 & \\
 & & & & & & \\
 Gupta~\cite{gupta} & 1994 & -0.9 & Cornell & 0.392 & 1.2 & V + S \\
 & & & & & & \\
 Zeng~\cite{zeng} & 1995 & +6 & Cornell$^{\dag}$ & & 1.53 & S \\
 & & & & & & \\
 Chen~\cite{oakes} & 1996 & $-5 \pm 1$ & QCD & runs & 1.478 & \\
 & & & & & & \\
 Bali~\cite{bali1997} & 1997 & $-1.5 \pm 2.5$ & Lattice & 0.183 & 1.33 & S$^{*}$ \\
 & & & & & & \\
 Okamoto~\cite{okamoto} & 2002 & $-1.5 \pm 2.6$ & Lattice & & & \\
 & & & & & & \\
 Lahde~\cite{lahde} & 2002 & +12 & QCD & 0.38 & 1.50 & V \\
 & & & & & & \\
 Ebert~\cite{ebert} & 2003 & -0.7 & Cornell & 0.314 & 1.55 & V + S$^{*}$ \\
 & & & & & & \\
 Recksiegel~\cite{recksiegel} & 2003 & $-0.8 \pm 0.8$ & QCD & runs & 1.243 & \\
 & & & & & & \\
\hline
\end{tabular}
\end{center}
\caption[A summary of potential model predictions for the mass and width of the $^1P_1$ (part 2).]{A summary of potential model predictions for the mass and width of the $^1P_1$ (part 2). The hyperfine mass splitting $\Delta M_{hf} \equiv <M(\chi_J)> - M(^1P_1)$ where $<M(\chi_J)>_{expt}$ = 3525.3 MeV. Asterisks denote inclusion of some relativistic corrections. Daggers denote modified potentials.}
\end{table}

\begin{table}
\begin{center}
\begin{tabular}{cccccc}
\hline
 & & & & & \\
Authors & $\Gamma(\eta_c\gamma)$ & $\Gamma(J/\psi \pi^0)$ & $\Gamma(J/\psi \pi^0 \pi^0)$ & $\Gamma(hadrons)$ & $\Gamma(total)$ \\
 & & & & & \\
 & (keV) & (keV) & (keV) & (keV) & (keV) \\
 & & & & & \\
\hline
 & & & & & \\
 Renard~\cite{renard} & 240 & & & 370 & 500-1000 \\
 & & & & & \\
 Novikov~\cite{novikov} & 975 & & & 60-350 & \\
 & & & & & \\
 McClary~\cite{mcclary} & 485 & & & & \\
 & & & & & \\
 Kuang~\cite{kuang} & & 2 & 4-8 & 54 & 395-400 \\
 & & & & & \\
 Galkin~\cite{galkin} & 560 & & & & \\
 & & & & & \\
 Chemtob~\cite{chemtob} & & 0.006 & 53 & & \\
 & & & & & \\
 Bodwin~\cite{bodwin} & 450 & & & 530 & 980 \\
 & & & & & \\
 Chen~\cite{chen3} & & 0.3-1.2 & 4-14 & 19-51 & 360-390 \\
 & & & & & \\
 Chao~\cite{chao} & 385 & & & & \\
 & & & & & \\
 Casalbuoni~\cite{casalbouni} & 450 & & & & \\
 & & & & & \\
 Ko~\cite{ko} & 400 & $>$1.6 & & & \\
 & & & & & \\
 Gupta~\cite{gupta} & 341.8 & & & & \\
 & & & & & \\
\hline
\end{tabular}
\end{center}
\caption{A summary of predictions for the partial widths of various $^1P_1$ decays.}
\end{table}

%\include{newchap2}

%Chapter 3--Experimental aspects of the study of charmonium
\baselineskip=24pt
\chapter{Experimental Apparatus}

The experimental set-up for the Fermilab E835 experiment, which follows the original E760 experiment, consists of a hydrogen gas-jet acting as a fixed proton target in the path of a circulating beam of antiprotons in the Fermilab Antiproton Accumulator. Decay products from the resulting $p\bar p$ annihilations are detected in a detector system with cylidrical geometry surrounding the interaction region. It has no magnetic field, and is designed to optimally detect and identify electrons, positrons, and gammas. As such, it is particularly suitable for the study of those states whose decay involves the vector $(1^{--})$ states of charmonium, $J/\psi$ and $\psi^{\prime}$ which have significant branching fractions into $e^{+}e^{-}$ final state, and/or hadrons which can decay into multiple gammas, in particular $\pi^0$ and $\eta$, which have large branching fractions for decay into two photons. Thus, E835 is particularly suited to the search for $h_c$, the $^1P_1$ resonance of charmonium, one of whose principal decay modes is expected to be $p\bar p \rightarrow ^1P_1 \rightarrow J/\psi + \pi^0 \rightarrow e^{+}e^{-} + \gamma\gamma$. The study of this decay is the primary objective of this dissertation. Several other possible decay channels of $h_c$ were also studied, and are also described. In this chapter we describe the experimental set-up in some detail. A fuller discussion of the E835 experimental setup can be found in~\cite{nimpaper}. E835 had two runs, one in 1997 with 141.4 pb$^{-1}$ of luminosity, and one in 2000 (also called E835p) with 113.2 pb$^{-1}$ of luminosity. This dissertation is devoted to the search for $h_c$ in the year 2000 data, although we also refer to the E760 and E835 (1997) data.

The excitation of a charmonium resonance may be described in terms of the Breit Wigner formula for resonance cross sections:

\begin{equation}
\sigma_{BW} = {\pi (2J + 1) \over s - 4m^2_p} {\Gamma^2_R B_{in}B_{out} \over (\sqrt{s} - M_R)^2 + \Gamma^2_R/4}
\end{equation}

where $M_R$ is the mass of the resonance, $\Gamma_R$ is the total width of the resonance, $B_{in} \equiv B(p\bar p \rightarrow R)$ and $B_{out} \equiv B(R \rightarrow final state)$ are the branching ratios to the initial and final states, $J$ is the spin of the resonance, $m_p$ is the proton/antiproton mass, and $\sqrt{s}$ is the energy in the center of mass frame.

Charmonium resonances at Fermilab E835 are studied by sweeping the $p\bar p$ center of mass energy through the resonance region, and measuring the yield of various decay channels as a function of center of mass energy. An example of such a scan is shown in Figure 3.1. 

\begin{figure}[htb]
\begin{center}
\includegraphics[width=8cm]
%%%{chi2_paper2.eps}
{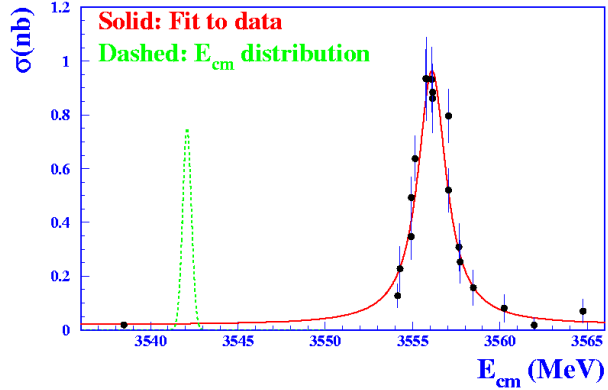}
\caption[A scan of the $\chi_{c2}$ resonance of charmonium using the E835 detector.]{A scan of the $\chi_{c2}$ resonance of charmonium using the E835 detector. The half width of the $\chi_{c2}$ resonance is $\sim 1.8$ MeV. The center of mass energy distibution of the beam, with typical half width of $\sim 500$ keV, is shown by the dotted curve.}
\label{fig:chi2}
\end{center}
\end{figure}

The antiproton beam is stochastically cooled so as to be as monochromatic as possible, but it still maintains a small but finite momentum spread, which must be convoluted with the Breit Wigner resonance cross section and multiplied by the efficiency and acceptance of the detector to obtain the measured cross section:

\begin{equation}
\sigma_{meas}(\sqrt{s}) = \epsilon\alpha{\int_0^{\infty} \sigma_{BW}\sqrt{s^{\prime}}G(\sqrt{s^{\prime}} - \sqrt{s})d\sqrt{s^{\prime}}}
\end{equation}

where $G(\sqrt{s^{\prime}}-\sqrt{s})$ is the function describing the center of mass energy distribution of the antiproton beam as a function of the energy $\sqrt{s^{\prime}}$, when $\sqrt{s}$ is the center of mass energy at which the cross section is being measured, and $\epsilon$ and $\alpha$ are the detector efficiency and acceptance respectively. In general, there is also a background cross section $\sigma_{bkg}$ with a relatively slow variation with center of mass energy, and the Breit Wigner resonance sits on top of this background. 

The number of events $N_i$ observed at a given energy point $\sqrt{s_i}$ is given by:

\begin{equation}
N_i = {\int_t{\cal L}_i(t) \times [\sigma(\sqrt{s_i}) + \sigma_{bkgd}(\sqrt{s_i})]dt}
\end{equation}

where ${\cal L}_i(t)$ is the instantaneous luminosity at energy point $i$, $\sigma_{bkgd}$ is the background cross section at that energy, and $\sigma(\sqrt{s_i})$ is given by equation 3.1. The background cross section is generally measured with data taken several half-widths away from the resonant energy.  Resonance parameters such as mass, width, and branching ratios may be calculated from measurements of events $N_i$ for a particular set of decay products at various energy points $\sqrt{s_i}$ across the resonance region. 

The plot of cross section versus energy is known as the excitation curve, and the area under this curve is given by:

\begin{equation}
A = {\int_0^{\infty}\sigma(\sqrt{s})d\sqrt{s}} = {\pi \over 2} \sigma_{peak} \Gamma_R
\end{equation}

where, 

\begin{equation}
\sigma_{peak} = {4\pi(2J+1)B_{in}B_{out}\alpha\epsilon \over M^2_R - 4m^2_p}
\end{equation}

For relatively broad resonances which have a width much larger than the spread of the beam, $\Gamma_R$ may be measured directly from the shape of the excitation curve.  Even when the width of the beam affects the shape of the excitation curve, causing it to differ from a pure Breit-Wigner for narrow resonances, the area under the curve is conserved.

\section{The Fermilab Antiproton Accumulator}

Antiprotons for experiment E835 are produced in $pp$ collisions using protons from the Fermilab Main Injector. Protons are initially injected into the accelerator system as $H^{-}$ ions and boosted to 750 keV by a Cockroft-Walton accelerator. They are then injected into the Fermilab Linac, where they are accelerated to 400 MeV. At this point they are made to pass through a carbon foil and stripped of their electrons, transforming them from $H^{-}$ ions to bare protons. These protons are then accelerated in two rings. The first, known as the Booster ring, brings them to an energy of 8 GeV, and the second, the Fermilab Main Injector, which brings them to an energy of 120 GeV.

These 120 GeV protons from the Main Injector are then focused onto a nickel target, where they produce a wide variety of secondary particles. For every million protons hitting the target, approximately 20 antiprotons are produced, with energies which have a distribution with a peak around 8.9 GeV. These, along with other negatively charged particles, are focused by a lithium lens towards a magnet, which then directs them to the Debuncher Ring. Negatively charged pions and muons which accompany the antiprotons decay on the way to the debuncher, while electrons miss the beam aperture as they lose energy due to synchrotron radiation. This results into an essentially pure antiproton beam reaching the Debuncher, the primary purpose of which is to reduce the initial large divergence of the captured antiprotons, and use the stochastic beam-cooling systems to create a brighter beam. This beam is injected into the Antiproton Accumulator Ring which is located beside the Debuncher Ring. The various elements of the Fermilab accelerator system relevant to the production of antiprotons are shown in Fig. 3.2.

\begin{figure}[htb]
\begin{center}
\includegraphics[width=8cm]
{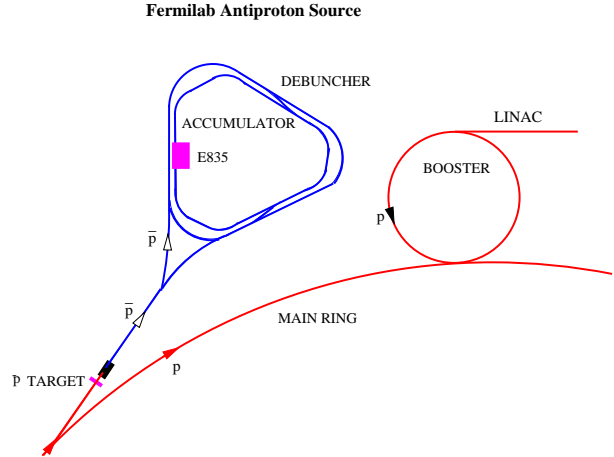}
%%%{chapter2/fig/pbar.xfig.eps}
\caption[Schematic of the Fermilab Antiproton Source and its position relative to the FNAL Main Ring.]{Schematic of the Fermilab Antiproton Source and its position relative to the FNAL Main Ring. The location of the E835 experiment in the Antiproton Accumulator is also shown.}
\label{fig:pbar_chain}
\end{center}
\end{figure}

The Antiproton Accumulator is designed to accumulate antiprotons, a process which is called stacking, for use in the Tevatron~\cite{fnal1}. Antiprotons are stacked in the accumulator at a rate of $\sim 2-4 \times 10^{10} \enskip \bar p$/hour, until a beam of approximately $6 \times 10^{11} \enskip \bar p$ has been collected. The Antiproton accumulator also reduces the momentum spread of the beam by a process known as stochastic cooling~\cite{stochastic}~\cite{stochastic2}. Once enough beam has been accumulated and cooled, it can either by extracted for the Tevatron, or, when the E835 experiment is running, decelerated from the momentum of 8.9 GeV to the momenta needed by E835, which ranged from $\sim 5-7$ GeV/c for the E835 year 2000 run. This deceleration is done with an RF cavity operating at the second harmonic of the beam revolution frequency. This cavity has a maximum RF voltage of 3 kV, allowing for a deceleration rate of 20 MeV/s. The deceleration is performed in several steps constituting a 'ramp'. The momenta to which the $\bar p$ beam must be brought for the various charmoinum states are shown in Table 3.1. The beam parameters are measured and corrected at the end of each ramp, and the magnet settings of the dipoles and focusing quadrupoles of the Antiproton Accumulator are appropriately adjusted. 

\linespread{2.4}
\begin{table}[htb]
\begin{center}
\begin{tabular}{|c|c|c|}
\hline
 & & \\
$c\bar c$ State & $E_{cm}$ (GeV/c$^2$) & $p_{beam}$ (GeV/c)\\
 & & \\
\hline
 & & \\
$\eta_c$ & 2.9797 & 3.6919 \\
 & & \\
$J/\psi$ & 3.0969 & 4.0657 \\
 & & \\
$\chi_0$ & 3.4151 & 5.1931 \\
 & & \\
$\chi_1$ & 3.5105 & 5.5502 \\
 & & \\
$^1P_1$ & 3.5262 & 5.6009 \\
 & & \\
$\chi_2$ & 3.5562 & 5.7246 \\
 & & \\
$\psi^{\prime}$ & 3.6860 & 6.2321 \\
 & & \\
\hline
\end{tabular}
\caption[Masses of charmonium states and the necessary $\bar p$ beam momenta.]{Masses of charmonium states and the necessary $\bar p$ beam momenta. Masses are from PDG2002~\cite{PDG2002}.}
\end{center}
\end{table}

There are 48 horizontal and 42 vertical Beam Position Monitors (BPMs) positioned around the Accumulator Ring. 

There are 38 dipole magnets that bend the beam horizontally around the ring. There are also 48 horizontal and 42 vertical Beam Position Monitors (BPMs) positioned around the Accumulator Ring. They are positioned so that the orbit displacement created by each individual magnet is measured by at least one BPM.

\subsection{Stochastic Cooling}

Once the desired beam momentum is reached the beam is again stochastically cooled. There are two types of cooling, transverse and longitudinal. The transverse cooling reduces the growth of the beam which occurs due to multiple scattering of the antiprotons with the target and with residual gas in the Accumulator. The transverse cooling reduces the size of the beam so that at the gas jet target 95\% of the beam is contained in an approximately circular region of radius 2.45 mm. The longitudinal cooling reduces the momentum spread in the beam and achieve the small beam energy spread shown in Figure 3.1. A distribution of the center of mass energy spread of the beam for all runs in the year 2000 E835 $^1P_1$ data stacks is shown in Figure 3.3. It ranges from $\sigma = 0.2 - 0.55$ MeV, with an average of 0.32 MeV. Both the transverse and longitudinal cooling systems are explained in the following.

\begin{figure}[htb]
\begin{center}
\includegraphics[width=8cm]
{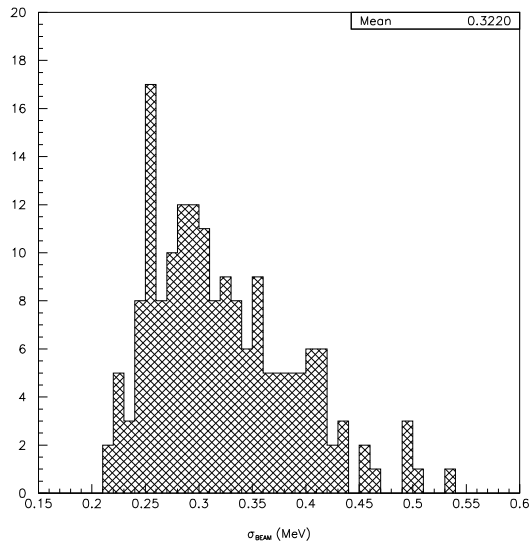}
%%%{chapter2/fig/pbar.xfig.eps}
\caption{Distribution of the center of mass energy spread (r.m.s.) for all runs of the year 2000 E835 $^1P_1$ data stacks.}
\label{fig:pbar_chain}
\end{center}
\end{figure}

\subsubsection{Transverse Stochastic Cooling}

As a $\bar p$ passes one of the several pickup electrodes positioned around the ring, its deviation $x$ from the central orbit position is detected. A correction can then be applied by transmitting a signal to a kicker electrode, which is located an odd number of quarter-wavelengths of betatron oscillation downstream from the pickup. The kick is timed so that it is delivered when the particle detected by the pickup arrives at the kicker. It causes the antiproton to have a transverse position $x - gx$ at the pickup, where $g$ is the system gain. For a ``beam'' which consists of a single particle, a single kick would be enough to correct its orbit. However, since we are dealing with a beam of $N$ antiprotons, each of which affects the motion of the others, the effect of each kick which is delivered by the kicker is smaller. Furthermore, the presence of the other particles means that the pickup detects the {\it mean} deviation of a portion of the beam, and delivers an appropriate kick. The effect of this, along with the fact that the system gain $g$ cannot be exactly 1, is that cooling the beam requires not one, but many, kicks. The cooling principle, though, is applicable for a beam of any size.

\subsubsection{Longitudinal Stochastic Cooling}

Transverse cooling, as described above, decreases the physical size of the beam by decreasing the amplitude of the betatron oscillations, but this has only a marginal effect on the momentum distribution of the beam. For that purpose, longitudinal, or momentum, cooling must be done.

In momentum cooling, it is necessary to detect variations, $\Delta p$, from the mean, or central, beam momentum $<p>$. The mechanism for momentum cooling is similar in nature to that used for transverse cooling. In this case, however, a band-pass or ``notch'' filter is used in the pickup-kicker network, so that particles nearest the central momentum, which corresponds to the central frequency, are the least affected. That is, the presence of the filter allows a positive correction to the slightly low frequency particles, and a negative kick to the slightly high frequency ones, while leaving the particles near the central orbit frequency alone. 

Transverse cooling in the Debuncher reduces the emittance of the $\bar p$ beam from $\sim 20\pi$ to $\sim 7\pi$ mm-rad. Longitudinal cooling reduces $\Delta p/p$ from the $\sim 0.2\%$ achieved by the RF to $\sim 0.09\%$~\cite{todd}. At this point, the $\bar p$ beam is transferred into the Accumulator Ring, and the Debuncher is ready to accept a new $\bar p$ batch. A similar cooling processes is then performed in the Accumulater Ring.

The vacuum in the Accumulator Ring is typically of the order $3 \times 10^{-10}$ torr, giving a fully stacked beam a lifetime $(t_{1/2})$ of approximately 1000 hours with the jet target off. With the jet target running, the beam lifetime is reduced substatially, to $\sim 50$ hours. The E835 target and detector is located in a low dispersion region of the Accumulator, where a momentum spread of $\Delta p/p$ of $10^{-4}$ corresponds to a longitudinal displacement of only 50 $\mu$m. 

\subsection{Measurement of Beam Center of Mass Energy}

The Lorentz-invariant center of mass energy of the beam/target interaction is determined from the antiproton energy and momenta in the lab frame by the relation:

\begin{equation}
E_{cm}^2 = (E_{\bar p(lab)} + m_{p}c^2)^2 - c^2p_{lab}^2
\end{equation}

Since,  

\begin{equation}
E_{\bar p(lab)} = \sqrt{m_{p}^2c^4 + c^2p_{lab}^2}
\end{equation}

we obtain

\begin{equation}
E_{cm}^2 = 2m_{p}c^2(m_{p}c^2+E_{\bar p(lab)}).
\end{equation}

$E_{\bar p(lab)}$ can also be written as:

\begin{equation}
E_{\bar p(lab)} = \gamma m_pc^2 = m_pc^2/\sqrt{1 - (v_{\bar p}^2/c^2)}
\end{equation}

%Here $\gamma = 1/\sqrt{1 - (v_{\bar p}^2/c^2)} = 1.8771$ for the $^1P_1$ energy. 

Thus, the measurement of the center of mass energy depends only on the mass of the proton/antiproton and the velocity of the antiproton in the lab. Since the proton/antiproton mass is known to be 938.271998 MeV$ \pm $38 eV~\cite{PDG2002}, the precision of the energy measurement rests depends only on the precision of the determination of the antiproton velocity. This velocity is obtained by multiplying the frequency of circulation of the antiprotons in the accumulator with their orbit length:

\begin{equation}
v_{\bar p} = f_{\bar p} \times L_{orbit}
\end{equation}

Thus 

\begin{equation}
E_{\bar p(lab)} = m_pc^2/\sqrt{1 - (f_{\bar p}L_{orbit}^2/c^2)}
\end{equation}

The uncertainty in the $E_{\bar p lab}$ measurement can then be calculated by dfferentiating this expression with respect to $f_{\bar p}$ and $L_{orbit}$ to give:

\begin{equation}
{\delta E_{\bar p(lab)} \over E_{\bar p(lab)}} = \beta_{\bar p}^2\gamma_{\bar p}^2({\delta f_{\bar p} \over f_{\bar p}} + {\delta L_{orbit} \over L_{orbit}})
\end{equation}

where $\beta_{\bar p}$ and $\gamma_{\bar p}$ are the relativistic factors for the antiproton in the lab frame, and $\delta E{\bar p(lab)}$ is the uncertainty on $E_{\bar p(lab)}$. 

The corresponding expression for $\delta E_{cm}$ can be similarly be written as:

\begin{equation}
{\delta E_{cm} \over E_{cm}} = {\beta_{\bar p}^2\gamma_{\bar p}^3 \over 2(1 + \gamma_{\bar p})}({\delta f_{\bar p} \over f_{\bar p}} + {\delta L_{orbit} \over L_{orbit}})
\end{equation}

As an example of the effect of the uncertainties on the $L_{orbit}$ and $f_{\bar p}$ measurements on the $E_{cm}$ uncertainty, at the $\psi^{\prime}$ center of mass energy, $dE_{cm}/df$ = 113.2 keV/Hz and $dE_{cm}/dL$ = 149.3 keV/mm~\cite{werkema}. The $\psi^{\prime}$ center of mass energy is of crucial importance for this calculation, since this is the energy at which the orbit length is measured. Typical values for $f_{\bar p}$ and $L_{orbit}$ are 0.6 MHz and 474 m, while typical uncertainties are in the range of $\delta f \approx 0.1$ Hz and $\delta L \approx 1.2$ mm, which leads to:

\begin{equation}
{\delta f_{\bar p} \over f_{\bar p}} = 0.6 \times 10^{-7}, \quad {\delta L_0 \over L_0} = 2.5 \times 10^{-6}.
\end{equation}

Thus, the uncertainty in the measurement of the center of mass energy is dominated by the uncertainty determination of the orbit length. Measurement of $f$, $L$, $\delta f$, and $\delta L$ is described in the following subsections. Using these measurements we may constrain the uncertainty in the center of mass energy to within 180 keV for $\sqrt{s}$ in the $\psi^{\prime}$ region, and 60 keV for $\sqrt{s}$ in the $J/\psi$ region~\cite{werkema}. These uncertainties are further reduced by scanning the $\psi^{\prime}$ resonance and calibrating the orbit length using the accepted value for the $\psi^{\prime}$ center of mass energy, as will be described in Section 3.1.4.

\subsection{Measurement of $f_{\bar p}$}

The orbit frequency of the beam is determined using the Schottky noise spectrum, which is created by the sum of the pulses generated by the passage of each particle in the vicinity of the pickup. The Schottky noise power spectrum is measured by using a Schottky pickup and a network analyzer. This power spectrum $P(f)$, is proportional to the frequency spectrum of the orbiting antiprotons ($dN/df$), or one of the harmonics of that frequency spectrum, by the relation:

\begin{equation}
P(f) = 2\pi (ef)^2 dN/df
\end{equation}

where $e$ is the electric charge.

An example of the Schottky noise power spectrum is shown in Figure 3.4 for the 127th harmonic. This power spectrum is fit in order to determine the peak value to within 10 Hz, corresponding to an uncertainty in the fundamental frequency of less than 0.1 Hz. Dividing the central frequency and the width of the peak by 127 for this plot gives an orbit frequency of (625.366 $\pm$ 0.004) kHz, and thus $\delta f/f \approx 10^{-7}$~\cite{nimpaper}. Note that the y-axis in Fig. 2 is plotted in a logarithmic scale. 

\begin{figure}[htb]
\begin{center}
\includegraphics[width=10cm]
{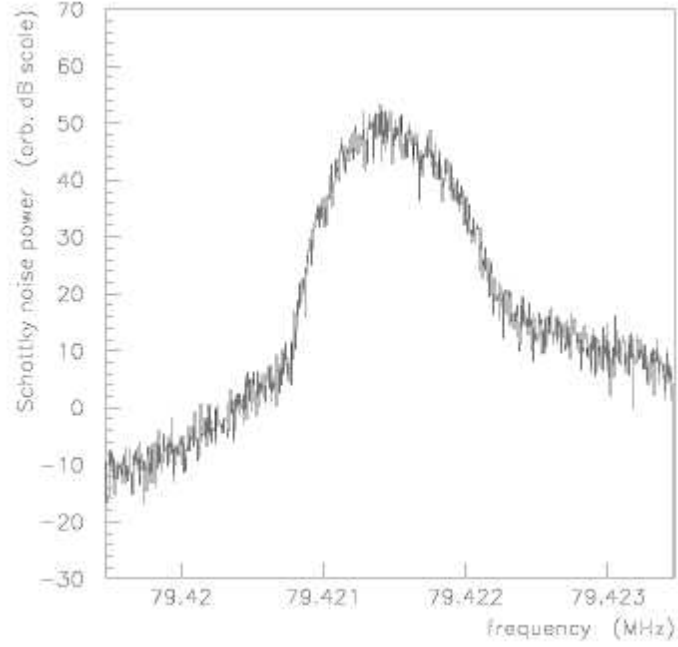}
\caption[Measured Schottky noise power spectrum of the $127^{th}$ harmonic 
of the antiproton beam with momentum 6232 MeV/c (the $\psi^{\prime}$ resonance).]{Measured Schottky noise power spectrum of the $127^{th}$ harmonic 
of the antiproton beam with momentum 6232 MeV/c (the $\psi^{\prime}$ resonance). The central frequency is 79.421474 MHz~\cite{nimpaper}. An uncertainty $\delta f$ of $\sim 10$ Hz in the $127^{th}$ harmonic corresponds to a $\delta f$ of $\sim 0.1$ Hz in the fundamental.}
\label{fig:schottky}
\end{center}
\end{figure}

The exact relation between the spectrum of the beam momenta and the frequency spectrum (measured via the Schottky noise power spectrum) is given by the relation~\cite{werkema}:

\begin{equation}
\label{eq:pbeam_f}
\frac{d p_{\bar p}}{p_{\bar p}} = 
- ~\frac{1}{\eta} \times \frac{df}{f}~. 
\end{equation}

where $\eta$ is a parameter known as the slip factor, and is given by:

\begin{equation}
\label{eq:slip}
\eta = \frac{1}{\gamma_t^2} - \frac{1}{\gamma_{\bar p}^2} ~, 
\end{equation}

where $\gamma_t$, the gamma factor at the transition energy, is a parameter determined by the machine lattice~\cite{acc1}~\cite{acc2}. Since $df/f$ can be well measured directly from the Schottky spectrum, to obtain the beam momenta we must determine $\eta$ as a function of beam energy. This can be done with two different techniques~\cite{nimpaper}.

The first technique is to determine $\eta$ from the synchrotron frequency as a function of peak RF voltage. With RF power on, the energy of the orbiting antiprotons will oscillate around the central energy with a characteristic synchrotron frequency $f_s$ given by:

\begin{equation}
f_s^2 = - {eV_{rf} \over E}{f^2_{RF} \over 2\pi h}{\eta cos(\phi_s) \over \beta^2}
\end{equation}

where $V_{rf}$ and $f_{RF}$ are the peak RF voltage and RF frequency, $E$ is the beam energy, $h$ is the harmonic number, and $\phi_s$ is the synchronous phase, which is 0 above transition $(\gamma > \gamma_t)$ and $\pi$ below transition. The synchrotron frequency may be determined to a precision of 1\%, and the uncertainty in the measurement of $\eta$ is dominated by the uncertainty in the RF voltage, which is of the order of 5\%. 

The second technique is to determine $\eta$ from $\gamma_t$ (eq. 3.17). $\gamma_t$ may be determined by varying the dipole magnetic fields in the absence of RF and measuring the resulting change in $f$. The gamma factor at the transition energy may then be determined by using the relation:

\begin{equation}
dB/B = \gamma_t^2 df/f
\end{equation}

with the uncertainty in $\eta$ related to the uncertainty in $\gamma_t$ by:

\begin{equation}
\delta \eta = (2/\gamma_t^3) \delta \gamma_t
\end{equation}

As this method requires knowledge of the magnetic fields in the dipoles, and these have large systematic errors, it is used primarily to cross check the result from the first technique.

\subsection{Measurement of $L_{orbit}$}

The Measurement of the orbit length; $L_{orbit}$ is done by measuring the difference in length between the current orbit and a reference orbit using a system of 48 Beam Position Monitors. The reference orbit is chosen to be that which has a center of mass energy at the peak of an easily detectable resonance whose mass is well known. E835 uses the $\psi^{\prime}$ resonance, which has a mass of $(3686.111 \pm 0.025 \pm 0.009)$ MeV~\cite{aulchenko}, and which is observed through its inclusive decays into $J/\psi$ followed by the subsequent decay $J/\psi \rightarrow e^{+}e^{-}$, as well as its direct decay $\psi^{\prime} \rightarrow e^{+}e^{-}$, which give clear signal $e^{+}e^{-}$ signals in the detector.

The reference length $L$ may then be determined from the mass of the $\psi^{\prime}$ resonance and the measured beam frequency at the $\psi^{\prime}$ energy using the relation:

\begin{equation}
\label{eq:lref}
M_{\psi^{\prime}}^2 = 2 \times (m_p c^2)^2 \times 
\biggl(1+\frac{1}{\sqrt{1-(f \times L_{ref}/c)^2}} \biggr)~,
\end{equation}

which may be derived from Equations 3.12 and 3.13 for the case where the center of mass energy is equal to the $\psi^{\prime}$ mass. The error in this measurement is then dominated by the error in the knowledge of the mass of the $\psi^{\prime}$, and generates an uncertainty in $L_{ref}$ which is given by:

\begin{equation}
\label{eq:dlref}
\delta L_{ref} = L_{ref} \times
\frac{M_{\psi^{\prime}}}
{\gamma_{\bar p}^3~\beta_{\bar p}^3~m_p^2} \times \delta M_{\psi^{\prime}}
~=~ 0.6~\mathrm{mm} times 0.28 ~=~ 0.17\mathrm{mm} .
\end{equation}

Once the reference orbit length is known, the difference between the length of the orbit at any given energy and the length of the reference orbit can be determined using the beam position monitors (BPMs). The data from these BPMs must be used in a piecewise manner to determine the change in the orbit length, or by using a constrained fit using a detailed model of the Antiproton Accumulator lattice. The overall uncertainty in the orbit length is of the order $\pm$ 1 mm, which is the largest contributor to the uncertainty of the center of mass energy in E835.

\section{The Gas Jet Target}

The E835 target is a hydrogen gas jet which ejects clusters of hydrogen perpendicularly to the antiproton beam axis at a rate of $\sim$ 1000 m/s~\cite{jettarget}. With the hydrogen jet target on, a typical 50 mA beam lasts for 2 or 3 days before being depleted by a factor $\sim$ 3; the beam lifetime is $\sim$10 times longer with the gas jet turned off.

In order to maintain a constant instantaneous luminosity during data taking, the density of the jet may be varied during operation to compensate for the loss of antiprotons in the beam. In a typical run the jet density is varied to keep the instantaneous luminosity in the range of $2 - 3 \times 10^{31}$ s$^{-1}$cm$^{-2}$. This corresponds to a minimum bias trigger rate of 3 MHz, which is close to the maximum sustainable by the detector and data acquisition systems, and thus optimizes the run time alloted to E835. The constant instantaneous luminosity also allowed for better understanding of effects such as event contamination due to pileup, which is strongly luminosity dependent. A plot showing the variation of the gas jet density, antiproton current, and instantaneous luminosity is shown in Fig. 3.5.

\begin{figure}[htbp]
\begin{center}
\includegraphics[height=14cm,width=14cm]
{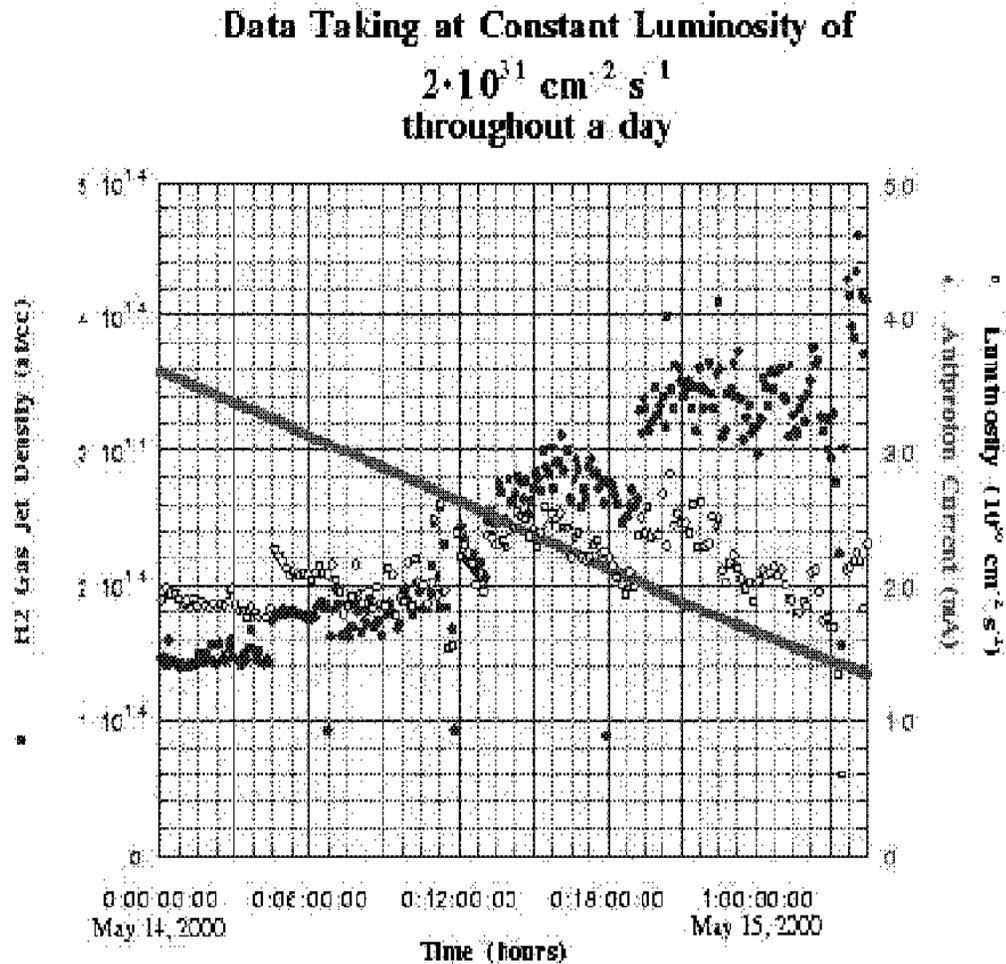}
\caption[Jet target density and antiproton beam current 
during $\sim$ 30 hours of data taking at $E_{cm}=3526$~MeV.]{Jet target density and antiproton beam current 
during $\sim$ 30 hours of data taking at $E_{cm}=3526$~MeV. 
As the beam current steadily decreases, the jet density is 
increased to maintain an approximately constant instantaneous luminosity. The jet density is shown by the solid blue points, the antiproton current by the solid red curve, and the instantaneous luminosity by the open black circles~\cite{paolo}.}
\label{fig:jetY2K_ConstLum}
\end{center}
\end{figure}

The gas jet used in E835 is of a type known as a cluster jet because the core of the jet is made up of small droplets, or clusters, of condensed hydrogen. This cluster jet is created by allowing hydrogen, which is kept at high pressure and low temperature, to expand through a convergent-divergent nozzle. The expansion of the gas through the nozzle is isentropic, and the the sudden decrease in pressure and temperature caused by the expansion leaves the gas in a supersaturated state, favoring the formation and growth of a jet of clusters whose size varies from $10^7$ to $10^8$ molecules.

A schematic of the E835 gas jet nozzle is shown in Fig. 3.6. The nozzle is trumpet shaped, and has an opening angle of $3.5^{\circ}$, a divergent length of 8 mm, and a throat diameter of 37 $\mu$m. As can be seen from the shape of the isentropes on the P-T diagram of hydrogen (shown in Fig. 3.7) the density of the jet may be maximized by allowing the expansion to begin at the highest possible pressure and the lowest possible temperature, which puts the hydrogen as close to the $H_2$ saturation curve as possible. In order to keep the temperature low, a helium cryo-cooler was installed at the final stage of the hydrogen line, allowing operation with hydrogen gas temperatures as low as 20$^{\circ}$ K. Further manipulation of the point on the P-T curve where the expansion starts may also be done by changing the pressure; reduced pressure at the nozzle allows for correspondingly lower densities. The open circles in Fig. 3.7 show the operating points used by the E835 gas-jet target; these lie directly above the $H_2$ saturation curve, and give the range of jet densities shown in the upper curve.

\begin{figure}[htb]
\begin{center}
\includegraphics[height=6cm]
{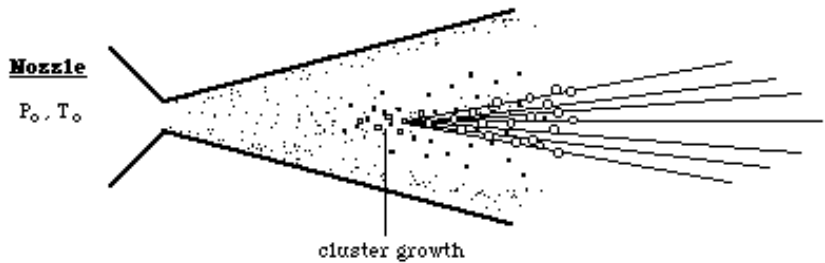}
\caption{Detail of the gas jet nozzle and the formation of hydrogen clusters.}
\label{fig:jet_ClusForm}
\end{center}
\end{figure}

\begin{figure}[htbp]
\begin{center}
\includegraphics[height=10cm]
{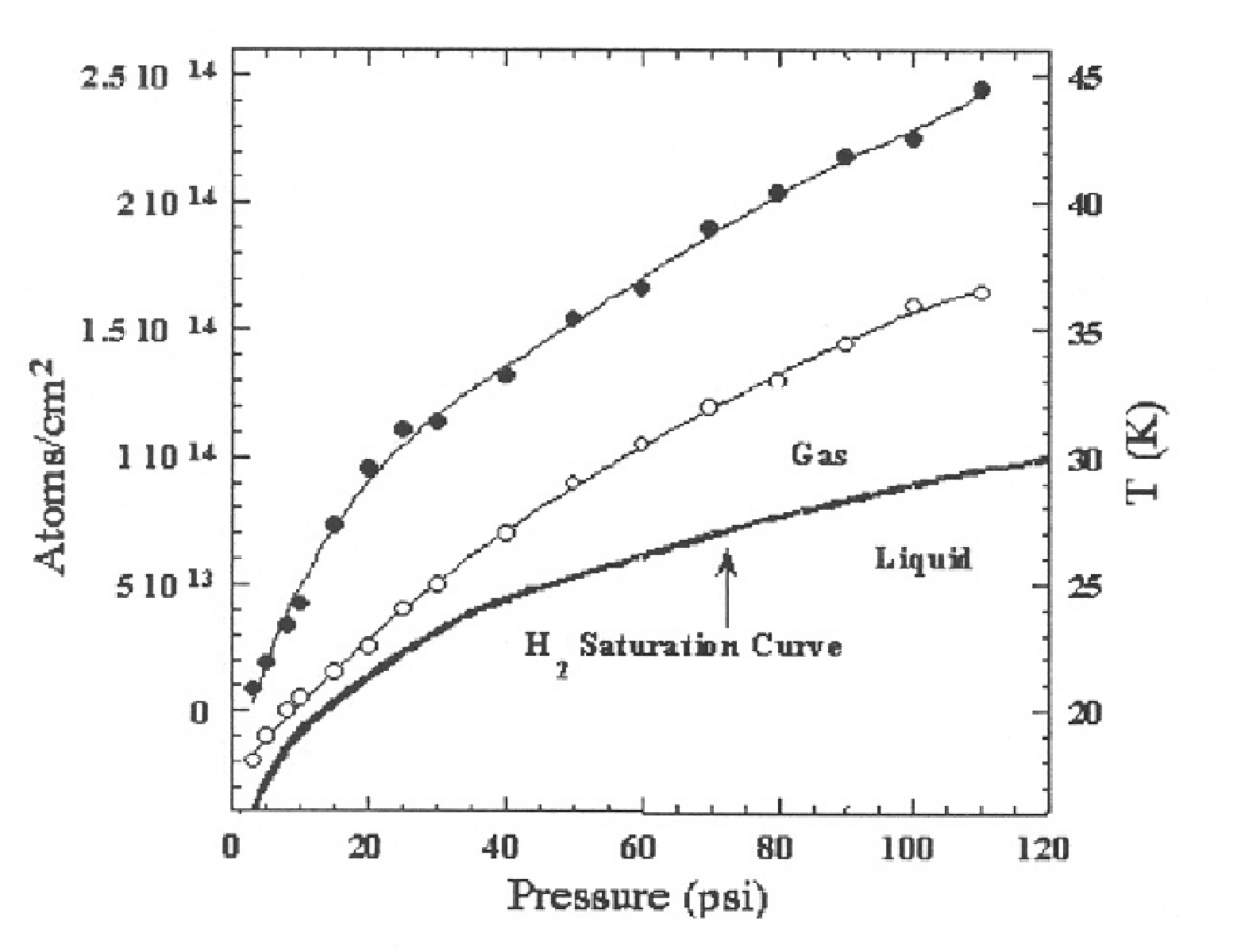}
\caption[The $P-T$ phase diagram for H$_2$.]{The $P-T$ phase diagram for H$_2$. The operating points utilized by E835 are shown by open circles and the jet density corresponding to those operating points are shown by filled circles~\cite{nimpaper}.}
\label{fig:jetY2K_JDpath}
\end{center}
\end{figure}

\begin{figure}[htbp]
\begin{center}
\includegraphics[height=15cm]
{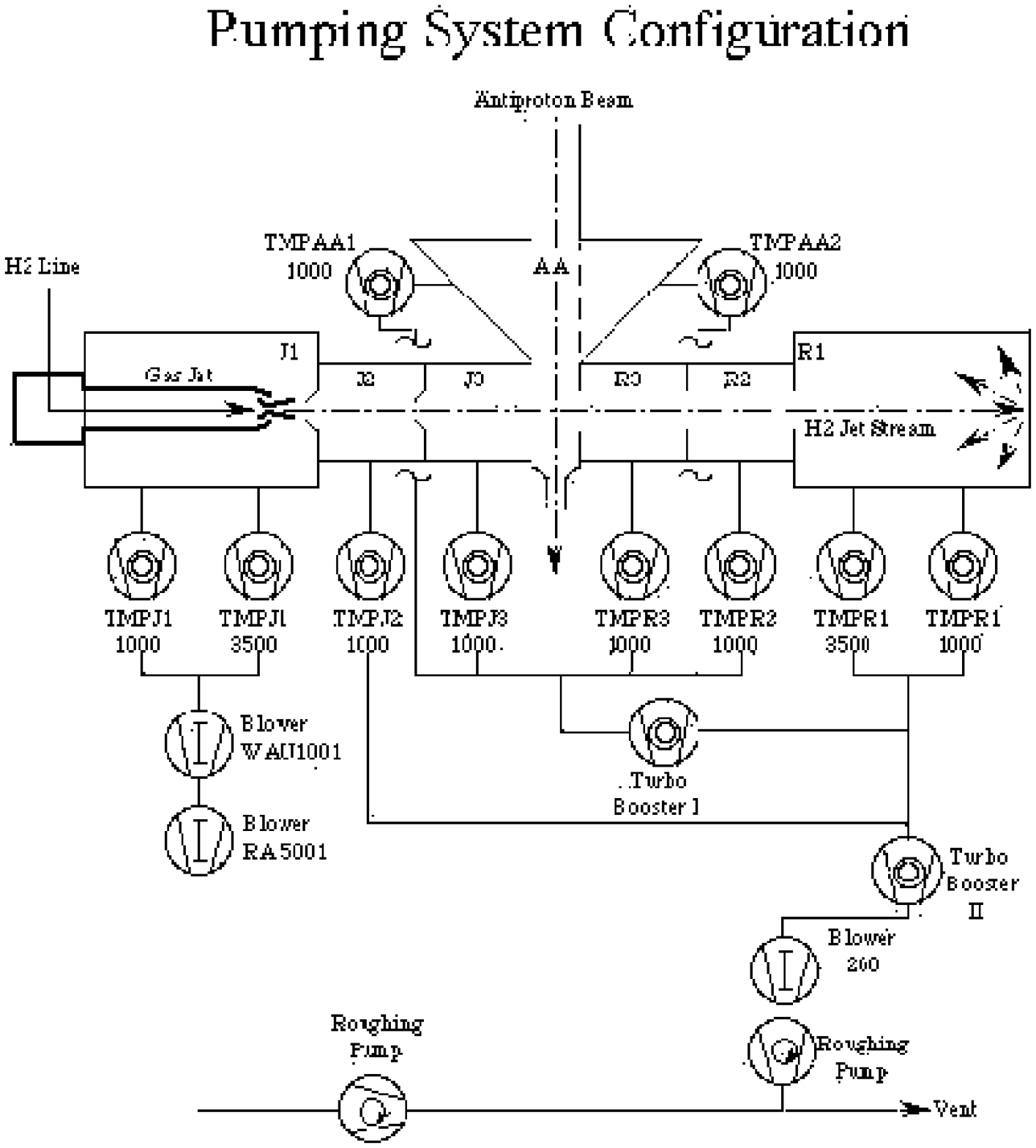}
\caption{Layout of the hydrogen jet target and pumping system.}
\label{fig:jet_PumpingSystem}
\end{center}
\end{figure}

Using the gas jet as a target inside the Antiproton Accumulator required setting up a system of turbo-molecular pumps in order to maintain the high vacuum in the beam pipe which is required to preserve a high quality $\bar p$ beam. The ability of this pumping system to maintain high vacuum is further enhanced by separating the cluster jet stream from the remaining gas exiting the nozzle. A differential pumping scheme is therefore used (as shown in Fig. 3.8), in which the jet crosses a series of chambers which are independently evacuated. These chambers have installed into them a series of ten turbo-molecular pumps (TMPs), eight of which have a capacity of 1000 liters per second, and two of which have a capacity of 3500 liters per second. As TMPs have a low compression ratio for hydrogen, two additional TMPs, three positive displacement blowers and two roughing pumps were arranged in a cascade configuration in over to avoid limiting the pressure in the high vacuum zone of each pump due to the rough vacuum (see bottom of Fig. 3.8). The chambers which are upstream from the interaction zone are labelled J1, J2, and J3, and the TMPs in these chambers remove the part of the gas which does not clusterize into the core of the jet. The chambers which are downstream from the interaction zone are labelled R1, R2 and R3, and the TMPs in these chambers are used to remove the core jet once it has crossed the interaction zone. These must be removed as only a small percentage of the protons in the clusters interact with the antiprotons in the beam. By using this system, we can reduce the number of interactions of the $\bar p$ beam outside of the interaction zone to 5\% of that inside the interaction zone.

In order to maximize the cluster density during data taking, the control system which regulated the temperature of the nozzle and the pressure of the hydrogen line was automated. This system allowed for regulation of the pressure to within 0.5 psi and the regulation of the temperature to within 0.05$^{\circ}$ K, with a response time of 10 s. Thus the densities of the hydrogen could be varied from $\sim 1 \times 10^{14}$ atoms/cm$^{-3}$ to $\sim 3 \times 10^{14}$ atoms/cm$^{-3}$ over the lifetime of an antiproton stack. The jet diameter in the interaction region is 6 mm, which is only slightly larger than the beam diameter at this point ($\sim 5$ mm). This diameter is regulated by the geometry of the skimmer between the second and third vacuum chambers. The system of collimators which cross the axis of the gas jet confine the direction of the jet to be perpendicular to the beam axis to within 2$^{\circ}$. The gas jet beam pipe is stainless steel with a thickness of 0.18 mm in the region where the secondary particles pass into the main detector, which will be described in the next section.

\section{The E835 Detector}

The E835 Detector is designed to detect electromagnetic final states, $\gamma, e^{+}$ and $e^{-}$, with two or more of these particles resulting from the decay of charmonium resonances. The detector is also designed to perform at a high interaction rate because $p\bar p$ total cross sections are very large, being $\sim$ 70 mb in the $\sqrt{s}$ = 3-4 GeV region. A complete and detailed description of the E835 detector may be found in the recently published in Nucl. Inst. Meth. A~\cite{nimpaper}; much of the following discussion is based on that paper.

The schematic of the E835 detector system is shown in Fig. 3.9. It has a cylindrical geometry about the antiproton beam axis, with full coverage of the azimuthal angle $\phi$ for all its central components. The system consists of the central and forward electromagnetic shower calorimeters to measure the energy and momentum of electrons, positrons and photons, a system of inner detectors to detect charged particles, a \v Cerenkov counter to discriminate electrons and positrons from heavier charged particles, and a luminosity monitor to measure the interaction luminosity. The inner detectors are made up of three plastic scintillator hodoscopes, four layers of drift tubes (straws), scintillation counters for forward angle veto, and two scintillating fiber detectors. The inner detectors are contained in a cylinder of radius 17 cm and length 60 cm; and their total thickness is less than 7\% of a radiation length for particles crossing at normal incidence. Other than the luminosity monitor, these detectors are all highly segmented to allow for a higher rate of signal and are equipped with time-to-digital converters (TDCs) to reject out of time signals.

The detector's polar acceptance for charged particles is $\theta = 15^{\circ} - 65^{\circ}$, and $10.6^{\circ} - 70^{\circ}$ for photons in the central calorimeter (CCAL). A forward calorimeter (FCAL) extends the photon acceptance down to $2^{\circ}$, but it is, as in the present measurements, used primarily to provide a veto.

\begin{figure}[htb]
\begin{center}
\vspace*{20pt}
\includegraphics[height=14cm,angle=270]
{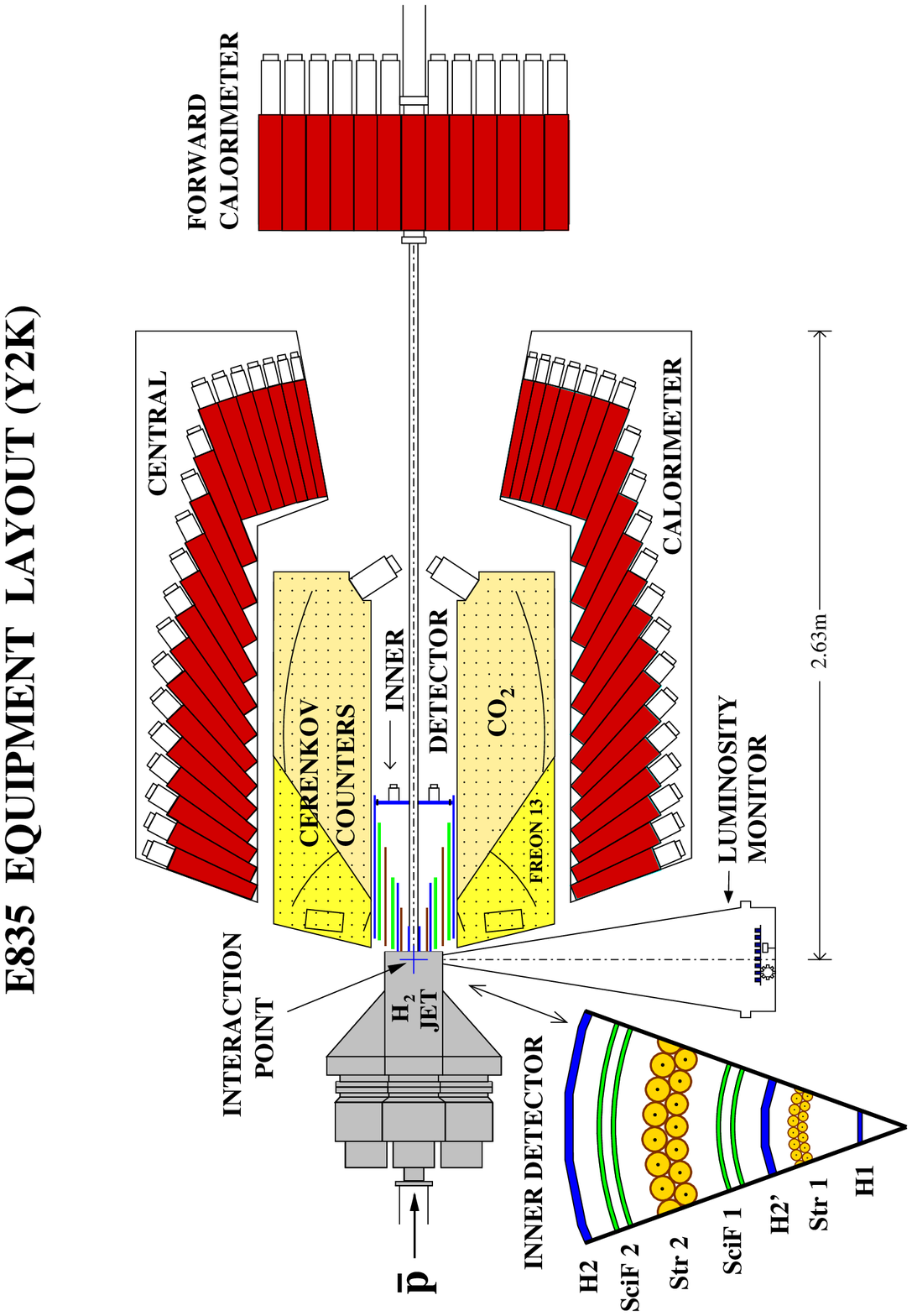}
\caption{Side view of the E835 detector system.}
\label{fig:detector}
\end{center}
\end{figure}

\subsection{The Hodoscopes and the Veto Counters}

There are three scintillator hodoscopes in the inner detector; these are labeled H1, H2, and H2$^{\prime}$, as well as a set of forward veto counters. These are all segmented, and each unit is made of a plastic scintillator connected by a light guide to a phototube. The Bicron 408 plastic which is used for the scintillators has an index of refraction of 1.58 and a density of 1.03 g/cm$^{-3}$. 

The three hodoscopes H1, H2, and H2$^{\prime}$ are all designed to be axially symmetric around the antiproton beam. They are segmented azimuthally into planes of constant angle $\phi$. The innermost hodoscope, H1, consists of 8 plastic scintillators which form a cone shaped structure around the segment of the beam pipe which is attached to the jet target body. The thickness of the H1 scintillators is 2 mm; they provide full coverage in the azimuthal angle $\phi$, and coverage from 9$^{\circ}$ to 65$^{\circ}$ in the polar angle $\theta$. The H1 light yeild is about 10-20 photoelectrons for a single minimum ionizing particle. H1 is located at distance of 2.2 cm from the beam axis. It is the innermost element of the entire E835 detector.

The outermost hodoscope, H2, is a cylindrical device of radius 17 cm, which is made of 32 segments. These segments are 60 cm long, 3 cm wide, and have a thickness of 4 mm. Like H1, H2 provides full geometric coverage in the azimuthal angle, and has an acceptance between 12$^{\circ}$ and 65$^{\circ}$ in the polar angle. The light yeild of H2 is on the order of 50-100 photoelectrons per minimum ionizing particle, much higher than the light yeild of H1. This allows H2 to give the best $dE/dx$ measurement of all the E835 hodoscopes. 

In between H1 and H2 is the third hodoscope, H2$^{\prime}$, which is generally similar to H2 in shape, but is made up of 24 segments rather than 32. These segments are 40.8 cm long and 4 mm thick, and are located at a distance of 7 cm from the beam axis. Like the other two hodoscopes, they provide complete azimuthal coverage, and coverage in the polar angle $\theta$ between 9$^{\circ}$ and 65$^{\circ}$. The cracks between the segments of H2$^{\prime}$ are deliberately not aligned with the cracks in H1 and H2 in order to reduce the leakage of particles. It was added primarily to improve the charge veto for neutral triggers, and to improve the $dE/dx$ measurement. 

A forward veto counter also exists which consists of a holed disk placed perpendicularly to the beam pipe as the end cap of the inner detector cylinder. It is made up of 8 trapezoidal scintillators of 2 mm thickness, which form an annulus. It provides full azimuthal coverage, and coverage in polar angle $\theta$ of between 2$^{\circ}$ and 12$^{\circ}$. It is used only as a charged particle veto in the forward acceptance region.

All three hodoscopes, H1, H2, and H2$^{\prime}$, as well as the forward veto counter FV, are used to trigger on charged particles, to act as a veto on neutral triggers. For example, the coincidence between H1 and H2 generates a first level trigger for charged events, and the coincidence between H1, H2$^{\prime}$ and FV generates a veto signal for neutral events. The hodoscopes are also used to measure $dE/dx$. Finally, they form the first piece of the electron identfication algorithm known as electron weight which will be described later.

\subsection{The Straw Chambers}

After the hodoscopes, the next components of the E835 inner detector system are the two straw chambers. These are cylindrical chambers placed with the antiproton beam along their axis, and are comprised of aluminized mylar straws which act as proportional drift tubes. These straw chambers are used to determine the azimuthal angle $\phi$ of charged particles passing through the inner detector. They give full coverage in the azimuthal, and coverage between 15$^{\circ}$ and 58$^{\circ}$ in the polar angle for the first chamber and between 15$^{\circ}$ and 65$^{\circ}$ for the second. Each of the two straw chambers is made of two layers of straws, which are staggered azimuthally with respect to each other in order to resolve left-right ambiguity. Each layer of the chamber is made up of 64 straws. The inner straw chamber has a radius of 54 mm and is labelled STR1, and the outer straw chamber has a radius of 120 mm and is labelled STR2. The length along the beam axis of STR1 is 182 mm, while the length of STR2 is 414 mm.

The straws are designed to have a low mass to minimize multiple scattering and photon conversions, and to have fine granularity to limit occupancy and increase the azimuthal angle resolution. Their thickness is 0.11\% of a radiation length at a polar angle of 90$^{\circ}$. These drift tubes are self supporting between two grooved flanges that allow gas to flow continously through the tubes. The gas in the tubes is a mixture of Argon, C$_4$H$_{10}$, and (OCH$_3$)$_2$CH$_2$ in the ration 82:15:3. The tubes themselves are made of mylar with a thickness of 80$\mu$m, and are coated on the inside with a layer of aluminum approximately 1000 atoms thick; these form the cathodes of the drift tubes. The anode of each drift tube is made up of a gold plated tungsten wire placed along its axis. These wires had diameters of 5.0-5.4 mm in STR1, and 11.1-12.1 mm in STR2. They are crimped at the end of each tube to gold plated copper pins. The voltage between the tubes and the wires was 1320 V and 1530 V for the inner and outer chambers, respectively. The drift velocity at these operating voltages was $\sim$ 40 $\mu$m/ns. 

The readout electronics of the straw chambers are designed to withstand high rates in the chambers. They use a custom analog bipolar integrated circuit, with an ASD-8B chip and an 8 channel amplifier-shaper-discriminator with fast peaking time (6-7 ns) and good double pulse resolution (25 ns) to avoid pile-up. The straw electronics have a signal amplitude of about 20 mV/fC, and the total power dissipation per channel is 23 mW. The front end electronics are mounted on the downstream flange of each chamber to minimize oscillations and pickup, and due to limited space only Surface Mounting Devices (SMDs) were used. Signals from the straw chambers are sent to the E835 counting room, which is located in the AP50 building directly above the Antiproton Accumulator. In the counting room the signals are processed by 32-channel LRS multihit Time-to-Digital (TDC) 3377 converters used in common-stop mode. 

The particle detection efficiency of a single straw goes from $\sim 100\%$ in the vicinity of the wire, to $\sim 80\%$ close to the aluminum cathode surface. The measured efficiency of track reconstruction with at least two layers of straws is 97\%, with an efficiency of about $\sim 90\%$ per layer. Using both chambers, the angular resolution of a track is approximately 9 mrad. The straw chambers are described fully in a dedicated paper published in Nucl. Inst. Meth A~\cite{straw}. The straw chambers were not used for the analysis in this thesis, and the azimuthal angles of the $e^{+}e^{-}$ pairs were determined using the Central Calorimeter (CCAL), which is described in a later section.

\subsection{The Scintillating Fiber Tracker}

The final elements of the E835 inner detector system are made up of two scintillating fiber trackers. The purpose of these trackers is to provide a measurement of polar angle $\theta$ for charged particles. The detectors are made of two concentric layers of scintillating fibers wound around two coaxial cylindrical supports. The inner tracker, SciF1, has 240 fibers per layer, gives full azimuthal coverage, and has a coverage of between 15$^{\circ}$ and 55$^{\circ}$ in the polar angle $\theta$. The outer tracker, SciF2, has 430 fibers per layer, gives full azimuthal coverage, and has a coverage of between 15$^{\circ}$ and 65$^{\circ}$ in the polar angle $\theta$. The radii of the two cylidrical supports around which each of the trackers are wound are 85.0 mm and 92.0 mm for SciF1 and 144.0 mm and 150.6 mm for SciF2. 

The scintillation light from these fibers is detected by solid state photosensitive devices called Visible Light Photon Counters (VLPCs). These VLPCs were chosen because of their very high quantum efficiency ($\sim 70\%$ for 550 nm photons). The fibers themselves are positioned onto the cylinders in a series of machined U-shaped grooves of pitches 1.10/1.19 mm, and 1.10/1.15 mm for the inner and outer SciF1 and SciF2 support cylinders, respectively. These grooves are machined so that their depth varies linearly with the azimuthal angle $\phi$ in order that the fiber can overlap itself after one turn without any change in polar angle $\theta$. The starting azimuthal angle of each fiber is offset from that of the neighboring fibers in order that the fibers do not overlap in any way as they are drawn axially away from the cylinder. The fibers are aluminized at one end to increase the light yield and reduce signal dependence on azimuthal position. On the other end they are thermally spliced to clear fibers which are 4 m long, and which bring the light to the VLPCs, which are kept in a cyrostat at a temperature of 6.5$^{\circ}$ K. 

The signals from the VLPCs are amplified by QPA02 cards and then sent to discriminator-OR-splitter modules which provide an analog and a digital output for each input channel, together with the digital OR of all inputs. The analog signal is then sent to an Analog-To-Digital (ADC) converter, while the digital signal is sent to a Time-To-Digital (TDC) converter. The outer scintillating fiber tracker signals were grouped together into 19 bundles of adjacent fibers, and the digital OR of the signals from each of these bundles is sent both to a TDC and to the first-level trigger logic of the experiment. The scintillating fiber tracker has an efficiency which is better than 99\% on average in the angular region between 15$^{\circ}$ and 50$^{\circ}$, and better than 90\% in the region between 50$^{\circ}$ and 65$^{\circ}$. The intrinsic angular resolution of the fibers is $(0.7 \pm 0.1)$ mrad.

The typical signal generated by a track through one fiber is 180 mV high and 80 ns wide, corresponding to a collected charge of $\approx$ 0.2 nC. During calibration, the one p.e. equivalent in ADC counts was measured using an LED test for each channel in the final readout configuration. The pulse charge in ADC counts generated by a minimum ionizing particle was obtained by studying a high statisics sample of hadronic tracks ($\approx 10^3$ events/fiber).

The detection efficiency of the scintillating fiber tracker was measured by using the $e^{+}e^{-}$ tracks from $J/\psi$ and $\psi^{\prime}$ decays and $p\bar p$ elastic scattering events. For each track an associated hit was sought in the scintillating fiber detector about a given software threshold ($\sim 0.2$ m.i.p.) and within a polar angle window of $\pm 50$ mrad. The results showed almost 100\% efficiency, with the exception of the backward region between 50$^{\circ}$ and 65$^{\circ}$ in polar angle $\theta$, where there is less redundancy as each track intercepts fewer fibers. 

One background which is particularly important for E835 is $e^{+}e^{-}$ pairs with a small opening angle, which are generated by photon conversion or by Dalitz decays of neutral pions, and which simulate single tracks. The scintillating fiber tracker can be used to help identify such $e^{+}e^{-}$ pairs in two ways; one using pulse height and the other using granularity. In the first case, when the opening angle of the $e^{+}e^{-}$ pair is so small that just one set of adjacent hit fibers is produced, the energy deposit is likely to be big. In the second case, when the pair separation is large, an extra set of adjacent hit fibers is produced. 

The intrinsic time resolution of the scintillating fiber tracker was evaluated by selecting tracks which hit two fibers belonging to adjacent bundles, and thus were read out by two separate TDC channels. The rms time resolution is the standard deviation divided by $\sqrt{2}$ of the distribution in time of the signals of tracks which registered in two bundles; it was determined to be approximately 3.5 ns, due mostly to the decay time of the scintillator. Further information on the scintillating fiber tracker may be found in Refs.~\cite{scifi}~\cite{scifi2}. The scintillating fiber tracker was not used in the analysis in this thesis for $e^{+}e^{-}$ polar angle determination. This was done using the CCAL, which will be described in a later section.

\subsection{The \v Cerenkov Counter}

The E835 \v Cerenkov Counter is used to identify the lightest charged particles, electrons and positrons, in a background of much more numerous charged hadrons. This counter works on the principle that \v Cerenkov light is emitted when a charged particle passes through a medium with a velocity greater than the velocity of light $(c/n)$ in the medium. Thus a particle must have a velocity $\beta$ greater than $1/n$ where $n$ is the index of refraction of the medium. Since the energy of a particle moving at speed $\beta$ is given by:

\begin{equation}
E = \gamma mc^2 = {mc^2 \over \sqrt{1-\beta^2}}
\end{equation}

the condition for \v Cerenkov light being produced is

\begin{equation}
E > {mc^2 \over \sqrt{1-1/n^2}}
\end{equation}

Since this threshold energy for the production of \v Cerenkov light is proportional to the mass of the particle, and there are more than two orders of magnitude between the mass of the electron/positron and that of the next heaviest particle, the pion, it is possible to choose a medium with an appropriate index of refraction $n$ which can discriminate between $e^{+}$ and $e^{-}$ and all other charged states over a large range of energies. The E835 \v Cerenkov counter uses two gases as its \v Cerenkov media. It is divided into two cells in polar angle $\theta$, with different gases in each cell to optimize the electron detection efficiency and the differentiation between electrons and the lightest charged hadrons, the $\pi^{\pm}$s. 

The overall structure of the \v Cerenkov detector is that of a cylindrical shell with an inner radius of 17 cm and an outer radius of 59 cm. It is designed to completely enclose the inner detector within its inner radius, and extends past it along the beam axis in the forward direction. The \v Cerenkov detector has full coverage in the azimuthal angle $\phi$ and coverage from 15$^{\circ}$ to 65$^{\circ}$ in the polar angle $\theta$. The detector is divided internally into eight wedges, each with two separate gas-tight cells separated by a septum which is positioned inside the \v Cerenkov Counter at a polar angle of 34$^{\circ}$ or 38$^{\circ}$, depending on the distance from the beam axis. Each of the two cells is divided interally into eight wedges, each subtending an azimuthal angle of 45$^{\circ}$. A cross section of one of these wedges is shown in Fig. 3.10.

\begin{figure}[htb]
\begin{center}
\includegraphics[width=5in]
{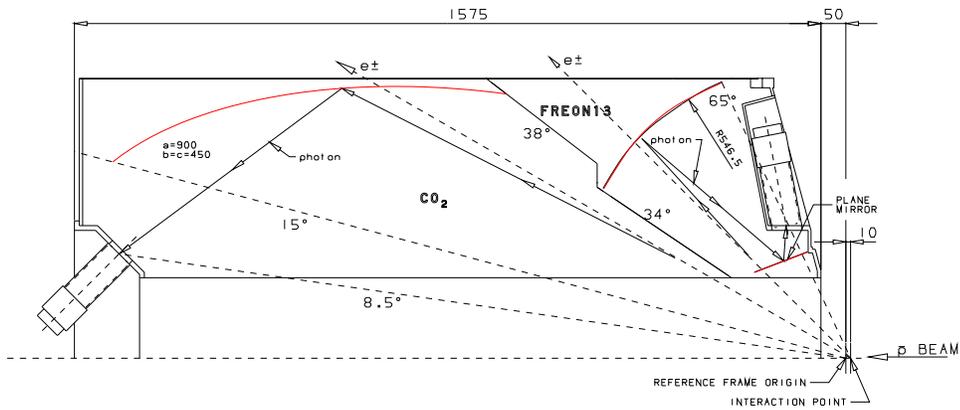}
\caption[Schematics of the \v Cerenkov counter.]{Schematics of the \v Cerenkov counter, showing one octant and its mirrors and angles. The septum at $34^\circ/38^\circ$ divides each chamber into its forward and backward-angle cells. The spherical, plane and ellipsoidal mirrors are shown in red.}
\label{fig:cerenkov}
\end{center}
\end{figure}

Each of the wedges in the small $\theta$ cell is filled with CO$_2$ gas, which has an index of refraction $n = 1.00041$. This corresponds to a \v Cerenkov energy threshold for charged pions of 4.875 GeV, which is greater than the largest pion energy observable by E835. The large $\theta$ cell is filled with either Freon 12 (CF$_2$Cl$_2$), which has an index of refraction of 1.00108, corresponding to an \v Cerenkov energy threshold for charged pions of 3.003 GeV, or with Freon 13 (CF$_3$Cl), which has an index of refraction of 1.0072, corresponding to a pion energy threshold of 3.677 GeV. Freon 13 was always used in the year 2000 run. The \v Cerenkov light which is emitted by charged particles travelling above the threshold velocity forms a cone around the direction of travel of the particle. The half angle $\theta_c$ of this cone is given by the relation:

\begin{equation}
\theta_c = cos^{-1}(1/n\beta)
\end{equation}

For $\beta \approx 1$ this gives a $\theta_c$ of 1.64$^{\circ}$ for CO$_2$, 2.17$^{\circ}$ for Freon 13, and 2.66$^{\circ}$ for Freon 12.

The 8 wedges in the large $\theta$ cell are each equipped with one plane and one spherical mirror to reflect \v Cerenkov light into a photomultiplier. The wedges in the small $\theta$ cell are each equipped with an ellipsoidal mirror which has as its two foci the interaction point and a photomuliplier for that wedge. The arrangement of the mirrors and photomultipliers for both cells are shown in Fig. 3.10. Both the small $\theta$ mirrors and the overall mechanical structure of the \v Cerenkov Counter are constructed of carbon fiber epoxy composites covered with a layer of plexiglass to improve their reflectance. This was done to minimize the overall weight of the structure and the amount of material traversed. The large $\theta$ mirrors, however, were made of glass. In these large $\theta$ cells light coming from the direction of the interaction region was reflected twice, first by the spherical mirror and then onto the plane mirror and from there into the photomultiplier. In both the geometry of the large $\theta$ and small $\theta$ cells, light coming from the direction of the interaction region, which is approximately a cube with side 5 mm, is focused by the mirrors onto the photomultiplier windows. 

The \v Cerenkov Counter mirrors are coated with a layer of aluminum which is approximately 100 nm thick, and are furthermore protected against oxidation by thin films which are MgF$_2$ on the ellipsoidal mirrors and the plane mirrors, and SiO$_2$ on the spherical mirrors. The thickness of the carbon-fiber/plexiglass mirrors was 4.3 mm; and each had a weight of 2 kg. The spherical glass mirrors each had a thickness of 3.0 mm and a weight of $\sim$ 1 kg. The plane glass mirrors were 1 mm thick. The \v Cerenkov Counter had an overall light collection efficiency of 0.84-0.90 for the CO$_2$ chamber, and an efficiency of 0.84-0.98 for the large $\theta$ chamber filled with Freon 13. For Freon 12 this efficiency dropped to the range 0.75-0.88. The reflectance of these mirrors was found to be approximately constant for wavelengths of light ranging from $\sim$200-300 nm.

The signals from the \v Cerenkov photomultipliers were directly amplified by a factor of approximately 10. These outputs were split in two, with one half sent to the trigger logic, and the other to an ADC to record the integrated charge. A signal from at least one of the \v Cerenkov wedges to the trigger is required for an event to be tagged as an electron candidate. The efficiency of this tagging was done by examining samples of the reactions $p\bar p \rightarrow J/\psi \rightarrow e^{+}e^{-}$ and $p\bar p \rightarrow J/\psi \rightarrow e^{+}e^{-} \gamma$ and comparing the numbers of events with at least one electron tagged (1e) with the number of events with both electrons tagged (2e). From the ratio $r = 2e/1e$ between these, the detection probability for a single electron is given by $\epsilon = 2r/(1+r)$ which was found to be $(98.1 \pm 0.5)\%$ for both the direct $J/\psi$ decay and the $\chi_{c2}$ decay.

The average number of photoelectrons per electron track was also measured using data from the reactions $p\bar p \rightarrow J/\psi \rightarrow e^{+}e^{-}$ and $p\bar p \rightarrow \chi_{c2} \rightarrow J/\psi \gamma \rightarrow e^{+}e^{-} \gamma$. The \v Cerenkov signal charge distribution was used to estimate the light yield. The average charge may be converted to an average number of photoelectrons provided one knows the photomultiplier gain, or the number of ADC counts per photoelectron. This was determined by measuring the pedestal and dark current separation for each photomultiplier; the separation between the dark current peak and the pedestal corresponds to one photoelectron. For the small $\theta$ cell the average number of photoelectrons per electron track was found to be dependent on the polar angle $\theta$ of the electron, ranging from $19.8 \pm 2.3$ photoelectrons for electron tracks in the $\theta = 15^{\circ}-20^{\circ}$ region, to $11.8 \pm 1.1$ photoelectrons for tracks in the $\theta = 32^{\circ}-34^{\circ}$ region. For the large $\theta$ cell, the average number of photoelectrons per track showed a much lower variation in polar angle. It ranged from $8.2 \pm 1.6$ in the $\theta = 34^{\circ}-38^{\circ}$ region to $6.8 \pm 1.1$ in the $\theta = 58^{\circ}-65^{\circ}$ region. A fuller description of the \v Cerenkov detector may be found in Refs.~\cite{cerenkov}~\cite{cerenkov2}.

\subsection{The Central Calorimeter}

The Central Calorimeter (CCAL) is one of the most important elements in the E835 detector system. It is composed of a cylindrical array of 1280 lead-glass \v Cerenkov counters, and is used to measure the energy and position of photons, electrons and positrons. It has full coverage in the azimuthal angle $\phi$ and coverage in the polar angle $\theta$ between 10.6$^{\circ}$ and 70.0$^{\circ}$. The main features of the detector are described below; more details may be found in a number of refereneces~\cite{bartoszek}~\cite{gollwitzer}~\cite{paolo}~\cite{michelle}. 

The 1280 lead-glass blocks which make up the CCAL are shaped and oriented to form a projective geometry in which each block is pointed towards the interaction region. This cylindrical array of blocks may be grouped into rings and wedges. A CCAL ring is made up of 64 blocks positioned at the same polar angle $\theta$, while each of the 64 CCAL wedges covers 5.625$^{\circ}$ in azimuth, and is made up of 20 blocks positioned at the same azimuthal angle $\phi$. A diagram of the CCAL showing its ring-and-wedge structure may been seen in Figure 3.11. Each CCAL wedge is contained in a light-weight stainless steel container with exterior surfaces which are 1.735 mm thick, while the separators between counters between each block in the wedge are 0.254 mm thick stainless steel. The thicker material separating the wedges leads to 2\% of the surface being inactive in the azimuthal direction; in the polar direction it is only 0.5\% due to the thinner separations. A detailed description of the orientation of the CCAL blocks may be found in Table 3.2 ~\cite{paolo}~\cite{michelle}.

\begin{figure}[htb]
\begin{center}
\includegraphics[height=6cm]
{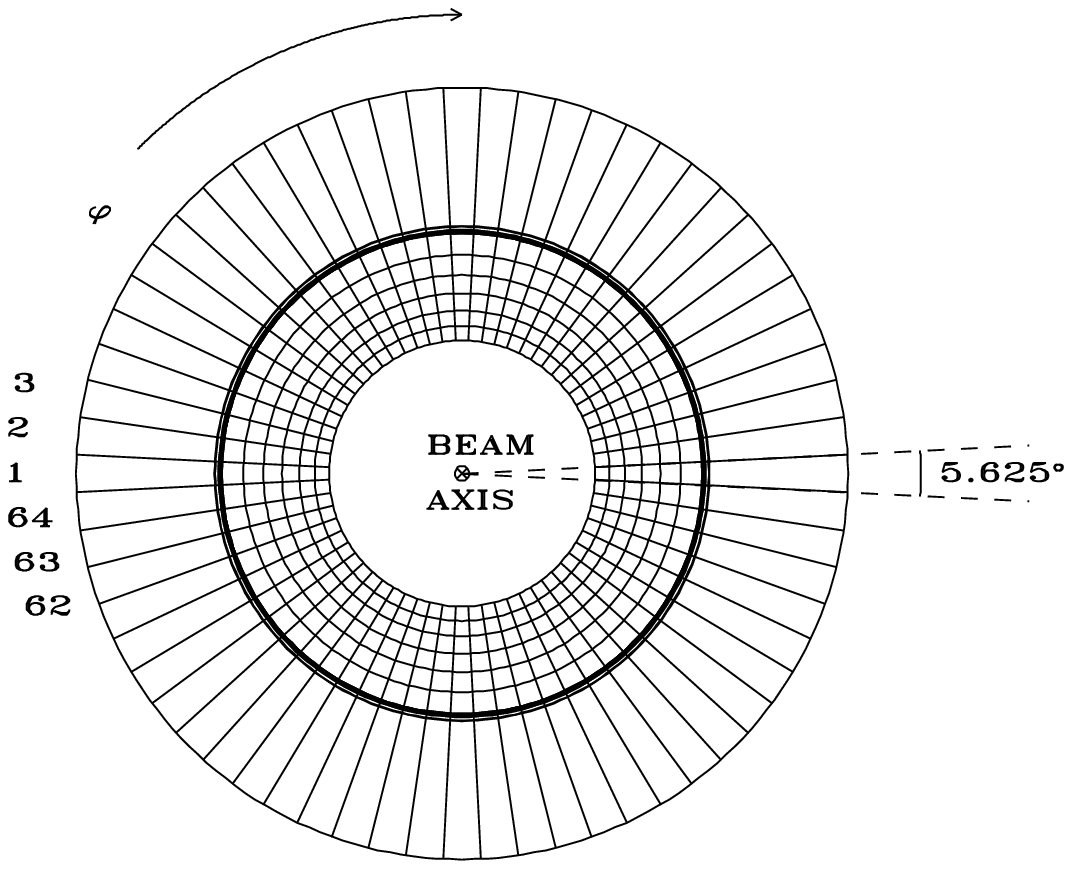}
\includegraphics[height=6cm]
{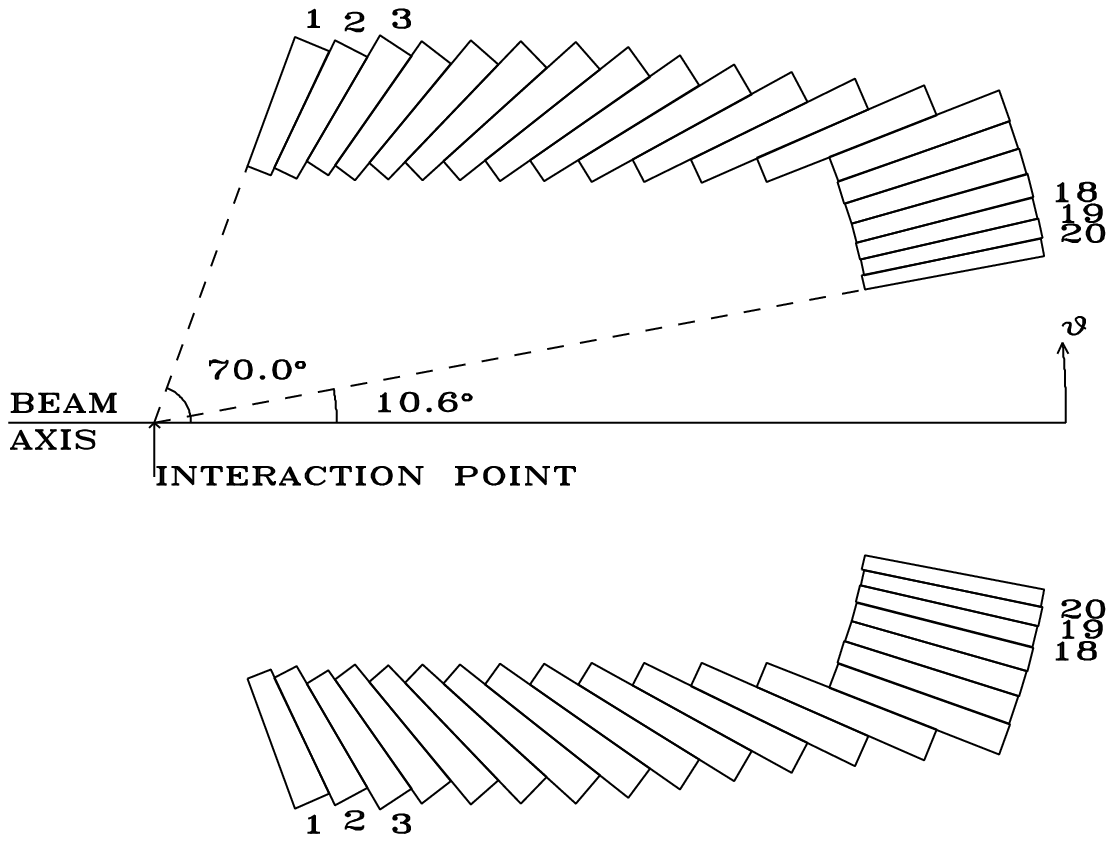}
\caption{View of a ring (left) and two opposite wedges (right) of the central 
calorimeter.}
\label{fig:ccal_ring}
\end{center}
\end{figure}

\linespread{2.4}
\begin{table}[htbp]
\begin{center}
\begin{tabular}{|ccccc|}
\hline\hline
 & & & & \\
       &     	    &  ~~Central~~   	 &                  & Distance from  	   \\
 & & & & \\
~Ring~~& ~~Length~~ &  $\theta$          &~~$\Delta\theta$~~& Interaction Region  \\
 & & & & \\
       &  (cm)	    &  (degree)	         &  (degree)        & (cm)		   \\
 & & & & \\
\hline\hline
 & & & & \\
 1     & 37.80      &  67.387		 & 5.226            &  72.44	    \\
 & & & & \\
 2     & 38.65      &  62.259		 & 5.031            &  75.87	    \\
 & & & & \\
 3     & 39.88      &  57.342		 & 4.803            &  80.07	    \\
 & & & & \\
 4     & 41.50      &  52.664		 & 4.552            &  85.08	    \\
 & & & & \\
 5     & 43.54      &  48.246		 & 4.284            &  90.96	    \\
 & & & & \\
 6     & 46.06      &  44.101		 & 4.007            &  97.79	    \\
 & & & & \\
 7     & 48.98      &  40.234		 & 3.728            &  105.62	    \\
 & & & & \\
 8     & 50.00      &  36.644		 & 3.451            &  114.54	    \\
 & & & & \\
 9     & 50.00      &  33.327		 & 3.183            &  124.66	    \\
 & & & & \\
 10    & 50.00      &  30.273		 & 2.925            &  136.07	    \\
 & & & & \\
 11    & 50.00      &  27.472		 & 2.679            &  148.89	    \\
 & & & & \\
 12    & 50.00      &  24.908		 & 2.449            &  163.26	    \\
 & & & & \\
 13    & 50.00      &  22.567		 & 2.233            &  179.34	    \\
 & & & & \\
 14    & 50.00      &  20.434		 & 2.033            &  197.28	    \\
 & & & & \\
 15    & 50.00      &  18.493		 & 1.848            &  197.29	    \\
 & & & & \\
 16    & 50.00      &  16.730		 & 1.678            &  197.29	    \\
 & & & & \\
 17    & 50.00      &  15.130		 & 1.552            &  197.30	    \\
 & & & & \\
 18    & 50.00      &  13.679		 & 1.380            &  197.30	    \\
 & & & & \\
 19    & 50.00      &  12.364		 & 1.250            &  197.30	    \\
 & & & & \\
 20    & 50.00      &  11.174		 & 1.131            &  197.30	    \\
 & & & & \\
\hline\hline
\end{tabular}
\caption[Geometrical characteristics of the CCAL blocks.]{Geometrical characteristics of the CCAL blocks. The 64 blocks contained in a ring share the same features.}
\label{tab:blocks}
\end{center}
\end{table}

The CCAL blocks are made of a Schott F2 type lead glass. They have an index of refraction of 1.651 for light with wavelength 404.7 nm, and a density of 3.61 g/cm$^{3}$. Their composition by weight is 42.2\% lead, 29.5\% oxygen, 21.4\% silicon, 4.2\% potassium, 2.3\% sodium, and 0.15\% arsenic. The glass has a radiation length of 3.141 cm, and its transmittance through 10 cm varies by wavelength from 95.5\% for light in the 385-394 nm region to 99.4\% for light in the 585-594 nm region. For light with smaller wavelengths the transmittance is significantly lower; it is 56.9\% for the 335-344 nm wavelength region.

A Hammamatsu photomultiplier was glued to the end of each block to collect the \v Cerenkov light produced by electromagnetic showers. These photomultipliers were chosen for efficient photoelectron collection rather than fast timing, and also for their relative insensitivity to magnetic fields. Photomultipliers with four different diameters were used in different rings of the CCAL in order to maximize the efficiency of light collection from the different sizes of blocks. 

The output of the photomultipliers is first sent to a summer box, where 5\% of the signal from each photomultiplier was diverted to the CCAL summer circuit to be used for triggering purposes. The remaining signal was sent over a 310 ns coaxial RG-58 delay cable to an electronic board that reshapes the signal to fit into a 100 ns gate for ADC integration. The long cable allows the trigger electronics time to make a decision about the signal. The reshaping is required in part because the long cable disperses and attenuates the pulse tail to over 600 ns. This was done by means of the Splitter-Shaper-Discriminator Circuit (SSD), or simply, the shaper, whose schematic is presented in Figure 3.12. There are three versions of the shaper circuit for rings 1-16, rings 17-18, and rings 19-20. By narrowing the pulses, the charge may be collected in the 100 ns ADC gate. After ADC integration, a portion of the reshaped signal is sent to a discriminator, followed by a TDC. The short ADC gate and the TDC time information are needed to reduce the event contamination due to pile-up. 

\begin{figure}[htbp]
\begin{center}
\includegraphics[width=5in]{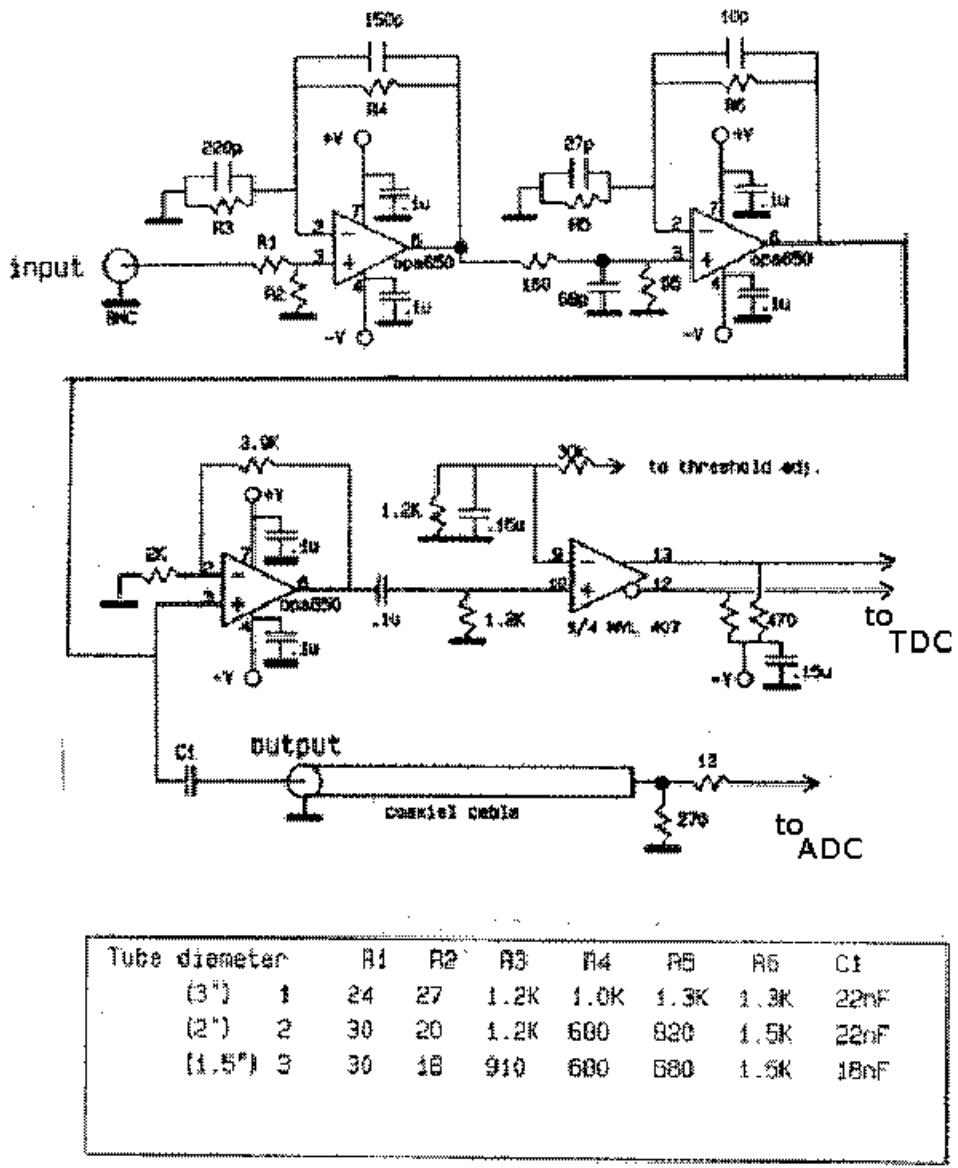}
\caption{\baselineskip=18pt
Schematic diagram of an SSD channel for the shaping of the 
input signal from a CCAL block.\label{circuit}}
\end{center}
\end{figure}

As photons and electrons entering the CCAL are likely to produce a signal in more than one block, signals from the CCAL are analyzed by a software procedure called a clusterizer, the purpose of which is to determine the energy and position of a photon or electron that generates a signal in a group of neighboring blocks known as a cluster. The clusterizer first searches for cluster seeds, whcih are defined as groups of 3 $\times$ 3 CCAL blocks where the central block has an energy deposit which is greater than the deposit in any of its eight neighbors. The central energy deposit must also exceed a set seed threshold, generally 25 MeV. The sum of all energy deposits in the 3 $\times$ 3 grid is called the cluster energy, and this must also exceed a set threshold to be considered a cluster, generally 50 MeV. These thresholds may be lowered to 5 and 20 MeV for the seed and cluster thresholds respectively for analyses which are particularly dependent on the detection of low energy photons. Clusters may be grouped into three categories: isolated, shared, and split. 

\subsubsection{Isolated Clusters}

Isolated clusters are defined as those which do not have a second seed block within a 5 $\times$ 5 grid of blocks around the original seed. For an isolated cluster, the position of the incoming particle is calculated as the energy-weighted average of the centers of the blocks in the 3 $\times$ 3 cluster grid. Defining $(x,y)$ as the cluster coordinates in units of blocks, in the ($\theta$(wedge),$\phi$(ring)) directions, where the origin is the center of the seed block, we have:

\begin{equation}
x=\Sigma^9_{i=1}f_iE_ix_i/\Sigma^9_{i=1}f_iE_i
\end{equation}

\begin{equation}
y=\Sigma^9_{i=1}f_iE_iy_i/\Sigma^9_{i=1}f_iE_i
\end{equation}

where $E_i$ is the energy of the $i$th block, $f_i$ is the fraction of the energy of the $i$th block taken by the $j$th cluster, and $x_i = -1,0,1$ and $y_i = -1,0,1$ are the coordinates of the centers of the nine blocks of the cluster in block units.

In order to take into account the energy which is lost by the cracks between the blocks, a correction then is made which takes into account block and crack positions, and the electromagnetic shower profile, which is described by a function whose parameters were measured with a test beam at the Brookhaven National Laboratory. The correction to the shower position is computed as follows:

\begin{equation}
x^{\prime} = N_x[A_wa_w(1-e^{-x/a_w}) + B_wb_w(1-e^{-x/b_w})]
\end{equation}

\begin{equation}
N_x = 0.5/[A_wa_w(1-e^{-0.5/a_w}) + B_wb_w(1-e^{-0.5/b_w})]
\end{equation}

\begin{equation}
y^{\prime} = N_y[A_ra_r(1-e^{-y/a_r}) + B_rb_r(1-e^{-y/b_r})]
\end{equation}

\begin{equation}
N_y = 0.5/[A_ra_r(1-e^{-0.5/a_r}) + B_rb_r(1-e^{-0.5/b_r})]
\end{equation}

where $x^{\prime} (y^{\prime})$ is the distance in the wedge (ring) direction, in block units, from the center of the seed block. The constants $A_w$, $A_r$, $a_w$, $a_r$, $b_w$, and $b_r$ were measured empirically with $e^{+}e^{-}$ decays of the $J/\psi$ resonance and are listed in Table 3.3. The effectiveness of this correction can be seen in Fig. 3.13.

\linespread{2.4}
\begin{table}[htb]
\begin{center}
\begin{tabular}{cccc}
\hline
 & & & \\
$A_r$ & 724.4 & $a_r$ & 0.03208 \\
 & & & \\
$A_w$ & 706.5 & $a_w$ & 0.03969 \\
 & & & \\
$B_r$ & 123.6 & $b_r$ & 0.1860 \\
 & & & \\
$B_w$ & 102.6 & $b_w$ & 0.1715 \\
 & & & \\
$C_l$ & 0.0614 & $c_l$ & 7.367 \\
 & & & \\
$C_h$ & 0.0857 & $c_h$ & 19.690 \\
 & & & \\
$D_1$ & 0.14736 & $d_1$ & 48.908 \\
 & & & \\
$D_2$ & 0.15935 & $d_2$ & 12.761 \\
 & & & \\
\hline
\end{tabular}
\caption{Constants used in calculating the position and energy of CCAL blocks.}
\label{tab:CCALcorr}
\end{center}
\end{table}

The corrected energy of the cluster is given by:

\begin{equation}
E_{corr} - E_{meas}/[(1-C_{h(l)}e^{-\vert x^{*} \vert/c_{h(l)}})\times(1-D_1e^{-\vert y^{*} \vert/d_1} - D_2e^{-\vert y^{*} \vert/d_2})]
\end{equation}

where $x^{*}, y^{*}$ are the distances from the cluster center to the closest edges of the seed block and the constants are listed in Table 3.3. The ring faces are staggered, and $c_h(c_l)$ corresponds to the high(low) $\theta$ block edge. The effectiveness of this correction can be seen in Fig. 3.13, where the dip in the ratio of measured to expected energies in the region around a CCAL crack is removed and there are fewer misidentified events after the correction is applied.

\begin{figure}[htbp]
\begin{center}
\includegraphics[height=12cm]
{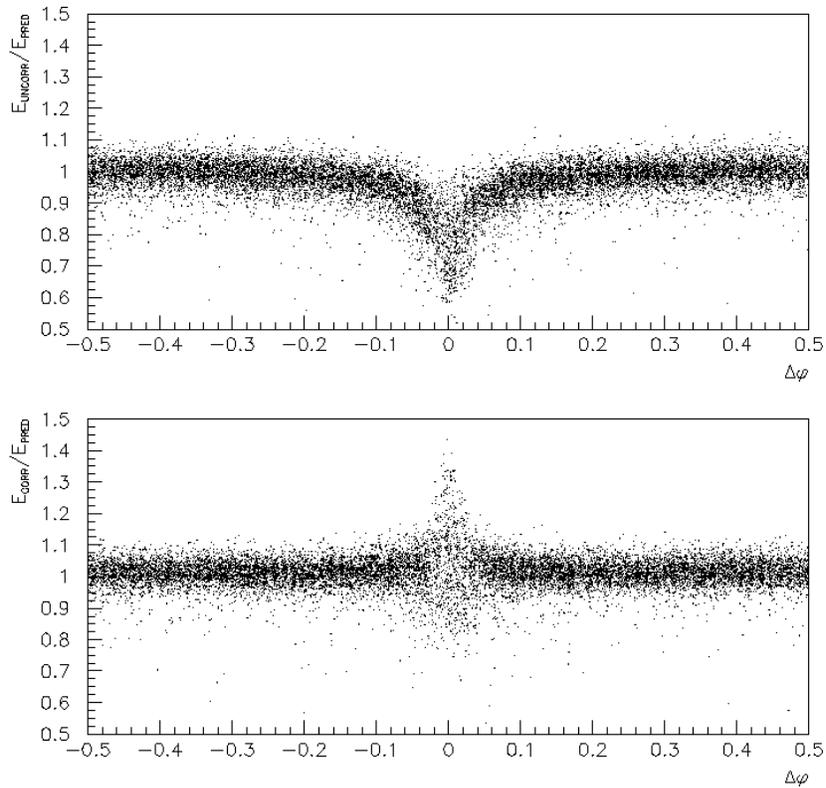}
\caption{Ratio between the measured energy of the cluster and the energy predicted by the
2 body kinematics in $J/\psi \rightarrow e^{+}e^{-}$ events, before (top) and after (bottom) energy correction, as a
function of the distance from a CCAL crack.}
\label{fig:crackcorrection}
\end{center}
\end{figure}

\subsubsection{Shared Clusters}

Shared clusters occur when a second seed is found in a 5 $\times$ 5 grid centered on the seed of the original cluster. When this occurs, the clusterizer assumes that two particles have impacted near each other in the CCAL, and it generates two clusters, with one 3 $\times$ 3 cluster around each seed. As some of the blocks in each 3 $\times$ 3 cluster will overlap, the energy of those blocks which belong to both clusters is shared between the two. The amount of energy which is assigned to each cluster from the shared blocks is determined by an iterative process which includes a method for correcting the position and the energy of the two clusters. This method is an extension of the position and energy determination algorithm described above for isolated clusters.

\begin{figure}[htb]
\begin{center}
\includegraphics[height=8cm]
{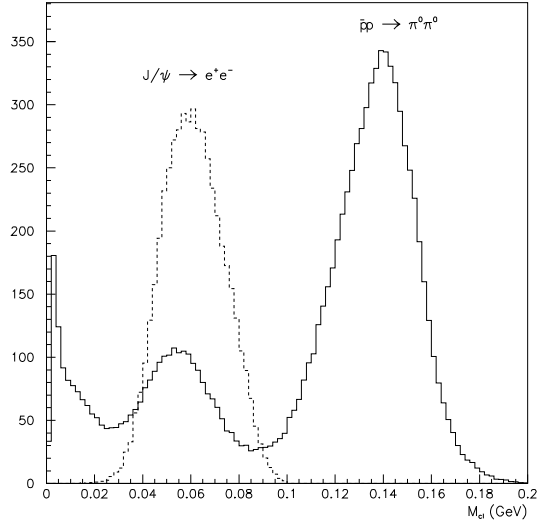}
\caption[Distribution of the cluster mass for electrons and positrons using data from 
$J/\psi$~decays (dashed),
and from photons from a $\pi^0\pi^0$ sample (solid).]{Distribution of the cluster mass for electrons and positrons using data from 
$J/\psi$~decays (dashed),
and from photons from a $\pi^0\pi^0$ sample (solid).
The large peak at $~M_{cl}>100~$MeV~ is due to coalesced photons
from the symmetric decay of the $\pi^0$.}
\label{fig:clustermass}
\end{center}
\end{figure}

\begin{figure}[htbp]
\begin{center}
\includegraphics[height=12cm]
{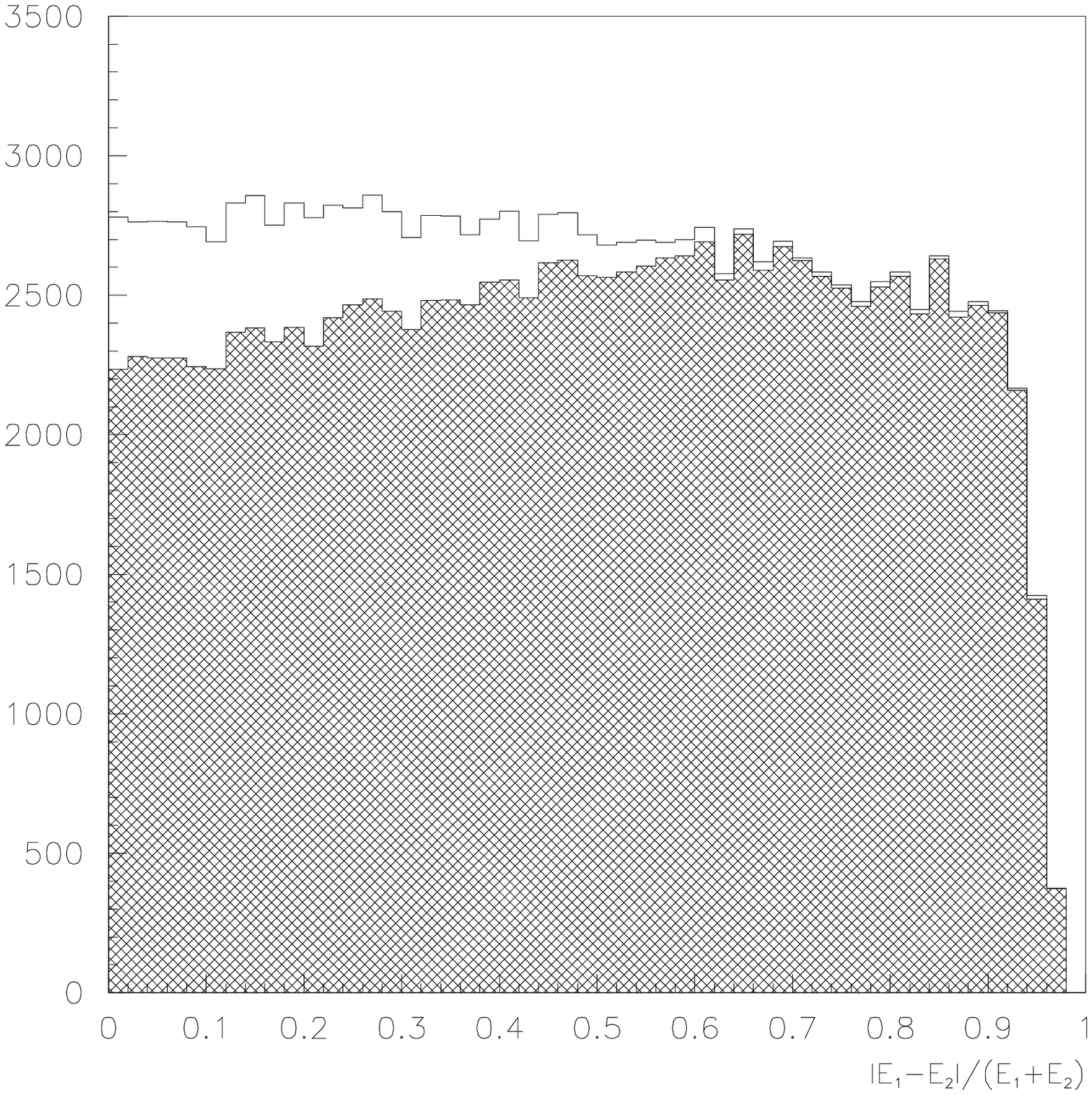}
\caption[Distribution of the $\pi^0$ asymmetry (data), 
defined as $~(E_1-E_2)/(E_1+E_2),~$ 
where $E_1$ and $E_2$ are the energies of the two photons.]{Distribution of the $\pi^0$ asymmetry (data), 
defined as $~(E_1-E_2)/(E_1+E_2),~$ 
where $E_1$ and $E_2$ are the energies of the two photons. 
The distribution is plotted without the split clusters (shaded) and including the split clusters (open). The distributions are expected to be uniform.}
\label{fig:asymmetry}
\end{center}
\end{figure}

There is also the possibility that a cluster may be due to two adjacent showers but may not have a second seed in the neighboring blocks. We then have to examine other features indicating that it has likely been formed by two different showers so close to each other so as not to create two distinguishable seeds. One such case occurs when a neutral pion decays into two photons of approximately equal energies in the laboratory frame, also known as symmetric decay, in which case the opening angle between the two photons will be a minimum. This minimum opening angle can be as low as $\sim$ 1.5 block widths at the upper energy range for pions which can be produced in E835. As this separation is too small to produce shared clusters a different clusterizing procedure must be used. This involves the cluster mass, which is defined as

\begin{equation}
\label{eq:clustermass}
M_{cl}\equiv \sqrt{\Big( \sum_{i=1}^{25}E_i \Big)^2 
           - \Big( \sum_{i=1}^{25}\vec{p}_i \Big)^2}~,
\end{equation}

where the sum runs over the 25 blocks of the $5\times5$ grid 
around the seed, $E_i$ is the energy deposited in the i-th block, and 
$\vec{p}_i$ is equal to $E_i\hat{r}_i$, $\hat{r}_i$ being the unit vector from
the interaction region to to the center of the i-th block. This gives a measure of the invariant mass of the cluster, which should be minimal (ideally $\le 0.5$ MeV) for clusters produced by single photons or electrons. By examining electron and positron showers from $J/\psi \rightarrow e^{+}e^{-}$ decays, it can be shown that the cluster mass of a single electron shower in the CCAL does not exceed $\sim$ 100 MeV, and the same is generally true for single photon showers. This is illustrated in Fig. 3.14, where the cluster mass distributions of electron clusters from the reaction $J/\psi \rightarrow e^{+}e^{-}$ are compared with cluster mass distributions from neutral pions from the reaction $p\bar p \rightarrow \pi^0\pi^0$, where the higher cluster mass (peak of $\sim$ 140 MeV) from symmetric pion decay is clearly observable. 

These split clusters have seeds which are too close together to be resolved with the shared cluster algorithm, but use instead a slightly different method in which the second seed is chosen as the block with the largest deposit among the four blocks at the corner of the original seed. The clusterizer then proceeds to calculate a total energy and position for each block using a similar technique to that used for shared clusters. Figure 3.15 shows the effectiveness of the split cluster algorithm. The $\pi^0$ asymmetry, defined as $(E_1 - E_2)/(E_1 + E_2)$, is expected to be uniform in the data over the range of the asymmetry from 0 to 1. As can be seen from the figure, however, the unformity of this distribution is recovered only by including the shared clusters. 

By taking into account cluster shape and correcting for cracks and split/shared clusters, the CCAL provides an angular resolution of $\sigma_{\theta} \approx$ 6 mrad in the polar angle $\theta$ and $\sigma_{\phi} \approx$ 11 mrad in the azimuthal angle $\phi$. The angular resolution is measured using a clean sample of electrons and positrons from $J/\psi \rightarrow e^{+}e^{-}$ decays, which are selected using the hodoscopes and the \v Cerenkov counter. The energy resolution is measured using a similar $J/\psi \rightarrow e^{+}e^{-}$ sample, where one can predict the energy of each electron from its direction. The rms energy resolution is found to be:

\begin{equation}
\label{eq:ccal_energyres}
\frac{\sigma_E(E)}{E}=\frac{6\%}{\sqrt{E(\mathrm{GeV})}}+1.4\%~.
\end{equation}

The CCAL is calibrated with both $e^{+}e^{-}$ events from $J/\psi$ and $\psi^{\prime}$ decays, and with $\pi^0\pi^0$ events. $\pi^0\pi^0$ calibration is generally superior because the yield of $\pi^0\pi^0$ is copious for all $p\bar p$ center of mass energies unlike $e^{+}e^{-}$ which is only copious at $J/\psi$ and $\psi^{\prime}$ energies, and also because the gammas from these decays populate all counters of the E835 detector. 

Using $\pi^0\pi^0$ events, a clean sample is selected for each stack using the following criteria: given four photon $\pi^0\pi^0$ events the combinations of two gammas to form each $\pi^0$ are taken by minimizing the value of $\sqrt{(2\Delta\theta)^2 + (\Delta\phi)^2}$, where $\Delta\phi \equiv \pi - \vert\phi_1 - \phi_2\vert$ and $\Delta\theta \equiv \theta_{1 pred} - \theta_{1 meas}$, where $\theta_{1 meas} \le \theta_{2 meas}$ and $\theta_{1 pred}$ is obtained from $\theta_{2 meas}$ by kinematics. Here $\theta$ and $\phi$ are the laboratory frame polar and azimuthal angles. Once these two two-photon combinations have been chosen to make up the two $\pi^0$s in the sample, further criteria are applied. These include:

$\bullet$ Requiring minimum energy and seed thresholds for all clusters. For energy and vertex calibration these must be 50 and 25 MeV respectively, while for timing calibration they must be 20 and 5 MeV.

$\bullet$ Requiring all four photons to be completely detected within the CCAL.

$\bullet$ Allowing a maximum $\Delta\phi$ (also called acoplanarity) accepted for both $\pi^0$s. This maximum $\Delta\phi$ is 10 mrad for energy and vertex calibration and 15 mrad for timing calibration.

$\bullet$ Allowing a maximum $\Delta\theta$ (also called akinematics) accepted for both $\pi^0$s. This maximum $\Delta\theta$ is 30 mrad for energy and vertex calibration, and 32 mrad for timing calibration.

$\bullet$ Accepting only a certain range of invariant masses, which for $\pi^0 \rightarrow 2\gamma$ decays may be expressed simply as $M_{\pi} = (2E_{\gamma1}E_{\gamma2}(1 - cos\theta_{open}))^{1/2}$ where $\theta_{open}$ is the angle between the two photons in the lab frame, and $E_{\gamma1}, E_{\gamma2}$ are their energies in the lab frame. For energy calibration $\vert M_{\pi} - 135$ MeV $\vert < 40$ is required, while for vertex and timing calibration $\vert M_{\pi}$ - 135 MeV $\vert < 35$ is required

$\bullet$ Accepting only events with exactly four CCAL clusters.

Since the event selection depends weakly on energy calibration, it is repeated after the energy calibration, and then recalculated.

The CCAL energy calibration is done by finding the gain constant $g_i$ of each one of the 1280 CCAL channels. The gain constant is defined as the ratio between the energy deposited in the block, and the value determined by the ADC. The gain constants are determined by the following procedure.

The predicted energy of each $\pi^0$, $E_j$, is calculated from its polar angle using two-body kinematics. This is then comapred with the measured energy, $M_{j}$\,, of each one of the two $\pi^0$'s, which is given by 
the sum of the energy deposited in all the blocks belonging to the clusters 
of the two decay photons:

\begin{equation}
\label{eq:ccal_pi0energy}
M_{j}=\sum_{i=1}^n~g_i~A_{ij}~,
\end{equation}

where $j$ is the index of one of the two pions, $i$ runs over the blocks of the
$j$-th pion, and $A_{ij}$ is the ADC value recorded for every involved block. 
Using $E_{j}~$, the energy calculated by using the measured polar angle of the 
pion, a $\chi^2$ is defined as

\begin{equation}
\label{eq:ccal_chisq}
\chi^2=\sum_{j=1}^{N}\frac{(M_{j}-E_{j})^2}{\sigma_j^2}~,
\end{equation}

where $N$ is the number of $\pi^0$'s and $\sigma_j$ the estimated uncertainty on
$E_j$. The values of the gain constant that minimize the $\chi^2$, calculated
analytically, are inserted back in Equation \ref{eq:ccal_chisq}. The procedure
is iterated until convergence is reached. The gain constants $g_k$ that minimize the $\chi^2$ are given by:

\begin{equation}
g_k = [\Sigma^N_{j=1}(A_{kj}/\sigma^2_j)(E_j - \Sigma^n_{i=1,j\ne k}A_{ij}g_i)]/[\Sigma^N_{j=1}(A^2_{kj}/\sigma^2_j)]
\end{equation}

For the thin, marginal ring 20 which cannot be calibrated with the $\pi^0\pi^0$ sample 
a punch-through method is used instead. It takes into account minimum ionizing particles that do not shower and release \v Cerenkov light in a single block, proportionally to the block length.

Once the above CCAL energy calibration is completed, the measured mass of the $\eta$ is compared with the actual mass. A scale correction, typically 1-2\%, is then applied to all gain constants to match the measured and actual values. This scale correction is found to be $2\%$ for electron clusters by examining the mass of the $J/\psi$ and comparing it to its expected value. A similar $1\%$ correction for gamma clusters is found using a GEANT Monte Carlo simulation examining the mass of the $\eta_c$.

Timing information is added to every CCAL cluster by considering the TDC channels (used in common stop mode) of the two blocks of the cluster with the highest ADC counts. If either of these channels has recorded a hit which is within 10 ns from the mean event time, then the cluster is classified as in-time. If the recorded hits are not within 10 ns from the mean event time, then the cluster is classified as out-of-time. If neither TDC channels recorded a hit at all, then the cluster is classified as undetermined.

The timing algorithm is tested with a $\pi^0\pi^0$ sample from which split clusters have been excluded, and for which the cluster energy threshold is set to 20 MeV, rather than 50 MeV, in order to examine the algorithm response at low as well as high energies. The number of clusters for which time information is determined rises sharply as the cluster energy increases from 20 to 50 MeV, since the higher cluster energies make it more likely that the TDC hit threshold is exceeded. The fraction $f$ of ADC signals with a corresponding TDC signal can be expressed as a function of ADC counts as:

\begin{equation}
f = 0.5 \times (erf(slope \times (ADC - thresh)) + 1.0), \quad erf(x) = {2 \over \sqrt{\pi}} \int^x_0 e^{-u^2} du
\end{equation}

where $slope$ and $thresh$ are empirically determined parameters. The result of the timing algorithm is that over 99.5\% of the $\pi^0\pi^0$ clusters with cluster energy above 50 MeV are found to be in-time, and the root mean square deviation in time for in-time clusters is found to be 1.3 ns.

When all of the pulses associated with a cluster are below discriminator threshold, a pulse from a prior interaction will cause it to be mistakenly identified as out-of-time. Thus the frequency of out-of-time clusters increases with instantaneous luminosity, and this loss of CCAL efficiency must be corrected for later during data analysis (see Sec. 3.4.3).

For monitoring and testing purposes, a network of plastic-polymer optical fibers was installed to transmit light from a nitrogen laser to the back of each lead glass block. The laser produces 3 ns wide pulses with wavelength 337.1 nm which are directed to a scintillator that produces light at $\sim$ 430 nm, a wavelength which is transmitted efficiently by the fibers. This scintillation light is passed to a rectangular lucite mixing bar and distributed to optical fibers coupled to the lucite which are directed to each wedge of the CCAL. Within the wedge container, the light encounters a secondary mixing bar which further distributes it to 20 more fibers, one coupled to the back of each counter. A light-tight aluminum box holds the laser, the scintillator, the major mixing bar, and two PIN diodes used to measure the intensity of the laser pulse. This system was used throughout the setting up and running of E835 to test the CCAL counters and to monitor their gain.

\newpage

\subsection{The Forward Calorimeter}

A second electromagnetic shower calorimeter known as the forward calorimeter, or FCAL, covers the forward region around the beam pipe. It has full coverage in the azimuthal angle $\phi$ and coverage in the polar angle $\theta$ between 3$^{\circ}$ and 11$^{\circ}$. 

The FCAL is made up of an array of 144 rectangular SF2 lead-glass blocks of three different sizes arranged around the beam pipe as shown in Figure 3.16. The SF2 lead-glass used in the FCAL has an index of refraction of 1.673, and a radiation length of 2.76 cm. The blocks vary in thickness from 13-21 radiation lengths. Each FCAL block is glued to a photomultiplier, which is then connected to electronic shaper boards similar to those used by the CCAL. The FCAL signals are sent to both ADCs and TDCs. The spatial resolution of the FCAL is approximately 2 cm. 

\begin{figure}[htb]
\begin{center}
\includegraphics[height=10cm]
{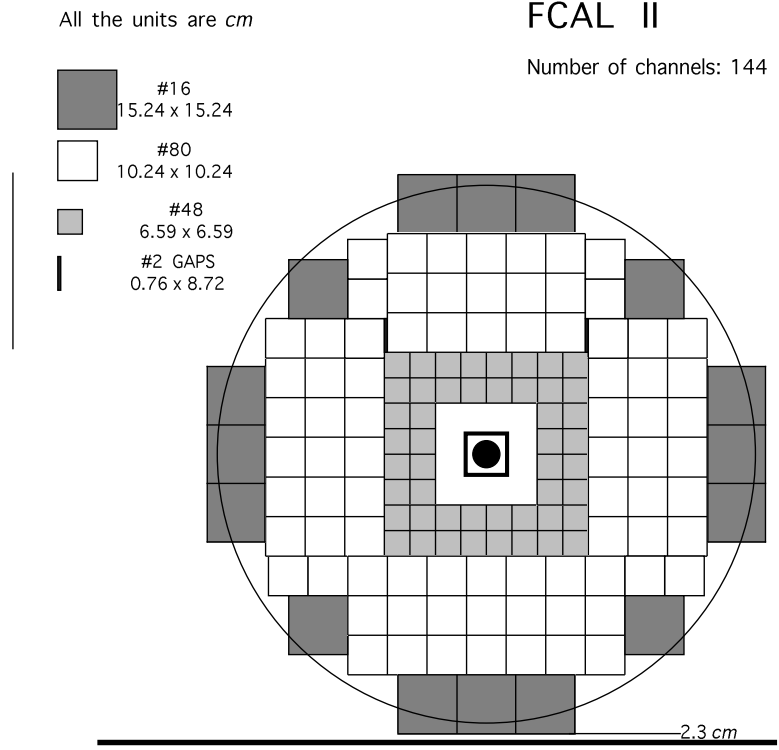}
\caption{Front face (perpendicular to the beam axis) of the forward calorimeter.}
\label{fig:fcal}
\end{center}
\end{figure}

To obtain an initial calibration for the FCAL, each block was placed in a muon beam downstream from the Fermilab MTEST area. For every block, the high voltage was adjusted so that the signals from through-going muons, measured using an LRS QVT and weighted by the length of each block, were equal.

After this initial calibration had been done, a sample of six-photon events from one of the reactions $p\bar p \rightarrow \pi^0\pi^0\pi^0 \rightarrow 6\gamma$, $p\bar p \rightarrow \pi^0\pi^0\eta \rightarrow 6\gamma$, or $p\bar p \rightarrow \pi^0\eta\eta \rightarrow 6\gamma$, taken at the center of mass energy of the $\chi_{c1}$ resonance, was used to calibrate the FCAL. For this calibration 5 of the 6 gammas were required to be detected in the CCAL, with the final gamma detected in the FCAL. Two pairs of CCAL clusters were required to form an invariant mass consistent with either a $\pi^0$ or an $\eta$, and the remaining CCAL cluster is paired with the single FCAL cluster. Without using the FCAL cluster energy, the event is fit using SQUAW to one of the hypotheses $\pi^0\pi^0\pi^0$, $\pi^0\pi^0\eta$, or $\pi^0\eta\eta$. The energy resolution of the FCAL, $\sigma_E(E)$ for a photon of energy E was found to be:

\begin{equation}
\frac{\sigma_E(E)}{E}=\frac{6\%}{\sqrt{E(\mathrm{GeV})}}+4\%~.
\end{equation}

For the analysis in this thesis, the FCAL is used primarily as a veto. The FCAL is described more fully in Ref.~\cite{fcal}.

\subsection{The Luminosity Monitor}

For an experiment such as E835, which measures excitation curves of resonances by varying the center of mass energy of the $p\bar p$ interaction and measuring the cross sections at each energy, it is crucial to get an accurate measure of the luminosity. To take full advantage of the excellent beam energy resolution afforded by the Antiproton Accumulator, it is necessary that these luminosities be measured with an accuracy in the relative luminosity better than 1\%, and an accuracy in absolute luminosity with an uncertainty of $\sim$2-3\%.
This is done with the E835 luminosity monitor, which was built
and designed by our Northwestern research group.  A full description
of the previous E760 version of the luminosity monitor is found in Ref.~\cite{lummon}. The present version is described in Ref.~\cite{nimpaper}. 

The luminosity monitor is based on the detection of proton recoils from low momentum transfer
($t \approx  0.0077$-$0.0106$ (GeV/c)$^2$) $p\bar p$ 
elastic scattering events. This typically corresponds to a polar angle $\theta$ of the scattered $\bar p$ of $\sim 0.20^{\circ}$, and a polar angle $\theta$ of the scattered proton of $\sim 89.2^{\circ}$. At these tiny momentum transfers, the elastic scattering is overwhelmingly Coulombic, allowing accurate absolute normalization. The recoil proton energies for these scattering events is also small (for $\alpha \equiv 90^{\circ} - \theta \le 6^{\circ}, T(p) \le 15$ MeV for the highest $\bar p$ energies used in E835), so that they may be detected in solid state detectors of excellent energy resolution and stability.

The luminosity monitor is made up of three solid state detectors mounted in a steel vacuum vessel below the interaction point, at a polar angle  of $\theta_r\approx 86.5^{\circ}$.  
One of these detectors lies directly
beneath the beam axis, while the other two lie symmetrically 
on either side.  The central detector is movable in the direction of the $\bar p$ beam.

The solid angles of the detectors were precisely defined 
by a machined tungsten mask, with rectangular openings,
with each dimension known to 
$\pm 0.0001$ inch. The masks for the three
detectors had dimensions
\begin{eqnarray*}
\mbox{(beam left)}: 
0.3886'' 
&\times& 
0.7889''
\\
\mbox{(central)}: 
0.2752'' 
&\times& 
1.7713''
\\
\mbox{(beam right)}: 
0.3884'' 
&\times& 
0.7886''.
\\
\end{eqnarray*}
The detector areas were thus known to better than
$0.04\%$.
The mask specifications were confirmed 
with measurements using an $\alpha$ source.
The three detectors provided not
only for threefold redundancy in luminosity monitoring 
to safeguard against detector failure, but they 
also allow precise and continuous monitoring of 
horizontal displacements of the $\bar p$ beam
by a comparison of the event rates in the 
three detectors. A schematic of the luminosity monitor is shown in Figure 3.17, and its orientation with respect to the interaction region and attachment to the gas-jet target is shown in Figure 3.18. The luminosity monitor is attached to the gas-jet by means of a conical vacuum enclosure called the horn. This is suspended just below the interaction region perpendicularly to the beam. Attached to the bottom of the horn is a pan which holds the solid state detectors, the central one of which may be moved along the beam axis from the outside, and the left and right detectors which were fixed at angles $\alpha_L = 3.496 \pm 0.005^{\circ}$ and $\alpha_R = 3.511 \pm 0.005^{\circ}$ during the running of E835, where $\alpha \equiv 90^{\circ} - \theta$.

\begin{figure}[htb]
\begin{center}
\includegraphics[height=7cm]
{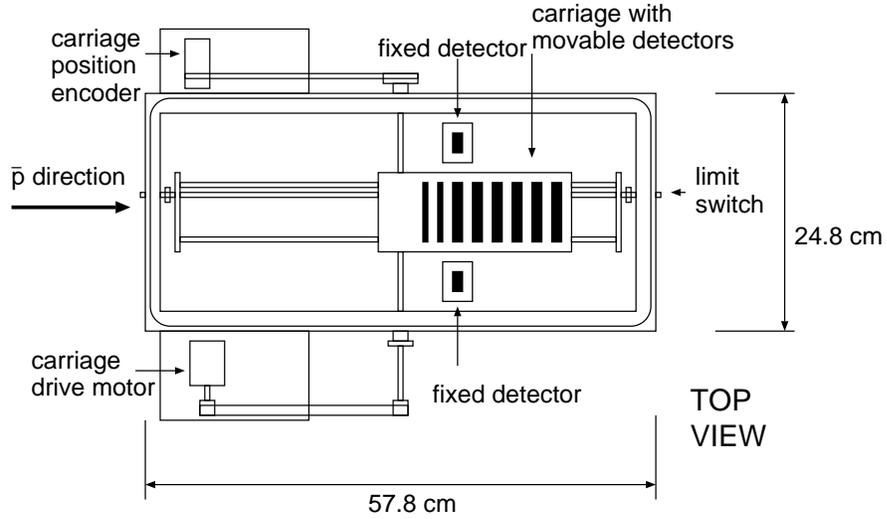}
\caption{Schematic of the E835 luminosity monitor.}
\label{fig:lum_mon}
\end{center}
\end{figure}

\begin{figure}[htb]
\begin{center}
\includegraphics[height=7cm]
{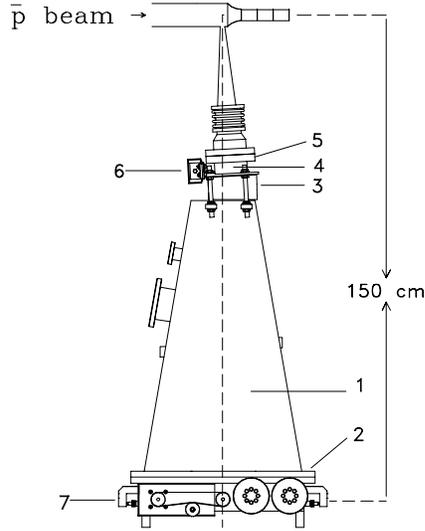}
\caption{The E835 luminosity monitor as connected to the gas-jet target.}
\label{fig:lum_mon_port}
\end{center}
\end{figure}

The luminosity monitor detectors are silicon surface barrier (500 $\mu$m deep) and Si-Li drift (3000 $\mu$m deep), with areas $\sim 1$ cm $\times 5$ cm, so that the area exposed by each mask is fully contained within each detector. 

The number of protons, N, passing through each one of the luminosity monitor detectors is given by

\begin{equation}
N=\mathcal{L} \int_{\Delta \Omega} \frac{d\sigma}{d\Omega}d\Omega~,
\end{equation}

where $\mathcal{L}$ is the luminosity, $\Delta\Omega$ is the solid
angle of acceptance, and $d\sigma/d\Omega$ is the differential cross section for $p\bar p$ elastic scattering.
The detectors' silicon surface barriers are sufficiently thick ($\ge$ 500 $\mu$m)
to completely stop the recoil protons, which have an energy of just a few MeV.

The absolute luminosity can be calculated by using the relation,
\begin{equation}
{\cal{L}} = \frac{N}{(d \sigma / dt)(dt / d \Omega)(d \Omega)}
\label{lumformula}
\end{equation} 
where N is the number of proton recoils counted, $(d \sigma / dt)$ 
is the known 
differential cross section(~\cite{trokenheim})
and $d\Omega$ is the solid angle subtended
by the detector.

The differential cross-section for the $p\bar p$ elastic scattering is given by:

\begin{equation}
d\sigma/dt = \vert F_c(t)e^{i\delta} + F_n(t) \vert^2
\end{equation}

The Coulomb amplitude $F_c(t)$ and phase $\delta$ are exactly calculable. The nuclear amplitude is given by the relation:

\begin{equation}
F_n(t) = \sigma_T(\rho + i)e^{-b\vert t\vert/2}/(4\sqrt{\pi\hbar})
\end{equation}

Thus, in order to determine absolute luminosity, three parameters must be determined. These are the total cross section $\sigma_T$, the 'slope' parameter $b$ of forward nuclear scattering, and the parameter $\rho \equiv$ Re$f(0)$/Im$f(0)$.

In its E760 version, the luminosity monitor was designed to also determine the parameters $\sigma_T$, $b$, and $\rho$ for $p\bar p$ elastic scattering in the energy region of interest. Absolute luminosity for all E835 measurements could then be obtained by using these parameters. The elastic scattering 
differential cross sections $d\sigma/d|t|$, 
versus squared momentum transfer, $|t|$, were
measured using the movable detector with better than $\pm 0.5\%$ precision in 
E760~\cite{lummon2}, the predecessor experiment to E835, for six values of the $\bar p$ beam momentum. These cross sections are shown in Fig. 3.19. Analysis of these cross-sections led to the determination of the $\sigma_T$, $b$, and $\rho$ parameters at various antiproton momenta. The cross sections shown were fitted with smooth curves, resulting in the following parametrization:

\begin{equation}
\sigma_T = 34.48 + 89.7p^{-0.70}, \quad b = 13.64 - 0.2p, \quad \rho = -0.12 + 0.03p
\end{equation}

where $\sigma_T$ is measured in mb, and $b$ in (GeV/c)$^{-2}$.

These parameters, which are valid for lab momentum $p$ from 2-8 GeV, are used to calculate $d\sigma/dt$ for any beam momentum at momentum transfer squared, $\vert t \vert$, corresponding to the acceptance of the luminosity monitor detector. Once $d\sigma/dt$ is known, the number of protons $N$ incident on the luminosity monitor leads to an absolute luminosity according to equation 3.41. 

\begin{figure}[htb]
\begin{center}
\includegraphics[height=10cm]
{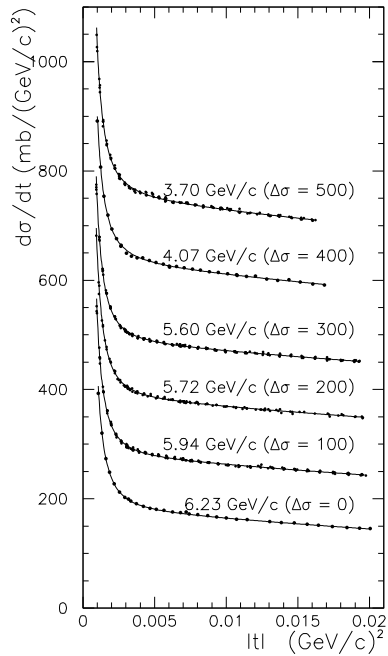}
\caption[Measured $p\bar p~$ differential cross section, 
$d\sigma/dt$, versus squared momentum transfer, $|t|$, 
at six values of the $\bar p$-beam momentum.]{Measured $p\bar p~$ differential cross section, 
$d\sigma/dt$, versus squared momentum transfer, $|t|$, 
at six values of the $\bar p$-beam momentum 
(shifted vertically by an amount $\Delta\sigma$ to distinguish them on the plot).
Fits to the data are shown.}
\label{fig:ppbar_dsigmadt}
\end{center}
\end{figure}

An example of the proton recoil spectra
obtained by one of the luminosity monitor detectors plotted in a logarithmic scale is shown in Fig. 3.20. $N$ is determined from the spectra by taking the number of counts in the signal region and subtracting the background, which is well approximated by an exponential (which appears as a straight line in the logarithmic plot). The signal region is defined separately for each center of mass energy. A luminosity measurement is carried out every 2 minutes during the running of E835, so that a 120 second average of instantaneous luminosity is available during data taking.

\begin{figure}[htb]
\begin{center}
\includegraphics[height=7cm]
{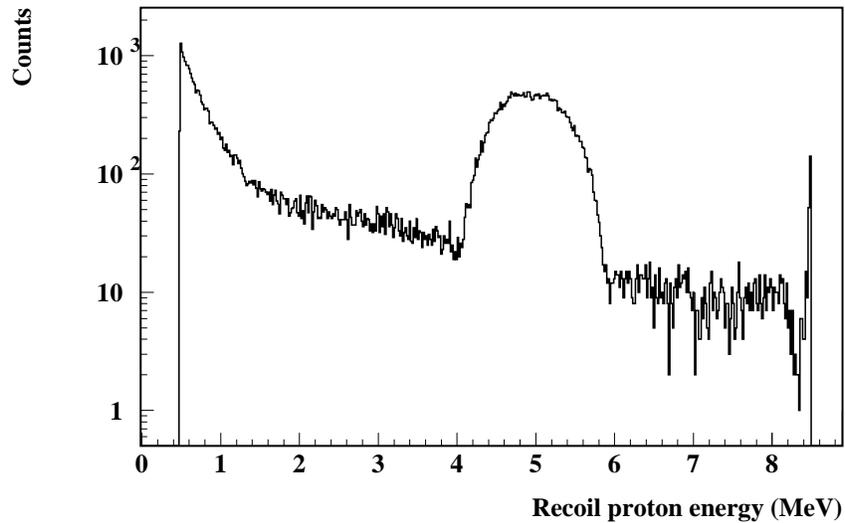}
\caption{Typical spectrum recorded by one of the three solid state detectors
of the luminosity monitor.}
\label{fig:proton_spectrum}
\end{center}
\end{figure}

The luminosity detector response was highly
stable throughout the year 2000 run. As can be seen from Fig. 3.20, the smooth exponential background occurs at the level of $\sim 2-4\%$, and it can be reliably subtracted from the spectrum. The typical count rates in the detectors during data taking were of the order of 10-20 Hz.  
Statistical error in a typical luminosity measurement was
very small, of the order of $0.3\%$ for 100 nb$^{-1}$ of data. The dominant source of error was the
error in the differential cross sections paramatrization of equation 3.44, which was 
estimated to be $\leq \pm$2.5$\%$~\cite{trokenheim}. 

The FORTRAN code used to read the counts from the luminosity monitor and calculate a luminosity off-line is found in Appendix B. The algorithm used in this software to determine the background level samples 21 channels above and 21 channels below the signal window. A fit is then made to an exponential and this is subtracted from the total number of counts detected in the signal window. A simplified version of the luminosity monitor code was also used for on-line monitoring, in order to give readings of the instantaneous luminosity every two minutes. In this version, the readings from the 21 background channels were averaged for each side of the window, and a straight line was drawn between them which was used as the background level. The on-line code also gave readings of the left-right beam asymmetry: by comparing relative rates in the ``beam-left'' and ``beam-right''
detectors, it was possible to determine whether the $\bar p$ beam
was displaced with respect to the nominal axis. This was important because during running it was found that the beam could occaisionally undergo horizontal shifts of several millimeters perpendicular to the beam axis. These shifts lead to errors in the luminosity measurement due to shadowing of the intersection volume by the aperture at the entrance to the luminosity monitor horn. Displacement of the beam from its central orbit causes asymmetry in the two fixed detectors. For orbit displacements of magnitude $<$ 1.5 mm, the asymmetry $(L-R)/(L+R)$ is calculated to be less than $2\%$, and no corrections were applied to the average luminosity. For displacements of magnitudes between 1.5 mm and 3.0 mm, the corrections to the luminosity were calculated to be of the order 10\%, but these corrections were generally not used; instead the beam orbit was adjusted back to its central value if the asymmetry was found to be greater than $2\%$. 

The detectors were regularly calibrated over the course of the experiment with a $^{244}$Cm alpha source, which is inserted pneumatically into the horn between the interaction region and the detector pan. The typical energy resolution for the detectors was found to be $\sim$ 40 keV for 5.742 MeV alphas. Before the pan assembly was attached to the horn, the surface area of each detector exposed by the mask was determined by mounting a calibrated $(\pm 0.4\%) ^{241}$Am alpha source supplied by the National Bureau of Standards. The solid angles subtended by the areas of the detectors exposed by the masks were then determined by surveying the distance between the effective $p\bar p$ interaction point and the detector surfaces; this was measured to be $149.60 \pm 0.03$ cm. The best determination of the recoil angle $\alpha$ for each detector was obtained from the centroid of the recoil peak.

\section{The Trigger System}

The E835 detector has independent triggers for both charged and neutral final states. The charged trigger is primarilly used to identify $e^{+}e^{-}$ pairs from the electromagnetic decay of the $J/\psi$ or the $\psi^{\prime}$, and is used in this thesis in the study of the reactions $p\bar p \rightarrow (h_c) \rightarrow J/\psi + X \rightarrow e^{+}e^{-} + X$. The neutral trigger is primarilly used to identify reactions in which the final decay products are entirely photons. It is used in this thesis for the study of the reaction $p\bar p \rightarrow (h_c) \rightarrow \eta_c \gamma \rightarrow 3\gamma$. There is also a specific trigger for the completely hadronic reaction $p\bar p \rightarrow c\bar c \rightarrow \phi\phi \rightarrow K^{+}K^{-}K^{+}K^{-}$, and, of course, this trigger was not used for any analysis in this thesis. In addition to these, several monitoring and efficiency study triggers were used in E835. Most important of these is the random trigger, also called Random Gate, which is run at a frequency of 10 Hz to study accidental signals and event pileup. The random gate was used in the analyses in this thesis primarilly in order to correct for dead-time due to event pileup. The triggers for both the charged and neutral final states are formed independently, and then fanned into a final trigger state along with the random gate. A schematic illustrating the general trigger layout is presented in Figure 3.21.

\begin{figure}[htb]
\begin{center}
\includegraphics[width=5in]{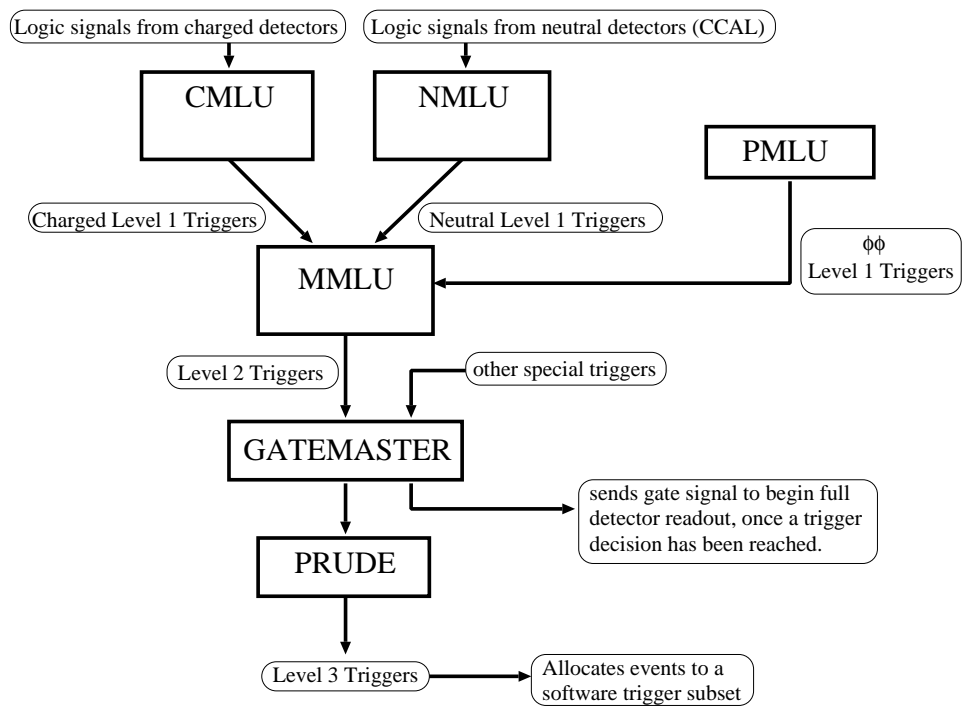}
\caption{\baselineskip=16pt
Schematic showing the process of 
E835 trigger construction.}
\end{center}
\end{figure}

\subsection{The Neutral Trigger}

The neutral trigger is formed by using the CCAL, complemented by the hodoscopes to detect and veto on the presence of charged tracks. Two distinct event categories are selected by the neutral trigger; the first one is characterized by two large deposits of energy in the CCAL at opposite or almost opposite positions in the azimuthal angle $\phi$, and the second one is characterized by a large fraction of the energy in the center of mass system being deposited in the CCAL. The triggers PBG1 and PBG3 belong to the first category, and the triggers ETOT-HI and ETOT-LO belong to the second category.

The large number of CCAL blocks (1280) is reduced for trigger purposes to a more managable number of signals by associating neighboring blocks into groups. The number of CCAL elements is first reduced from 1280 to 160 signals consisting of the sum of 9 counters in the same ring (forming what is called a super-wedge) overlapping with adjacent sums within the same ring. These super-wedge signals are then split with one path being discriminated at $\sim$ 100 MeV and the other path being summed in the ring direction to form 40 superblocks. Each superblock consists of nine wedges and five rings as shown in Fig. 3.22, with the exception of the 8 superblocks at high $\theta$, which span only four rings. These 40 superblocks overlap each other in order to prevent trigger inefficiencies for particles hitting the peripheral elements of a superblock. The superblocks divide the full CCAL into eight parts in the azimuthal angle $\phi$, and five parts in the polar angle $\theta$. A small fraction of the energy from every block is diverted to the neutral trigger; these signals from all of the blocks within a superblock are then summed to form 40 superblock signals, which are in turn used as the inputs of discriminator modules.

\begin{figure}[htbp]
\begin{center}
\includegraphics[height=11cm, angle=0]
%{chapter2/fig/superblocks_unfold.eps}
{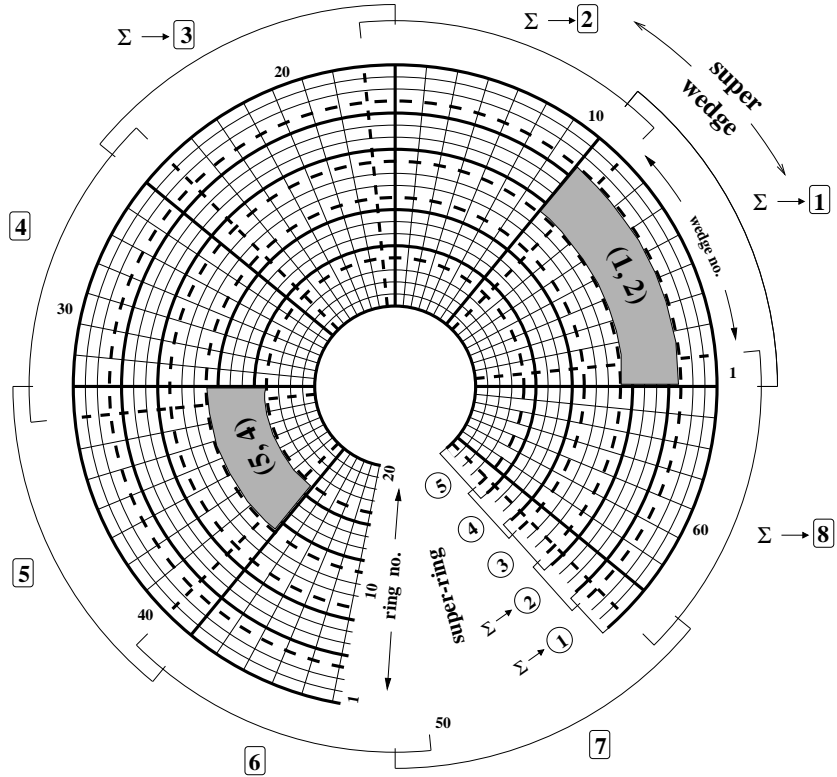}
\caption[View of the front face of the CCAL from the interaction region.]{View of the front face of the CCAL from the interaction region.
Successive superblocks, delimited by tick lines (solid and dashed, alternatively), 
overlap each other
to prevent inefficiencies when a shower hits the borders of the superblocks, as indicated by $\Sigma \rightarrow [1]$ and $\Sigma \rightarrow [2]$.}
\label{fig:superblocks}
\end{center}
\end{figure}

The 40 signals from the superblocks are discriminated and sent to custom programmable CAMAC memory-lookup units (MLUs). The Neutral MLU (N-MLU) then produces two level-1 triggers known as PBG1 and PBG3 for large invariant mass two-body events. Two additional triggers for multiphoton events known as ETOT-HI and ETOT-LO are also formed. All these modules are configured via CAMAC in such a way that each output can be any logical combination of the incoming signals. The output levels are determined at the trailing edge of a strobe signal in order to synchronize the output signals for different triggers. The strobe for the N-MLU is based upon having at least two of the super-wedge signals above the discriminator threshold. 

The various level-1 CCAL triggers used in the neutral trigger are defined as follows:

$\bullet$ The PBG1 trigger requires a signal in two superblocks whose position is consistant with two-body kinematics. It is designed for selecting out $\gamma\gamma$ and $e^{+}e^{-}$ final states. Events from decays such as $p\bar p \rightarrow \pi^0\pi^0 \rightarrow 4\gamma$ are also selected through PBG1 because of the small opening angle between the two photons from each $\pi^0$ decay.

$\bullet$ The PBG3 trigger has similar requirements to the PBG1 trigger, but with a looser requirement on coplanarity: the two hit superblocks do not need to be exactly opposite each other in the azimuthal angle $\phi$; rather, the superblock positions adjacent to the opposite position are also allowed. PBG3 can thus select events from such reactions as $p\bar p \rightarrow J/\psi \gamma \rightarrow e^{+}e^{-}\gamma$, or in the neutral case $p\bar p \rightarrow \eta_c \gamma \rightarrow 3\gamma$.

$\bullet$ The ETOT-HI trigger requires that 80\% of the energy in the center of mass is deposited into the CCAL. It is designed to select multiple photon events. 

$\bullet$ The ETOT-LO trigger requires that 70\% of the energy in the center of mass is deposited into the CCAL; as such it is a looser version of ETOT-HI.

Once an event passes one of these CCAL triggers, the signals from the hodoscopes are used to determine whether the event is neutral or not. The hodoscope trigger signals are defined according to the following logics:

$\bullet$ The FV-OR trigger is defined as the logical ``OR'' between the eight elements of the forward veto counter.

$\bullet$ The H1$\times$H2$^{\prime}$ trigger is defined as the logical ``AND'' between any element of the H1 hodoscope and one of the four corresponding channels of H2$^{\prime}$. Four elements of H2$^{\prime}$ are required to competely cover one element of H1 as the cracks in the two hodoscopes are not aligned. This trigger also allows cases where more than one H1 element is matched with a signal from one of its corresponding H2$^{\prime}$ elements.

$\bullet$ The H2$\ge$2 trigger indicates that two or more elements of H2 are hit.

$\bullet$ Finally the charged veto CV is defined as the logical ``OR'' between FV-OR and H1$\times$H2$^{\prime}$; CV = FV-OR $\times$ H1$\times$H2$^{\prime}$. Its negation is used to declare an event as neutral for triggering purposes.

Once any of the PBG1, PBG3, ETOT-HI, or ETOT-LO triggers are formed in the CCAL, they are then associated with the hodoscope trigger logics into the four final neutral triggers:

\begin{description} 
\item[Neutral-PBG1~]
$\equiv~$~PBG1~$\times~\overline{\mathrm{H1}\times\mathrm{H2}'}~\times\overline{\mathrm{FV-OR}}~\equiv$~PBG1~$\times~\overline{\mathrm{CV}}$
\item[Neutral-ETOT~]
$\equiv~$~ETOT-HI~$\times~\overline{\mathrm{H1}\times\mathrm{H2}'}~\times\overline{\mathrm{FV-OR}}$
\item[ETOT-NoVeto~]
$\equiv~$~ETOT-HI~$\times~\overline{\mathrm{H2}\ge2}$
\item[Neutral-ETOT-LO~]
$\equiv~$~ETOT-LO~$\times~\overline{\mathrm{H1}\times\mathrm{H2}'}~\times\overline{\mathrm{FV-OR}}$~$\equiv~$~ETOT-LO~$\times~\overline{\mathrm{CV}}$
\end{description} 

where the $\times$ symbol denotes the logical ``AND'' operator, and the overline denotes the logical ``NOT'' operator.

The Neutral-PBG1 and Neutral-ETOT triggers are the ones which are primarilly used to record neutral events, while the ETOT-NoVeto and Neutral-ETOT-LO triggers are mostly used for monitoring and efficiency purposes. The events selected by these triggers are then further processed by the software trigger PRUDE (Program Rejecting Unwanted Data Events) for online reconstruction, filtering, sorting, and logging.

The formation time of the PBG triggers is 200 ns and that of the ETOT triggers is 155 ns. The formation time of the neutral-veto triggers alone is approximately 40 ns. The efficiencies of these various triggers are summarized in Table 3.4 for various physics channels studied. The hardware for the E835 neutral trigger is shown schematically in Fig. 3.23. A detailed description of the neutral trigger may be found in Refs.~\cite{ntrig1}~\cite{ntrig2}.

\linespread{2.4}
\begin{table}[htb]
\begin{center}
\begin{tabular}{cccc}
\hline
 & & & \\
Res. & $\sqrt{s}$ (GeV) & PBG1 eff. (\%) & ETOT-H1 eff. (\%) \\
 & & & \\
\hline
 & & & \\
$\eta_c$ & 2.99 & 99.5 & 99.8 \\
 & & & \\
$J/\psi$ & 3.10 & 100 & 100 \\
 & & & \\
Cont. & 3.25 & 98.8 & 99.8 \\
 & & & \\
$\chi_0$ & 3.41 & 100 & 99.8 \\
 & & & \\
$\chi_1$ & 3.51 & 100 & 99.1 \\
 & & & \\
{\bf $^1P_1$} & {\bf 3.53} & {\bf 99.4} & {\bf 99.3} \\
 & & & \\
$\chi_2$ & 3.55 & 99.7 & 99.0 \\
 & & & \\
$\eta_c^{\prime}$ & 3.60 & 100. & 98.2 \\
 & & & \\
$\psi^{\prime}$ & 3.67 & 100. & 97.7 \\
 & & & \\
Cont. & 3.70 & 100. & 95.0 \\
 & & & \\
\hline
\end{tabular}
\caption{Efficiencies of the neutral trigger for charmonium (PBG1) and multiphoton (ETOT-H1) final states as a function of center of mass energy.}
\label{tab:trigeff}
\end{center}
\end{table}

\begin{figure}[htbp]
\begin{center}
\includegraphics[height=6.5in]{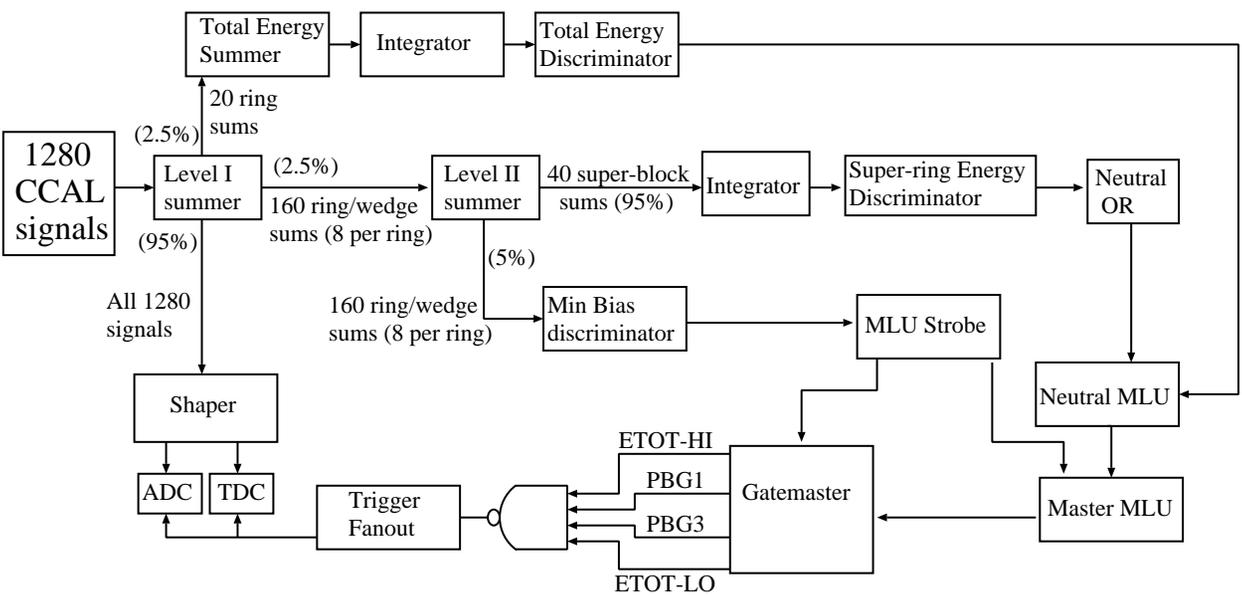}
\caption[Schematic of the E835 neutral trigger.]{\baselineskip=18pt
Schematic of the E835 neutral trigger.  The numbers in 
parentheses denote the fraction of the signal input to each 
summer which is used for the indicated operation.  For 
example, $2.5\%$ of the output from the Level I summers
is used as input (after amplification by a factor 20)
to the Level II summers.}
\end{center}
\end{figure}

\subsection{The Charged Trigger}

The charged trigger is designed mainly to accept $e^{+}e^{-}$ events from the decays of charmonium states such as $J/\psi$ and $\psi^{\prime}$~\cite{ctrig}. As stated earlier, there is a separate trigger to detect $\phi\phi$ states; it was not used in the analysis in this thesis and will not be discussed here.

The charged trigger selects events with one or two electron tracks and generates a neutral veto. The input signals to the charged trigger are the outputs from the hodoscopes H1, H2, and H2$^{\prime}$, the \v Cerenkov counter, and the scintillating fiber detector, although the latter is not used in the present analysis. The crucial signal for the charged trigger for $e^{+}e^{-}$ events is that from the \v Cerenkov counter. A schematic of the charged trigger is shown in Figure 3.24.

\begin{figure}[htb]
\begin{center}
\includegraphics[width=5.5in]{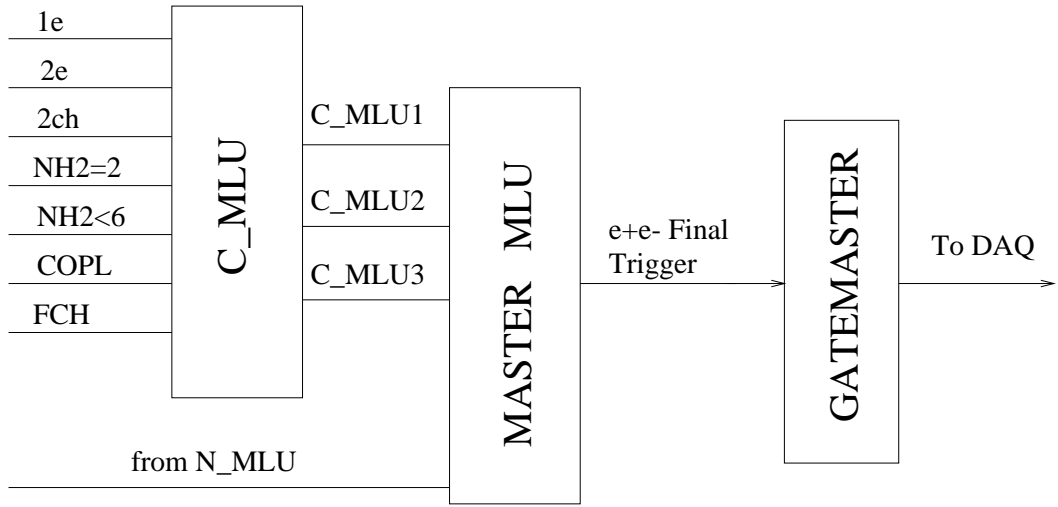}
\caption{\baselineskip=18pt
Schematic of the E835 charged trigger.}
\end{center}
\end{figure}

The charged trigger hardware consists of both CAMAC and NIM modules, divided into three successive stages; the discrimination stage, the single logic stage, and the final trigger stage. In the discrimination stage, the signals from the hodoscopes and the \v Cerenkov are discriminated by six LRS 4413 discriminators, and the 80 output discriminated signals are sent to the trigger logic and to TDCs. The outputs from the scintillating fiber detector are discriminated seperately with 32 discriminator modules. The second stage of the charged trigger consists of the single-logic modules; track reconstruction, multiplicity, coplanarity and also the $\phi\phi$ logic for the $\phi\phi$ trigger which outputs to its own separate MLU. The neutral-veto and forward-veto modules which also use information from the hodoscopes have been described in the previous section.

The track reconstruction module gives the number of charged tracks and the number of electrons. A charged track is defined as the coincidence between one element of H1 and a corresponding element of H2. An electron for trigger purposes is defined as a charged track where there is also a coincidence with a signal from the corresponding \v Cerenkov octant. The outputs of the track reconstruction module are the signals 1e, 2e, and 2ch, which represent the identifcation of one electron, two electrons and two charged tracks. The track reconstruction logic is done with LRS NIM modules in approxmately 70 ns.

The multiplicity module provides the numbers of H2 scintillators and fiber bundles which recorded signals; these are labelled NH2 and HSF respectively. The multiplicity module is comprised of LRS 4532 logic modules that give output signals proportional to the number of active input signals. The multiplicities that are selected from this module for the final charged trigger are NH2=2, NH2$\le$4, NH2=3 or 4, NSF=3 or 4, and the combined multiplicities NH2 + NSF = 7 or 8. The mulitplicity module requires approximately 65 ns to output a signal.

The coplanarity module requires at least two hits in H2 which are back to back within 22.5$^{\circ}$, i.e. $180^{\circ} \pm 22.5^{\circ}$. This module uses only the signals coming from the H2 hodoscope and requires that if an element of H2 is hit then at least one of the 3 opposite elements of H2 are also hit. The coplanarity module requires approximately 50 ns to output a signal.

Once the track reconstruction, multiplicity and coplanarity signals have been formed by the single-logic modules, they are sent to a custom CAMAC programmable MLU known as the C-MLU (charged memory-lookup unit). There is also a separate $\phi$-MLU for the $\phi\phi$ trigger signals. The outputs from these MLUs are combined with the neutral trigger output from the N-MLU and sent to the master memory-lookup unit (M-MLU). The output of the M-MLU is then strobed into the data acquisition system (DAQ), which writes the event. 

The strobe signal for the C-MLU is the logical OR of the H2 hodoscope signals. Since the timing of this signal is critical, it is reshaped using an Ortec 934 constant fraction discriminator, which leads to a jitter of 2 ns or less. Input signals that arrive at least 8 ns before the strobe are successfully latched. To allow for a 5 ns jitter, the inputs are set to reach the MLUs 15 ns before the strobe and are 30 ns wide. The strobe itself is 10 ns wide.

The transit time through the single-logic stage of the charged trigger is 96 ns for the $e^{+}e^{-}$ branch. The transit time through the final trigger stage to the output of the M-MLU is about 90 ns (including some delay needed for synchronization of the neutral logic). The total transit time through the whole trigger is thus about 200 ns. The total efficiency of the charged trigger is $\epsilon = (90 \pm 1)\%$; this is the number which is used as the trigger efficiency for all charged analyses in this thesis~\cite{nimpaper}. This efficiency was calculated by examining a set of $\psi^{\prime} \rightarrow e^{+}e^{-}$ events which were collected using a simplified trigger requiring only one electron track, and then checking to see how many times the second electron track passed the full charged trigger~\cite{wandertrigger}.
 
\subsection{The Random Gate Trigger}

Because of the high instantaneous luminosity at which E835 is run, one of the main sources of inefficiency is the dead-time due to the pileup of different events. The event pileup is difficult to deal with because it varies with the instantaneous luminosity, which, despite the efforts made to limit its fluctuations during the data taking, may be different at various center-of-mass energy points. Thus when performing an energy scan, it is critically important to estimate the amount of the luminosity dependent pileup contamination and correct for it according to the instananeous luminosity of each data taking run. This is done primarilly by using information recorded through the random gate trigger.

During the data taking, this random gate trigger is run at a frequency of 10 Hz with priority over the physical triggers. While it is active the gate of the data acquisition is opened and the event signals from all channels of the CCAL, FCAL and Charged Veto (hodoscopes) are recorded. Off-line a Monte Carlo simulation is run and the efficiency of the detector system is determined by overlapping random gate events with generated Monte Carlo events; during this overlap the Monte Carlo content of each ADC and TDC channel of the detectors is summed to the random gate content of the same channel, and afterwards the clusterization and event selection is performed on the composite Monte Carlo/random gate signal. In this way it can be determined how likely a random gate event is to cause a genuine event to be lost. That information, combined with the overall rate of random gate events, is then used to determine the efficiency.

For charged channels the detection efficiency as a function of instantaneous luminosity was found to be:

\begin{equation}
\epsilon \approx (1 - 0.04\times{\cal L}_{inst}) 
\end{equation}

where the instantaneous luminosity ${\cal L}_{inst}$ is measured in units of $10^{-31}$cm$^{-2}$s$^{-1}$.

This efficiency was found using the random gate data collected at $\chi_0$ energies, overlapped with GEANT Monto Carlo generated $p\bar p \rightarrow J/\psi \gamma \rightarrow e^{+}e^{-} \gamma$ events~\cite{margherita}. This correction includes the effect of random gate events on both the $e^{+}e^{-}$ signals, and the $\gamma$ signals.

For all-neutral channels, such as $p\bar p \rightarrow \pi^0\pi^0 \rightarrow 4\gamma$, the luminosity dependent correction to the efficiency is greater:

\begin{equation}
\epsilon \approx (1 - 0.075\times{\cal L}_{inst}) 
\end{equation}

Fig 3.25 shows the result of the random gate analysis for the charged channel described above. The luminosity dependent corrections to the efficiency in the above equations are calculated from the slope of the efficiency vs. luminosity plot for that reaction.

\begin{figure}[htbp]
\begin{center}
\includegraphics[width=15cm]
{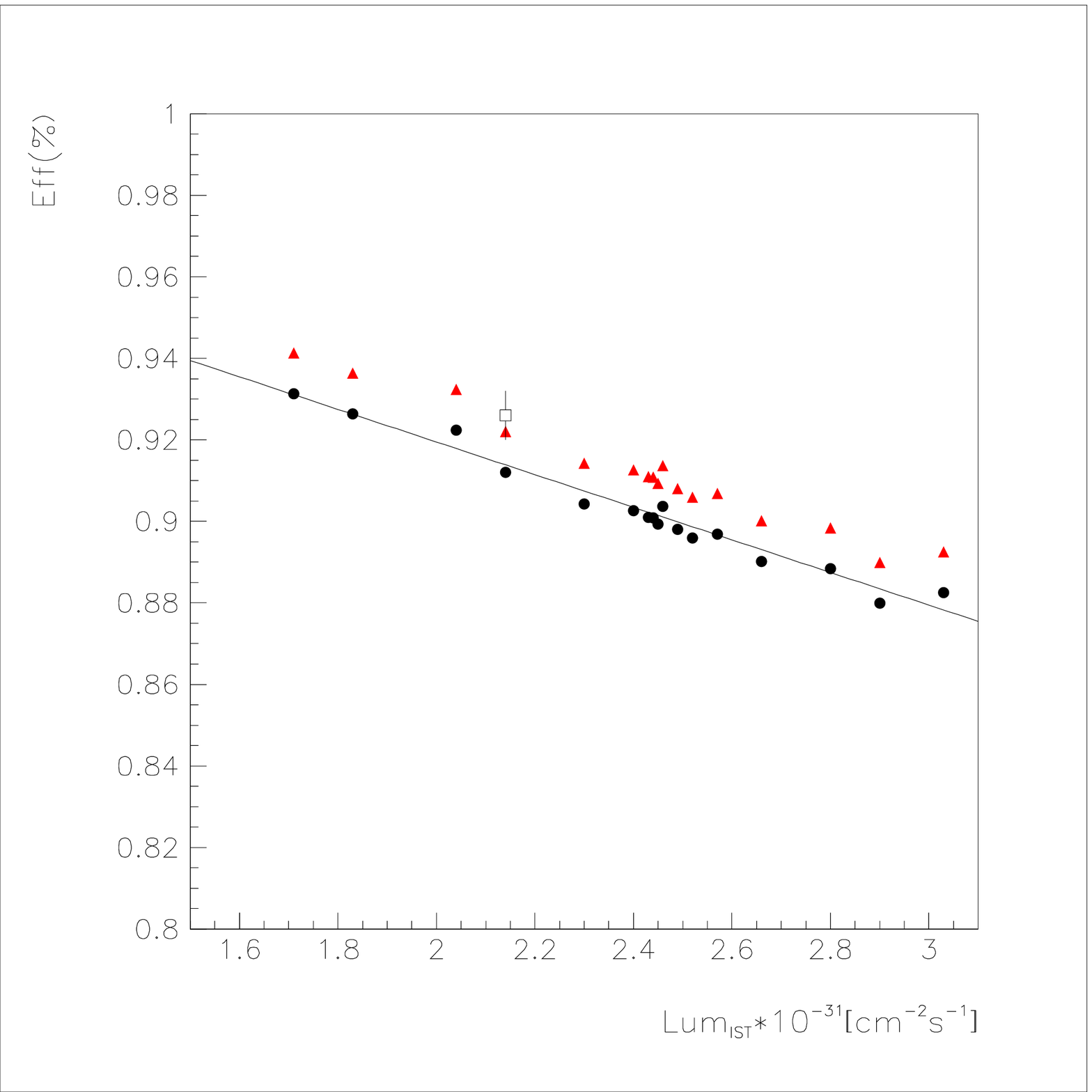}
\caption[The probability that an $e^{+}e^{-}$ event survives the pileup contamination versus the instantaneous luminosity, on a run by run basis.]{The probability that an $e^{+}e^{-}$ event survives the pileup contamination versus the instantaneous luminosity, on a run by run basis~\cite{margherita}. The probability was calculated using random gate events collected at the $\chi_0$ energy region. The black circles represent the efficiencies calculated using GEANT, and the red triangles represent the same efficiencies scaled to account for the difference between GEANT event and data events.}
\label{fig:pileup_runbyrun}
\end{center}
\end{figure}

The overall efficiency 
at each energy point is determined by averaging the Random Gate
efficiency of its runs, $(\epsilon)_{run}$, 
and taking a weighted average of the runs according to their integrated luminosity $\mathcal{L}_{run}$:
\begin{equation}
\label{eq:averagepileup}
a\times\epsilon=
\frac{1}{\sum_{run}~\mathcal{L}_{run}}~\times~
\sum_{run}\mathcal{L}_{run}\times (\epsilon)_{run}~.
\end{equation}
The uncertainty on $\epsilon$ is
\begin{equation}
\sigma_{\epsilon}=
\frac{1}{\sum_{run}~\mathcal{L}_{run}}~\times~
\sqrt{\sum_{run}~
\left[\mathcal{L}_{run}\times\sigma_{(\epsilon)_{run}}\right]^2}~.
\end{equation}

\section{The Data Acquisition System}

Data acquisition and recording in E835 is done from four independent data acquisition streams running in parallel: the event data stream, the data stream for parameters of the $\bar p$ beam, the luminosity monitor data stream, and the scaler data stream to monitor the rates of selected counters and triggers. Fig. 3.26 illustrates the schematic of the system.

\begin{figure}[htbp]
\begin{center}
\vspace*{-50pt}
\includegraphics[width=4.5in]
{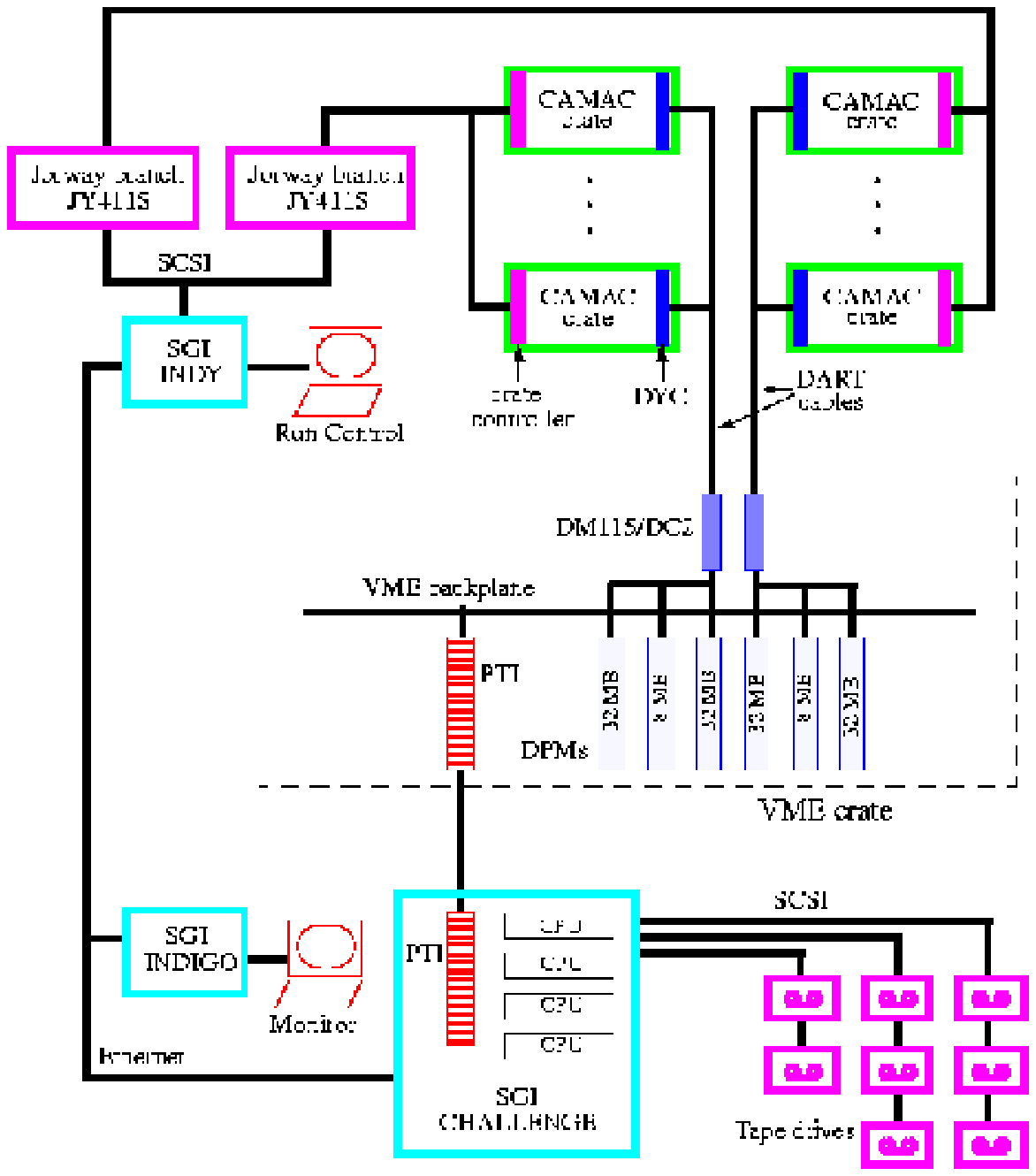}
\vspace*{-50pt}
\caption[Layout of the hardware of the event data acquisition.]{Layout of the hardware of the event data acquisition. The ADC and TDC readouts, where the signals first enter the data acquisition system, are located in the CAMAC crates on the upper right of the diagram. Here, DYC stands for Damn Yankee Controller, DPM's are the Dual Ported Memory units, SGI Indy and SGI Indigo are two of the Silicon Graphics computers used for run control and monitor display respectively, and PTI are Performance Technologies Interfaces~\cite{paolo}.}
\label{fig:daq}
\end{center}
\end{figure}

The event data stream is based on the DART~\cite{dart} data acquisition system, which was developed by a collaboration between the Fermilab Online System Department and several experiments that operated during the 1996-97 fixed target run. DART provides a common system of hardware and software which has been configured and extended to meet the specific requirements of the E835 experiment. The beam parameter information data stream is received from the FNAL Beams Division via ethernet. Both the luminosity monitor stream and the scaler data streams are input and received via standard CAMAC.

Three Silicon Graphics (SGI) computers were used to coordinate, process, filter, log, and monitor the data streams. The run-control computer, called SGI-Indy, coordinates the independent data streams, communicates with the CAMAC branches, and performs the slow data acquisition functions. The event data acquisition, which consists of event building, filtering, and logging, is performed by an SGI Challenge-L computer with 12 processors. Finally, the monitor-display computer is an SGI Indigo, which monitors the detectors and displays events.

\subsection{The Event Data Acquisition}

The layout of the event data acquisition is illustrated in Fig. 3.26. The readout consists of 163 ADC (LRS4300) modules and 114 TDC (LRS3377) modules organized in 14 CAMAC crates. They are further arranged into 3 CAMAC branches, two parallel and one serial, and addressed by the run-control computer through two SCSI 411 Jorway Interfaces.

Data from the detectors are read out by the ADC and TDC modules, and transferred through their front-end ECL ports to the DART-developed Damn Yankee Controller (DYC) modules. The DYC is an intermediate data buffer which stores the 16-bit input data in a 32-bit wide FIFO (first-in first-out). The DYC adds a header word including a word count, an error bit in the case of event overflow, and 4 bits of event synchronization. Data is sent out to two pairs of Access Dynamics DC2/DM115 modules over two RS-485 cables.

Each ADC or TDC CAMAC crate is served by a DYC. The DYC responds to the request from the readout electronic modules it serves, which are connected in a daisy-chain, by issuing a readout enable to the first module, starting the data transfer. When the first readout electronic module has finished sending data, it enables the following module in the chain by sending a pass signal. The end of the data transfer is signified by the receipt of the pass signal from the last module.

While receiving data, the DYC modules issue a busy signal. The logical ``OR'' of the busy signals from all the DYC modules is sent to the trigger logic to inhibit triggers during the data reception. 

The DYCs are connected in a daisy-chain via two RS-485 cables (labelled as DART cables in Fig. 3.21) in two groups, each of them connected to one of the DC2/DM115 pairs. Data transmission to the DC2/DM115 pairs is controlled by a wait signal, established by the destination buffer, and a permit token. On receipt of a permit token, and in the absence of a wait signal, the DYC transmits the header word and the data from one event over the RS-485 cable. When the DYC has completed its transmission, it passes the permit token to the following DYC in the daisy-chain. The DYC module is able to simultaneously perform the data reception from the readout electronic modules and the data transmission to the DC2/DM115.

Each DC2 is connected, via the VSBus of the VME crate, to two 32 MB and one 8 MB Dual Ported Memory (DPM) MM-6390 modules. Each DC2 fills the two 32 MB DPMs with the data it receives from the DYCs using a {\it ping-pong} algorithm. According to this algorithm the DC2 writes to the {\it ping} DPM with exactly N (the number which is set in the configuration database) events before it switches to the {\it pong} DPM.

A process running on the Challenge, called the gateway process, reads data from the {\it ping} DPM via PTI(PT-VME64) while the DC2 writes to the {\it pong} DPM. When all the data from the {\it ping} DPM are read, gateway allows the DC2 to write to the {\it ping} DPM and turns to read out the {\it pong} DPM. Both the DC2 and the gateway poll each other to see if each has finished using a DPM. The communication between the DC2 and the gateway is done using the 8 MB DPM as a mailbox. Neither the DC2 nor the gateway interrupt a data transfer while polling. 

The gateway writes the two DC2 sets of data, which comprise N events, to two buffers in a reserved part of the memory of the Challenge-L, which has been designated as memory to be shared among processes. When finished writing a buffer, the gateway releases it to a filter queue before writing into the other buffer in the shared memory. An online filter process runs continuously on each CPU of the Challenge-L. Each filter queues for a gateway buffer; when one is available, an online filter processes the N events. First, the events are built taking data from both of the two DC2 data sets. Then, synchronization numbers of the DPMs and the DYCs are checked and the data integrity is verified for 0.1\% of the events processed. Finally, an event header is added which includes a unique event number, date, and time of processing and pointers to each DYC header word for later use by the offline event reconstruction. Some basic event trigger classification and online analysis bit flags are also included.

The trigger information is decoded to classify the event. This classification determines which online filter analyses to perform, if any, and which logging stream the event is sent to if it is not rejected.

The following online analyses are performed: a simplified CCAL determination of energies and angles for electromagnetic showers and formation of invariant mass pairs to identify $\pi^0$, $\eta$, and $c\bar c$ candidates, simple charged track reconstruction with electron identification and association with CCAL energy deposits for identifying events that fit exclusive $e^{+}e^{-}$ final states or $J/\psi + X \rightarrow e^{+}e^{-} + X$ decay, and four-track identification for $\phi\phi$ events.

The online filters set analysis bit flags in the event header word and write summary event information which is added to the end of the event. The analysis bit flags are compared to a list of trigger masks to determine acceptance or rejection of the event.

Accepted events are copied to logging buffers reserved in the shared memory for eventual recording to tape and transfer to the monitor-display computer. There are three logging streams; which stream an event is written to is based on the event trigger classification and the analysis bit flags. Events for which the filter determines a large invariant mass and those to be used for calibration purposes are written to other logging buffers. When the logging buffers are full (they have a maximum size of 64 KB; the average event size is $\sim$ 1 KB), they are released to the appropriate logging queues and to the monitor-display computer. There are ten logging processes, each one with its own queue and tape drive. Every logging process receives buffers from all of the filter processes. Generally, five tape drivers record data in parallel to 8-mm EXABYTE tapes with a 5.0 GB capacity, while the other five frivers are left ready to be able to start as soon as one of the tapes in the first five drivers becomes full. The switching of tape drivers signifies the end of a single run of data taking and the beginning of a new one. Two of the five simultaneously running tape drivers were dedicated to neutral events and another two were dedicated to charged events; thus there were two charged tapes and two neutral tapes for each run. The final tape driver is used for efficiency study events.

For a characteristic instananeous luminosity of $2.5 \times 10^{31}$ cm$^{-2}$s$^{-1}$, the trigger rate is $\sim$ 4 KHz. The typical event size from the readout electronics is $\sim$ 1.2 KB. Thus the two daisy chains and the gateway typically transfer $\sim$ 4.8 MB per second to the online filters. The filters reduce the output event rate to $\sim$ 1.2 KHz, which corresponds to writing $\sim$ 1.4 MB per second to the logging streams. Each tape drive is capable of 1000 KB per second; the tape drives typically record data at a rate of 350, 300, and 250 KB per second for the neutral, charged, and efficiency study events, respectively. The resulting lifetime of the data acquisition as a whole is typically 93\%, at the instantaneous luminosity of $2.5 \times 10^{31}$ cm$^{-2}$s$^{-1}$. The inefficiency comes from inhibiting the triggers during the ADC conversion time and from the data transfer to the DYC modules.

\subsection{Non-Event Data Streams}

The control and monitoring of the FNAL accelerator complex is performed through the Accelerator Control Network System, known as ACNET. Accelerator data for E835 is acquired by a Fermilab Beams Division ACNET computer, which is interfaced to the accelerator instrumentation and sends a data stream to the run-control computer. This data stream consists of the positions and intensities recorded by the BPMs, the longitudinal frequency spectrum of the beam, the gas-jet density, pressures within the accelerator, magnet currents, magnetic field strenghts, and the online determination of the beam energy, emittances and current. The run-control computer writes the accelerator data to disk depending on the status of the run control. The online beam momentum and beam current are unpacked and made available for monitoring purposes every two minutes.

The CAMAC readout electronics of the luminosity monitor detector consists of a multiplexed ADC converter (EGG ORTEC-AD413A)and a FERAbus Histogramming Memory (EGG ORTEC-HM413). A self gating ADC channel is used to read each of the three silicon detectors. Each channel is read and histogrammed via the front-end ECL port by the HM413 module. Every two minutes a process running on the run-control computer reads the three data sets from the HM413 module via the CAMAC backplane, resets the memory, computes the luminosity by the algorithm described in Section 3.3.7, writes the result on disk, and sends it to the monitor-display computer.

The readout of the luminosity monitor data stream shares a CAMAC crate with the readout from the scaler data stream. Three LRS4434 scaler modules receive data via their front-end ECL ports from the trigger and detector electronics. A Fermilab CAMAC pulser (RFD01) module sends a signal to the scalers to update every 10 seconds. The data stream is sent every ten seconds via the CAMAC backplane to the run-control computer which writes it to disk and displays it on a monitor.

\section{The Offline Event Reconstruction}

Tracking of charged particles by the inner detectors is done separately for the $r/\phi$ and $r/z$ projections. The straw chambers are used for the $r/\phi$ projection and and scintillating fibers together with the nominal position of the interaction vertex for the $r/z$ projection. These processes are then associated using CCAL information.

The azimuthal angle $\phi$ for charged tracks is determined using the signals from the straw chambers. First the drift distance is determined within the hit straw and a fit to a straight line through the layers is made using the geometrical parameters of the straws. For each straw the discriminator threshold is set at one primary ionization. 

The drift time, which is measured with 500 ps (rms) resolution by an LRS multihit TDC 3377 used in Common Stop mode, is defined as the delay of the straw signal with respect to the event time, the latter of which is defined by a strobe constructed from CCAL signals. For each straw, a reference time $T_0$ is measured which depends on the propagation delays through the electronics and cables. The drift time distributions $(T-T_0)$ should be uniform for constant drift velocity and full efficiency, since the particle density is uniform in the azimuthal angle $\phi$. Non-constant drift velocity is corrected for to obtain the distance $X(T-T_0)$ for each layer using the relation:

\begin{equation}
X(T-T_0)/R = dN/dX \int_{T_0}^{T_1} (dX/dt)\times(dt/N)
\end{equation}

where $N$ is the total number of tracks, $dN/dx$ (= constant) is the track density, and $X$ is the perpendicular distance between the particle line of flight and the wire, which determines the azimuthal angle $\phi$ and is independent of $\theta$. For each track a straight line is fit through the hits to determine $\phi$ with a single-track rms resolution of $\sim$ 9 mrad. This is comparable to the angular resolution of the CCAL, which was measured to be $\sim$ 6 mrad in the polar angle and $\sim 11$ mrad in the azimuthal angle~\cite{bartoszek}. 

The polar angle $\theta$ is determined for charged tracks using the signals from the scintillating fiber detector. A charged particle traversing the scintillating fibers hits up to three consecutive fibers in each layer, depending on the polar angle $\theta$. Clusters are then defined as sets of adjacent hit fibers; those with more than three fibers are split into two or more smaller clusters. The $z$ coordinate of the intersection between the track and the layer is taken as the mean of the $z$ coordinates of the fibers in a cluster, weighted by the energy deposit in each fiber. This value of $z$ is then converted to a $\theta$ coordinate using the nominal coordinates of the interaction point. Once this is done, the clusters found in each layer are associated to form tracks, by requiring that their polar angles $\theta$ be within a specified range. The clusters for each track and the interaction point determined using the CCAL, as previously described, are then fit to a straight line. 

A nominal interaction point for the CCAL analysis is required in order to compute the laboratory angles of the clusters, and thus the momentum of each particle. The interaction region is located at the intersection of the beam and the gas jet. While the jet location does not change, the beam position and size varies from stack to stack. For every stack, the ($x,y$) coordinate of the center of the interaction region is determined from the $\pi^0\pi^0$ sample. The desired precision of 0.1 mm is obtained by using 10,000 events per stack.

The laboratory coordinate system is defined as follows: The $z$ axis is the beam direction, the $z$ coordinate of the gas-jet center is zero, the intersection point of the (pointing) counter axes of the CCAL is $(x,y,z) = (0,0,z_c)$, and the $x-z$ plane passes through the axes of all of the counters of CCAL wedge number 1.

For a two-body reaction, trigonometry gives the following formula for the acoplanarity, $\Delta \phi \equiv \pi - \vert \phi_1 - \phi_2 \vert$, for a vertex at $(x_0,y_0,0)$ where $\phi$ is the azimuthal angle of one of the $\pi^0$s and $R$ is the counter radius in the $x-y$ plane:

\begin{equation}
tan(\Delta\phi/2) = {x_0 sin\phi - y_0 cos\phi \over R - x_0 cos\phi - y_0 sin\phi}
\end{equation}

The distribution of $\Delta \phi$ for each stack is fit to the above equation to obtain $x_0$ and $y_0$. As these parameters depend only on the $\pi^0$ directions, $(x_0,y_0)$ is relatively insensitive to the energy calibration, and we determine it using the $g_i$ (described previously in the CCAL section) from the previous stack, prior to the energy calibration, which depends strongly on $(x_0,y_0)$. 

\section{Electron Identification}

Proper electron identification is crucially important for most of the analyses performed in this thesis; in particular the identifcation of electrons from $J/\psi$ decay. For all analyses of interest to the E835 experiment it is necessary to identify electromagnetic decays of charmonium in the presence of a hadronic background which is between $10^8$ and $10^{11}$ times larger, requiring not only excellent electron identification, but also the ability to distinguish between $\gamma$ and $\pi^0$ signals. Individual electrons must be distinguished from other ``tracks'' producing \v Cerenkov light, including $e^{+}e^{-}$ pairs due to photon conversion in the beam pipe and from Dalitz decays of $\pi^0$s.

The detectors that provide useful electron identification information are the H1, H2, and H2$^{\prime}$ scintillator hodoscopes, which measure $dE/dx$; the scintillating fiber detector, which also measures $dE/dx$ and can identify small angle electron pairs; the threshold \v Cerenkov counter, which is designed to count only electrons and can identify electron pairs using pulse height; and the CCAL, in which cluster shapes are different for single electrons, hadrons, and merged electron pairs.

Using all the available information, a likelihood ratio has been developed for E835 analyses. It is known as the electron weight (EW), which combines signals from all these detectors. A second method, known as the k-nearest neighbor rule, is used to check the assumptions of the EW method. It is the electron weight method which is used primarilly in this thesis to identify electrons.

The electron weight of any given charged track is calculated as follows. For each charged track, characterized by measured quantities ${\bf x} = (x_1, ... ,x_n)$, we seek to distinguish between the hypothesis $e$ that it is an electron vs. the hypothesis $b$ that it is not an electron. The optimal (Neyman-Pearson) method is the construction of the likelihood ratio $\rho({\bf x})$, derived from the probability density functions $F_e({\bf x})$ and $F_b({\bf x})$, under the two hypotheses respectively. For a single track we thus have:

\begin{equation}
\rho({\bf x}) = {F_e({\bf x}) \over F_b({\bf x})}
\end{equation}

We define $P_e$ and $P_b$ as the usually-unknown overall efficiencies for electrons $e$ and non-electrons $b$ respectively, resulting from the trigger and selection efficiencies. It is thus convenient to rewrite the above equation as:

\begin{equation}
\rho({\bf x}) = {f_e({\bf x})P_e \over f_b({\bf x})P_b}
\end{equation}
 
If $P_e$ and $P_b$ were known, the cut $\rho > 1$ would give the optimal discrimination criterion. Since they are unknown, the reduced likelihood ratio is used, which is defined as:

\begin{equation}
\rho^{\prime}({\bf x}) = {f_e({\bf x}) \over f_b({\bf x})}
\end{equation}

Since it is desirable to maximize the signal to background ratio rather than to minimize the absolute number of misassignments, a suitable minimum value of $\rho^{\prime}$ is empirically determined, depending on the analysis being performed. This reduced likelihood ratio $\rho^{\prime}$ is refered to as the electron weight. The measured quantities ${\bf x}$ which are included in the electron weight are:

$\bullet$ H1 pulse height corrected by polar angle $\theta$.

$\bullet$ H2 pulse height corrected by polar angle $\theta$.

$\bullet$ H2$^{\prime}$ pulse height corrected by polar angle $\theta$.

$\bullet$ \v Cerenkov counter pulse height corrected by polar angle $\theta$ and mirror photoelectron yield, with the correction done independently for all 16 mirrors.

$\bullet$ $s_{\theta}$ and $s_{\phi}$, the CCAL cluster second moments in wedge and ring directions, calculated according to the relations:

\begin{equation}
s_{\theta} = {\Sigma^3_{r,w=1} E(r,w) \times (w-w_0)^2 \over \Sigma^3_{r,w=1} E(r,w)}
\end{equation}

\begin{equation}
s_{\phi} = {\Sigma^3_{r,w=1} E(r,w) \times (r-r_0)^2 \over \Sigma^3_{r,w=1} E(r,w)}
\end{equation}

where the index $r$ runs over the CCAL rings, the index $w$ runs over the CCAL wedges, and $E(r,w)$ is the energy in block $(r,w)$.

$\bullet$ $F_{35}$, the ratio of the energy deposited in a 3 $\times$ 3 block matrix divided by that in a 5 $\times$ 5 block matrix, both surrounding the highest energy block in the cluster.

$\bullet$ $F_{24}$, the ratio defined similarly to the $F_{35}$ ratio, but using matrices of sizes 2 $\times$ 2 and 4 $\times$ 4.

$\bullet$ $M_{cl}$, the cluster mass as defined previously in the discussion of the CCAL (see Section 3.3.5).

For the CCAL variables $s_{\phi}$, $s_{\theta}$, $F_{35}$, $F_{24}$, and $M_{cl}$, all of the blocks surrounding that with the highest energy are used, whether or not the cluster is split.

To compute the electron weight, $f_e({\bf x})$ and $f_b({\bf x})$ are taken as the normalized products of the measured distributions of each variable, for clean samples of electron and non-electron tracks. Although components of the electron weight, such as hodoscope signals, are obviously correlated in the presence of a real electron track, it is assumed that the electron weight variables ${\bf x}$ are uncorrelated, in order to factor $f_e({\bf x})$ and $f_b({\bf x})$ into individual probability distribution functions. The electron weight EW is then calculated as:

\begin{equation}
EW = {\prod_i f^i_e(x_i) \over \prod_i f^i_b(x_i)} = \prod_i {f^i_e(x_i) \over f^i_b(x_i)} = \prod_i W_i
\end{equation}

where the $W_i$ are the individual weights of which the final electron weight is the product.

The individual probability distribution functions are determined from events selected by the trigger. The tracks must consist of a CCAL cluster associated with at least two out of three hodoscope hits and a \v Cerenkov signal. For the electron probability distribution functions a clean sample of $J/\psi$ and $\chi_{c2}$ events is taken with consistent event topology and acceptable kinematic fit probability. For the non-electron background probability distribution functions events are used from runs taken well outside the regions of the known charmonium resonances, specifically at energies between 3590 and 3660 MeV (which were taken during the $\eta_c^{\prime}$ search in the 1996/7 E835 run). The probability distribution functions themselves are described in detail in Ref.~\cite{nimpaper}. An illustration of the ability of the electron weight to distinguish between $e^{+}e^{-}$ events and non-electron background events is shown in Figure 3.27. The unshaded histogram in this figure shows the electron weight distribution of events from the reaction $p\bar p \rightarrow \chi_1 \rightarrow J/\psi \gamma \rightarrow e^{+}e^{-} \gamma$. The shaded histogram shows the electron weight distribution of background events from the same runs. Background events were defined as those in which $M(e^{+}e^{-}) < 2.8$ GeV.

\begin{figure}[htbp]
\begin{center}
\vspace*{-10pt}
\includegraphics[width=5.5in]
{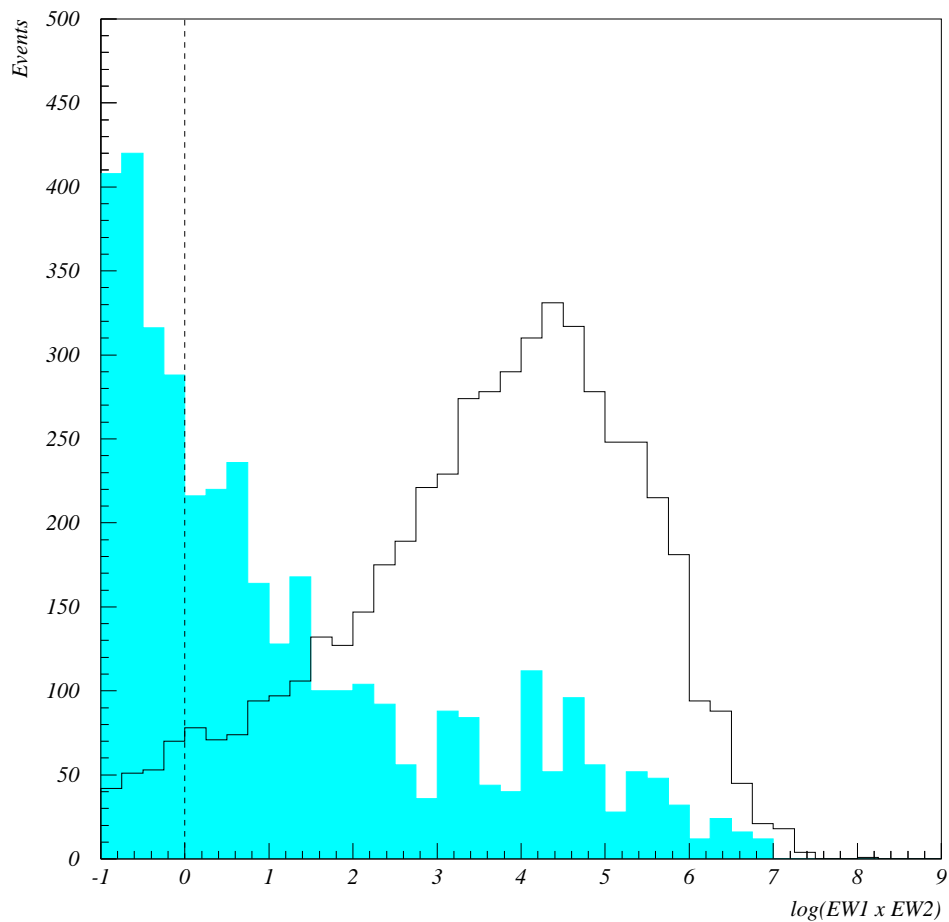}
\vspace*{-10pt}
\caption[Comparison of electron weights of $e^{+}e^{-}$ events and non-electron background events.]{Comparison of electron weights of $e^{+}e^{-}$ events and non-electron background events for all data taken in the $\chi_1$ region. The dashed line represents the electron weight cut at EW1 $\times$ EW2 $>$ 1, which was used in the analysis of most of the decay channels in this thesis.}
\label{fig:daq}
\end{center}
\end{figure}

Since all events of interest have either zero or two electrons in the final state the product of electron weights is most often used in E835 analyses which examine reactions containing the decays $J/\psi \rightarrow e^{+}e^{-}$ and $\psi^{\prime} \rightarrow e^{+}e^{-}$. Some of the analyses in this thesis also use requirements on the electron weights of individual electron tracks in order to give a particularly clean sample. The level of the electron weight cut for each particular analysis is done empirically to yield the maximum signal to background ratio; examples of such analyses are done in the next chapter for the various $e^{+}e^{-}$ channels. Once a cut level has been established an independent $J/\psi$ sample is used to measure the effiency of the electron weight cut, and a sample of $\eta_c^{\prime}$ search events are used to determine the background rejection power. The efficiencies of the various electron weight cuts used in the analyses in this theses are discussed in the next chapter.

\section{Simulations}

Several simulations are used in the analysis of data from E835. The most important of these is the GEANT 3.2 simulation used for simulated data sets and to compute efficiencies and response properties for individual detectors. When very large simulated data sets are required, mainly for all-neutral channels, events may also be simulated using a parametric description of the CCAL response which runs 10-20 times faster than GEANT and allows for enough event generation to study large backgrounds such as the $\pi^0\pi^0$ feeddown background to $\gamma\gamma$ events. Further discussion of this technique may be found in Ref.~\cite{nimpaper}; for the purposes of this thesis the GEANT simulation was used exclusively to compute the efficiencies of various cuts. A separate set of kinematic simulation programs was also written in FORTRAN specifically for the analyses performed in this thesis, in order to calculate geometric acceptances with large statistics.

%Chapter 4--Fermilab E835
\baselineskip=24pt
\chapter{Data Selection and Analysis}

\section{Decay Channels}

The present search for the $h_c (^1P_1)$ resonance was done by scanning in the range $\sqrt{s}$ = 3523 - 3529 MeV in the following decay channels:

% \vspace*{10pt}

(1) $p\bar p \rightarrow (^1P_1) \rightarrow J/\psi + X, J/\psi \rightarrow e^{+}e^{-}$ 

(2) $p\bar p \rightarrow (^1P_1) \rightarrow J/\psi + \pi^0, J/\psi \rightarrow e^{+}e^{-}, \pi^0 \rightarrow \gamma\gamma$

(3) $p\bar p \rightarrow J/\psi + \gamma, J/\psi \rightarrow e^{+}e^{-}$ 

(4) $p\bar p \rightarrow (^1P_1) \rightarrow J/\psi + \pi^0\pi^0, J/\psi \rightarrow e^{+}e^{-}, \pi^0\pi^0 \rightarrow 2(\gamma\gamma)$ 

(5) $p\bar p \rightarrow (^1P_1) \rightarrow \eta_c \gamma \rightarrow 3\gamma$

The decay channels (1-4) containing $J/\psi$ are the most promising because our detector is optimized for $e^{+}e^{-}$ and photon detection. The first channel $J/\psi + X$ includes all $p\bar p$ events which produce a $J/\psi$ and any number of gammas with an appropriate total energy. It is sensitive to any resonance which could produce $J/\psi$ as a decay product, and thus has the largest statistics of all the individual $J/\psi$ containing decay channels. The second channel, the isospin violating reaction $J/\psi + \pi^0$, is important because it was the channel in which E760 claimed observation of the $^1P_1$ in its scan of the 3523 - 3529 MeV energy range. The third channel is used only as a control channel, since the decay $^1P_1 \rightarrow J/\psi \gamma$ is strictly forbidden by C-parity conservation. The fourth channel, $^1P_1 \rightarrow J/\psi + 2\pi^0$ does not violate isospin, but we note that the phase space space for this reaction is small. The final channel ($p\bar p \rightarrow ^1P_1 \rightarrow\eta_c \gamma$) was investigated because it is the most promising of all decays which do not involve a $J/\psi$. This channel will be discussed separately at the end of this chapter.

% \vspace*{10pt}

The data used in this analysis was taken by experiment E835p (year 2000 run) at eighteen energy points in the region 3524.79 $\le \sqrt{s}$ MeV $\le$ 3527.29, which is approximately $\pm$ 1 MeV around the value of $\sqrt{s} = 3526.2$ MeV at which the resonance enhancement was reported by the E760 experiment. In order to constrain the background, data were also taken at three energies ($\sqrt{s} = 3523.3, 3528.7$ and 3529.1 MeV) which were about 1 MeV outside these limits. The total luminosity invested in the present $^1P_1$ search in E835p was 50.5 pb$^{-1}$. This is to be compared with 38.9 pb$^{-1}$ invested in the E835 $^1P_1$ search, and 15.9 pb$^{-1}$ invested in the E760 measurement. The typical mass resolution was $\sigma(\sqrt{s}) \approx$ 0.3 MeV, or FWHM$(\sqrt{s}) \approx$ 0.7 MeV, which was due entirely to the momentum spread of the antiproton beam which was measured by its Schottky noise spectrum. Table 4.1 lists the parameters of the present scan, and those of E835 (year 1997 run) and E760 are listed in Table 4.2. Energies shown in these tables are calculated by taking an average over an entire stack of the measured energies in each run, weighted by their respective luminosities. These weighted averages generally differ from the nominal energies of the stack, being higher by $\sim$ 300-500 keV.

\linespread{1.5}
\begin{table}[htbp]
\begin{center}
\begin{tabular}{|c|c|c|c|c|c|}
\hline
 & & & & & \\
 Stack No. & Nominal $\sqrt{s}$ & Measured $\sqrt{s}$ & $\cal{L}$ & $\cal{L}$(corr) & $\sigma_{beam}$ \\
 & & & & & \\
 & (MeV) & (MeV) & (nb$^{-1}$) & (nb$^{-1}$) & (MeV)  \\
 & & & & & \\
\hline
 & & & & & \\
 II-23 & 3523.3 & 3523.33 & 3058.6 & 2811.0 & 0.390 \\
 & & & & & \\
 II-39 & 3524.7 & 3524.79 & 2033.0 & 1820.7 & 0.297 \\
 & & & & & \\
 II-21 & 3525.1 & 3525.17 & 3709.6 & 3303.2 & 0.300 \\
 & & & & & \\
 II-20 & 3525.4 & 3525.46 & 4305.1 & 3840.4 & 0.309 \\
 & & & & & \\
 II-16 & 3525.7 & 3525.74 & 3316.0 & 2807.9 & 0.376 \\
 & & & & & \\
 II-37 & 3525.8 & 3525.88 & 1668.1 & 1506.5 & 0.256 \\
 & & & & & \\
 II-38 & 3525.8 & 3525.89 & 1638.3 & 1475.6 & 0.288 \\
 & & & & & \\
 II-18 & 3525.95 & 3526.02 & 3742.4 & 3353.0 & 0.294 \\
 & & & & & \\
 II-15 & 3526.2 & 3526.21 & 3674.5 & 3233.2 & 0.442 \\
 & & & & & \\
 II-24 & 3526.25 & 3526.25 & 510.5 & 464.5 & 0.346 \\
 & & & & & \\
 II-22 & 3526.2 & 3526.28 & 2627.9 & 2384.6 & 0.280 \\
 & & & & & \\
 II-36 & 3526.2 & 3526.29 & 1647.5 & 1458.5 & 0.268 \\
 & & & & & \\
 II-35 & 3526.2 & 3526.29 & 1488.5 & 1333.1 & 0.291 \\
 & & & & & \\
 II-41 & 3526.2 & 3526.30 & 1947.3 & 1775.3 & 0.253 \\
 & & & & & \\
 II-25 & 3526.25 & 3526.32 & 952.7 & 875.3 & 0.304 \\
 & & & & & \\
 II-26 & 3526.25 & 3526.42 & 2469.5 & 2238.0 & 0.328 \\
 & & & & & \\
 II-17 & 3526.5 & 3526.57 & 3707.5 & 3249.5 & 0.291 \\
 & & & & & \\
 II-19 & 3526.8 & 3526.89 & 3036.7 & 2720.2 & 0.311 \\
 & & & & & \\
 II-40 & 3527.2 & 3527.29 & 1223.3 & 1098.7 & 0.265 \\
 & & & & & \\
 II-28 & 3528.5 & 3528.61 & 1279.3 & 1135.7 & 0.364 \\
 & & & & & \\
 II-27 & 3528.9 & 3529.11 & 2467.0 & 2233.4 & 0.386 \\
 & & & & & \\
\hline
 & & & & & \\
 Total & & & 50503.3 & 45118.3 & \\
 & & & & & \\
\hline
\end{tabular}
\caption[Stack numbers, nominal energies, measured center of mass energies, luminosities, dead-time corrected luminosities and beam widths for each E835p data stack.]{Stack numbers, nominal energies, measured center of mass energies, luminosities, dead-time corrected luminosities and beam widths for each E835p data stack. The dead-time corrected luminosity is calculated for each stack using the random gate method described in Sec. 3.4.3.}
\label{tab:1}
\end{center}
\end{table}

\linespread{1.5}
\begin{table}[htbp]
\begin{center}
\begin{tabular}{|ccc|cccc|}
\hline
 & & & & & & \\
 E760 & & & E835 & & & \\
 & & & & & & \\
 Stack & $\sqrt{s}$ & $\cal{L}$ & Stack & $\sqrt{s}$(meas) & $\cal{L}$ & $\cal{L}$(corr)\\
 & & & & & & \\
 Number & (MeV) & (nb$^{-1}$) & Number & (MeV) & (nb$^{-1}$) & (nb$^{-1}$) \\
 & & & & & & \\
\hline
 & & & & & & \\
 3 & 3522.6 & 980 & I-61 & 3524.6 & 3717 & 3441 \\
 & & & & & & \\
 5 & 3523.5 & 490 & I-58 & 3525.2 & 2903 & 2768 \\
 & & & & & & \\
 2 & 3524.0 & 783 & I-62 & 3525.5 & 3532 & 3259 \\
 & & & & & & \\
 1 & 3524.3 & 823 & I-55 & 3525.7 & 3477 & 3237 \\
 & & & & & & \\
 9 & 3525.0 & 1041 & I-60 & 3525.8 & 2976 & 2750 \\
 & & & & & & \\
 13 & 3525.6 & 1310 & I-63 & 3525.8 & 3821 & 3509 \\
 & & & & & & \\
 18 & 3525.9 & 885 & I-59 & 3525.9 & 1196 & 1103 \\
 & & & & & & \\
 14 & 3526.1 & 1364 & I-19 & 3526.1 & 1992 & 1841 \\
 & & & & & & \\
 20 & 3526.1 & 980 & I-56 & 3526.1 & 2291 & 2117 \\
 & & & & & & \\
 11 & 3526.2 & 1337 & I-57 & 3526.2 & 3309 & 3041 \\
 & & & & & & \\
 22 & 3526.2 & 876 & I-64 & 3526.5 & 3234 & 2976 \\
 & & & & & & \\
 19 & 3526.2 & 940 & I-51 & 3526.5 & 976 & 904 \\
 & & & & & & \\
 16 & 3526.3 & 1017 & I-54 & 3526.6 & 649 & 601 \\
 & & & & & & \\
 21 & 3526.5 & 911 & I-52 & 3526.9 & 3094 & 2871 \\
 & & & & & & \\
 15 & 3526.6 & 1137 & I-53 & 3527.5 & 1396 & 1310 \\
 & & & & & & \\
 17 & 3527.2 & 1016 & I-50 & 3529.1 & 2328 & 2177 \\
 & & & & & & \\
\hline
 & & & & & & \\
Total & & 15890 & & & 38899 & 36064 \\
 & & & & & & \\
\hline
\end{tabular}
\caption[Center of mass energies and luminosities for each E760 data stack. Center of mass energies, luminosities, and dead-time corrected luminosities for each E835 data stack.]{Center of mass energies and luminosities for each E760 data stack. Center of mass energies, luminosities, and dead-time corrected luminosities for each E835 data stack. The dead-time corrected luminosity is calculated for each stack using the random gate method described in Sec. 3.4.3.}
\label{tab:1}
\end{center}
\end{table}

% \vspace*{10pt}

The hardware trigger for the first four decay channels was designed to accept all events with a $J/\psi \rightarrow e^{+}e^{-}$ decay. It required exactly two charged particle hits in the corresponding elements of the inner and outer scintillator hodoscopes, with at least one particle identified as an electron by the associated Cerenkov signal. Independently, it was required that two large energy clusters in the CCAL be separated by more than 90$^{\circ}$ in azimuth, and have an invariant mass greater than 2.2 GeV. Additional cuts were imposed in off-line event selection. 

\begin{figure}[htbp]
\begin{center}
\includegraphics[width=15cm]
{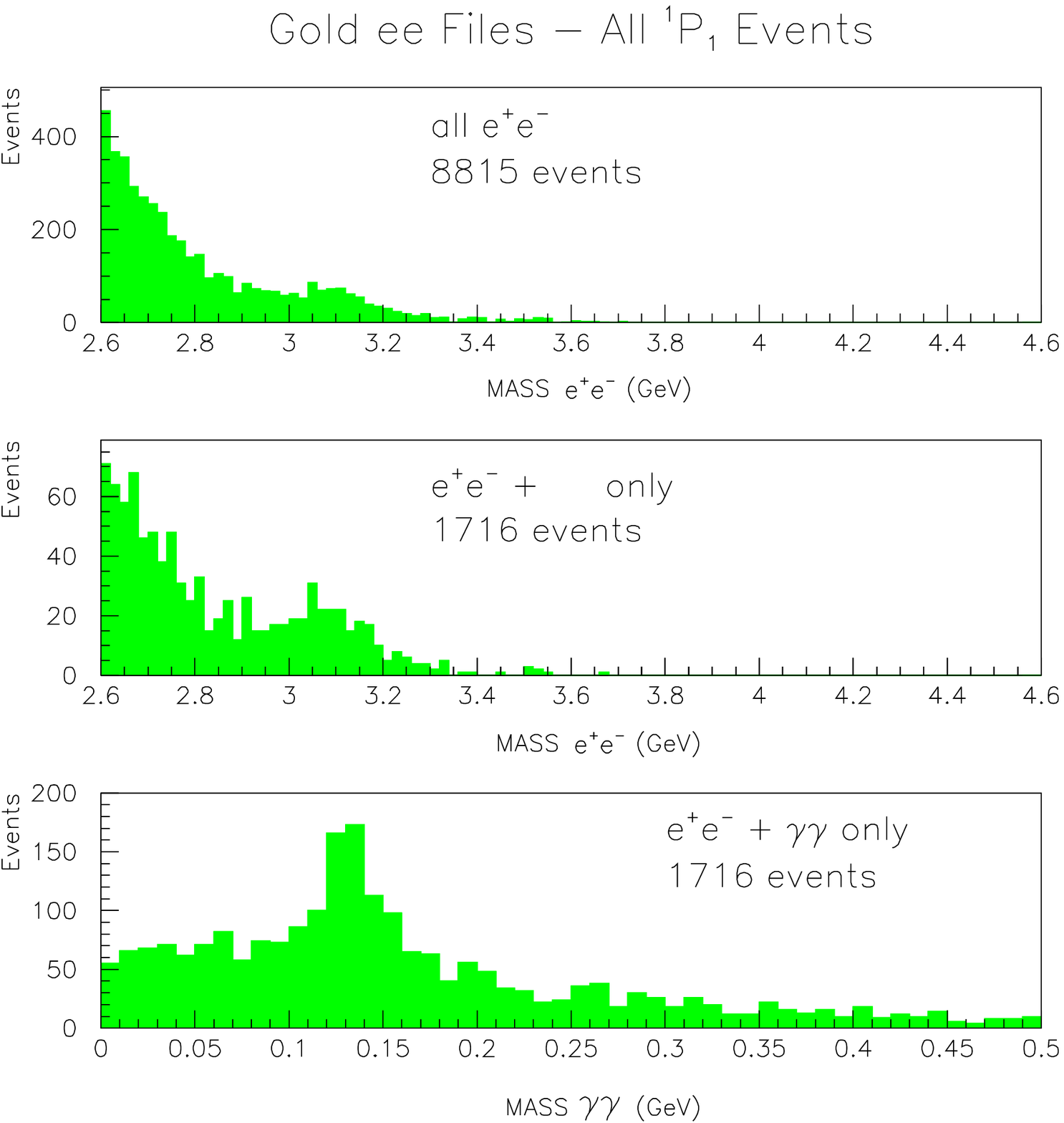}
\caption[All $e^{+}e^{-}$ events from the 21 stacks taken during the E835p search.]{All $e^{+}e^{-}$ events from the 21 stacks taken during the E835p $^1P_1$ search. The top figure shows the invariant mass of the two electrons for all $e^{+}e^{-}$ events. The $J/\psi$ peak is barely visible. The middle figure shows the $e^{+}e^{-}$ invariant mass for all $e^{+}e^{-} + \gamma\gamma$ events. The $J/\psi$ peak is observable. The bottom figure shows the $\gamma\gamma$ invariant mass for all $e^{+}e^{-} + \gamma\gamma$ events. The $\pi^0$ peak is clearly visible. In all cases the need for further cuts to define the $e^{+}e^{-}$ and $\gamma\gamma$ peaks clearly is indicated.}
\label{fig:1}
\end{center}
\end{figure}

The total $e^{+}e^{-}$ data set from all 50.5 pb$^-1$ taken in the E835p $^1P_1$ search is shown in the top panel of Figure 4.1. As can be seen from that figure, additional cuts are required to identify $J/\psi$, as well as $\pi^0$, cleanly. These cuts are discussed below.

\section{Analysis Cuts}

The cuts used for this analysis fall into three main categories; cuts for electron identification, cuts for gamma identification, and probability cuts for overall hypothesis testing. 

All channels involving $J/\psi$ require cuts for clean identification of electrons. This is done by the use of the electron weight cut. As described in Sec. 3.7, the electron weight takes into account pulse heights in the hodoscopes and Cerenkov counters, and the transverse energy distributions in the CCAL clusters. 

Several levels of electron weight cuts were examined in this analysis. The results are shown in Figure 4.2 (with any number of gammas), Figure 4.3 (with two gammas), Figure 4.4 (with one gamma), and Figure 4.5 (with four gammas). As can be seen from these figures, a clear $J/\psi$ peak at 3.1 GeV is visible with the minimum electron weight cut generally used in E835 and E835p, i.e. EW1 $\times$ EW2 $>$ 1. As can be seen from Figures 4.2-4.5, tightening the selection to require EW1 $\times$ EW2 $>$ 10 has the effect of making the $J/\psi$ peak sharper, but at some cost to the total number of events. Requiring both electron weights to each be separately greater than one is an even stronger cut. This is shown as the third choice of electron weight cuts in these figures. As shown later, for most channels, choosing the combined EW(1) $\times$ EW(2) $>$ 1 proved to be most {\it efficient} at giving a clean $J/\psi$ sample when used together with probability cuts as described later.

The gamma cuts were based mostly on the number of on-time clusters observed in the CCAL which were not associated with one of the two electrons. For the $J/\psi + X$ channel, no number-of-clusters cuts were imposed, and any number of gammas were allowed. The data for this channel is shown in Figure 4.2. For the $J/\psi + \pi^0$, $J/\psi + \gamma$, and $J/\psi + \pi^0\pi^0$ channels, 2, 1, and 4 gamma clusters were required. These data are shown in Figures 4.3, 4.4, and 4.5, respectively. Figure 4.3 also shows the invariant mass spectrum of the two gammas; the $\pi^0$ peak is already clearly observable even before additional cuts are made. One additional on-time cluster was also allowed in the $J/\psi + \pi^0$, $J/\psi + \gamma$, and $J/\psi + \pi^0\pi^0$ to account for a possible bremsstrahlung gamma as long as it made an angle of $< 10^{\circ}$ with one of the charged particles, had an energy $<$ 100 MeV, and did not make a $\pi^0$ (i.e. have invariant mass within the limits $135 \pm 35$ MeV) with any of the other gammas.

\newpage

\clearpage

\begin{figure}[htbp]
\begin{center}
\includegraphics[width=15cm]{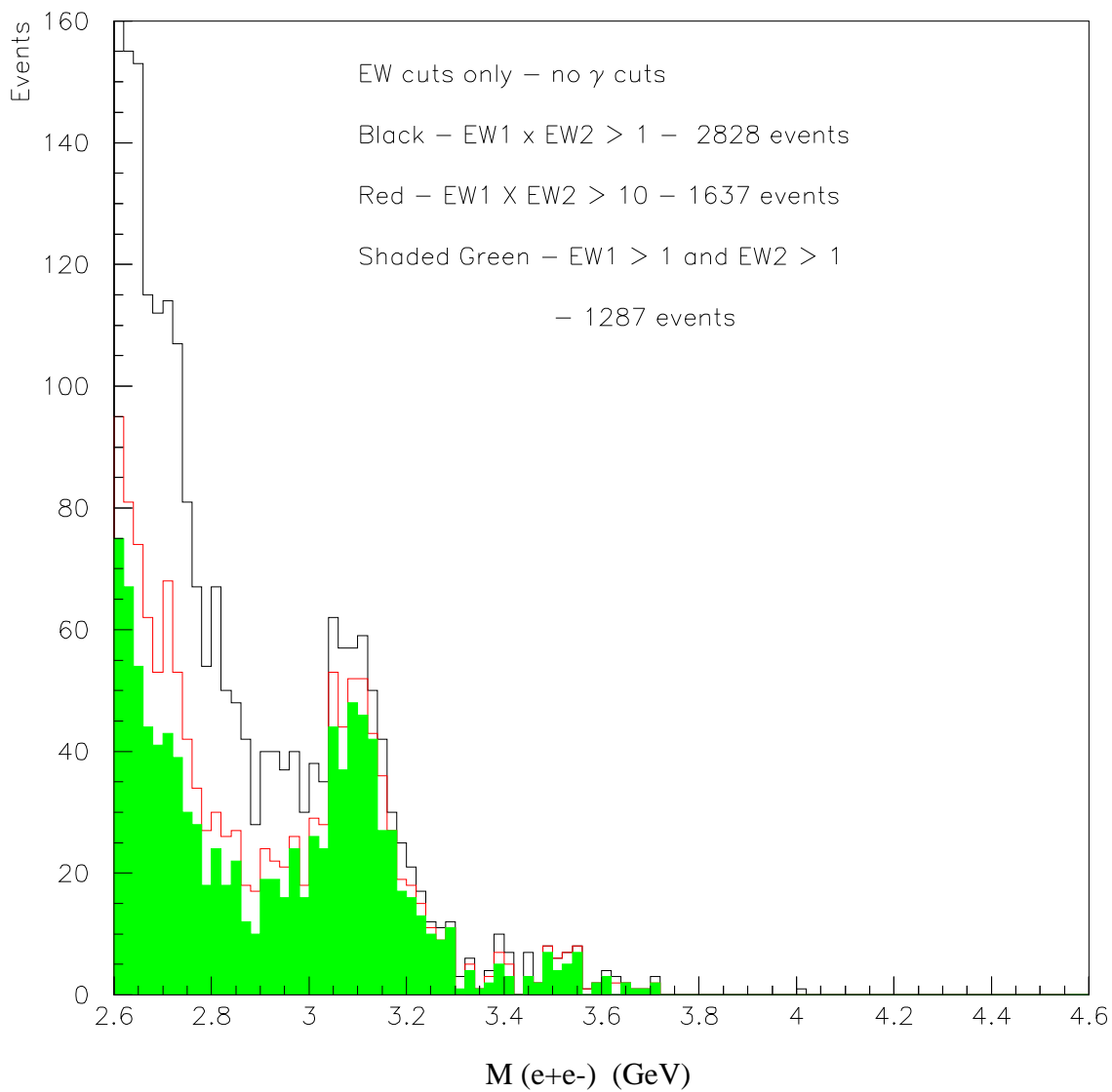}
\caption[Effect of electron weight cuts on the $M(e^{+}e^{-})$ distribution for $e^{+}e^{-}$ events with any number of gammas.]{Effect of electron weight cuts on the $M(e^{+}e^{-})$ distribution for $e^{+}e^{-}$ events with any number of gammas. The tightest cuts (shaded green) are those requiring both EW1 and EW2 to be separately greater than one. The effect of the tighter cuts is to reduce the number of non-$J/\psi$ events at the lower end of the $e^{+}e^{-}$ invariant mass spectrum. The small number of $e^{+}e^{-}$ events with an invariant mass in the 3.4-3.6 GeV region are ascribable to direct production or the proton form factor.}
\label{fig:2}
\end{center}
\end{figure}

\newpage

\clearpage

\begin{figure}[htbp]
\begin{center}
\includegraphics[width=15cm]{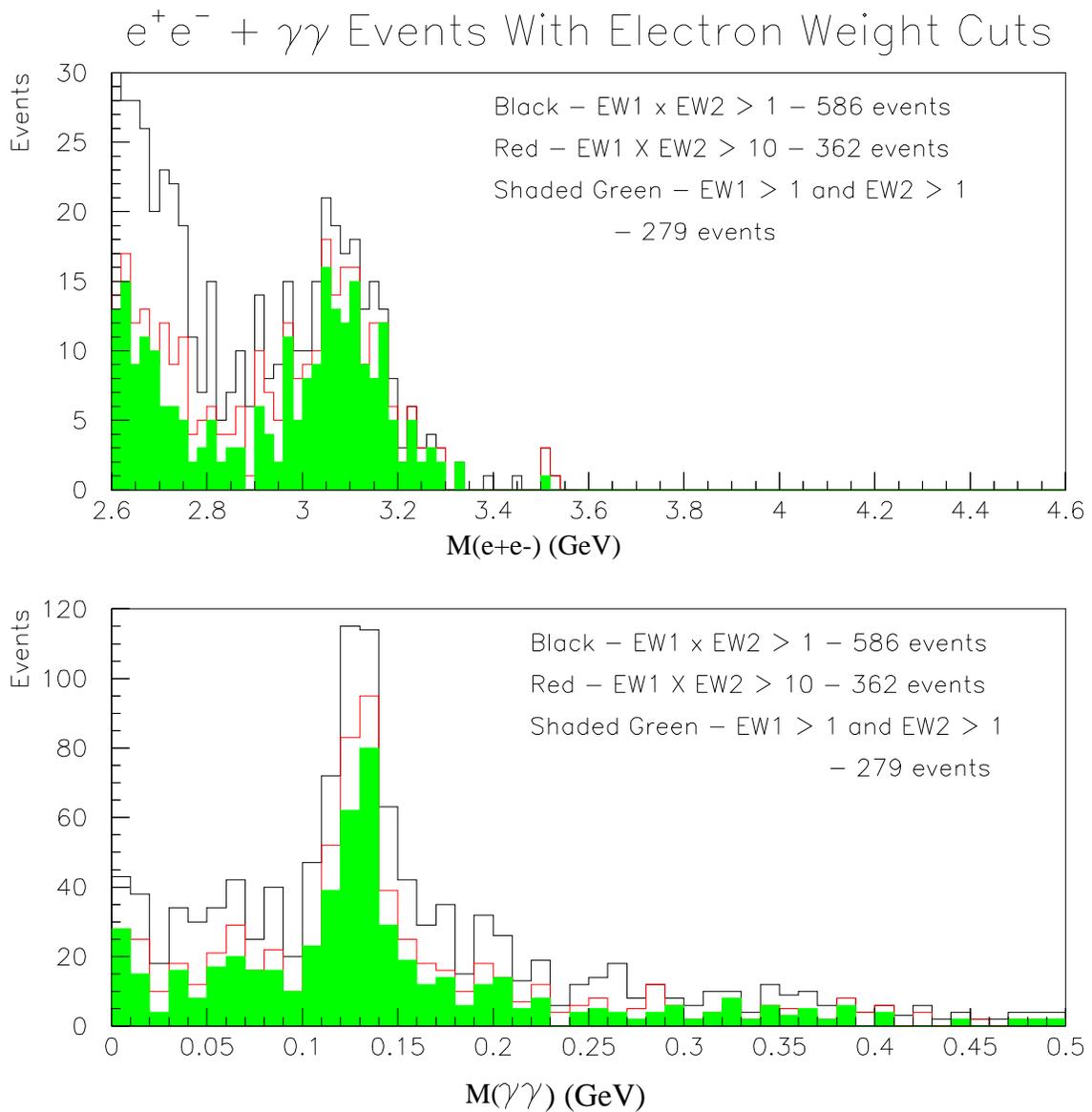}
\caption[Effect of electron weight cuts on the $M(e^{+}e^{-})$ distribution for $e^{+}e^{-} + \gamma\gamma$ events.]{Effect of electron weight cuts on the $M(e^{+}e^{-})$ distribution for $e^{+}e^{-} + \gamma\gamma$ events. Top figure shows the $e^{+}e^{-}$ invariant mass for three choices of electron weight cuts. Bottom figure shows the corresponding $\gamma\gamma$ invariant mass distributions.}
\label{fig:3}
\end{center}
\end{figure}

\newpage

\clearpage

\begin{figure}[htbp]
\begin{center}
\includegraphics[width=15cm]{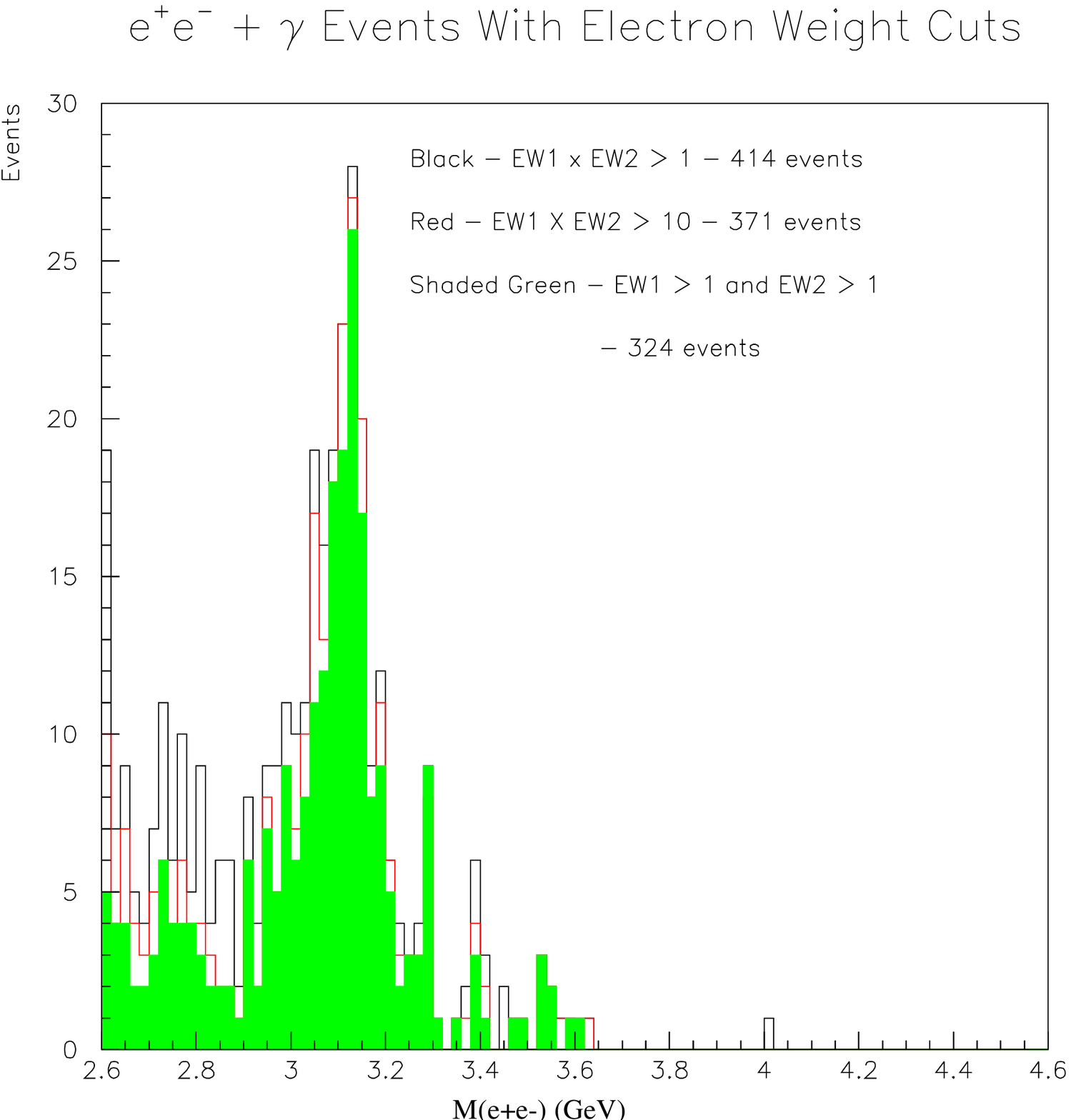}
\caption{Effect of electron weight cuts on the $M(e^{+}e^{-})$ distribution for $e^{+}e^{-} + \gamma$ events.}
\label{fig:11}
\end{center}
\end{figure}

\newpage

\clearpage

\begin{figure}[htbp]
\begin{center}
\includegraphics[width=15cm]{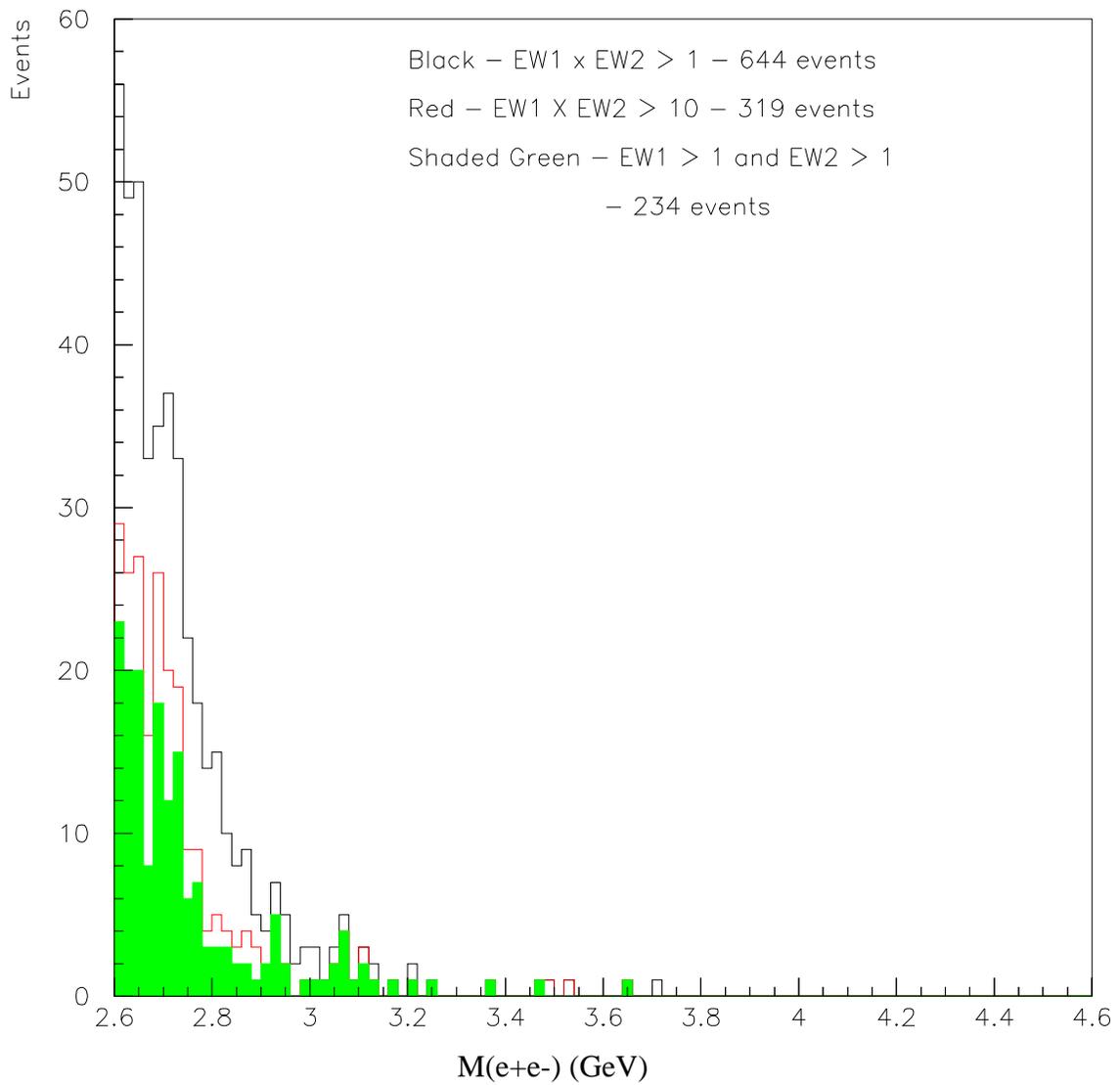}
\caption{Effect of electron weight cuts on the $M(e^{+}e^{-})$ distribution for $e^{+}e^{-} + 4\gamma$ events.}
\label{fig:23}
\end{center}
\end{figure}

\newpage

\clearpage

\subsection{Probability Cuts}

Probability distributions were constructed for each final state hypothesis and these are shown in Figures 4.6, 4.7, 4.8, and 4.9 for the $J/\psi + X$, $J/\psi + \pi^0$, $J/\psi + \gamma$, and $J/\psi + \pi^0\pi^0$ channels respectively. As can be seen from the first three figures, the probability distribution begins to flatten out at $\sim P = 0.05$, and thus in the final selection events with $P > 5\%$ were accepted for each of these three decay channels. The probability plot is also shown for the $J/\psi \pi^0\pi^0$ (4 cluster) channel, but in this case only two events survive the $P > 5\%$ cut.

The effect of various combinations of electron weight and probability cuts are shown in Figures 4.10-14 for the $J/\psi + X$, $J/\psi + \pi^0$, $J/\psi + \gamma$, and $J/\psi + \pi^0\pi^0$ channels, respectively. Figures 4.11 - 4.14 illustrate the fact that when combined with probability cuts, EW(1) $\times$ EW(2) $>$ 10 does no better in isolating $J/\psi$ or $\pi^0$ than EW(1) $\times$ EW(2) $>$ 1, but costs some in net counts. In the $J/\psi + X$ case, however, the tighter electron weight cut (EW1 $\times$ EW2 $>$ 10) is found to give a significantly cleaner $J/\psi$ mass spectrum, and was therefore used.

The probability distributions for the $J/\psi + X$ and $J/\psi + \gamma$ decays were calculated using a five constraint fit; the constraints being the total energy and momentum of the event and the invariant mass of the $e^{+}e^{-}$ pair which makes up the $J/\psi$. For the $J/\psi + \pi^0$ channel, the probability distributions had six constraints, the additional constraint being the invariant mass of the two gammas which make up the $\pi^0$.

\newpage

\clearpage

\begin{figure}[htbp]
\begin{center}
\includegraphics[width=15cm]
{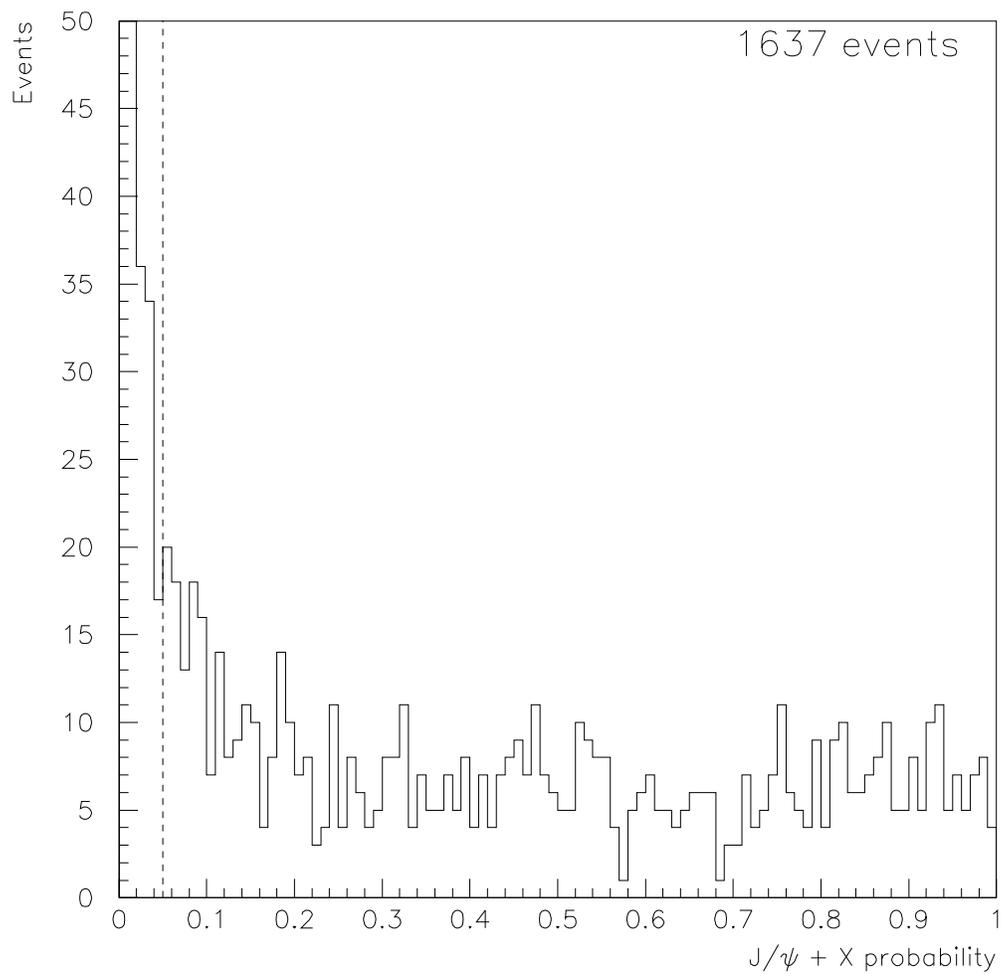}
\caption[Probability distribution for $J/\psi + X$ for all $e^{+}e^{-}$ events with EW1 $\times$ EW2 $>$ 10.]{Probability distribution for $J/\psi + X$ for all $e^{+}e^{-}$ events with EW1 $\times$ EW2 $>$ 10. Events with $P(J/\psi + X) \ge 0.05$ (dotted line) were used in the final analysis.}
\label{fig:2a}
\end{center}
\end{figure}

\newpage

\clearpage

\begin{figure}[htbp]
\begin{center}
\includegraphics[width=15cm]
{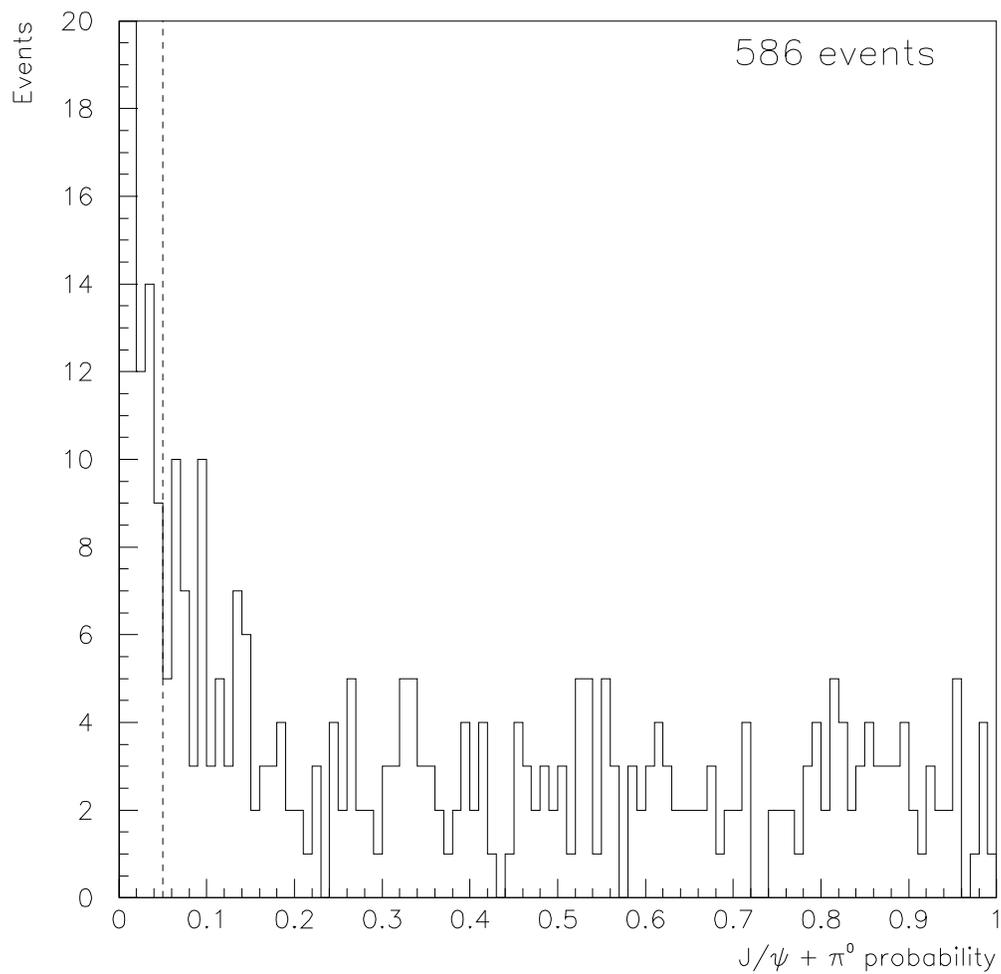}
\caption[Probability distribution for $J/\psi + \pi^0$ for all $e^{+}e^{-} + \gamma\gamma$ events with EW1 $\times$ EW2 $>$ 1.]{Probability distribution for $J/\psi + \pi^0$ for all $e^{+}e^{-} + \gamma\gamma$ events with EW1 $\times$ EW2 $>$ 1. Events with $P(J/\psi + \pi^0) \ge 0.05$ (dotted line) were used in the final analysis.}
\label{fig:3a}
\end{center}
\end{figure}

\newpage

\clearpage

\begin{figure}[htbp]
\begin{center}
\includegraphics[width=15cm]
{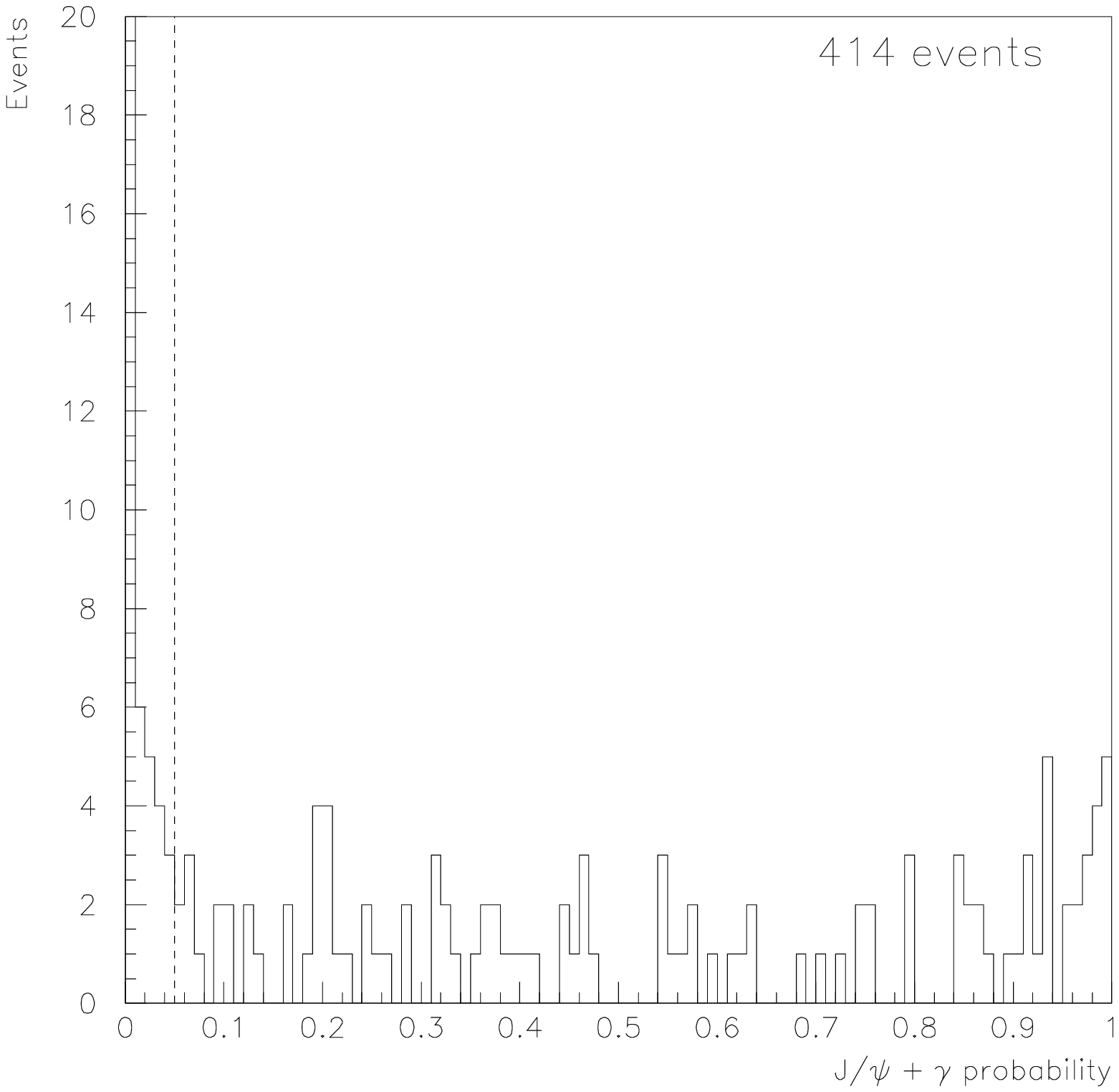}
\caption[Probability distribution for $J/\psi + \gamma$ for all $e^{+}e^{-} + \gamma$ events with EW1 $\times$ EW2 $>$ 1.]{Probability distribution for $J/\psi + \gamma$ for all $e^{+}e^{-} + \gamma$ events with EW1 $\times$ EW2 $>$ 1. Events with $P(J/\psi + \gamma) \ge 0.05$ (dotted line) were used in the final analysis.}
\label{fig:12}
\end{center}
\end{figure}

\newpage

\clearpage

\begin{figure}[htbp]
\begin{center}
\includegraphics[width=15cm]
{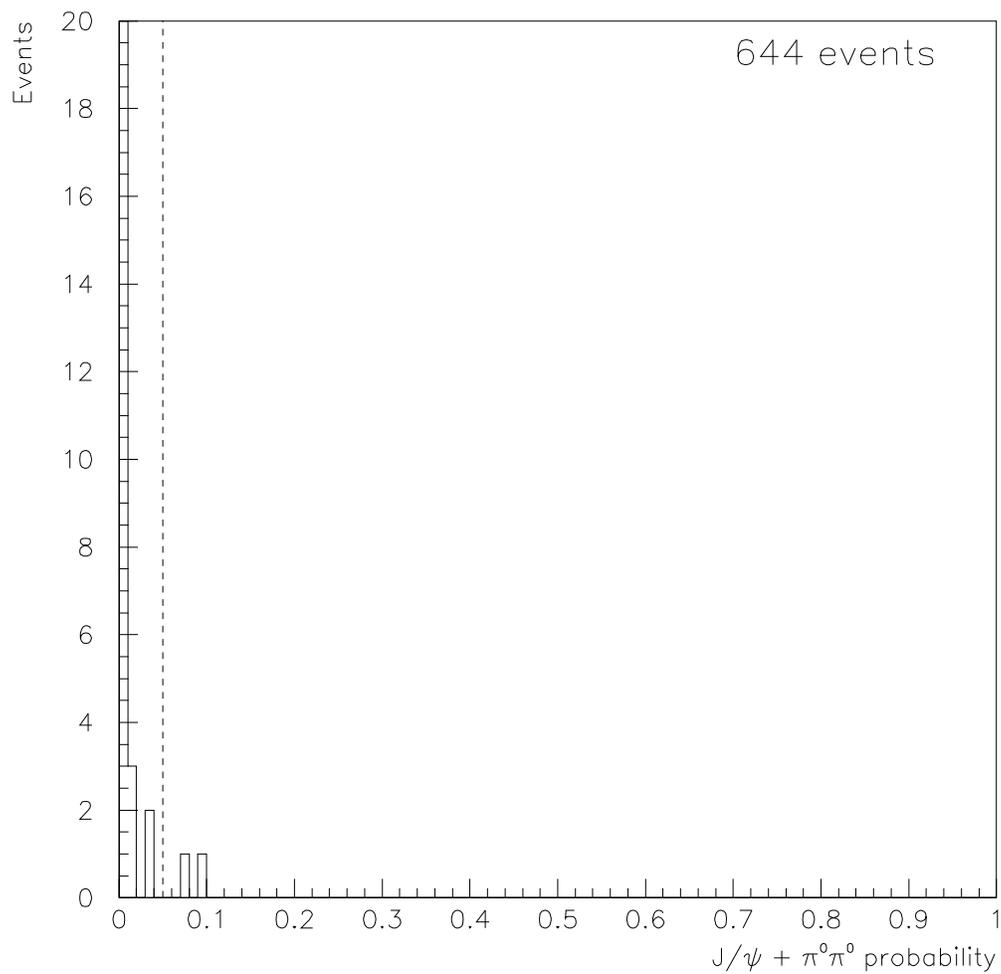}
\caption[Probability distribution for $J/\psi + \pi^0\pi^0$ for all $e^{+}e^{-} + 4\gamma$ events with EW1 $\times$ EW2 $>$ 1.]{Probability distribution for the channel $J/\psi + \pi^0\pi^0$ for all $e^{+}e^{-} + 4\gamma$ events with EW1$\times$ EW2 $>$ 1. Events with $P(J/\psi + \pi^0\pi^0) \ge 0.05$ (dotted line) were used in the final analysis. Only two events survive this cut.}
\label{fig:24}
\end{center}
\end{figure}

\newpage

\clearpage

\begin{figure}[htbp]
\begin{center}
\includegraphics[width=15cm]
{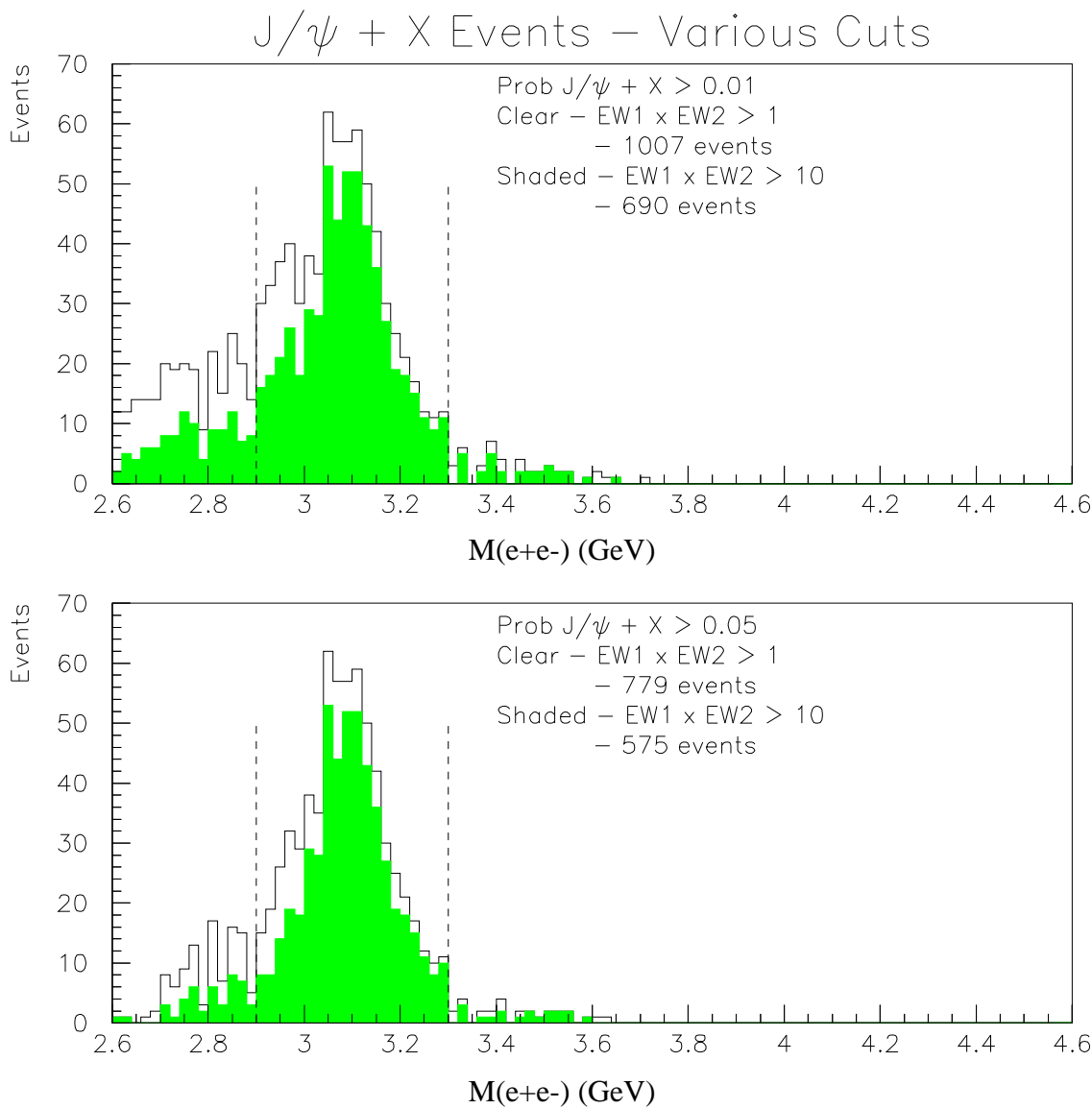}
\caption[Effect of various electron weight and probability cuts on the $M(e^{+}e^{-})$ distribution for $J/\psi + X$ events.]{Effect of various electron weight and probability cuts on the $M(e^{+}e^{-})$ distribution for $J/\psi + X$ events. The top plot shows the effect on the $e^{+}e^{-}$ invariant mass distribution for two choices of electron weight cuts with $p(J/\psi + X) > 0.01$. The bottom plot shows the same for $p(J/\psi + X) > 0.05$. The tightest cuts (EW1 $\times$ EW2 $>$ 10, $p(J/\psi + X) > 0.05)$ show the cleanest $J/\psi$ peak and are used in the final analysis. In addition, a $J/\psi$ mass cut is made to remove events outside the region 2.9 $< M(e^{+}e^{-}) <$ 3.3 GeV (dotted lines).}
\label{fig:4}
\end{center}
\end{figure}

\newpage

\clearpage

\begin{figure}[htbp]
\begin{center}
\includegraphics[width=15cm]
{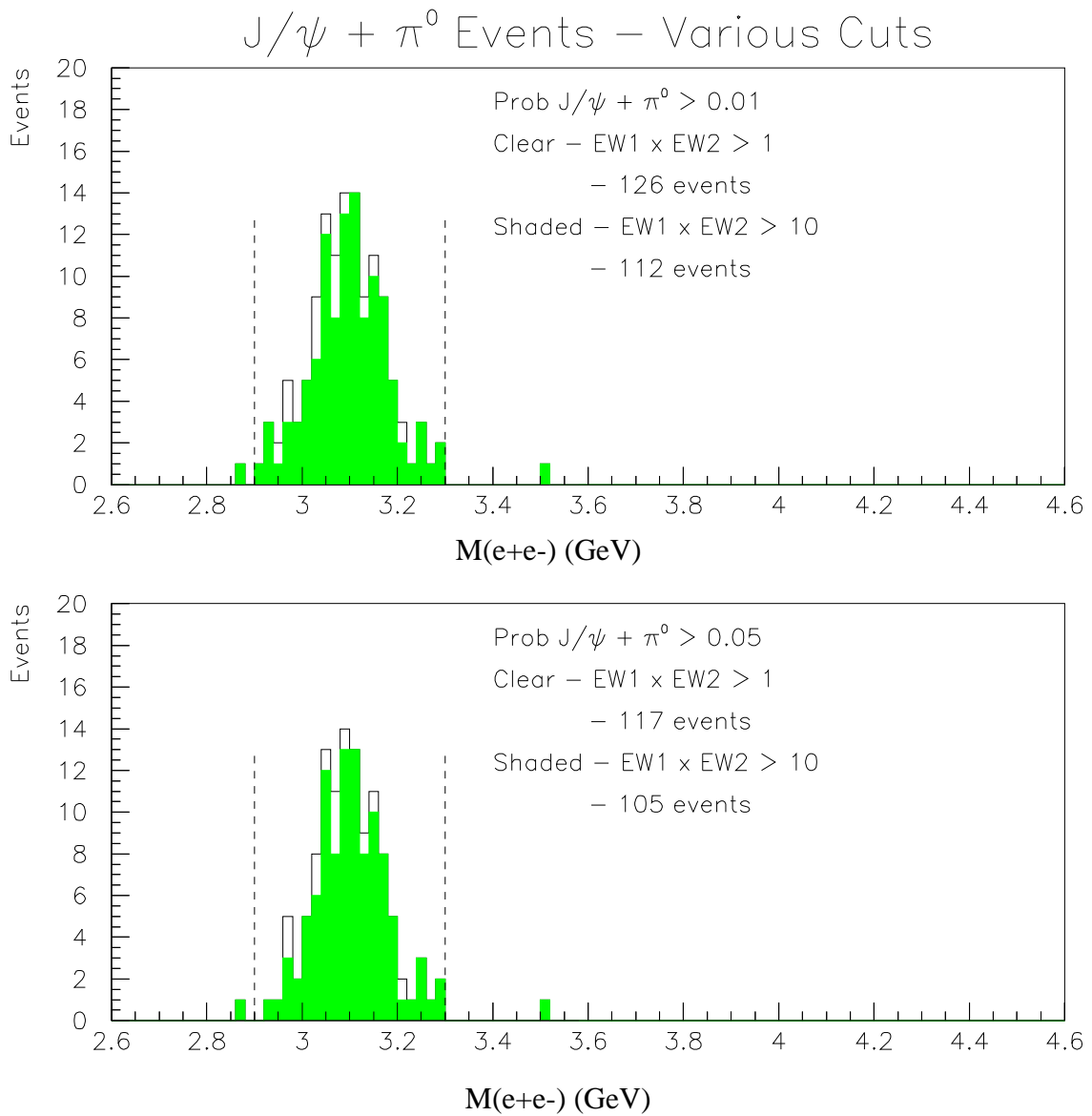}
\caption[Effect of various electron weight and probability cuts on the $M(e^{+}e^{-})$ distribution for $J/\psi + \pi^0$ events.]{Effect of various electron weight and probability cuts on the $M(e^{+}e^{-})$ distribution for $J/\psi + \pi^0$ events. The top plot shows the effect on the $e^{+}e^{-}$ invariant mass distribution for two choices of electron weight cuts with probability $P(J/\psi + \pi^0) > 0.01$. The bottom plot shows the same for probability $P(J/\psi + \pi^0) > 0.05$. The cuts (EW1 $\times$ EW2 $>$ 1, $P(J/\psi + \pi^0) > 0.05$) are used in the final analysis. $J/\psi$ mass cuts are made to remove events outside the region 2.9 $< M(e^{+}e^{-}) <$3.3 GeV (dotted lines).}
\label{fig:5}
\end{center}
\end{figure}

\newpage

\clearpage

\begin{figure}[htbp]
\begin{center}
\includegraphics[width=15cm]
{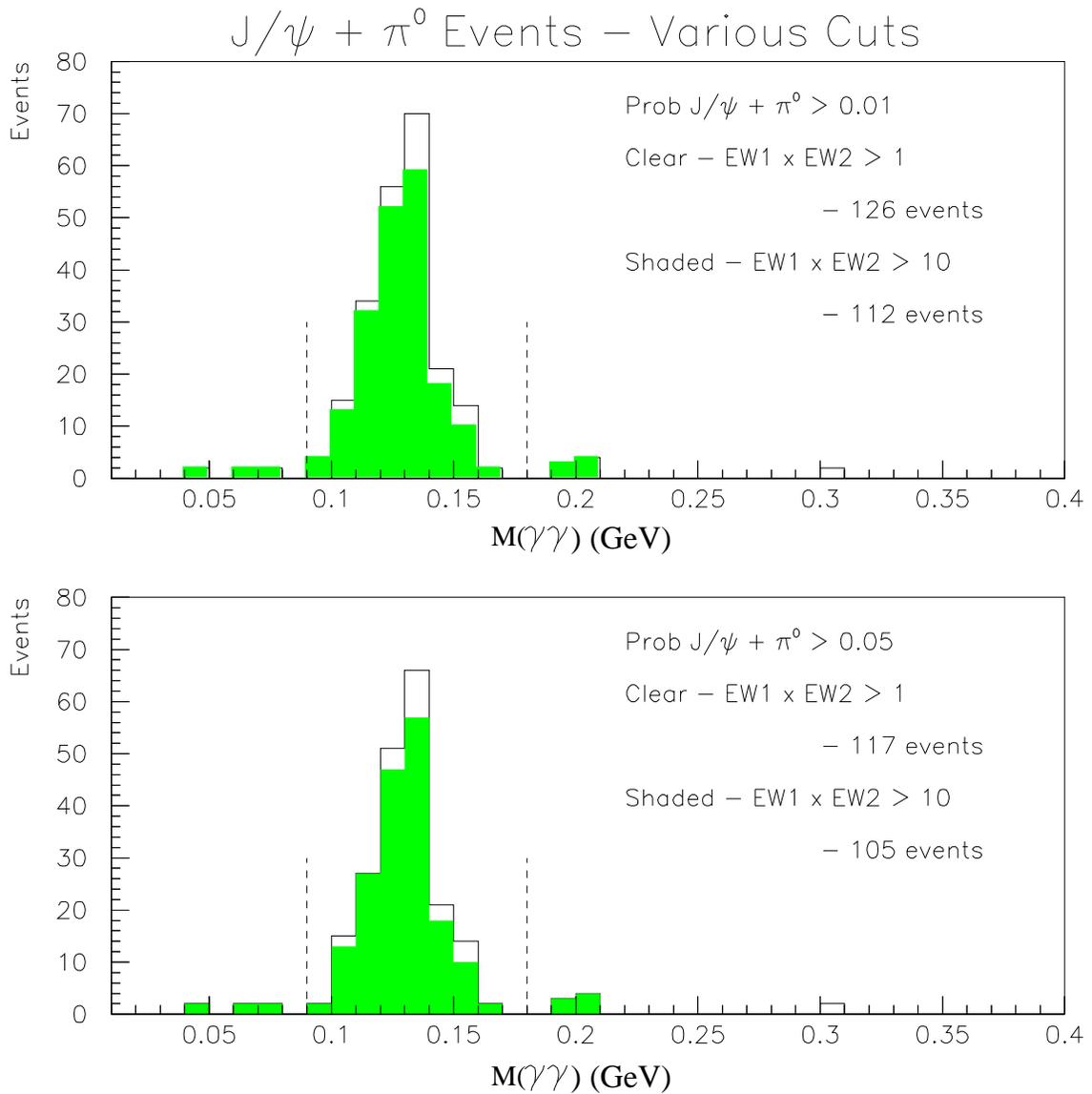}
\caption[Effect of various electron weight and probabililty cuts on the $M(\gamma_1\gamma_2)$ distribution for $J/\psi + \pi^0$ events.]{Effect of various electron weight and probabililty cuts on the $M(\gamma_1\gamma_2)$ distribution for $J/\psi + \pi^0$ events. The top plot shows the effect on the $\gamma\gamma$ invariant mass distribution for two choices of electron weight cuts with $P(J/\psi + \pi^0) > 0.01$. The bottom plot shows the same for $P(J/\psi + \pi^0) > 0.05$. $\pi^0$ mass cuts are made to remove events outside the region 90 $< M(gamma\gamma) <$ 180 MeV (dotted lines).}
\label{fig:6}
\end{center}
\end{figure}

\newpage

\clearpage

\begin{figure}[htbp]
\begin{center}
\includegraphics[width=15cm]
{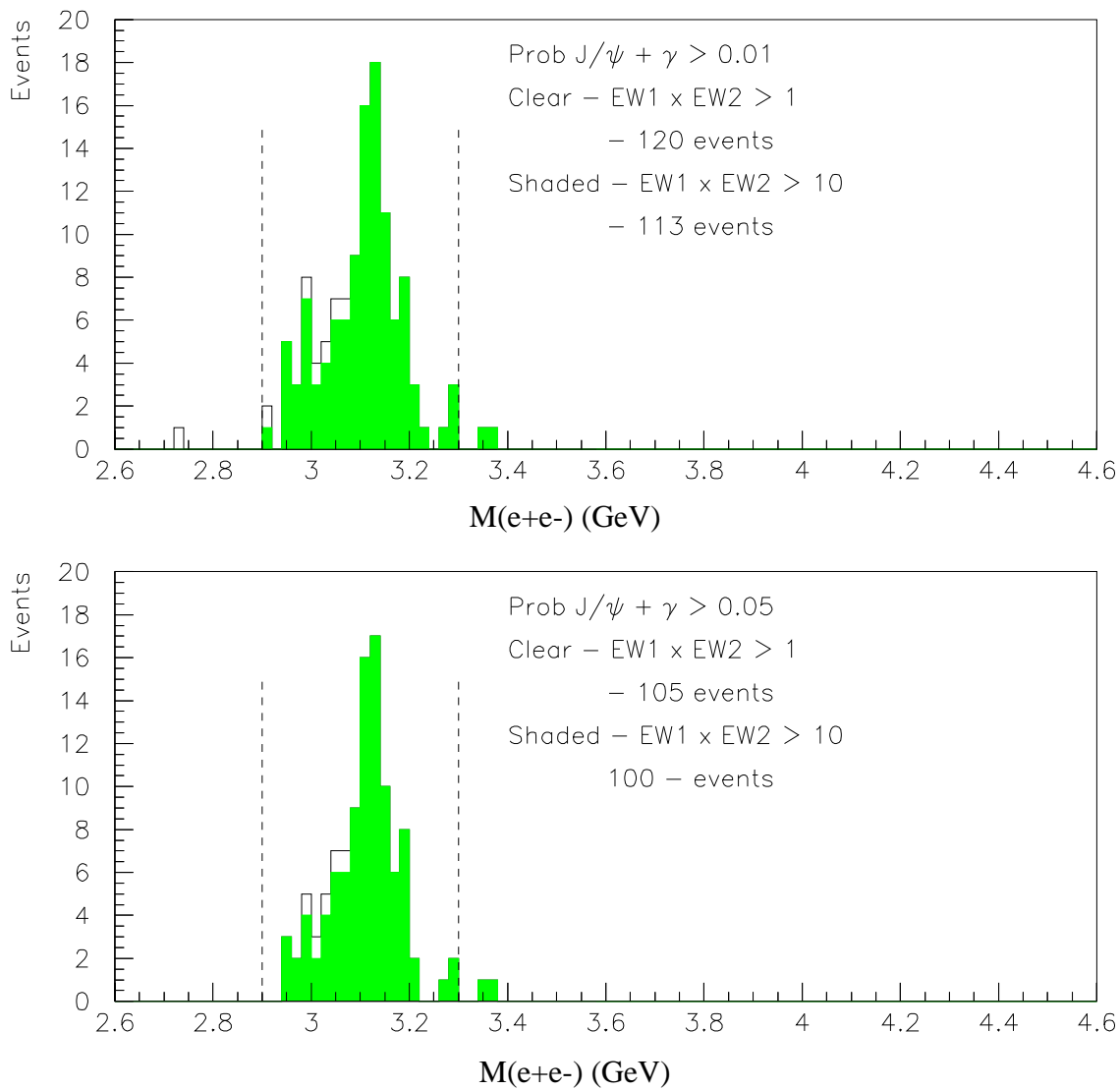}
\caption[Effect of various electron weight and probability cuts on the $M(e^{+}e^{-})$ distribution for $J/\psi + \gamma$ events.]{Effect of various electron weight and probability cuts on the $M(e^{+}e^{-})$ distributions for $J/\psi + \gamma$ events. The top plot shows $M(e^{+}e^{-})$ for $P(J/\psi + \gamma > 0.01)$, and the bottom plot for $P(J/\psi + \gamma > 0.05)$. A final mass cut is made to remove events outside the region 2.9 $< M(e^{+}e^{-}) <$ 3.3 GeV (dotted lines).}
\label{fig:13}
\end{center}
\end{figure}

\newpage

\clearpage

\begin{figure}[htbp]
\begin{center}
\includegraphics[width=15cm]
{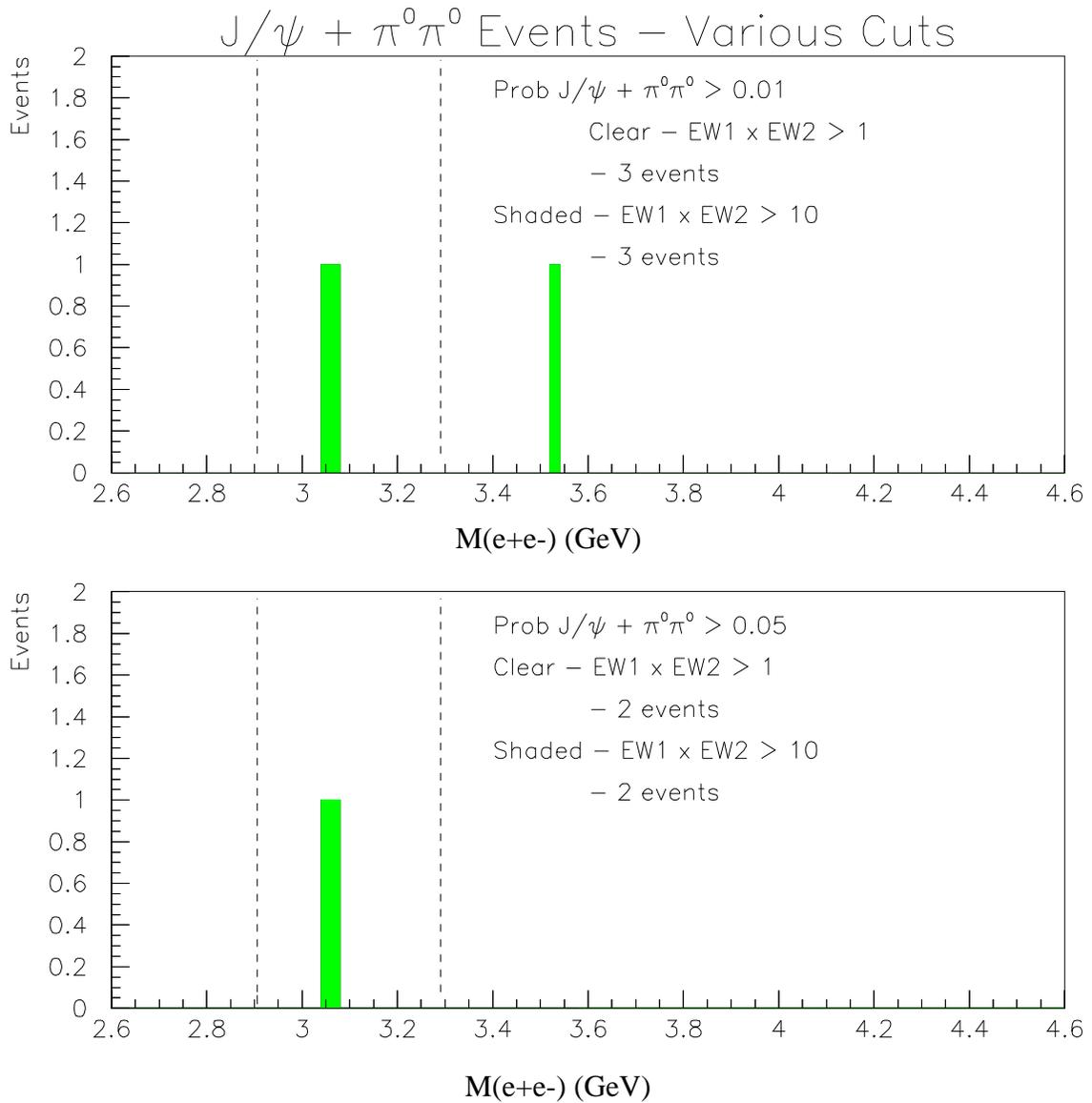}
\caption[Effect of various electron weight and probability cuts on the $M(e^{+}e^{-})$ distribution for $J/\psi + \pi^0\pi^0$ events.]{Effect of various electron weight and probability cuts on the $M(e^{+}e^{-})$ distribution for $J/\psi + \pi^0\pi^0$ events. The top plot shows the $M(e^{+}e^{-})$ spectrum for $P(J/\psi + \pi^0\pi^0) > 0.01$ and the bottom plot for $P(J/\psi + \pi^0\pi^0) > 0.05$. As in the previous channels, a final mass cut (dotted line) is made to remove events outside the region 2.9 $< M(e^{+}e^{-}) <$ 3.3 GeV.}
\label{fig:25}
\end{center}
\end{figure}

\newpage

\clearpage

Finally, motivated mainly by the tails present in the $M(e^{+}e^{-})$ distribution for the $J/\psi + X$ decay channel, a mass cut of 2.9 GeV $< m_{e^{+}e^{-}} <$ 3.3 GeV was made to define $J/\psi$. Although this cut is almost unnecessary in the $J/\psi + \pi^0$ and $J/\psi + \gamma$ channels, where it removes very few events, we also apply it for these decay channels for the sake of consistency with the $J/\psi + X$ decay channel. For the $J/\psi + \pi^0$ channel, an additional invariant mass cut was used for $\pi^0$. As indicated in Figure 4.12, the cut used was 90 MeV $< m_{\gamma\gamma} <$ 180 MeV. The same mass cuts were imposed on the $J/\psi + \pi^0\pi^0$ channel (Figure 4.14), although they had no effect, since only two events remained in this channel once the electron weight and $5\%$ probability cuts were used.

\section{Excitation Curves}

Having decided on the best cuts for event selection in each decay channel, the data for each stack was analyzed separately, in order to plot the excitation curves. In Table 4.3 the counts obtained in each final state are listed for each E835p energy point, along with the corresponding luminosities and beam widths. Table 4.4 shows the energy by energy event listing of the E760 experiment in the $J/\psi + \pi^0$ and $J/\psi \gamma$ channels, as published in {\it Phys. Rev. Lett.}[?]. Table 4.5 shows the same for the $J/\psi + \pi^0$ channel for the E835 (1997) run, using the same final cuts as were used for the current E835p analysis. Cross sections at each energy point were obtained by dividing the event counts $N_i$ at each energy by the luminosity ${\cal L}_i$ and the overall efficiency $\epsilon$ assumed to be constant over the $\le$ 5 MeV total range of $\sqrt{s}$ scanned. Account was taken of dead time corrections due to instantaneous luminosity variations using the random gate corrections described in Sec. 3.4.3. The efficiencies of the various cuts on the four $e^{+}e^{-}$ channels are shown in Table 4.6. In all cases, a trigger efficiency of $90\%$ is taken, as descibed in Sec. 3.4.2. The efficiency of the probability cuts for each final state was obtained by GEANT MC simulation, as were the acceptances for each decay channel. For the $J/\psi + X$ channel, which may have several final states, the efficiency was calculated as the average of those for the various individual final states. For example, the acceptance of the two electrons was 57.4$\%$ for the $J/\psi + \pi^0$, 56.3$\%$ for the $J/\psi + \pi^0$ channel, and $59.4\%$ for the $J/\psi + \pi^0\pi^0$ channel. The $J/\psi + X$ acceptance of $57.7\%$ is the average of these three. The electron weight and probability cut efficiencies for the $J/\psi + X$ channel were calculated similarly to obtain the overall efficiency for the $J/\psi + X$ channel. The overall efficiency (including acceptance) was as follows $\epsilon(J/\psi + X) = 39.1\%$, $\epsilon (J/\psi + \pi^0) = 17.2\%$, $\epsilon(J/\psi + \gamma) = 25.6\%$, and $\epsilon (J/\psi + \pi^0 + \pi^0) = 5.3\%$. 

Using the total efficiencies, the cross sections $\sigma_i = N_i/{\cal L}_i\epsilon)$ are obtained. These correspond to $\sigma(final state) \times B(J/\psi \rightarrow e^{+}e^{-})$, with final state $\equiv J/\psi + X, J/\psi + \gamma,$ or $J/\psi + \pi^0\pi^0$. These cross sections are listed for the $J/\psi + X$, $J/\psi + \pi^0$ and $J/\psi + \gamma$ channels in Table 4.7. Luminosities listed in Table 4.7 are after they have been corrected for dead time using the random gate. The table does not list $\sigma(J/\psi + \pi^0\pi^0$), since only two counts were observed in the entire scan. For $J/\psi + X$ the results from the tight EW selection are listed. The total efficiency for the $J/\psi + \pi^0$ channel in the E835 (year 1997 run) data was 16.2$\%$. For the E760 data, the total efficiencies (as reported in Ref. [?]) are 32.4$\%$ and 44.6$\%$ for the $J/\psi + \pi^0$ and $J/\psi + \gamma$ channels respectively.

The excitation curves corresponding to the cross sections in Table 4.7 for the decay channels $J/\psi + X$, $J/\psi + \pi^0$ and $J/\psi + \gamma$ are shown in Figures 4.15-4.18. For the sake of clarity the data for points with $\Delta \sqrt{s} \le 0.15$ MeV have been combined in all plots, although the subsequent determination of upper limits for branching ratios was done with the uncombined data. Figures 4.15 shows the excitation curve for the $J/\psi + X$ channel, using the two different choices of electron weight cuts described before. In neither case any enhancement is observed. A constant cross section fit gives $\sigma = 26.6 \pm 1.4$ pb ($\chi^2$/d.o.f. = 1.30) for the loose EW selection, and $\sigma = 25.2 \pm 1.4$ pb ($\chi^2$/d.o.f. = 1.75) for the tight EW selection.

Figure 4.16 shows the excitation curve for the $J/\psi + \pi^0$ channel, in which E760 found evidence for the observation of $^1P_1$ with a mass of $3526 \pm 0.15 \pm 0.2$ MeV. As can be seen from the figure, we do not find any evidence for an enhancement in this channel anywhere in the scan region. The data cross sections are fitted with a constant of 13.6 $\pm$ 1.4 pb with an excellent $\chi^2$/d.o.f. of 0.31. Figure 4.17 shows a comparison of the excitation curves for the $J/\psi + \pi^0$ channel for E835p, E835, and E760 data. The excitation curve for the $^1P_1$ for the forbidden decay channel $J/\psi + \gamma$ is shown in Figure 4.17. Again, no enhancement is observed, and $J/\psi + \gamma$ cross sections are fitted with a constant 8.8 $\pm$ 0.9 pb ($\chi^2$/d.o.f. = 0.66). Figure 4.18 shows a comparison of the excitation curves for the $J/\psi + \pi^0$ channel for E835p, E835, and E760 data.

\newpage

\clearpage

\linespread{1.5}
\begin{table}[htbp]
\begin{center}
\begin{tabular}{|c|c|c|ccccc|}
\hline
 & & & & & & & \\
 $\sqrt{s}$(meas) & $\cal{L}$(corr) & $\sigma_{beam}$ & Number & of events & observed & with & $J/\psi +$\\
 & & & & & & & \\
 (MeV) & (nb$^{-1}$) & (MeV) & $X$ (loose) & $X$ (tight) & $\pi^0$ & $\gamma$ & $\pi^0\pi^0$ \\
 & & & & & & & \\
\hline
 & & & & & & & \\
3523.33 & 2811.0 & 0.309 & 27 & 22 & 6  & 6  & 0 \\
 & & & & & & & \\
3524.79 & 1820.7 & 0.297 & 22 & 18 & 3  & 8  & 0 \\
 & & & & & & & \\
3525.17 & 3303.2 & 0.300 & 27 & 21 & 6  & 6  & 0 \\
 & & & & & & & \\
3525.46 & 3840.4 & 0.309 & 48 & 43 & 11 & 11 & 0 \\
 & & & & & & & \\
3525.74 & 2807.9 & 0.376 & 33 & 26 & 8  & 8  & 0 \\
 & & & & & & & \\
3525.88$^{*}$ & 1506.5 & 0.256 & 16 & 15 & 2  & 4  & 0 \\
 & & & & & & & \\
3525.89$^{*}$ & 1475.6 & 0.288 & 27 & 25 & 7  & 2  & 0 \\
 & & & & & & & \\
3526.02$^{*}$ & 3353.0 & 0.294 & 29 & 28 & 6  & 4  & 0 \\
 & & & & & & & \\
3526.21$^{\dag}$ & 3233.2 & 0.442 & 25 & 21 & 4  & 7  & 0 \\
 & & & & & & & \\
3526.25$^{\dag}$ & 464.5 & 0.346  & 5  & 6 & 1  & 0  & 0 \\
 & & & & & & & \\
3526.28$^{\dag}$ & 2384.6 & 0.280 & 29 & 29 & 6  & 4  & 0 \\
 & & & & & & & \\
3526.29$^{\dag}$ & 1458.5 & 0.268 & 22 & 20 & 7  & 6  & 0 \\
 & & & & & & & \\
3526.29$^{\dag}$ & 1333.1 & 0.291 & 19 & 18 & 5  & 3  & 0 \\
 & & & & & & & \\
3526.30$^{\dag}$ & 1775.3 & 0.253 & 29 & 26 & 5  & 6  & 0 \\
 & & & & & & & \\
3526.32$^{\dag}$ & 875.3 & 0.304  & 9  & 10 & 3  & 4  & 0 \\
 & & & & & & & \\
3526.42$^{\ddag}$ & 2238.0 & 0.328 & 27 & 24 & 5  & 4  & 0 \\
 & & & & & & & \\
3526.57$^{\ddag}$ & 3249.5 & 0.291 & 43 & 37 & 7  & 7  & 1 \\
 & & & & & & & \\
3526.89 & 2720.2 & 0.311 & 26 & 24 & 6  & 4  & 1 \\
 & & & & & & & \\
3527.29 & 1098.7 & 0.265 & 8  & 6 & 2  & 2  & 0 \\
 & & & & & & & \\
3528.61 & 1135.7 & 0.364 & 17 & 14 & 2  & 3  & 0 \\
 & & & & & & & \\
3529.11 & 2233.4 & 0.386 & 24 & 17 & 5  & 4  & 0 \\
 & & & & & & & \\
\hline
 & & & & & & & \\
Total & 45118.3 & & 512 & 450 & 107 & 103 & 2 \\
 & & & & & & & \\
\hline
\end{tabular}
\caption[Number of events observed for each E835p data stack in the channels $J/\psi + X$, $J/\psi + \gamma$, $J/\psi + \pi^0\pi^0$, and $J/\psi + \pi^0$.]{Number of events observed for each E835p data stack in the channels $J/\psi + X$, $J/\psi + \gamma$, $J/\psi + \pi^0\pi^0$, and $J/\psi + \pi^0$. Both tight and loose electron weight cuts for the $J/\psi + X$ channel are shown. Data stacks with $\Delta \sqrt{s} < 0.15$ MeV are combined for later analysis as follows: stacks marked $^{*}$ are combined to a single point at 3525.95 MeV, stacks marked $^{\dag}$ are combined to a single point at 3526.22 MeV, and stacks marked $^{\ddag}$ are combined to a single point at 3526.50 MeV.}
\label{tab:1}
\end{center}
\end{table}

\newpage

\clearpage

\linespread{1.5}
\begin{table}[htbp]
\begin{center}
\begin{tabular}{|c|c|cc|cc|}
\hline
 & & & & & \\
 $\sqrt{s}$ & $\cal{L}$ & Number of & events observed & $\sigma(J/\psi + \gamma)$ & $\sigma(J/\psi + \pi^0)$ \\
 & & & & & \\
 (MeV) & (nb$^{-1}$) & $J/\psi + \gamma$ & $J/\psi + \pi^0$ & (pb) & (pb) \\
 & & & & & \\
\hline
 & & & & & \\
3522.6 & 980 & 3 & 3 & 6.9 & 9.4 \\
 & & & & & \\
3523.5 & 490 & 5 & 0 & 22.9 & 0.0 \\
 & & & & & \\
3524.0 & 783 & 2 & 1 & 5.7 & 3.9 \\
 & & & & & \\
3524.3 & 823 & 4 & 3 & 10.9 & 11.3 \\
 & & & & & \\
3525.0 & 1041 & 8 & 1 & 17.2 & 3.0 \\
 & & & & & \\
3525.6 & 1310 & 3 & 4 & 5.1 & 9.4 \\
 & & & & & \\
3525.9 & 885 & 2 & 5 & 5.1 & 17.4 \\
 & & & & & \\
3526.1 & 1364 & 7 & 7 & 11.5 & 15.8 \\
 & & & & & \\
3526.1 & 980 & 2 & 7 & 4.6 & 22.0 \\
 & & & & & \\
3526.2 & 1337 & 2 & 9 & 3.4 & 20.8 \\
 & & & & & \\
3526.2 & 876 & 2 & 2 & 5.1 & 7.0 \\
 & & & & & \\
3526.2 & 940 & 4 & 6 & 9.5 & 19.7 \\
 & & & & & \\
3526.3 & 1017 & 4 & 9 & 8.8 & 27.3 \\
 & & & & & \\
3526.5 & 911 & 0 & 4 & 0.0 & 13.6 \\
 & & & & & \\
3526.6 & 1137 & 4 & 4 & 7.9 & 10.9 \\
 & & & & & \\
3527.2 & 1016 & 3 & 2 & 6.6 & 6.1 \\
 & & & & & \\
\hline
 & & & & & \\
Total & 15890 & 55 & 67 & & \\
 & & & & & \\
\hline
 & & & & & \\
Efficiency & ($\epsilon$) & & & $44.6\%$ & $32.4\%$ \\
 & & & & & \\
\hline
\end{tabular}
\caption[Number of events observed for each E760 data stack for the channels $J/\psi + \gamma$ and $J/\psi + \pi^0$.]{Number of events observed for each E760 data stack for the channels $J/\psi + \gamma$ and $J/\psi + \pi^0$. The cross sections have been obtained by using overall efficiencies $\epsilon = 32.4\%$ for the $J/\psi + \pi^0$ channel and $\epsilon = 44.6\%$ for the $J/\psi + \gamma$ channel, calculated as the products of the acceptances and analysis efficiencies given in the published paper.}
\label{tab:2}
\end{center}
\end{table}

\newpage

\clearpage

\linespread{1.5}
\begin{table}[htbp]
\begin{center}
\begin{tabular}{|c|c|c|c|}
\hline
 & & & \\
 $\sqrt{s}$(meas) & $\cal{L}$(corr) & Number of events observed & $\sigma(J/\psi + \pi^0)$ \\
 & & & \\
 (MeV) & (nb$^{-1}$) & $J/\psi + \pi^0$ & (pb) \\
 & & & \\
\hline
 & & & \\
 3524.6 & 3441 & 6 & 10.8\\
 & & & \\
 3525.2 & 2768 & 9 & 20.1\\
 & & & \\
 3525.5 & 3259 & 6 & 11.4\\
 & & & \\
 3525.7 & 3237 & 17 & 32.4\\
 & & & \\
 3525.8 & 2750 & 10 & 22.4\\
 & & & \\
 3525.8 & 3509 & 1 & 1.8\\
 & & & \\
 3525.9 & 1103 & 1 & 5.6\\
 & & & \\
 3526.1 & 2117 & 4 & 11.7\\
 & & & \\
 3526.2 & 3041 & 3 & 6.1\\
 & & & \\
 3526.5 & 2976 & 4 & 8.3\\
 & & & \\
 3526.5 & 904 & 2 & 13.7\\
 & & & \\
 3526.6 & 601 & 0 & 0.0\\
 & & & \\
 3526.9 & 2871 & 4 & 8.6\\
 & & & \\
 3527.5 & 1310 & 4 & 18.8\\
 & & & \\
 3529.1 & 2177 & 3 & 8.5\\
 & & & \\
\hline
 & & & \\
Total & 36064 & 74 & \\
 & & & \\
\hline
 & & & \\
Efficiency & ($\epsilon$) & & 16.2$\%$ \\
 & & & \\
\hline
\end{tabular}
\caption[Number of events observed for each E835 data stack for the $J/\psi + \pi^0$ channel.]{Number of events observed for each E835 data stack for the $J/\psi + \pi^0$ channel. The overall efficiency of this channel for the E835 data is $16.2\%$.}
\label{tab:4}
\end{center}
\end{table}

\newpage

\clearpage

\linespread{1.5}
\begin{table}[htbp]
\begin{center}
\begin{tabular}{|c|c|c|}
\hline
 & & \\
Channel & Cut & Efficiency \\
 & & \\
\hline
 & & \\
Common & Trigger & $90\%$ \\
 & & \\
 & $J/\psi$ mass cut & $98\%$ \\
 & & \\
(All but $J/\psi + X$) & EW1 $\times$ EW2 $>$ 1 & $95\%$ \\
 & & \\
\hline
 & & \\
$J/\psi + X$ & $p(J/\psi + X) > 5\%$ & $90\%$ \\
 & & \\
 loose & EW1 $\times$ EW2 $>$ 10 & 92$\%$ \\
 & & \\
 tight & EW1 $>$ 1, EW2 $>$ 1 & 85$\%$ \\
 & & \\
 & Acceptance & $57.7\%$ \\
 & & \\
 & Total (loose) & $42.1\%$ \\
 & & \\
 & Total (tight) & $39.1\%$ \\
 & & \\
\hline
 & & \\
$J/\psi + \pi^0$ & $p(J/\psi + \pi^0) > 5\%$ & $71\%$ \\
 & & \\
 & $N(\gamma) = 2$ & $93\%$ \\
 & & \\
 & $\pi^0$ mass cut & $95\%$ \\
 & & \\
 & Acceptance & $32\%$ \\
 & & \\
 & Total & $17.2\%$ \\
 & & \\
\hline
 & & \\
$J/\psi + \gamma$ & $p(J/\psi + \gamma) > 5\%$ & $77.5\%$ \\
 & & \\
 & $N(\gamma) = 1$ & $93.7 \%$ \\
 & & \\
 & Acceptance & $42.2\%$ \\
 & & \\
 & Total & $25.6\%$ \\
 & & \\
\hline
 & & \\
$J/\psi + \pi^0\pi^0$ & $p(J/\psi + \pi^0\pi^0) > 5\%$ & $56.8\%$ \\
 & & \\
 & $N(\gamma) = 4$ & $92.3\%$ \\
 & & \\
 & Acceptance & $12.0\%$ \\
 & & \\
 & Total & $5.3\%$ \\
 & & \\
\hline
\end{tabular}
\caption{Efficiencies of cuts used in decay channels containing $J/\psi$.}
\label{tab:3}
\end{center}
\end{table}

\newpage

\clearpage

\begin{table}[htbp]
\begin{center}
\begin{tabular}{|c|c|ccc|ccc|}
\hline
 & & & & & & & \\
 $\sqrt{s}$(meas) & $\Sigma{\cal L}$(corr) & $\Sigma N$ & for & $J/\psi +$ & $\sigma$ (pb) & for & $J/\psi (\rightarrow e^{+}e^{-})$ \\
 & & & & & & & \\
 (MeV) & (nb$^{-1}$) & $X$ & $\pi^0$ & $\gamma$ & $+ X$ & $+ \pi^0$ & $+ \gamma$ \\
 & & & & & & & \\
\hline
 & & & & & & & \\
3523.33 & 2811.0 & 22 & 6 & 6 & $20.0^{+4.3}_{-4.3}$ & $12.4^{+6.8}_{-4.5}$ & $8.3^{+4.6}_{-3.0}$\\
 & & & & & & & \\
3524.79 & 1820.7 & 18 & 3 & 8 & $25.2^{+6.8}_{-5.9}$ & $9.6^{+7.3}_{-6.1}$ & $17.2^{+7.1}_{-5.8}$ \\
 & & & & & & & \\
3525.17 & 3303.2 & 21 & 6 & 6 & $16.3^{+3.5}_{-3.5}$ & $10.6^{+5.8}_{-3.8}$ & $7.1^{+3.9}_{-2.6}$ \\
 & & & & & & & \\
3525.46 & 3840.4 & 43 & 11 & 11 & $28.6^{+4.4}_{-4.4}$ & $16.7^{+5.8}_{-4.8}$ & $11.2^{+3.9}_{-3.2}$ \\
 & & & & & & & \\
3525.74 & 2807.9 & 26 & 8 & 8 & $29.2^{+3.6}_{-3.6}$ & $17.1^{+4.8}_{-4.2}$ & $9.5^{+2.9}_{-2.5}$ \\
 & & & & & & & \\
3525.95 & 6335.1 & 68 & 15 & 8 & $22.4^{+4.2}_{-4.2}$ & $10.9^{+6.0}_{-4.0}$ & $4.9^{+3.4}_{-2.0}$ \\
 & & & & & & & \\
3526.22 & 11524.5 & 130 & 31 & 30 & $28.6^{+2.3}_{-2.3}$ & $15.2^{+2.5}_{-2.5}$ & $9.7^{+1.7}_{-1.7}$ \\
 & & & & & & & \\
3526.50 & 5487.5 & 61 & 12 & 11 & $29.1^{+4.8}_{-4.8}$ & $12.5^{+5.9}_{-4.9}$ & $8.4^{+4.0}_{-3.3}$ \\
 & & & & & & & \\
3526.89 & 2720.2 & 24 & 6 & 4 & $22.6^{+4.6}_{-4.6}$ & $12.8^{+7.0}_{-4.7}$ & $5.7^{+4.0}_{-2.4}$ \\
 & & & & & & & \\
3527.29 & 1098.7 & 6 & 2 & 2 & $14.0^{+7.6}_{-5.0}$ & $10.6^{+11.9}_{-6.7}$ & $7.1^{+8.0}_{-4.5}$ \\
 & & & & & & & \\
3528.61 & 1135.7 & 14 & 2 & 3 & $31.5^{+9.7}_{-8.3}$ & $10.2^{+11.5}_{-6.5}$ & $10.3^{+7.9}_{-6.5}$ \\
 & & & & & & & \\
3529.11 & 2233.4 & 17 & 5 & 4 & $19.5^{+5.5}_{-5.2}$ & $13.0^{+7.3}_{-5.9}$ & $7.0^{+4.9}_{-2.9}$ \\
 & & & & & & & \\
\hline
 & & & & & & & \\
Efficiency & ($\epsilon$) & & & & 39.1$\%$ & 17.2$\%$ & 25.6$\%$ \\ 
 & & & & & & & \\
\hline
%Average & \enskip $24.8 \pm 1.2$ & \enskip $13.9 \pm 1.6$ & \enskip $8.7 \pm 1%.0$\\
%\hline
\end{tabular}
\caption[Cross sections $\sigma(final state) \times B(J/\psi \rightarrow e^{+}e^{-})$ observed for the $final states$ $J/\psi + X$, $J/\psi + \pi^0$, and $J/\psi + \gamma$.]{Cross sections $\sigma(final state) \times B(J/\psi \rightarrow e^{+}e^{-})$ observed for the $final states$ $J/\psi + X$, $J/\psi + \pi^0$, and $J/\psi + \gamma$. Data taken at $\Delta \sqrt{s} < 0.15$ MeV have been combined. Errors in the individual cross sections are statistical (Poisson).}
\label{tab:test}
\end{center}
\end{table}

\newpage

\clearpage

\begin{figure}[htbp]
\begin{center}
\includegraphics[width=15cm]
{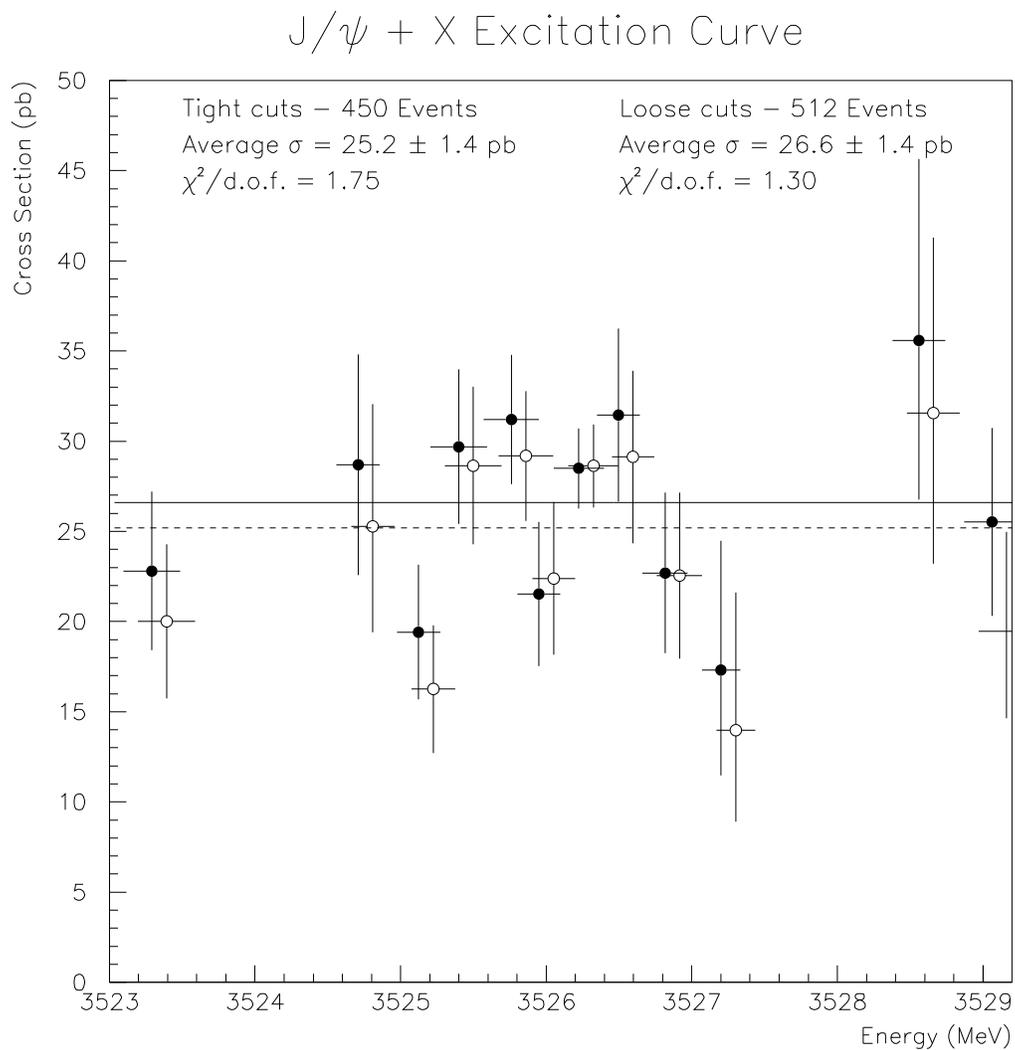}
\caption[Efficiency corrected excitation curve for the $J/\psi + X$ channel with loose electron weight cuts (EW1$\times$EW2$>$10) (closed circles) and tight electron weight cuts (EW1$>$1, EW2$>$1).]{Efficiency corrected excitation curve for the $J/\psi + X$ channel with {\bf loose} electron weight cuts (EW1$\times$EW2$>$10) (closed circles) and {\bf tight} electron weight cuts (EW1$>$1, EW2$>$1) (open circles - slightly displaced). The cross sections are consistent despite different cuts. Stacks with energies less than 0.15 MeV apart have been combined.}
\label{fig:8}
\end{center}
\end{figure}

\newpage

\clearpage

\begin{figure}[htbp]
\begin{center}
\includegraphics[width=15cm]
{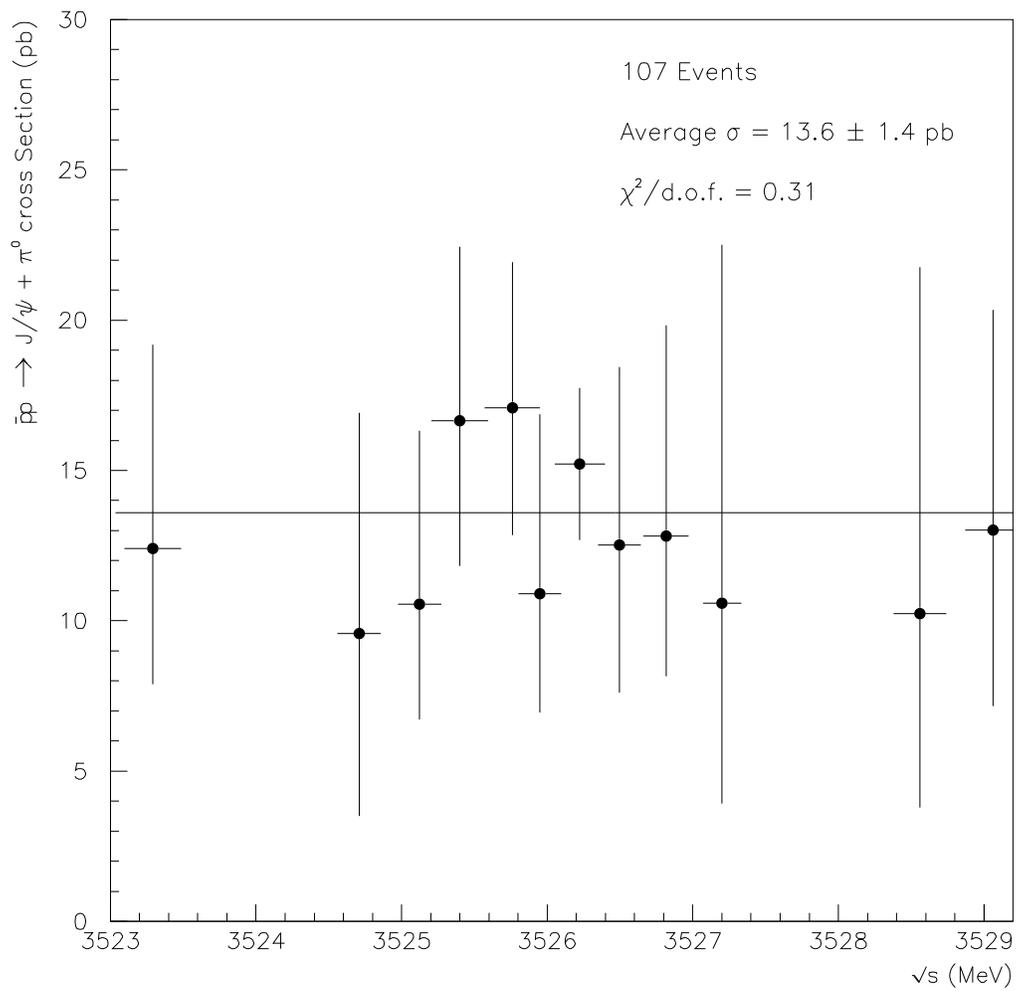}
\caption[Efficiency corrected excitation curve for the $J/\psi + \pi^0$ channel from the E835p data.]{Efficiency corrected excitation curve for the $J/\psi + \pi^0$ channel from the E835p data. Stacks with energies less than 0.15 MeV apart have been combined.}
\label{fig:9}
\end{center}
\end{figure}

\newpage

\clearpage

\begin{figure}[htbp]
\begin{center}
\includegraphics[width=15cm]
{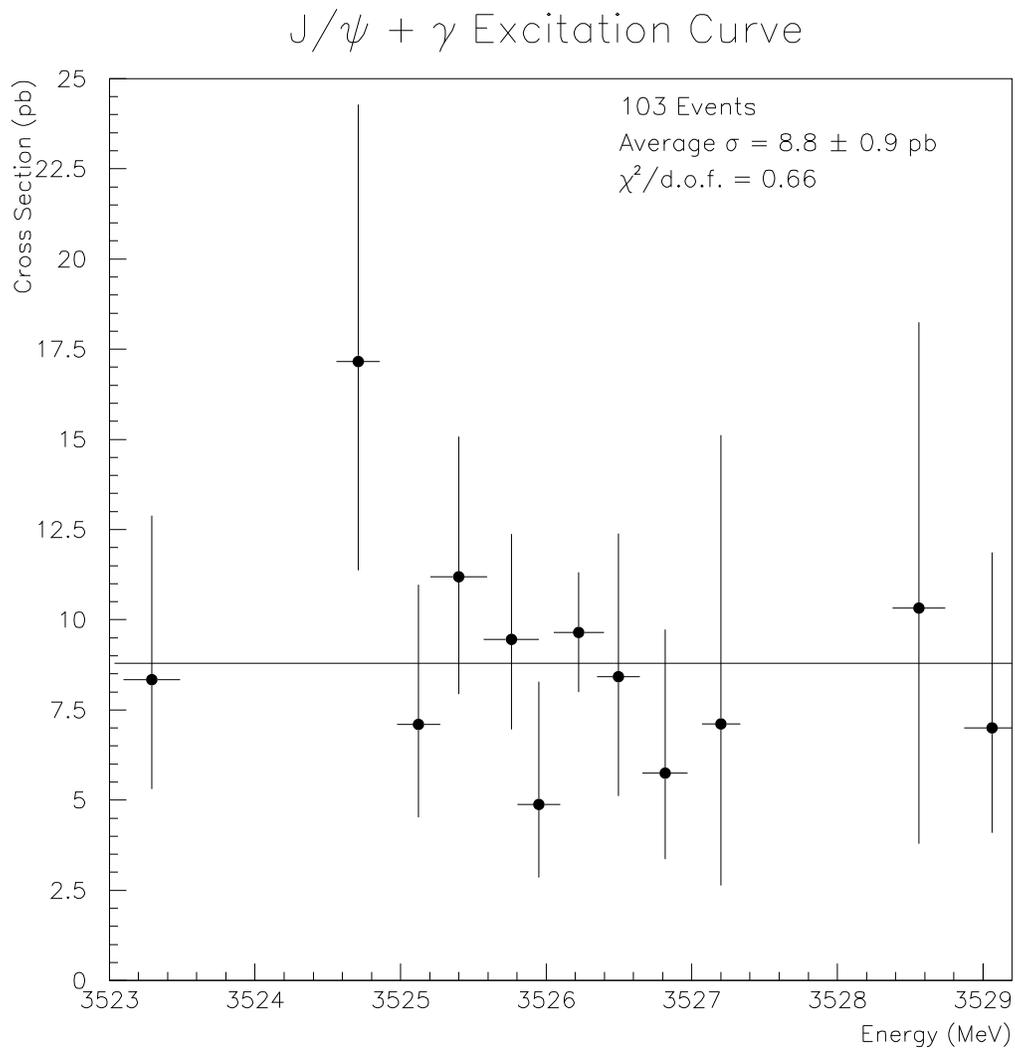}
\caption[Efficiency corrected excitation curve for the $J/\psi + \gamma$ channel from the E835p data.]{Efficiency corrected excitation curve for the $J/\psi + \gamma$ channel from the E835p data. Stacks with energies within 0.15 MeV of each other have been combined.}
\label{fig:14}
\end{center}
\end{figure}
\newpage

\clearpage

\begin{figure}[htbp]
\begin{center}
\includegraphics[width=12cm]
{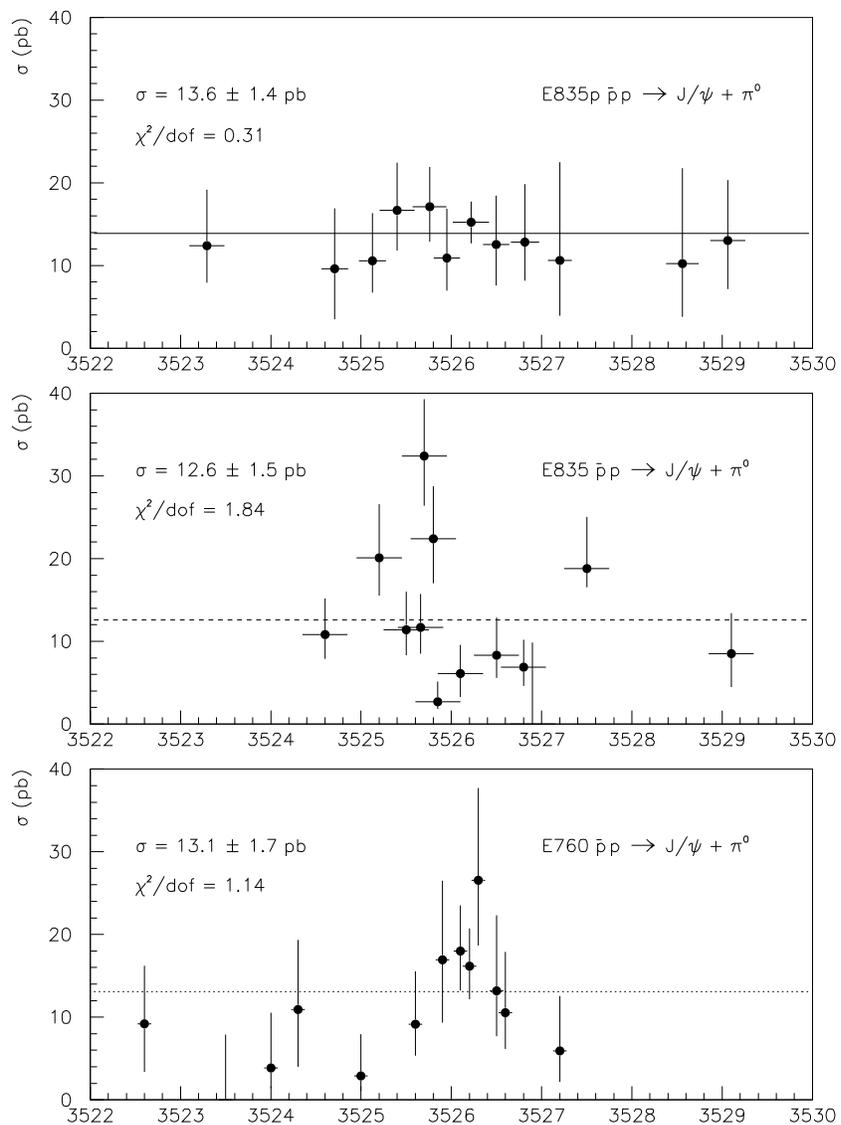}
\caption{Comparison of excitation curves for the $J/\psi + \pi^0$ channel for E835p (top plot), E835 (second plot), and E760 (third plot).}
\label{fig:10}
\end{center}
\end{figure}

\newpage

\clearpage

\section{Backgrounds}

As Table 4.7 and Figs. 4.15 - 18 show, there is no apparent enhancement in the excitation curves for the three reactions, and good fits are obtained assuming constant cross sections: $\sigma(J/\psi + X) = 25.2 \pm 1.4$ pb, $\sigma(J/\psi + \pi^0) = 13.6 \pm 1.4$ pb, and $\sigma(J/\psi + \gamma) = 8.8 \pm 0.9$ pb. The observed sum of cross sections $\sigma(J/\psi + \pi^0) + \sigma(J/\psi + \gamma) = 22.4 \pm 1.9$ pb essentially exhausts the observed $\sigma(J/\psi + X) = 25.2 \pm 1.4$ pb.

These observations suggest that the cross sections observed in all three decay channels arise primarilly from continuum production and tails of distant resonances. As illustrated by Table 4.8 and Fig. 4.19, an estimate of continuum $\sigma(J/\psi + \pi^0)$ is provided by its value of $11.9 \pm 1.9$ pb which we obtain from our scan in the region $\sqrt{s} = 3415 \pm 15$ MeV containing the $\chi_0$ resonance which is forbidden to decay in this channel by C-parity conservation. To obtain these results, cuts identical to those used in the $^1P_1$ scan were used. From the known branching ratios for $\chi_J \rightarrow J/\psi + \gamma$ we calculate the total contribution at 3526.2 MeV from the tails of the $\chi_J$ resonances as $\sigma_{\chi}(J/\psi + \gamma) = 5.5 \pm 0.9$ pb, of which $\sim 2.9$ pb comes from the tail of $\chi_2$, $\sim 2.1$ pb comes from the tail of $\chi_1$, and $\sim 0.5$ pb comes from the tail of $\chi_0$. It is estimated by Monte Carlo simulation that an additional contribution of $2.2 \pm 0.3$ pb arises from $J/\psi + \pi^0$ events in which a photon is lost, bringing the total estimated background to $\sigma(J/\psi + \gamma) = 7.7 \pm 1.0$ pb. Thus, both $\sigma(J/\psi + \pi^0)$ and $\sigma(J/\psi + \gamma)$ observed in the present measurements are indeed well accounted for by the 'background' processes described above, as is $\sigma(J/\psi + X)$ which is made up primarilly by the sum of the two channels. This leaves little room for an observable resonance enhancement in either $J/\psi + \pi^0$ or $J/\psi + X$ channels.

\newpage

\clearpage

\linespread{1.5}
\begin{table}[htbp]
\begin{center}
\begin{tabular}{|c|c|c|c|c|c|}
\hline
 & & & & & \\
 Stack & $\sqrt{s}$(meas) & $\cal{L}$ & $\cal{L}$(corr) & Events & $\sigma(J/\psi + \pi^0)$ \\
 & & & & & \\
 Number & (MeV) & (nb$^{-1}$) & (nb$^{-1}$) & observed & (pb) \\
 & & & & & \\
\hline
 & & & & & \\
 II-47-b & 3339.5 & 659 & 585 & 1 & 10.9 \\
 & & & & & \\
 II-47-a & 3365.0 & 1424 & 1287 & 0 & 0.0 \\
 & & & & & \\
 II-10 & 3384.4 & 1630 & 1464 & 2 & 8.6 \\
 & & & & & \\
 II-31 & 3384.8 & 3371 & 3038 & 4 & 8.4 \\
 & & & & & \\
 II-42 & 3392.0 & 1431 & 1276 & 0 & 0.0 \\
 & & & & & \\
 II-34-b & 3400.1 & 1483 & 1352 & 2 & 9.4 \\
 & & & & & \\
 II-12-b & 3404.7 & 76 & 68 & 0 & 0.0 \\
 & & & & & \\
 II-14-b & 3406.1 & 2481 & 2243 & 5 & 14.2 \\
 & & & & & \\
 II-32-b & 3409.1 & 1134 & 1057 & 0 & 0.0 \\
 & & & & & \\
 II-12-a & 3410.2 & 1613 & 1461 & 2 & 8.7 \\
 & & & & & \\
 II-33 & 3413.8 & 2959 & 2616 & 4 & 9.7 \\
 & & & & & \\
 II-9 & 3415.0 & 2352 & 2131 & 8 & 23.9 \\
 & & & & & \\
 II-13 & 3415.9 & 2801 & 2498 & 5 & 12.7 \\
 & & & & & \\
 II-7 & 3417.9 & 1456 & 1345 & 2 & 9.5 \\
 & & & & & \\
 II-32-a & 3422.1 & 2512 & 2250 & 3 & 8.5 \\
 & & & & & \\
 II-11 & 3425.9 & 1820 & 1631 & 6 & 23.4 \\
 & & & & & \\
 II-34-a & 3430.1 & 1438 & 1260 & 0 & 0.0 \\
 & & & & & \\
 II-8 & 3469.9 & 2513 & 2302 & 1 & 2.8 \\
 & & & & & \\
\hline
 & & & & & \\
Total & & 33153 & 29864 & 45 & $<\sigma> = 9.6 \pm 1.4$ \\
 & & & & & \\
\hline
 & & & & & \\
Efficiency & ($\epsilon$) & & & & 15.7$ \pm 0.4\%$ \\
 & & & & & \\
\hline
\end{tabular}
\caption[Number of events observed and corresponding cross sections for each E835 data stack in the $\chi_0$ region for the $J/\psi + \pi^0$ channel.]{Number of events observed and corresponding cross sections for each E835 data stack in the $\chi_0$ region for the $J/\psi + \pi^0$ channel. The 'a' and 'b' labels denote stacks which were split into two energies by decelerating the $\bar p$ beam between runs. The overall efficiency of this channel for the E835 data in this energy region is $15.7\% \pm 0.4\%$.}
\label{tab:4}
\end{center}
\end{table}

\newpage

\clearpage

\begin{figure}[htbp]
\begin{center}
\includegraphics[width=15cm]
{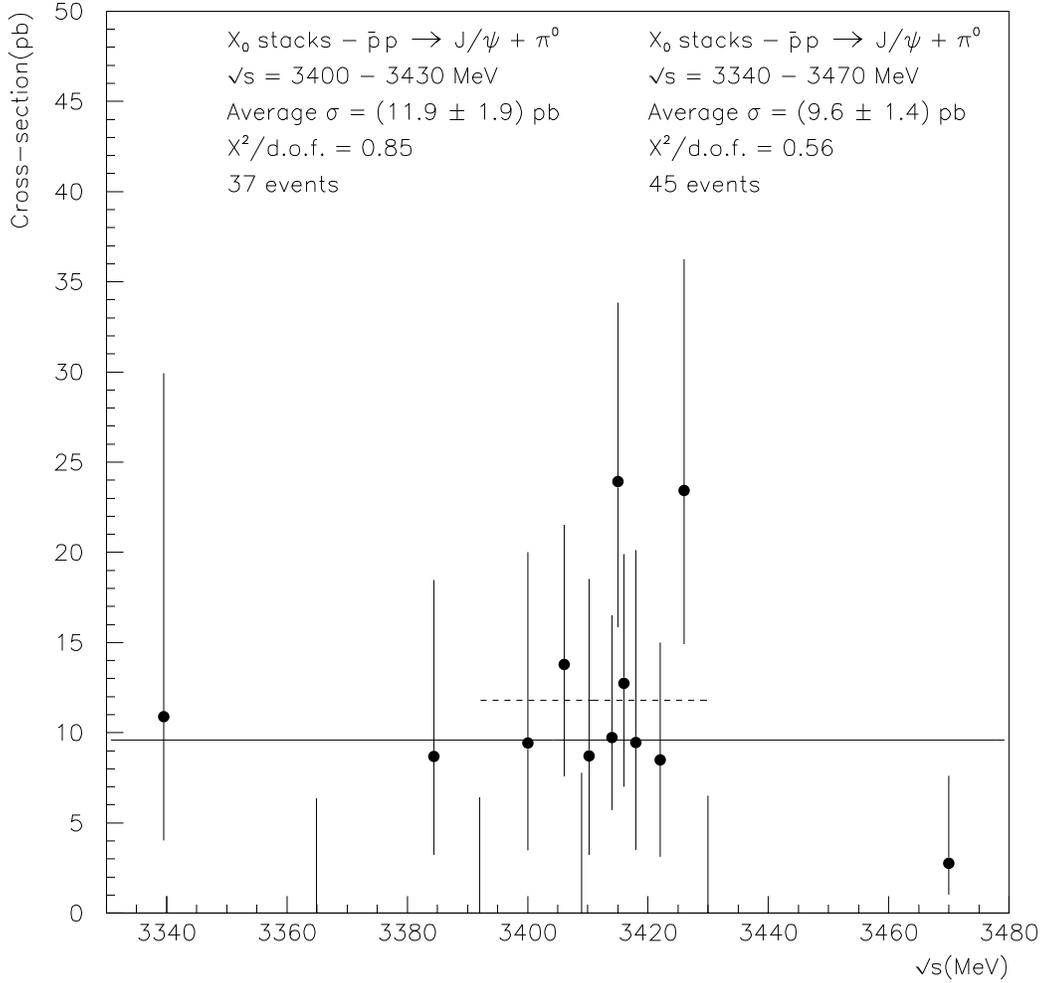}
\caption[Efficiency corrected excitation curve for the $J/\psi + \pi^0$ channel from the E835p data for the $\chi_0$ energy range.]{Efficiency corrected excitation curve for the $J/\psi + \pi^0$ channel from the E835p data for the $\chi_0$ energy range. The average cross section is 9.6 $\pm$ 1.4 pb over the whole $\chi_0$ region (solid line), and 11.9 $\pm$ 1.9 pb in the central $\chi_0$ region (dashed line), as compared to 13.6 $\pm$ 1.4 pb in the $^1P_1$ energy range. The total efficiency of this channel at $\sqrt{s} = 3415$ MeV is calculated using GEANT to be $15.7\% \pm 0.4\%$, as opposed to $17.2 \pm 0.4\%$ for $\sqrt{s} = 3526$ MeV.}
\label{fig:14}
\end{center}
\end{figure}

\newpage

\clearpage

\section{Upper Limits}

As shown in Figures 4.15-4.18, there is no apparent enhancement in the excitation curves for any of the three reactions $p\bar p \rightarrow J/\psi + X$, $p\bar p \rightarrow J/\psi \pi^0$ or $p\bar p \rightarrow J/\psi \gamma$, and fits assuming constant cross sections yield good $\chi^2$ per degree of freedom.

In order to quantify the above observations in terms of upper limits for the identification of $h_c (^1P_1)$, the data for $p\bar p \rightarrow (^1P_1) \rightarrow J/\psi + X$, $p\bar p \rightarrow (^1P_1) \rightarrow J/\psi + \pi^0$, and $p\bar p \rightarrow (^1P_1) \rightarrow J/\psi + \gamma$ (only for control purposes) have been analyzed by the maximum likelihood method for fit to a constant background and a Breit-Wigner resonance. The likelihood function was defined as

\begin{equation}
L = \prod\limits_{i=1}^N {\nu_i^{n_i}e^{-\nu_i} \over n_i !},
\end{equation}

 where

\begin{equation}
\nu_i(\sqrt{s}) = {\cal L}_i \cdot \epsilon \cdot [\sigma_{bkg} + \int f_i(\sqrt{s})\sigma_{BW}(\sqrt{s})d\sqrt{s}]
\end{equation}

The expected number of counts in the $i$th scan point is $\nu_i(\sqrt{s})$ with center of mass energy $\sqrt{s}$, while the observed number of events is $n_i$. ${\cal L}_i$ is the integrated luminosity and $\epsilon$ is the overall efficiency for the observation of the final state. The integral provides for the convolution of the Breit Wigner cross section with the (Gaussian) center of mass beam energy distribution function $f_i(\sqrt{s})$ as determined by the Schottky noise spectrum. For the Breit Wigner resonance, three different choices of width $\Gamma = 0.5$ MeV, $\Gamma = 1.0$ MeV, and $\Gamma = 2.0$ MeV were assumed. The data at $\sqrt{s} =$ 3523.33 MeV, 3528.61 MeV, and 3529.11 MeV were used to determine the level of the constant backgrounds which was $11.1 \pm 3.1$ pb for $J/\psi + \pi^0$, $22.1 \pm 3.0$ pb for $J/\psi + X$, and $8.2 \pm 2.3$ pb for $J/\psi \gamma$. These fixed backgrounds were therefore assumed and the mass of the resonance, $M_R$, was varied in 0.1 MeV steps. For each trial the $90\%$ confidence upper limit for the product branching ratios $B_{in} \times B_{out}$ were determined by the maximum likelihood fit, where $B_{in} = B(p\bar p \rightarrow R)$ and $B_{out} = B(R \rightarrow final state)$ (where $final state$ can be one of $J/\psi + X (J/\psi \rightarrow e^{+}e^{-})$, $J/\psi + \pi^0 (J/\psi \rightarrow e^{+}e^{-})$ or $J/\psi + \gamma (J/\psi \rightarrow e^{+}e^{-})$, all of which include the factor $B(J/\psi \rightarrow e^{+}e^{-}) = 0.06$. The resulting $90\%$ confidence upper limits for the three final states are plotted in various combinations in Figs. 4.20 - 4.27 for three choices of resonance width: $\Gamma = 0.5$ MeV, $\Gamma = 1.0$ MeV and $\Gamma = 2.0$ MeV. 

Figure 4.20 shows the upper limits associated with $\Gamma =$ 0.5, 1.0 and 2.0 MeV $^1P_1$ resonances for the $J/\psi + X$ channel from the E835p data. As can be seen from the plot, meaningful upper limits can only be set in the region $\sim$ 3525-3527 MeV, where the bulk of the data was taken. Outside these regions, the upper limits rise rapidly, as the fits are unconstrained by actual data (there being only one data point $\sim 2$ MeV lower, and two points $\sim$ 2 MeV higher). As may be expected, the smallest width ($\Gamma =$ 0.5 MeV) tends to give the largest upper limits.

Figures 4.21-4.23 show the upper limits for the $J/\psi + \pi^0$ channel separately for the three experiments E835p, E835, and E760, each for the three choices of resonance width. The background for the E835 scan was obtained from the data at 3524.6 MeV and 3529.1 MeV and that for E760 from data at 3522.6 MeV and 3523.5 MeV. The upper limit curves in all of these plots have the same basic shape as in the E835p $J/\psi + X$ case, but the overall level of the upper limits is lower in the E835p plot (Figure 4.21) than in either the E835 plot (Figure 4.22) or the E760 plot (Figure 4.23). This is essentially due to the larger luminosity invested in the E835p scan (50.5 pb$^{-1}$ in E835p as compared to 38.9 pb$^{-1}$ in E835 and 15.9 pb$^{-1}$ in E760). 

In Figures 4.24-4.26, we compare the upper limits for the $J/\psi + \pi^0$ channel from the three experiments for each of the three choices of the resonance width, 0.5 MeV in Figure 4.24, 1.0 MeV in Figure 4.25, and 2.0 MeV in Figure 4.26. These figures clearly demonstrate the superiority of the E835p scan over that of the E835 and E760 scans in establishing more stringent upper limits. Finally, Figure 4.27 shows the upper limits for the C-parity violating reaction $p\bar p \rightarrow ^1P_1 \rightarrow J/\psi \gamma$ for the E835p data. As in the previous plots, upper limits are shown for 0.5, 1.0, and 2.0 MeV resonance widths.

In summary, the present measurements do not show evidence for the presence of a resonance in the mass region 3524.8 MeV $\le M_R \le$ 3527.3 in any of the reactions studied: $p\bar p \rightarrow (h_c) \rightarrow J/\psi + X$, $p\bar p  \rightarrow (h_c) \rightarrow J/\psi + \pi^0$, and the C parity forbidden reaction $p\bar p \rightarrow (h_c) \rightarrow J/\psi + \gamma$. 

In terms of the 90$\%$ confidence upper limits, we find that for assumed $^1P_1$ total widths larger than 0.5 MeV, the product branching ratios are: $B(p\bar p \rightarrow h_c) \times B(h_c \rightarrow J/\psi + X (J/\psi \rightarrow e^{+}e^{-})) \le 2.3 \times 10^{-7}$ for $M(^1P_1)$ in the region 3524.6 MeV - 3527.0 MeV, and $\le 1.7 \times 10^{-7}$ for $M(^1P_1)$ in the smaller region $3526.2 \pm 0.25$ MeV. The corresponding limits for the $^1P_1 \rightarrow J/\psi + \pi^0$ decay channel are $1.8 \times 10^{-7}$ and $0.9 \times 10^{-7}$, and those for the $^1P_1 \rightarrow J/\psi + \gamma$ decay channel are $\le 1.6 \times 10^{-7}$ and $0.5 \times 10^{-7}$. These upper limits, along with upper limit values for $B_{in}\times B_{out}$ at the E760 claimed $^1P_1$ energy and at $<M(\chi_J)>$, the centroid of the $\chi$ states, are listed in Tables 4.9 and 4.10.

In the E760 experiment it was stated that for the claimed observation of $^1P_1$ at 3526.2 $\pm$ 0.2 MeV the product branching ratio, $B(p\bar p \rightarrow h_c) \times B(h_c \rightarrow J/\psi + \pi^0 (J/\psi \rightarrow e^{+}e^{-})) = (2.3 \pm 0.6) \times 10^{-7}$ for $\Gamma(h_c) = 0.5$ MeV and $(1.7 \pm 0.4) \times 10^{-7}$ for $\Gamma(h_c) = 1.0$ MeV. These correspond to $90\%$ confidence upper limits of $3.4 \times 10^{-7}$ and $2.3 \times 10^{-7}$. Our present limits are therefore nearly a factor two smaller in both cases.

\newpage

\clearpage

\begin{figure}[htbp]
\begin{center}
\includegraphics[width=15cm]
{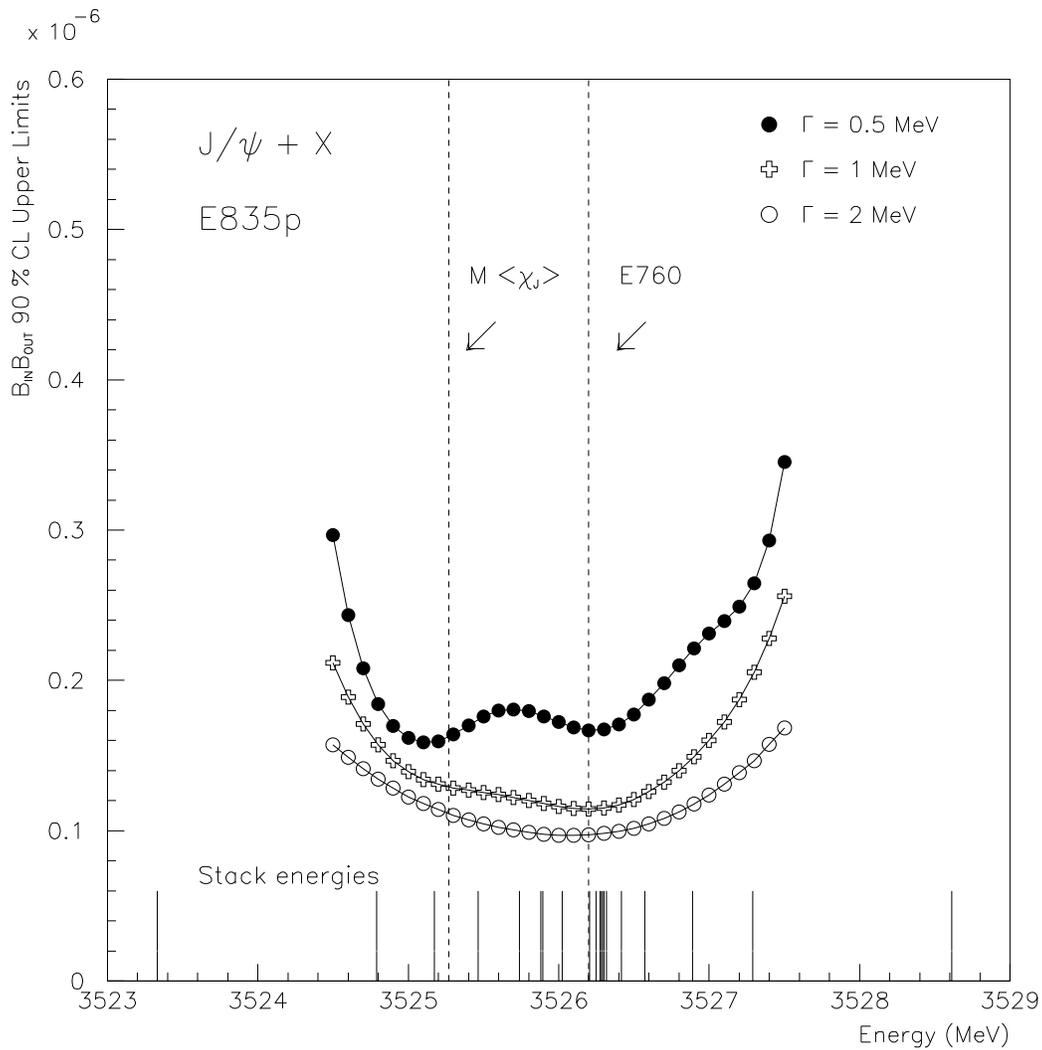}
\caption[$90\%$ CL upper limits for $B_{in}B_{out}(p\bar p \rightarrow ^1P_1 \rightarrow J/\psi + X, J/\psi \rightarrow e^{+}e^{-})$, calculated using the events from the tight $J/\psi + X$ data selection.]{$90\%$ CL upper limits for $B_{in}B_{out}(p\bar p \rightarrow ^1P_1 \rightarrow J/\psi + X, J/\psi \rightarrow e^{+}e^{-})$, calculated using the events from the tight $J/\psi + X$ data selection. The three curves show the results for $\Gamma_{tot}(^1P_1) = 0.5, 1$ and 2 MeV. The stack energies are marked by the short lines at the bottom of the plot. The first vertical dashed line represents the center of gravity of the $\chi$ states, and the second line shows the $^1P_1$ mass value claimed by E760.}
\label{fig:16}
\end{center}
\end{figure}

\newpage

\clearpage

\begin{figure}[htbp]
\begin{center}
\includegraphics[width=15cm]
{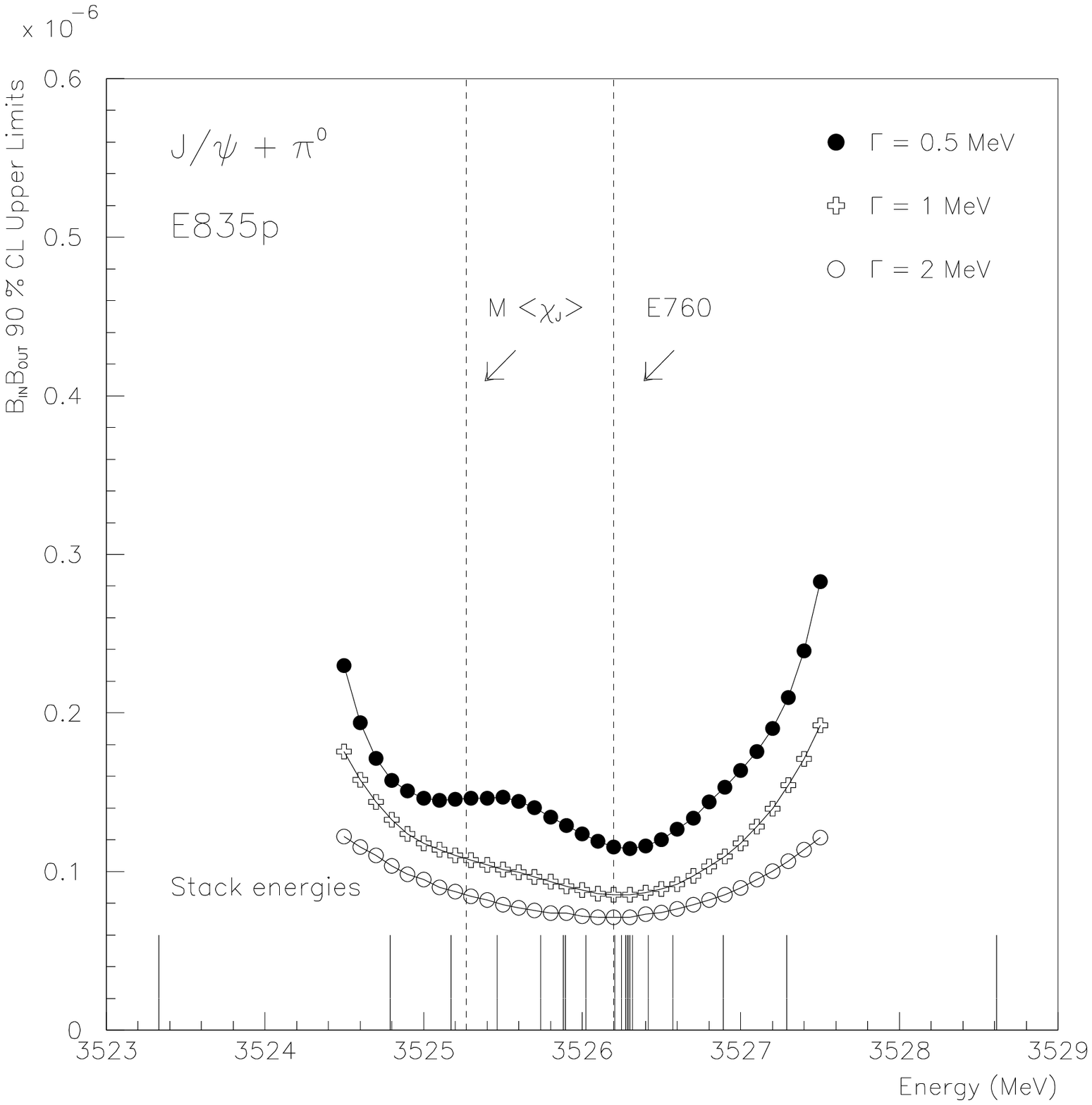}
\caption[$90\%$ CL upper limits for $B_{in}B_{out}(p\bar p \rightarrow ^1P_1 \rightarrow J/\psi + \pi^0, J/\psi \rightarrow e^{+}e^{-})$ for data from E835p.]{$90\%$ CL upper limits for $B_{in}B_{out}(p\bar p \rightarrow ^1P_1 \rightarrow J/\psi + \pi^0, J/\psi \rightarrow e^{+}e^{-})$ for data from E835p. The three curves show the results calculated for $\Gamma(^1P_1) = 0.5, 1$ and 2 MeV. The first vertical dashed line represents the center of gravity of the $\chi$ states, and the second line shows the $^1P_1$ mass value claimed by E760.}
\label{fig:20}
\end{center}
\end{figure}

\newpage

\clearpage

\begin{figure}[htbp]
\begin{center}
%\vspace*{100pt}
\includegraphics[width=15cm]
{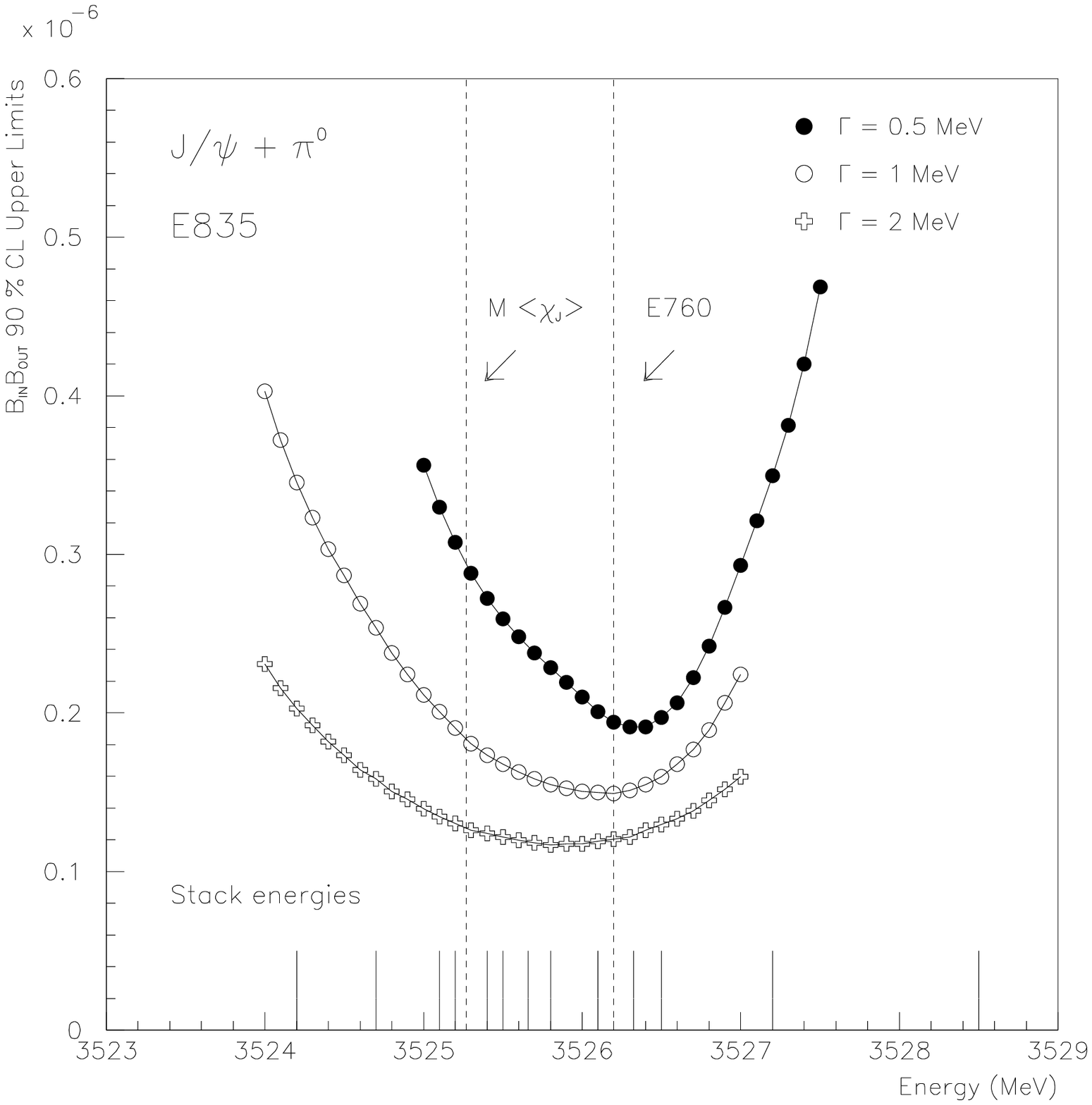}
%\vspace*{-100pt}
\caption[$90\%$ CL upper limits for $B_{in}B_{out}(p\bar p \rightarrow ^1P_1 \rightarrow J/\psi + \pi^0, J/\psi \rightarrow e^{+}e^{-})$ for data from E835.]{$90\%$ CL upper limits for $B_{in}B_{out}(p\bar p \rightarrow ^1P_1 \rightarrow J/\psi + \pi^0, J/\psi \rightarrow e^{+}e^{-})$ for data from E835. The three curves show the results calculated for $\Gamma(^1P_1) = 0.5, 1$ and 2 MeV. The first vertical dashed line represents the center of gravity of the $\chi$ states, and the second line shows the $^1P_1$ mass value claimed by E760.}
\label{fig:21}
\end{center}
\end{figure}

\newpage

\clearpage

\begin{figure}[htbp]
\begin{center}
%\vspace*{100pt}
\includegraphics[width=15cm]
{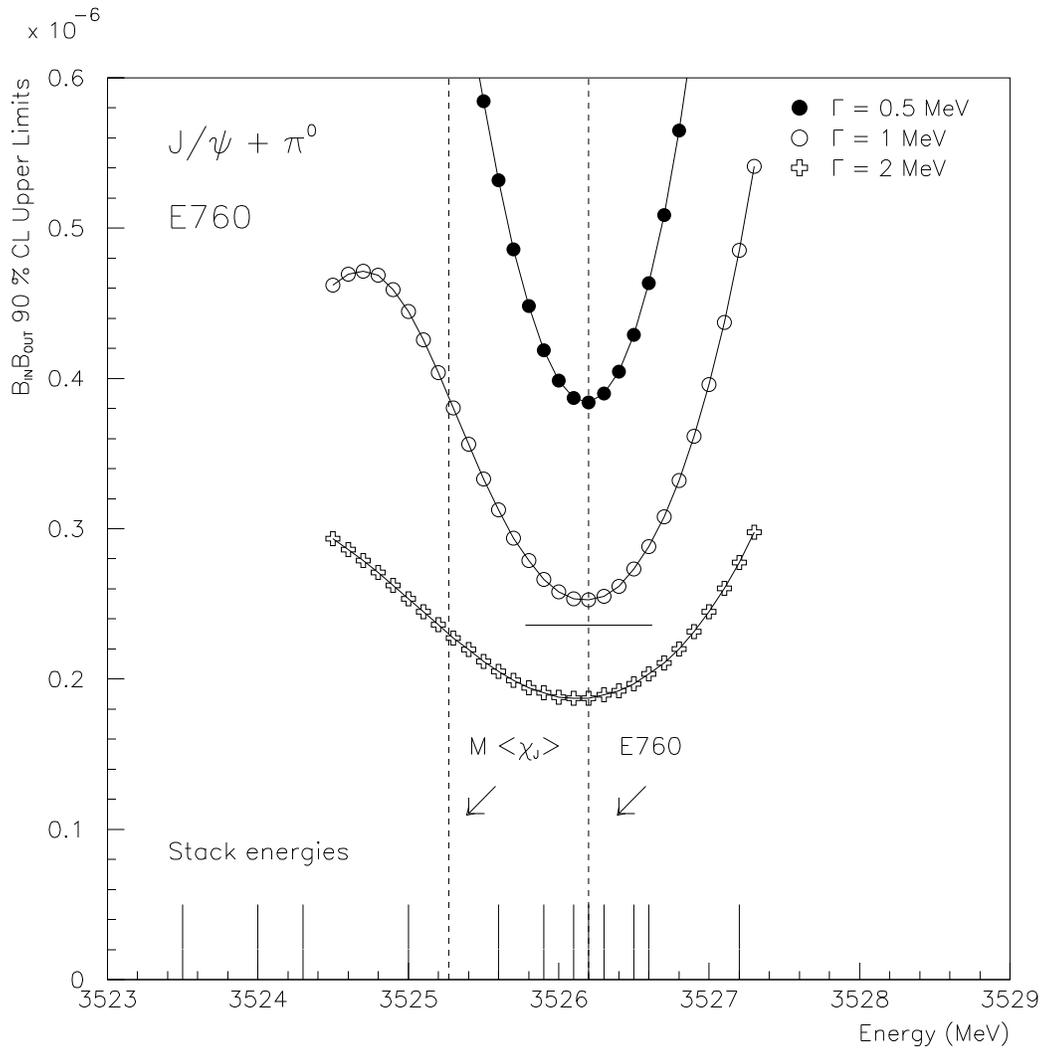}
%\vspace*{-100pt}
\caption[$90\%$ CL upper limits for $B_{in}B_{out}(p\bar p \rightarrow ^1P_1 \rightarrow J/\psi + \pi^0, J/\psi \rightarrow e^{+}e^{-})$ for data from E760.]{$90\%$ CL upper limits for $B_{in}B_{out}(p\bar p \rightarrow ^1P_1 \rightarrow J/\psi + \pi^0, J/\psi \rightarrow e^{+}e^{-})$ for data from E760. The three curves show the results calculated for $\Gamma(^1P_1) = 0.5, 1$ and 2 MeV. The first vertical dashed line represents the center of gravity of the $\chi$ states, and the second line shows the $^1P_1$ mass value claimed by E760. The horizontal line in the middle of the plot represents the E760 $90\%$ CL upper limit for $B_{in}B_{out}$.}
\label{fig:22}
\end{center}
\end{figure}

\newpage

\clearpage

\begin{figure}[htbp]
\begin{center}
%\vspace*{100pt}
\includegraphics[width=15cm]
{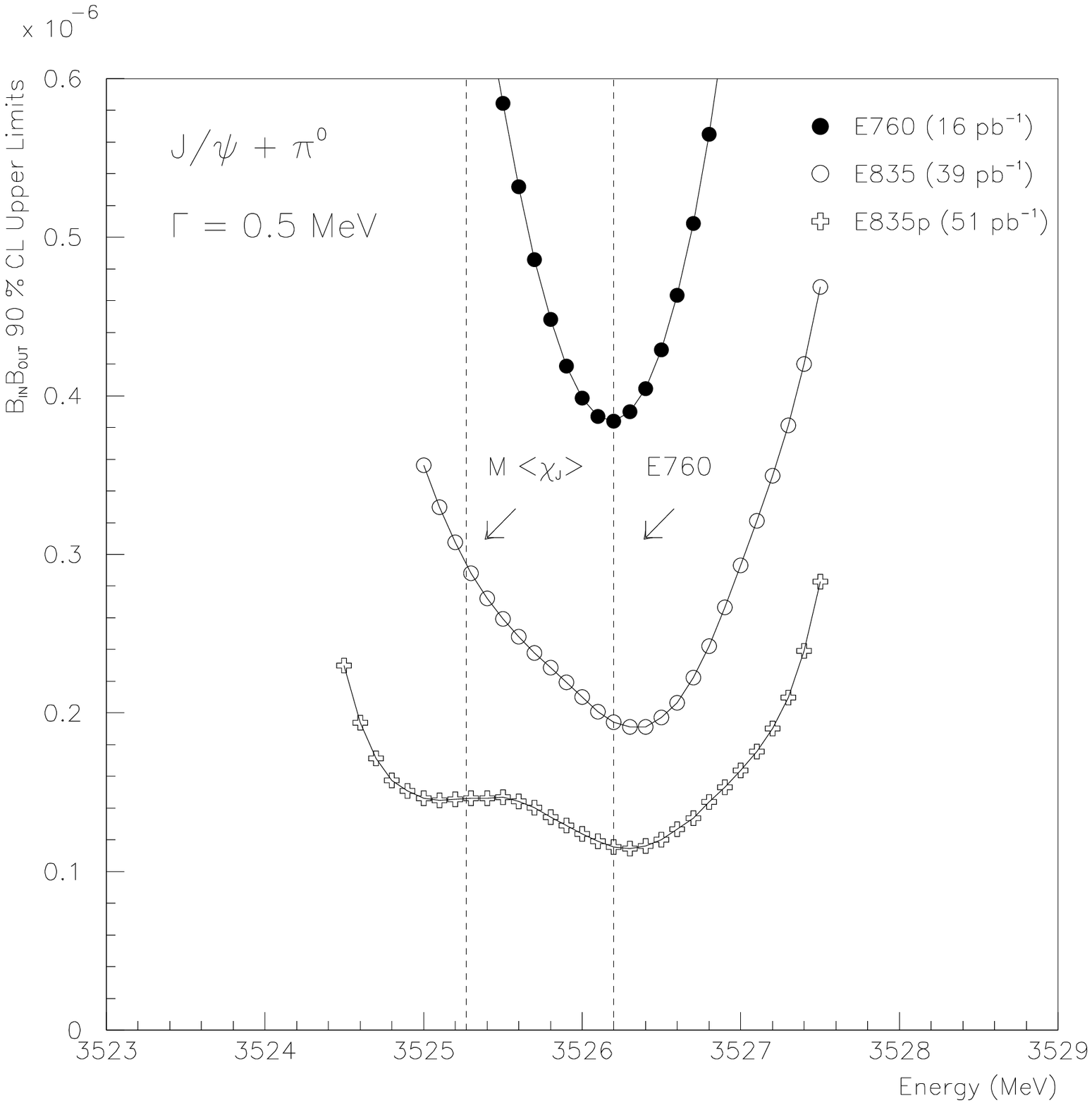}
%\vspace*{-100pt}
\caption[$90\%$ CL upper limits for $B_{in}B_{out}(p\bar p \rightarrow ^1P_1 \rightarrow J/\psi + \pi^0, J/\psi \rightarrow e^{+}e^{-})$ for $\Gamma(^1P_1) = 0.5$ MeV.]{$90\%$ CL upper limits for $B_{in}B_{out}(p\bar p \rightarrow ^1P_1 \rightarrow J/\psi + \pi^0, J/\psi \rightarrow e^{+}e^{-})$ for $\Gamma(^1P_1) = 0.5$ MeV. The three curves shown show the results from the data from E760, E835, and E835p. The first vertical dashed line represents the center of gravity of the $\chi$ states, and the second line shows the $^1P_1$ mass value claimed by E760.}
\label{fig:17}
\end{center}
\end{figure}

\newpage

\clearpage

\begin{figure}[htbp]
\begin{center}
\includegraphics[width=15cm]
{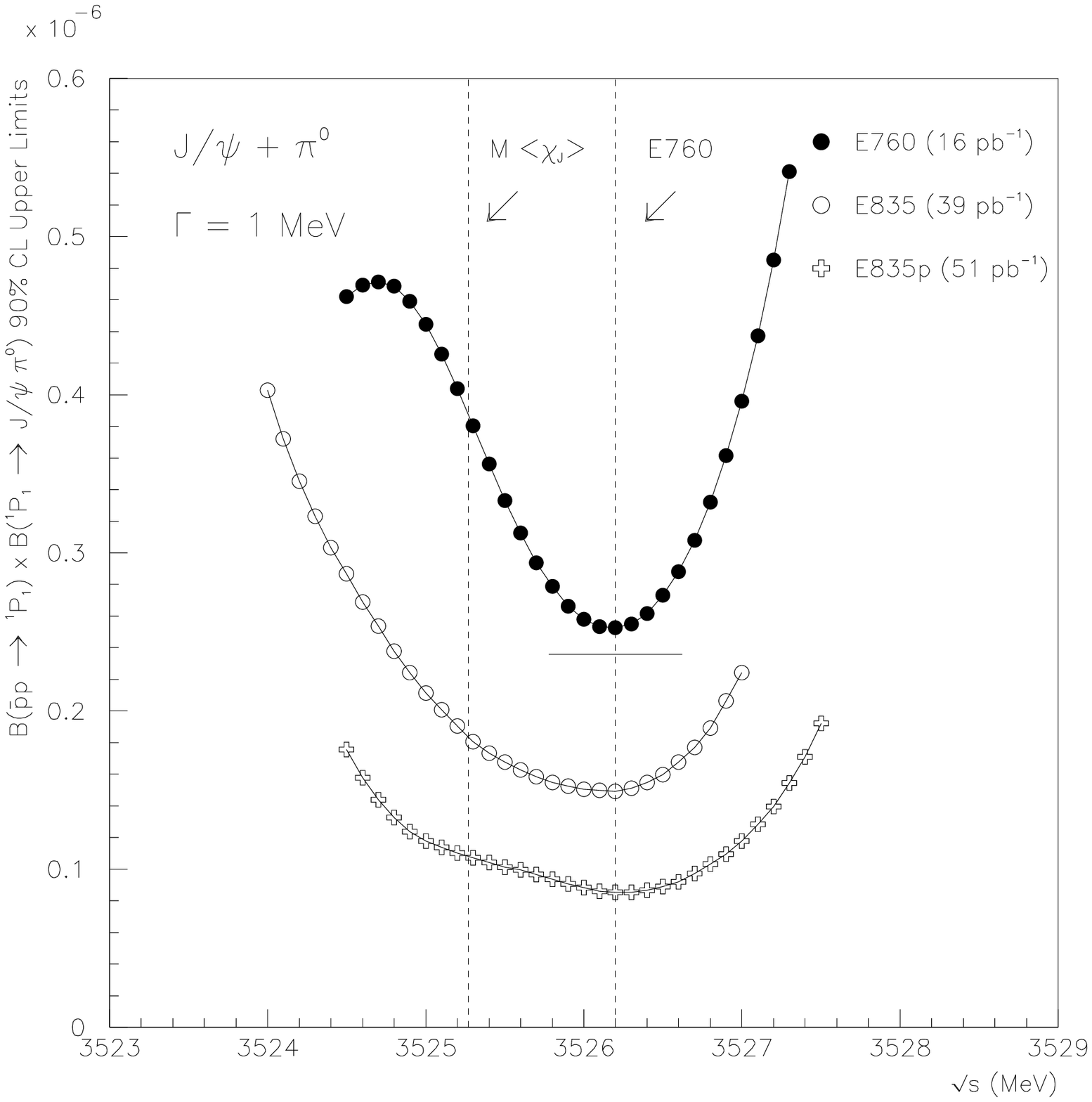}
\caption[$90\%$ CL upper limits for $B_{in}B_{out}(p\bar p \rightarrow ^1P_1 \rightarrow J/\psi + \pi^0, J/\psi \rightarrow e^{+}e^{-})$ for $\Gamma(^1P_1) = 1$ MeV.]{$90\%$ CL upper limits for $B_{in}B_{out}(p\bar p \rightarrow ^1P_1 \rightarrow J/\psi + \pi^0, J/\psi \rightarrow e^{+}e^{-})$ for $\Gamma(^1P_1) = 1$ MeV. The three curves shown show the results from the data from E760, E835, and E835p. The first vertical dashed line represents the center of gravity of the $\chi$ states, and the second line shows the $^1P_1$ mass value claimed by E760. The horizontal line in the middle of the plot represents the E760 $90\%$ CL upper limit for $B_{in}B_{out}$.}
\label{fig:18}
\end{center}
\end{figure}

\newpage

\clearpage

\begin{figure}[htbp]
\begin{center}
%\vspace*{100pt}
\includegraphics[width=15cm]
{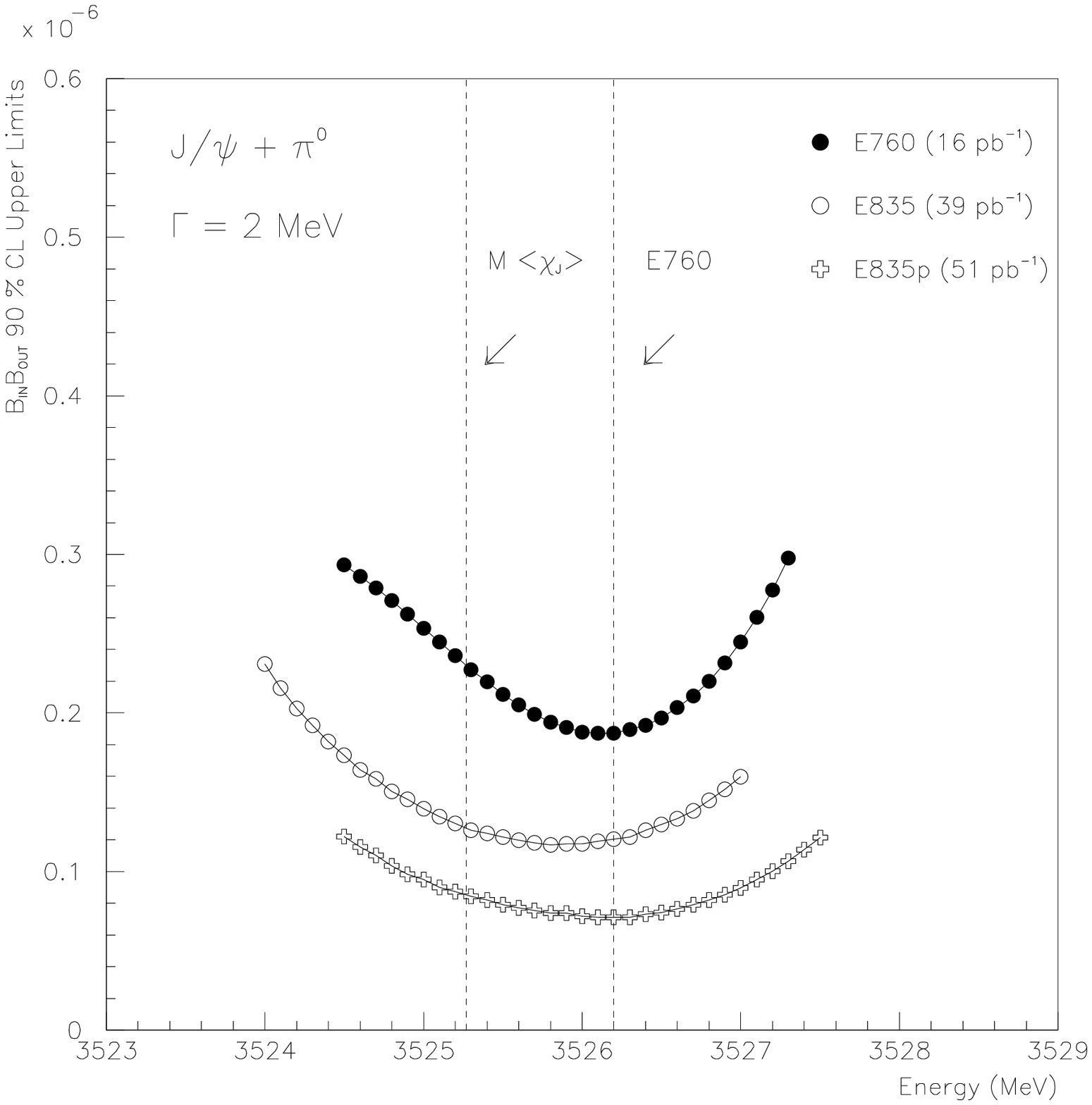}
%\vspace*{-100pt}
\caption[$90\%$ CL upper limits for $B_{in}B_{out}(p\bar p \rightarrow ^1P_1 \rightarrow J/\psi + \pi^0, J/\psi \rightarrow e^{+}e^{-})$ for $\Gamma(^1P_1) = 2$ MeV.]{$90\%$ CL upper limits for $B_{in}B_{out}(p\bar p \rightarrow ^1P_1 \rightarrow J/\psi + \pi^0, J/\psi \rightarrow e^{+}e^{-})$ for $\Gamma(^1P_1) = 2$ MeV. The three curves shown show the results from the data from E760, E835, and E835p. The first vertical dashed line represents the center of gravity of the $\chi$ states, and the second line shows the $^1P_1$ mass value claimed by E760.}
\label{fig:19}
\end{center}
\end{figure}

\newpage

\clearpage

\begin{figure}[htbp]
\begin{center}
\includegraphics[width=15cm]
{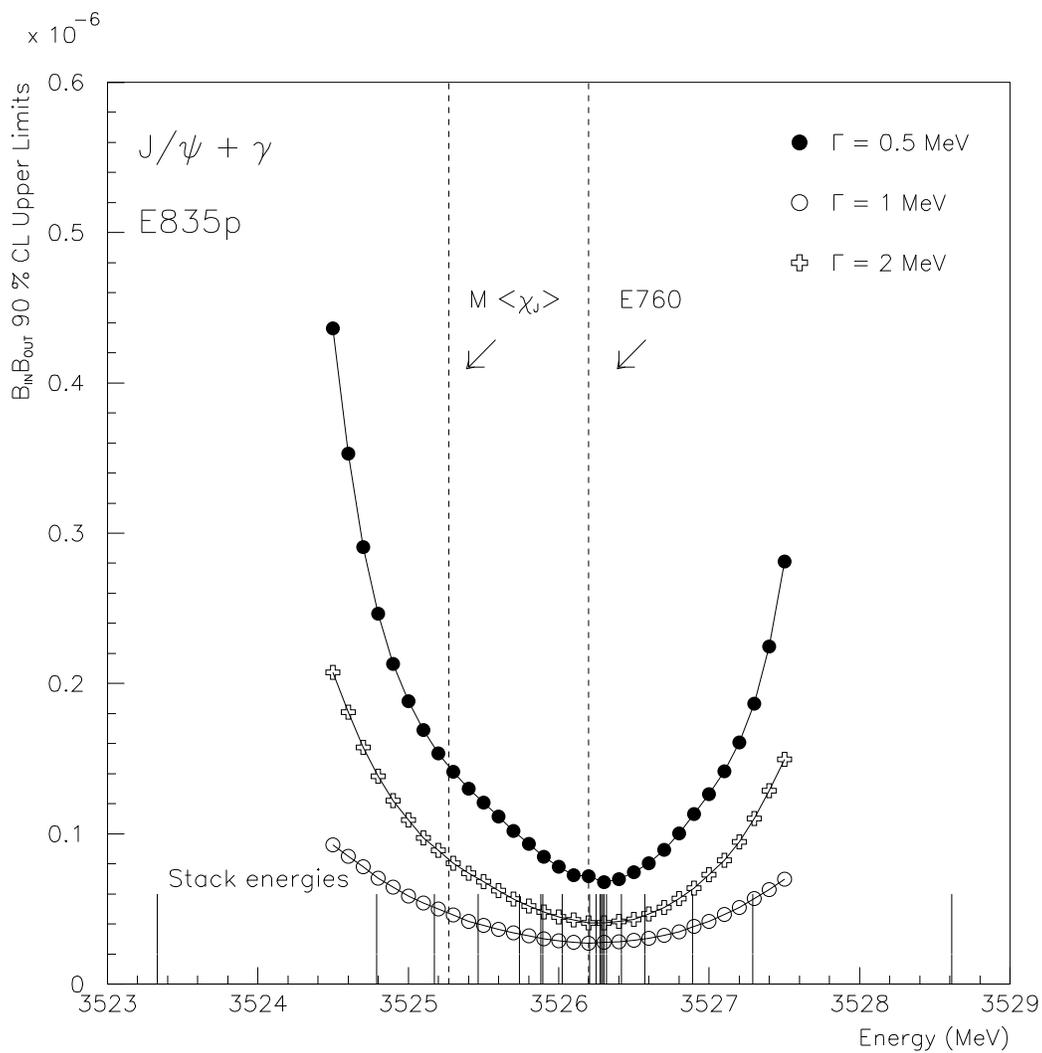}
\caption[$90\%$ CL upper limits for $B_{in}B_{out}(p\bar p \rightarrow ^1P_1 \rightarrow J/\psi + \gamma, J/\psi \rightarrow e^{+}e^{-})$.]{$90\%$ CL upper limits for $B_{in}B_{out}(p\bar p \rightarrow ^1P_1 \rightarrow J/\psi + \gamma, J/\psi \rightarrow e^{+}e^{-})$. The three curves show the results for $\Gamma(^1P_1) = 0.5, 1$, and 2 MeV. The stack energies are shown at the bottom of the plot. The first vertical dashed line represents the center of gravity of the $\chi$ states, and the second line shows the $^1P_1$ mass value claimed by E760.}
\label{fig:15}
\end{center}
\end{figure}

\newpage

\clearpage

\linespread{1.5}
\begin{table}[htbp]
\begin{center}
\begin{tabular}{|c|c|c|c|}
\hline
 & & & \\
Channel & Energy & $^1P_1$ Width & $B_{in}B_{out} 90\%$ \\
 & & & \\
 & (MeV) & (MeV) & CL Upper Limits \\
 & & & \\
\hline
 & & & \\
$J/\psi + X$ & 3525.27 & 0.5 & $1.6 \times 10^{-7}$\\
 & & & \\
 & & 1 & $1.3 \times 10^{-7}$\\
 & & & \\
 & & 2 & $1.1 \times 10^{-7}$\\
 & & & \\
$J/\psi + X$ & 3526.2 & 0.5 & $1.7 \times 10^{-7}$\\
 & & & \\
 & & 1 & $1.1 \times 10^{-7}$\\
 & & & \\
 & & 2 & $1.0 \times 10^{-7}$\\
 & & & \\
\hline
 & & & \\
$J/\psi + \pi^0$ & 3525.27 & 0.5 & $1.4 \times 10^{-7}$\\
 & & & \\
 & & 1 & $1.1 \times 10^{-7}$\\
 & & & \\
 & & 2 & $0.8 \times 10^{-7}$\\
 & & & \\
$J/\psi + \pi^0$ & 3526.2 & 0.5 & $1.1 \times 10^{-7}$\\
 & & & \\
 & & 1 & $0.8 \times 10^{-7}$\\
 & & & \\
 & & 2 & $0.7 \times 10^{-7}$\\
 & & & \\
\hline
 & & & \\
$J/\psi + \gamma$ & 3525.27 & 0.5 & $1.4 \times 10^{-7}$\\
 & & & \\
 & & 1 & $0.8 \times 10^{-7}$\\
 & & & \\
 & & 2 & $0.5 \times 10^{-7}$\\
 & & & \\
$J/\psi + \gamma$ & 3526.2 & 0.5 & $0.7 \times 10^{-7}$\\
 & & & \\
 & & 1 & $0.4 \times 10^{-7}$\\
 & & & \\
 & & 2 & $0.3 \times 10^{-7}$\\
 & & & \\
\hline
\end{tabular}
\caption{$B_{in}B_{out} 90\%$ CL Upper Limits for the $J/\psi + X$, $J/\psi + \pi^0$, and $J/\psi + \gamma$ channels, fit for 0.5, 1, and 2 MeV width $^1P_1$ resonances at the centroid of the $\chi_J$ states and the E760 observed $^1P_1$ energy.}
\label{tab:test}
\end{center}
\end{table}

\newpage

\clearpage

\begin{table}[htb]
\begin{center}
\begin{tabular}{|c|ccc|}
\hline
 & & & \\
 & $90\%$ Confidence & Upper Limits & $\times 10^7$ \\
 & & & \\
\hline
 & & & \\
Mass Region (MeV) & $B(p\bar p \rightarrow h_c) \times$ & $B(p\bar p \rightarrow h_c) \times$ & $B(p\bar p \rightarrow h_c) \times$ \\
 & & & \\
(Width) & $B(h_c \rightarrow J/\psi + \pi^0)$ \enskip & $B(h_c \rightarrow J/\psi + X)$ & $B(h_c \rightarrow J/\psi + \gamma)$ \\ 
 & & & \\
\hline
 & & & \\
3526.2 $\pm$ 1.1 & & & \\
 & & & \\
($\Gamma$ = 2.0 MeV) & $\le 0.9 $ & $\le 1.3 $ & $\le 0.5 $ \\
 & & & \\
($\Gamma$ = 1.0 MeV) & $\le 1.2 $ & $\le 1.8 $ & $\le 1.0 $ \\
 & & & \\
($\Gamma$ = 0.5 MeV) & $\le 1.8 $ & $\le 2.3 $ & $\le 1.6 $ \\
 & & & \\
\hline
 & & & \\
3526.2 $\pm$ 0.25 & & & \\
 & & & \\
($\Gamma$ = 2.0 MeV) & $\le 0.7 $ & $\le 1.0 $ & $\le 0.3 $ \\
 & & & \\
($\Gamma$ = 1.0 MeV) & $\le 0.8 $ & $\le 1.1 $ & $\le 0.5 $ \\
 & & & \\
($\Gamma$ = 0.5 MeV) & $\le 1.1 $ & $\le 1.7 $ & $\le 0.8 $ \\
 & & & \\
\hline
\end{tabular}
\caption{$90\%$ confidence upper limits from the present measurements for the  mass regions 3526.2 $\pm$ 1.1 MeV and $3526.2 \pm 0.25$ MeV.}
\end{center}
\end{table}

%Chapter 5--Data Acquisition
\baselineskip=24pt
\chapter{The $p\bar p \rightarrow ^1P_1 \rightarrow \eta_c\gamma \rightarrow 3\gamma$ Reaction}

\section{The $\eta_c \gamma$ Decay Channel of $h_c (^1P_1)$}

An additional $^1P_1$ decay channel which was examined in the E835p data was the reaction $p\bar p \rightarrow ^1P_1 \rightarrow \eta_c \gamma \rightarrow (\gamma\gamma) \gamma$. This channel was examined even though it lacks a $J/\psi \rightarrow e^{+}e^{-}$ signal for clear charmonium identification. This is because the radiative decay to $\eta_c$ is expected to be the dominant decay mode of the $^1P_1$ being the spin singlet version of the radiative decay $\chi_{c1} \rightarrow J/\psi \gamma$ for which the PDG gives a branching ratio of $\sim 32\%$. Unfortunately, $\eta_c$ ($0^{-+}$) cannot decay into $e^{+}e^{-}$, but decays almost entirely hadronically. This makes its identification difficult in a detector optimized for the detection of electrons and photons. There is, however, a two photon decay of the $\eta_c$, with a branching ratio of $\sim 3 \times 10^{-4}$ which can be identified in the E835 detector. In fact, if the products of the estimated branching ratios for both the $J/\psi \pi^0 \rightarrow e^{+}e^{-} \gamma\gamma$ and $\eta_c \gamma \rightarrow (\gamma\gamma) \gamma$ decays of the $^1P_1$ are compared, they should be of roughly the same order. This is because the $J/\psi \pi^0$ decay is suppressed by approximately two orders of magnitude relative to $\eta_c \gamma$, but $B(\eta_c \rightarrow \gamma\gamma) = 3 \times 10^{-4}$ is suppressed by roughly two orders of magnitude relative to $B(J/\psi \rightarrow e^{+}e^{-}) = 6 \times 10^{-2}$. Thus the expected decay rate of $^1P_1 \rightarrow \eta_c \gamma \rightarrow (\gamma\gamma)\gamma$ is comparable to $^1P_1 \rightarrow J/\psi \pi^0 \rightarrow (e^{+}e^{-}) \gamma\gamma$.

\section{GEANT Monte Carlo Study of the $\eta_c \gamma$ Decay Channel}

A study of the $\eta_c \gamma$ channel was done using the full E835p GEANT Monte Carlo to examine the feasibility of detecting a signal in this channel. Figure 5.1 shows the two photon invariant mass spectrum from the GEANT Monte Carlo simulation of the reaction $p\bar p \rightarrow ^1P_1 \rightarrow \eta_c \gamma \rightarrow 3\gamma$. The resulting $\eta_c$ signal shows a spread in $M(\gamma_1\gamma_2)$ over $2.98 \pm 0.20$ GeV due to detector effects. This range was eventually chosen as the $\eta_c$ mass cut for this channel.

Figure 5.2 shows the $\gamma_1\gamma_3$ invariant mass spectrum from the GEANT Monte Carlo generated events for the $\eta_c\gamma \rightarrow (\gamma_1\gamma_2)\gamma_3$ decay. The Monte Carlo, of course, does not contain $\pi^0$, $\eta$, and $\eta^{\prime}$ which are produced in great numbers in $p\bar p$ annihilations, as shown later. To remove them from the data, cuts were made in the data for $M(\gamma_1\gamma_3) < 1.1$ GeV and $M(\gamma_2\gamma_3) < 1.1$ GeV. This is indicated by the dashed line in Fig. 5.2. The expected angular distribution of photons from the $\eta_c$ decay as calculated by GEANT Monte Carlo simulation is shown in Figure 5.3, and as expected it is flat over the acceptance region. Although the angular distribution of the two gammas coming from the decay of the $\eta_c$ is expected to be essentially isotropic, the gamma background in the detector is known to be strongly peaked in the forward and backward regions. This forward-backward peaking is known to arise from the very large photon yield from $\pi^0$ decays from $p\bar p \rightarrow \pi^0\pi^0$ events. Cuts are made in the data to remove events with $\vert \cos(\theta^{*}_1)$, $\cos(\theta^{*}_2 \vert > 0.4$ in order to eliminate the large background from feeddown from $\pi^0\pi^0$ events. The cuts are indicated by dashed lines in Fig. 5.3.

Finally, the total energy of the GEANT Monte Carlo event in the laboratory frame, $E_{tot}$(lab), for the events which survive the invariant mass and angular cuts is plotted in Figure 5.4 with respect to the center of mass energy of the radiative gamma $(\gamma_3)$ which is expected to be monoenergetic with $E_{cm}(\gamma_3) = 504$ MeV. These two quantities are shown again in projections in Figure 5.5, where the detector effects can be clearly seen: they spread the events around the nominal values of total energy (6.626 GeV), and $E_{cm}(\gamma_3)$ = 504 MeV for the assumed $M(^1P_1) = 3.5262$ GeV. Cuts are therefore made in the data sample to remove events which do not have 6.376 GeV $ < E_{lab}(tot) <$ 7.000 GeV, and 454 MeV $< E_{cm}(\gamma_3) <$ 580 MeV. These cuts are non-symmetric about the nominal values due to the $1\%$ energy correction applied to each $\gamma$, which is described in chapter 3.

\clearpage

\newpage

\begin{figure}[htbp]
\begin{center}
%\vspace*{100pt}
\includegraphics[width=15cm]{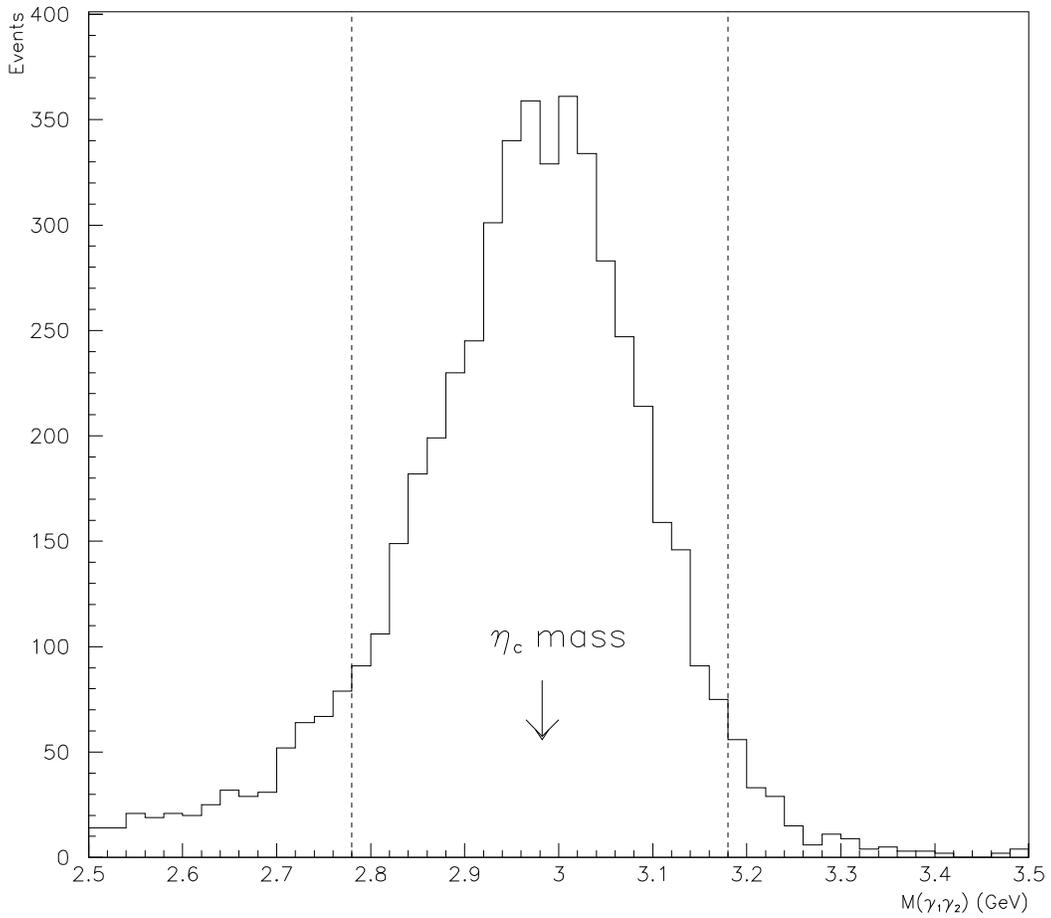}
%\vspace*{-100pt}
\caption[The $\gamma_1\gamma_2$ invariant mass distribution for all 3 $\gamma$ events in the GEANT monte carlo simulation of the $^1P_1 \rightarrow \eta_c\gamma \rightarrow 3\gamma$ decay channel.]{The $\gamma_1\gamma_2$ invariant mass distribution for all 3 $\gamma$ events in the GEANT monte carlo simulation of the $^1P_1 \rightarrow \eta_c\gamma \rightarrow 3\gamma$ decay channel. $\eta_c$ mass cuts are made for events with invariant mass $M(\gamma_1\gamma_2) <$ 2.78 and $M(\gamma_1\gamma_2) >$ 3.18 GeV.}
\label{fig:gg1}
\end{center}
\end{figure}

\clearpage

\newpage

\begin{figure}[htbp]
\begin{center}
%\vspace*{100pt}
\includegraphics[width=15cm]{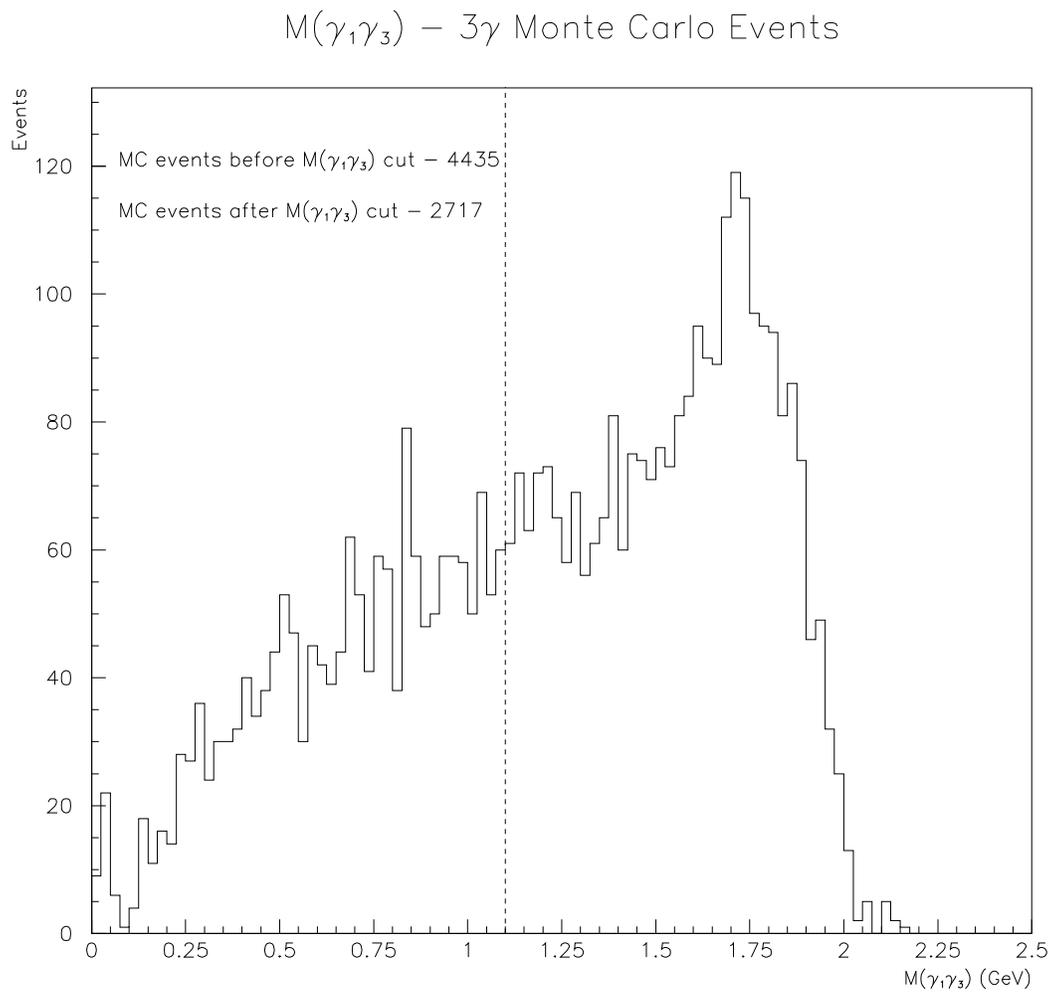}
%\vspace*{-100pt}
\caption{The $\gamma_1\gamma_3$ invariant mass distribution for all 3 $\gamma$ events in the GEANT Monte Carlo simulation with 2.8 GeV $< M(\gamma_1\gamma_2) <$ 3.2 GeV.}
\label{fig:gg1}
\end{center}
\end{figure}

\newpage

\clearpage

\begin{figure}[htbp]
\begin{center}
%\vspace*{100pt}
\includegraphics[width=15cm]{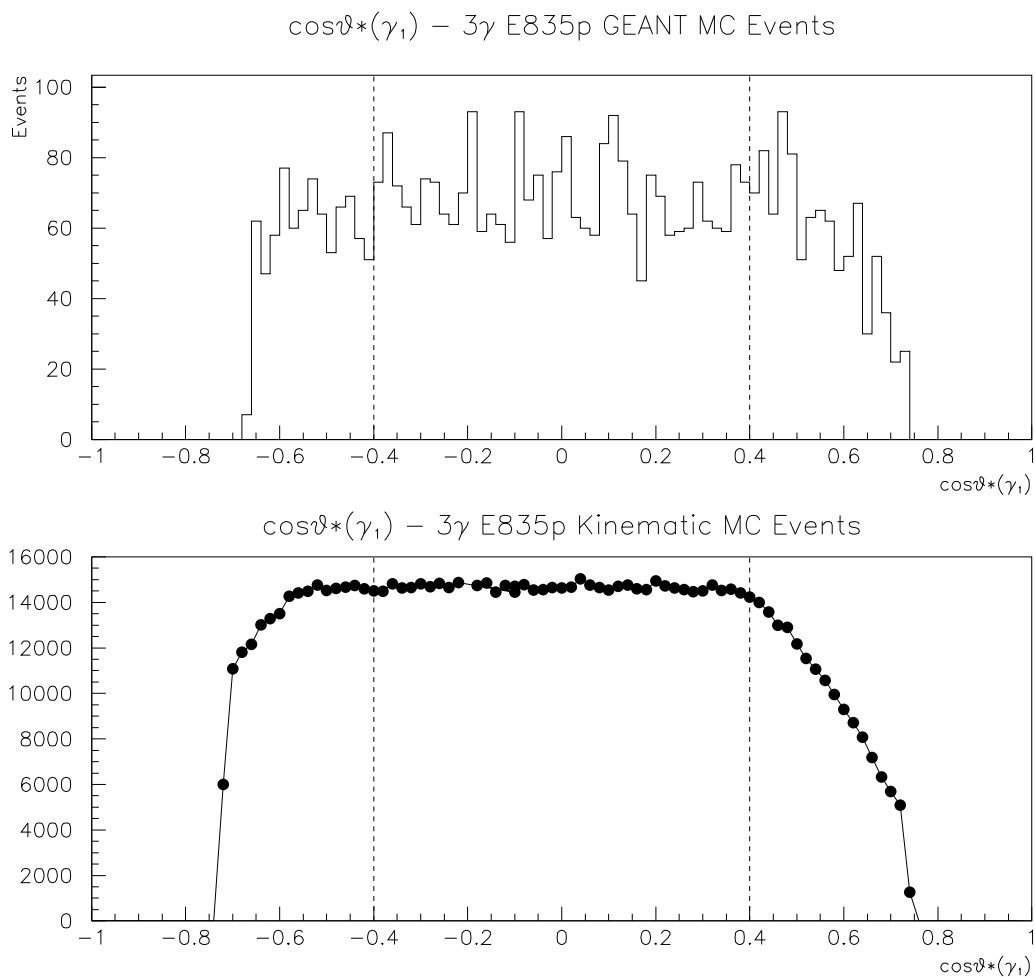}
%\vspace*{-100pt}
\caption[Top: The cos$(\theta^*)(\gamma_{1,2})$ distribution for all 3 $\gamma$ events in the GEANT Monte Carlo simulation of the reaction $p\bar p \rightarrow ^1P_1 \rightarrow \eta_c\gamma \rightarrow (\gamma_1\gamma_2)\gamma$. Bottom: The same plot made with a strictly kinematic Monte Carlo with 100 times the number of generated events.]{Top: The cos$(\theta^*)(\gamma_{1,2})$ distribution for all 3 $\gamma$ events in the GEANT Monte Carlo simulation of the reaction $p\bar p \rightarrow ^1P_1 \rightarrow \eta_c\gamma \rightarrow (\gamma_1\gamma_2)\gamma$. The vertical dashed lines show the cuts made to remove events with $\vert \cos(\theta^*)(\gamma_1) \vert > 0.4$. Bottom: The same plot made with a strictly kinematic Monte Carlo with 100 times the number of generated events.}
\label{fig:gg1}
\end{center}
\end{figure}

\newpage

\clearpage

\begin{figure}[htbp]
\begin{center}
%\vspace*{100pt}
\includegraphics[width=15cm]{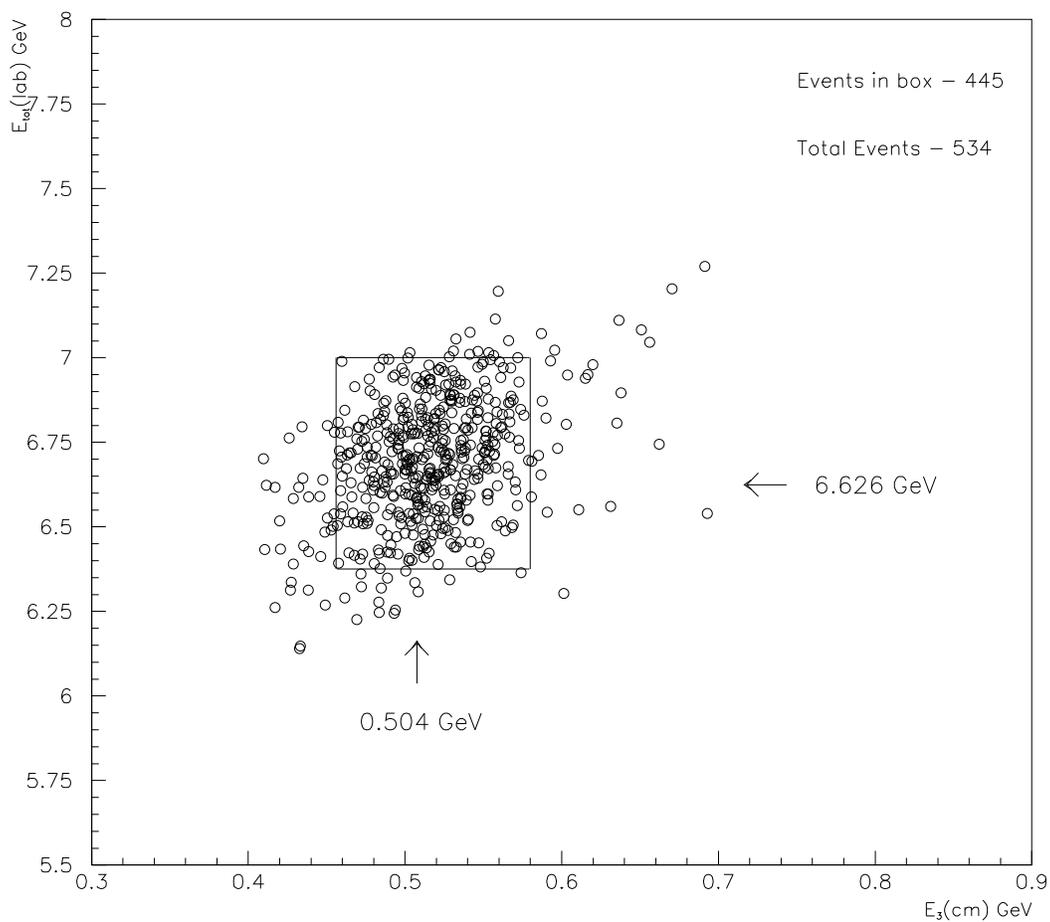}
%\vspace*{-100pt}
\caption[The $E_{tot}$(lab) vs. $E_{cm}(\gamma_3)$ distribution for all 3 $\gamma$ events in the GEANT Monte Carlo with 2.78 GeV $< M(\gamma_1\gamma_2) <$ 3.18 GeV, and $M(\gamma_1\gamma_3), M(\gamma_2\gamma_3) >$ 1.1 GeV and $\vert \cos(\theta^*)(\gamma_1,\gamma_2) \vert < 0.4$.]{The $E_{tot}$(lab) vs. $E_{cm}(\gamma_3)$ distribution for all 3 $\gamma$ events in the GEANT Monte Carlo with 2.78 GeV $< M(\gamma_1\gamma_2) <$ 3.18 GeV, and $M(\gamma_1\gamma_3), M(\gamma_2\gamma_3) >$ 1.1 GeV and $\vert \cos(\theta^*)(\gamma_1,\gamma_2) \vert < 0.4$. The box outlines the region in which 6.376 GeV $< E(tot) < 7.000$ GeV and 0.454 GeV $< E(\gamma_3) < 0.580$ GeV.}
\label{fig:gg1}
\end{center}
\end{figure}

\newpage

\clearpage

\begin{figure}[htbp]
\begin{center}
%\vspace*{100pt}
\includegraphics[width=15cm]{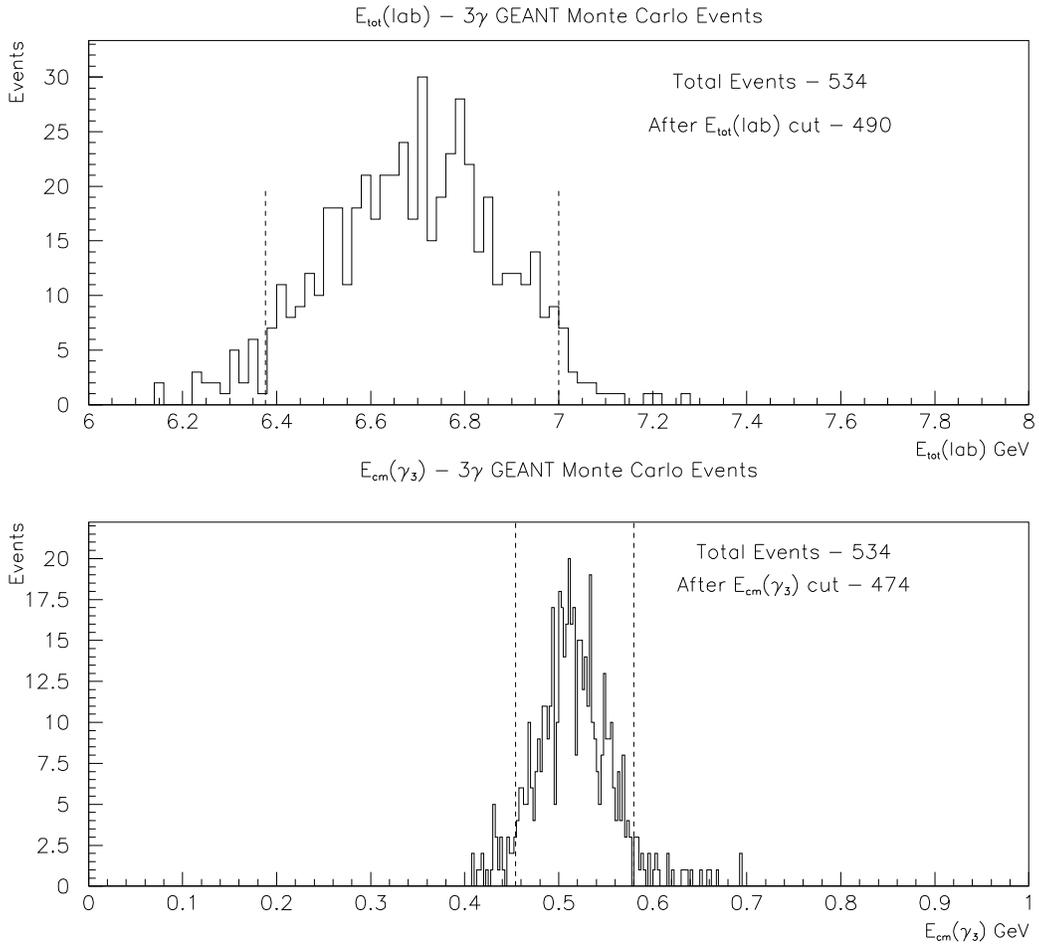}
%\vspace*{-100pt}
\caption{Top: The $E_{tot}$(lab) distribution for all 3 $\gamma$ events in the GEANT Monte Carlo with 2.78 GeV $< M(\gamma_1\gamma_2) <$ 3.18 GeV, and $M(\gamma_1\gamma_3), M(\gamma_2\gamma_3) >$ 1.1 GeV and $\vert \cos(\theta^*)(\gamma_1,\gamma_2) \vert < 0.4$. Bottom: The $E_{cm}(\gamma_3)$ distribution for the same GEANT Monte Carlo events.}
\label{fig:gg1}
\end{center}
\end{figure}

\clearpage

\newpage

\section{Data and Analysis Cuts for the $\eta_c \gamma$ Decay Channel} 

The $\eta_c \gamma$ decay channel was studied for the 21 energy points of the $^1P_1$ search using the neutral trigger data collected in E835p. Most of the cuts in the data are based on GEANT Monte Carlo simulations as described in Section 5.2. Here we refer only to the figures illustrating the Monte Carlo results. All three gamma events from this data sample were tested for their highest possible two gamma invariant mass combinations. These gammas were designated as $\gamma_1$ and $\gamma_2$. The remaining gamma was, of course, $\gamma_3$. Events with $M(\gamma_1\gamma_2) > 2.75$ GeV were used in the data analysis. The $\gamma_1\gamma_2$ invariant mass spectrum for the entire data sample is shown in Figure 5.6. (see Figure 5.1 for the corresponding Monte Carlo results) Note that at this stage no enhancement is visible at the $\eta_c$ region around 2.98 GeV. 

Figure 5.7 shows the $\gamma_1\gamma_3$ invariant mass spectrum. As can be clearly seen in the figure, a large fraction of these photons make a $\pi^0$ ($M = 135$ MeV) or an $\eta$ ($M = 547$ MeV). Figure 5.8 shows $M(\gamma_2\gamma_3)$ after a cut has been made for $M(\gamma_1\gamma_3) < 1.1$ GeV to remove $\pi^0$, $\eta$, and $\eta^{\prime}$. Now one can see $\eta^{\prime}$ ($M$ = 958 MeV) and some other residual structure. The cut at $M(\gamma_{1,2}\gamma_3) <$ 1.1 GeV rejects all of these. Figure 5.9 shows $M(\gamma_1\gamma_2)$ after the $M(\gamma_1\gamma_3)$ and $M(\gamma_2\gamma_3)$ cuts have been made. No $\eta_c$ enhancement is visible, even now.

Although the angular distribution of the two gammas coming from the decay of the $\eta_c$ is expected to be isotropic (see Figure 5.3 for the Monte Carlo results), the gamma background in the detector is known to be strongly peaked in the forward and backward regions because of mainly feed-down from the prolific yield of the reaction $p\bar p \rightarrow \pi^0\pi^0$. This can be seen in Figures 5.10 and 5.11, which show the angular distributions of $\gamma_1$ and $\gamma_2$ respectively. Cuts are made to remove background events with $-0.4 < \vert \cos(\theta^{*})(\gamma_1,\gamma_2) \vert < 0.4$. The invariant mass spectrum $M(\gamma_1\gamma_2)$ after the cuts are made to $\cos(\theta^{*})(\gamma_1,\gamma_2)$ is shown in Figure 5.12.

Lastly, the total energy of the event in the lab frame for the remaining events in the E835p data is plotted in Figure 5.13 with respect to the center of mass energy of the radiative gamma $(\gamma_3)$. Cuts are made in the data sample to remove events which do not have 6.376 GeV $< E_{lab}(tot) <$ 7.000 GeV, and further cuts on the radiative gamma are made to remove events which do not have 454 MeV $< E_{cm}(\gamma_3) <$ 580 MeV. The levels of these cuts and their efficiencies are determined from the GEANT Monte Carlo (see Figure 5.4 for Monte Carlo results).

\clearpage

\begin{figure}[htbp]
\begin{center}
%\vspace*{100pt}
\includegraphics[width=15cm]{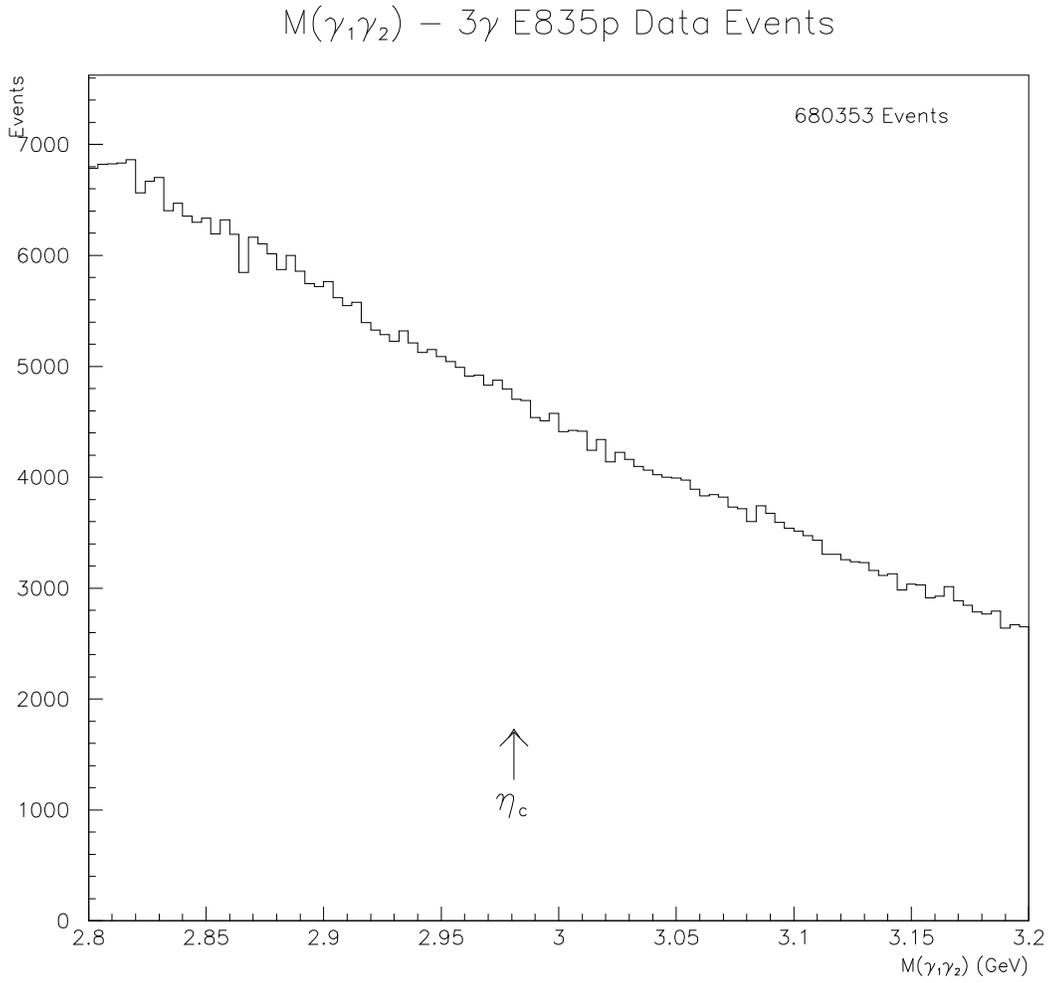}
%\vspace*{-100pt}
\caption[The $\gamma_1\gamma_2$ invariant mass distribution for 3 $\gamma$ events in the E835p $^1P_1$ stacks with 2.8 GeV $< M_{\gamma_1\gamma_2} < 3.2$ GeV.]{The $\gamma_1\gamma_2$ invariant mass distribution for 3 $\gamma$ events in the E835p $^1P_1$ stacks with 2.8 GeV $< M_{\gamma_1\gamma_2} < 3.2$ GeV. No enhancement is seen in the $\eta_c$ mass region. $\gamma_1$, $\gamma_2$, and $\gamma_3$ are numbered according to their energies in the center of mass frame, with $\gamma_1$ being the highest energy and $\gamma_3$ the lowest. Using this numbering system for the reaction $p\bar p \rightarrow ^1P_1 \rightarrow \eta_c\gamma \rightarrow 3\gamma$, the two $\gamma$s from the $\eta_c$ will be $\gamma_1$ and $\gamma_2$ while the radiative $\gamma$ will be $\gamma_3$.}
\label{fig:gg1}
\end{center}
\end{figure}

\clearpage

\newpage

\begin{figure}[htbp]
\begin{center}
%\vspace*{100pt}
\includegraphics[width=15cm]{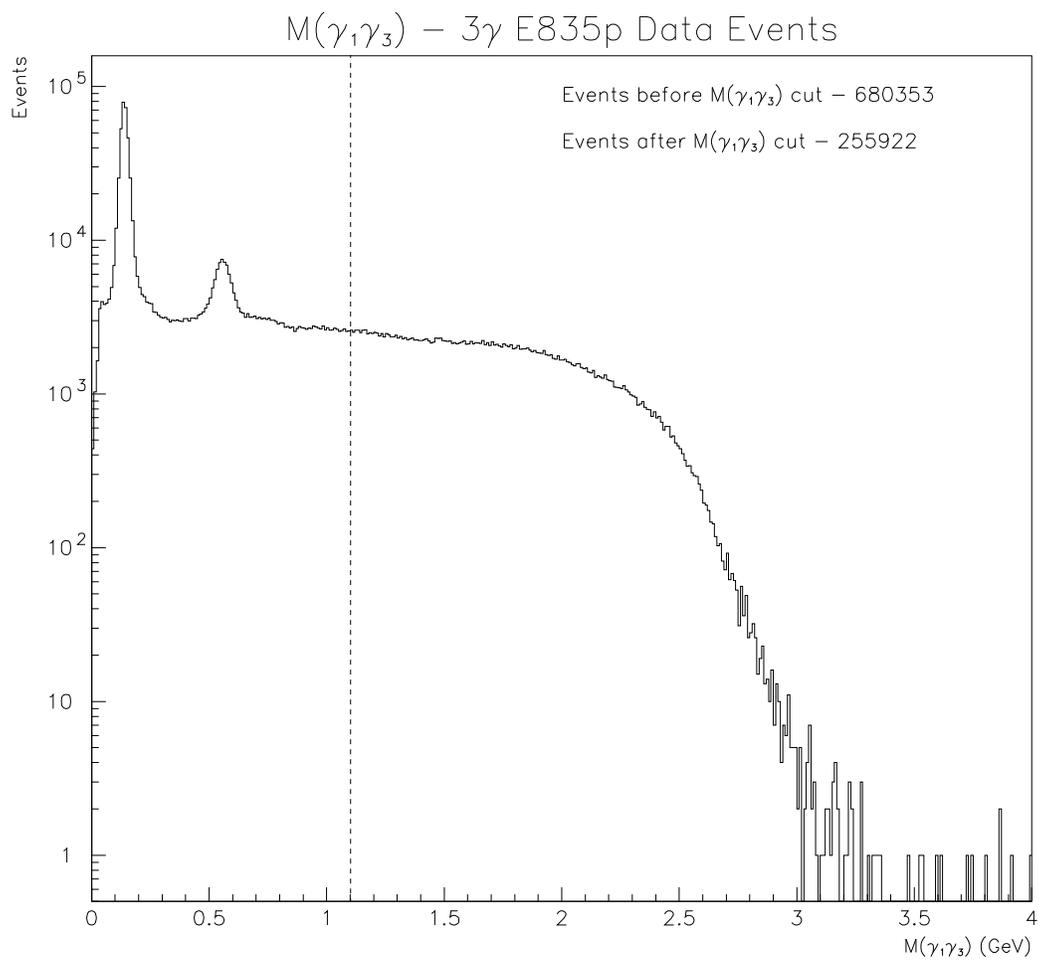}
%\vspace*{-100pt}
\caption[The $\gamma_1\gamma_3$ invariant mass distribution for 3 $\gamma$ events in the E835p $^1P_1$ data.]{The $\gamma_1\gamma_3$ invariant mass distribution for 3 $\gamma$ events in the E835p $^1P_1$ data. The $\pi^0$ and $\eta$ peaks are clearly visible. Mass cuts are made to remove events with invariant mass $M(\gamma_1\gamma_3) <$ 1.1 GeV.}
\label{fig:gg1}
\end{center}
\end{figure}

\newpage

\clearpage

\begin{figure}[htbp]
\begin{center}
%\vspace*{100pt}
\includegraphics[width=15cm]{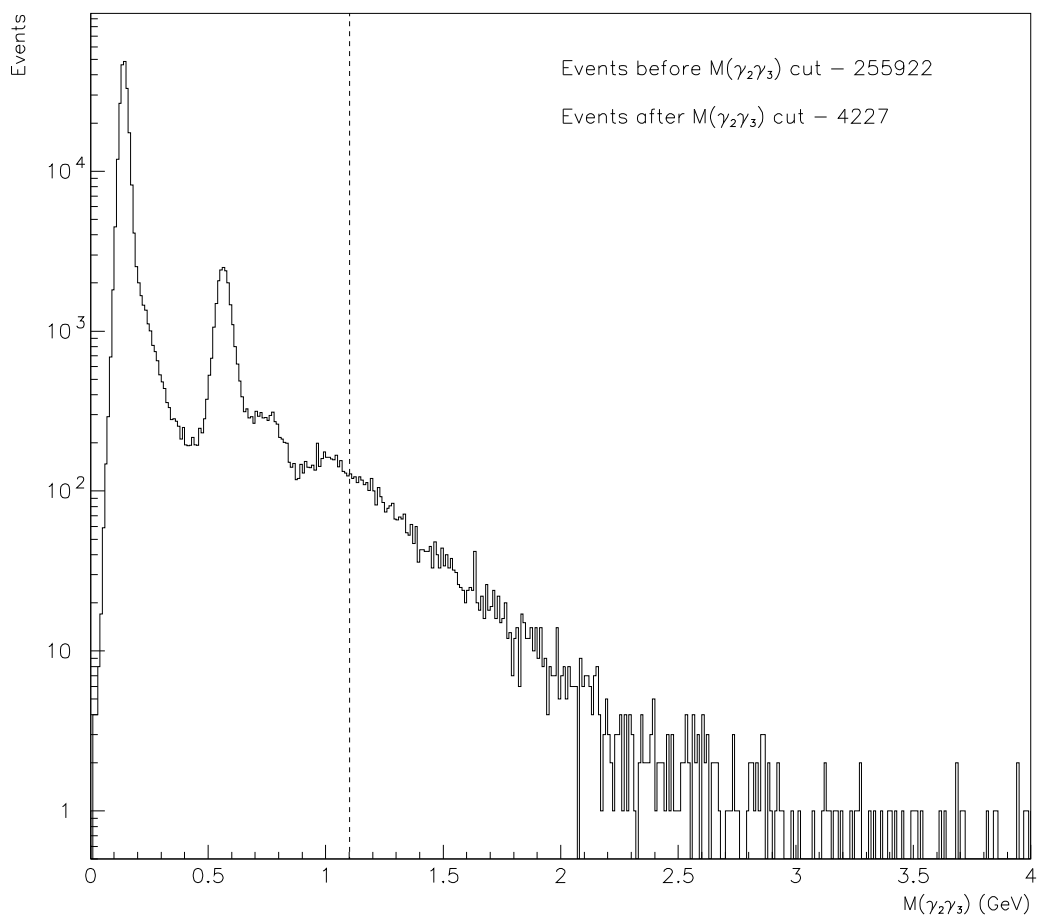}
%\vspace*{-100pt}
\caption[The $\gamma_2\gamma_3$ invariant mass distribution for 3 $\gamma$ events in the E835p $^1P_1$ data.]{The $\gamma_2\gamma_3$ invariant mass distribution for 3 $\gamma$ events in the E835p $^1P_1$ data. The vertical axis is plotted logarithmically so that the contamination from the $\eta^{\prime}$ is visible as well as the $\pi^0$ and $\eta$ peaks. Mass cuts are made to remove events with invariant mass $M(\gamma_2\gamma_3) <$ 1.1 GeV.}
\label{fig:gg1}
\end{center}
\end{figure}

\newpage

\clearpage

\begin{figure}[htbp]
\begin{center}
%\vspace*{100pt}
\includegraphics[width=15cm]{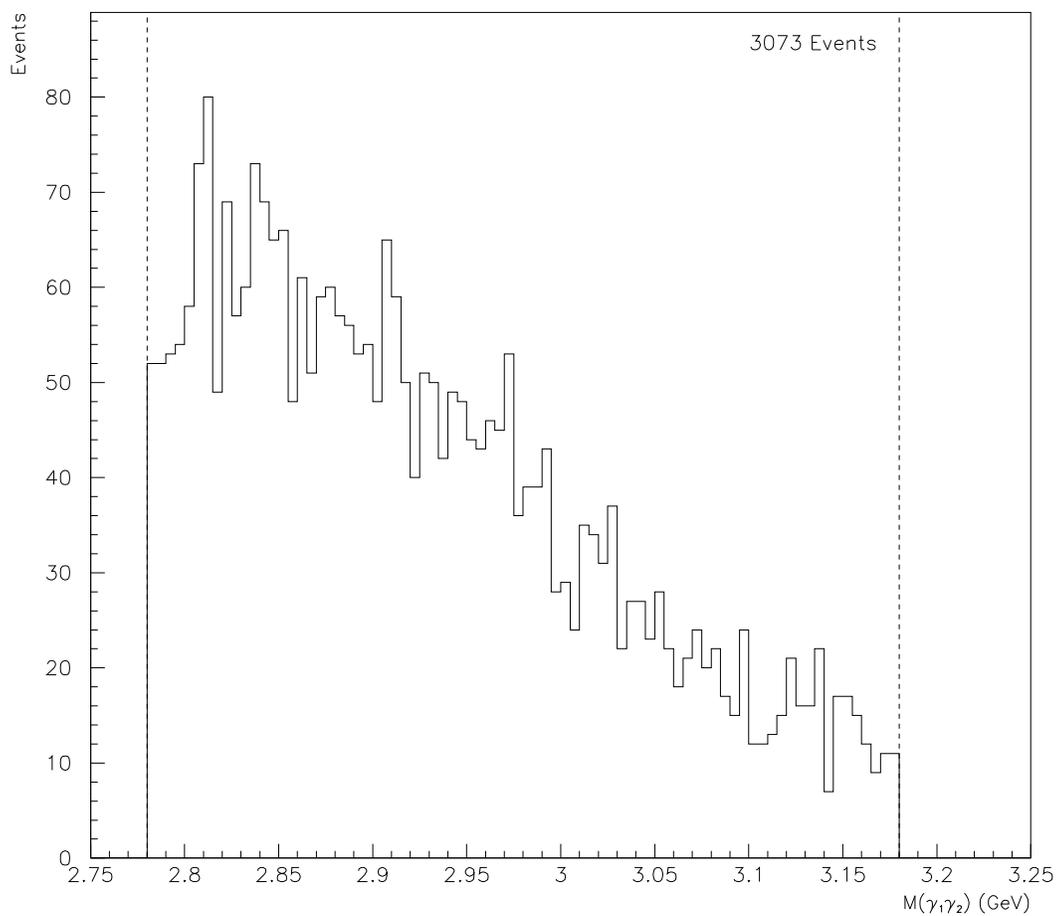}
%\vspace*{-100pt}
\caption[The $\gamma_1\gamma_2$ invariant mass distribution for 3 $\gamma$ events in the E835p $^1P_1$ data with 2.78 GeV $< M(\gamma_1\gamma_2) <$ 3.18 GeV, and $M(\gamma_1\gamma_3), M(\gamma_2\gamma_3) >$ 1.1 GeV.]{The $\gamma_1\gamma_2$ invariant mass distribution for 3 $\gamma$ events in the E835p $^1P_1$ data with 2.78 GeV $< M(\gamma_1\gamma_2) <$ 3.18 GeV, and $M(\gamma_1\gamma_3), M(\gamma_2\gamma_3) >$ 1.1 GeV. As before, no $\eta_c$ enhancement is visible at 2.98 GeV.}
\label{fig:gg1}
\end{center}
\end{figure}

\newpage

\clearpage

\begin{figure}[htbp]
\begin{center}
%\vspace*{100pt}
\includegraphics[width=15cm]{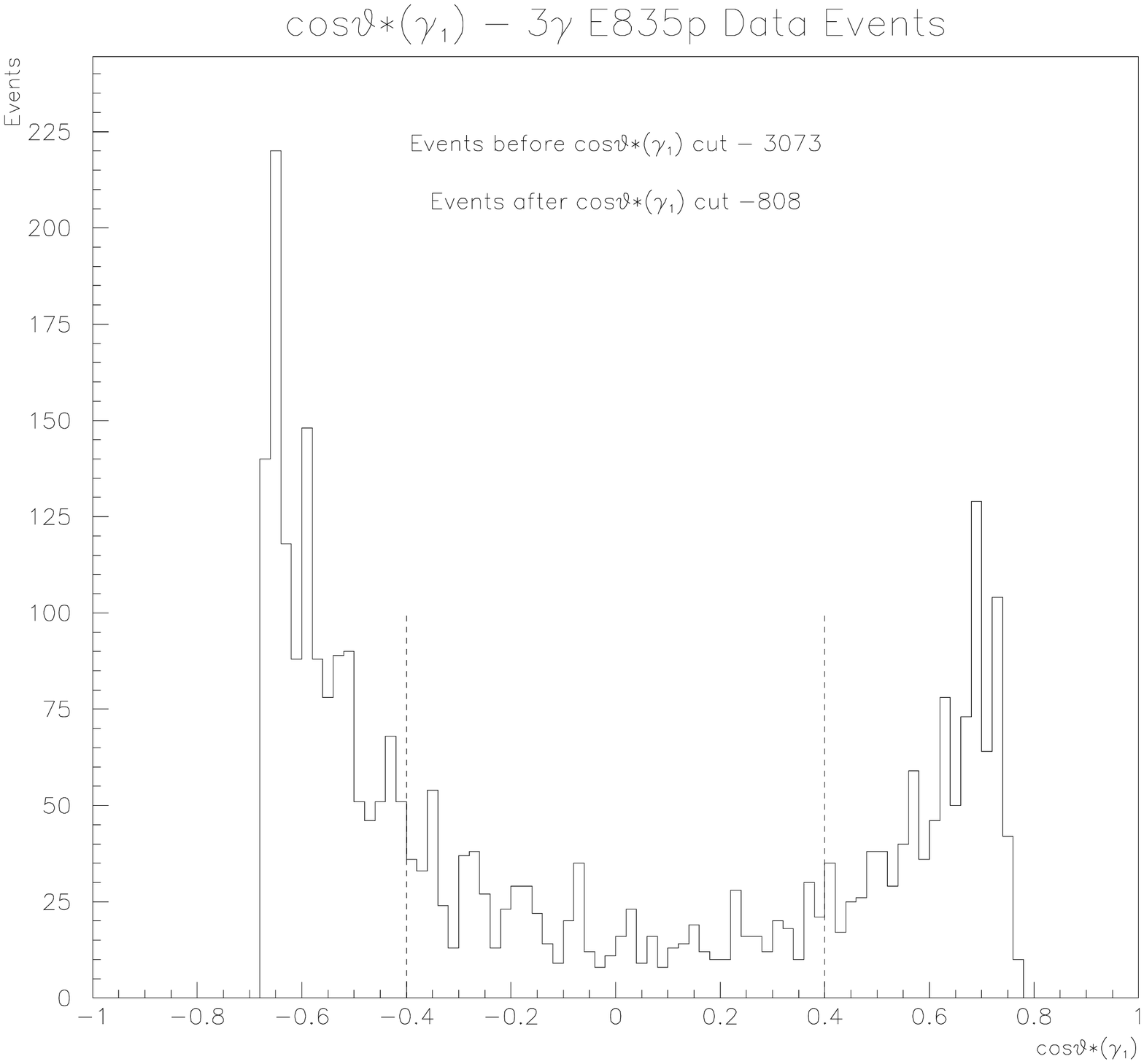}
%\vspace*{-100pt}
\caption[The cos$(\theta^*)(\gamma_1)$ distribution for 3 $\gamma$ events in the E835p $^1P_1$ data with 2.78 GeV $< M(\gamma_1\gamma_2) <$ 3.18 GeV, and $M(\gamma_1\gamma_3), M(\gamma_2\gamma_3) >$ 1.1 GeV.]{The cos$(\theta^*)(\gamma_1)$ distribution for 3 $\gamma$ events in the E835p $^1P_1$ data with 2.78 GeV $< M(\gamma_1\gamma_2) <$ 3.18 GeV, and $M(\gamma_1\gamma_3), M(\gamma_2\gamma_3) >$ 1.1 GeV. As shown in Fig. 5.3, the distribution is expected to be flat within the acceptance range of the detector. A cut is made to remove background events with $\vert \cos(\theta^*)(\gamma_1) \vert > 0.4$.}
\label{fig:gg1}
\end{center}
\end{figure}

\newpage

\clearpage

\begin{figure}[htbp]
\begin{center}
%\vspace*{100pt}
\includegraphics[width=15cm]{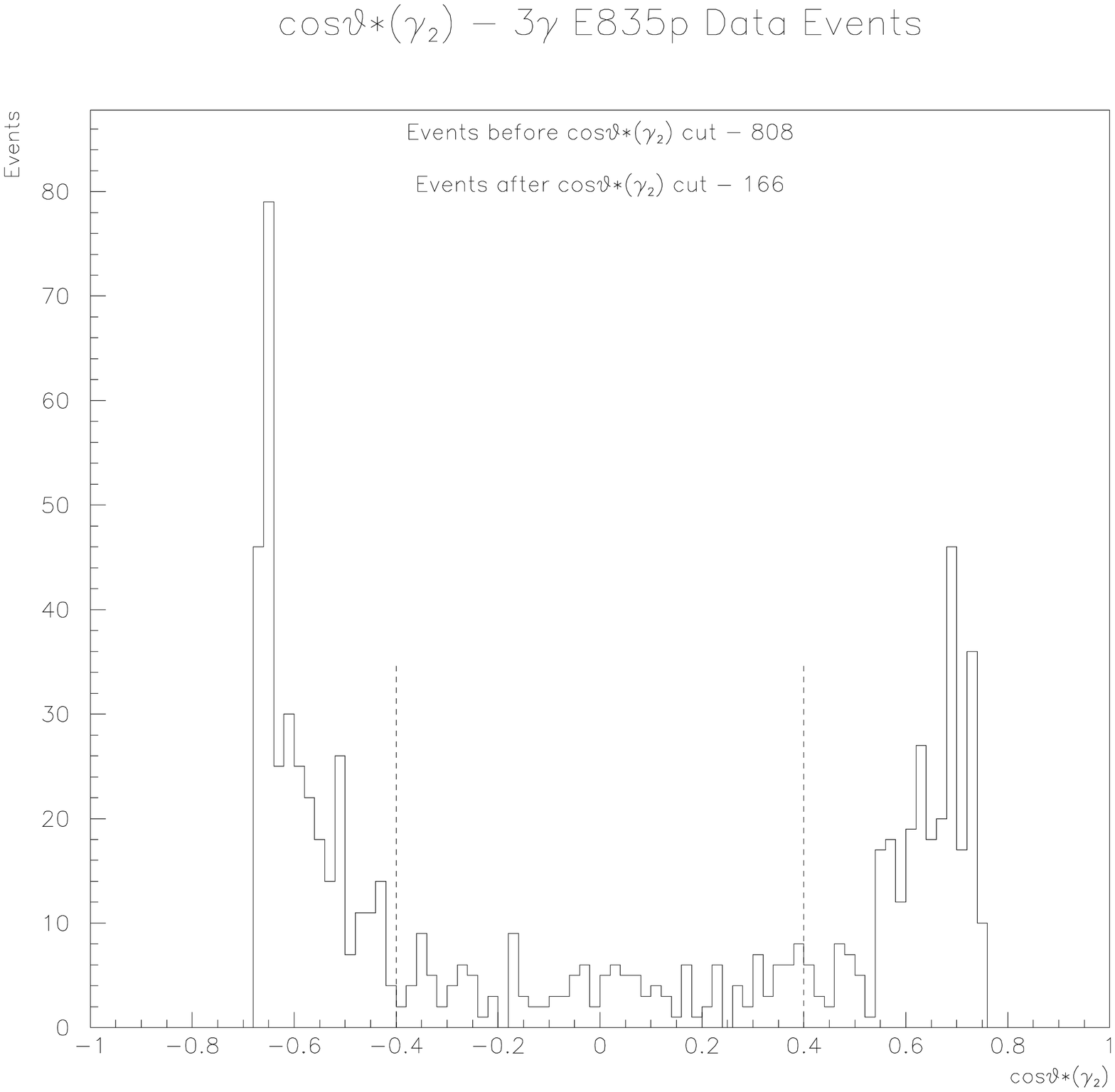}
%\vspace*{-100pt}
\caption[The cos$(\theta^*)(\gamma_2)$ distribution for 3 $\gamma$ events in the E835p $^1P_1$ data with 2.78 GeV $< M(\gamma_1\gamma_2) <$ 3.18 GeV, and $M(\gamma_1\gamma_3), M(\gamma_2\gamma_3) >$ 1.1 GeV and $\vert \cos(\theta^*)(\gamma_1) \vert < 0.4$.]{The cos$(\theta^*)(\gamma_2)$ distribution for 3 $\gamma$ events in the E835p $^1P_1$ data with 2.78 GeV $< M(\gamma_1\gamma_2) <$ 3.18 GeV, and $M(\gamma_1\gamma_3), M(\gamma_2\gamma_3) >$ 1.1 GeV and $\vert \cos(\theta^*)(\gamma_1) \vert < 0.4$. As shown in Fig. 5.3, the distribution is expected to be flat within the acceptance range of the detector. A cut is made to remove background events with $\vert \cos(\theta^*)(\gamma_1) \vert > 0.4$.}
\label{fig:gg1}
\end{center}
\end{figure}

\newpage

\clearpage

\begin{figure}[htbp]
\begin{center}
%\vspace*{100pt}
\includegraphics[width=15cm]{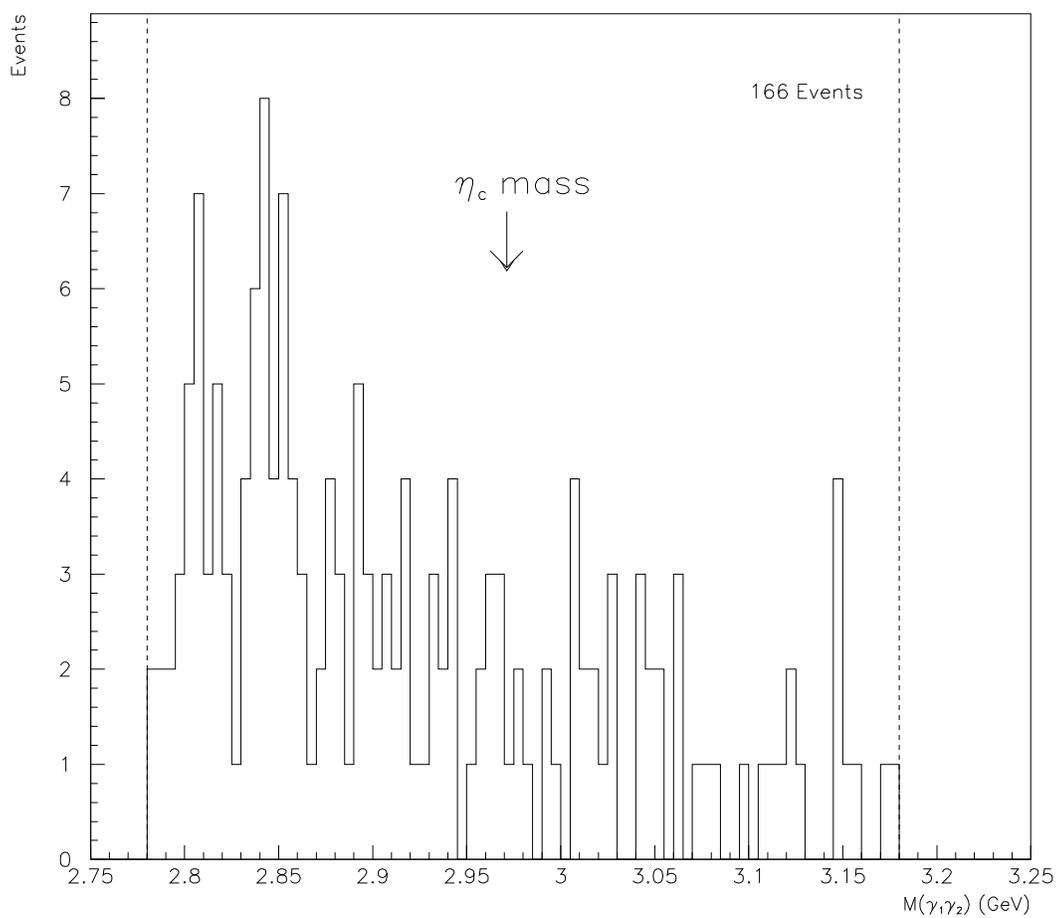}
%\vspace*{-100pt}
\caption{The $M(\gamma_1\gamma_2)$ distribution for 3 $\gamma$ events in the E835p $^1P_1$ data with 2.78 GeV $< M(\gamma_1\gamma_2) <$ 3.18 GeV, and $M(\gamma_1\gamma_3), M(\gamma_2\gamma_3) >$ 1.1 GeV and $\vert \cos(\theta^*)(\gamma_1,\gamma_2) \vert < 0.4$.}
\label{fig:gg1}
\end{center}
\end{figure}

\newpage

\clearpage

\begin{figure}[htbp]
\begin{center}
%\vspace*{100pt}
\includegraphics[width=15cm]{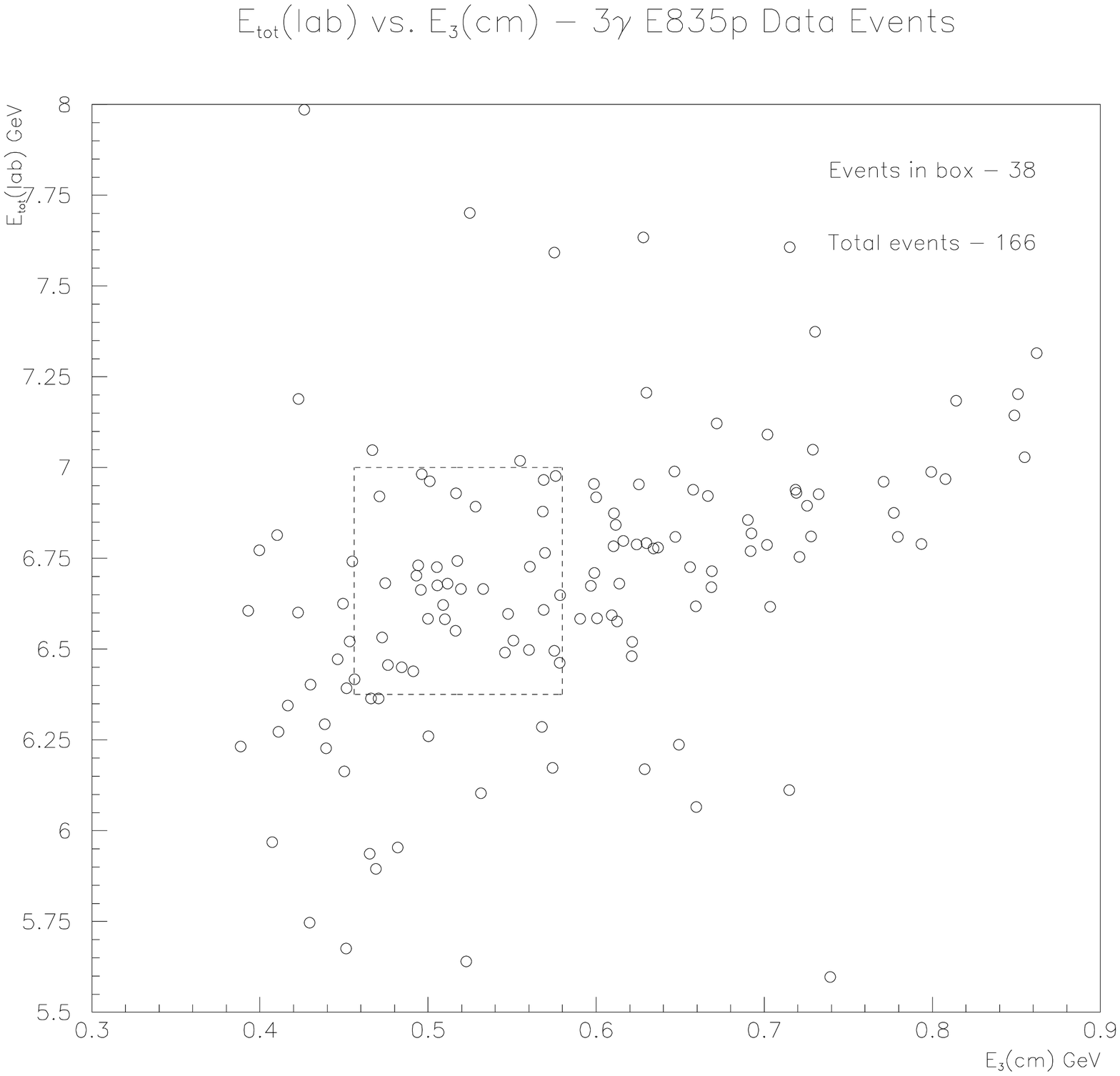}
%\vspace*{-100pt}
\caption[The $E_{tot}$(lab) vs. $E_{cm}(\gamma_3)$ distribution for 3 $\gamma$ events in the E835p $^1P_1$ data with 2.78 GeV $< M(\gamma_1\gamma_2) <$ 3.18 GeV, and $M(\gamma_1\gamma_3), M(\gamma_2\gamma_3) >$ 1.1 GeV and $\vert \cos(\theta^*)(\gamma_1,\gamma_2) \vert < 0.4$.]{The $E_{tot}$(lab) vs. $E_{cm}(\gamma_3)$ distribution for 3 $\gamma$ events in the E835p $^1P_1$ data with 2.78 GeV $< M(\gamma_1\gamma_2) <$ 3.18 GeV, and $M(\gamma_1\gamma_3), M(\gamma_2\gamma_3) >$ 1.1 GeV and $\vert \cos(\theta^*)(\gamma_1,\gamma_2) \vert < 0.4$. $E(tot)$ and $E(\gamma_3)$ cuts have not been made yet. The box outlines the region in which 6.376 GeV $< E(tot) < 6.876$ GeV and 0.454 GeV $< E(\gamma_3) < 0.554$ GeV.}
\label{fig:gg1}
\end{center}
\end{figure}

\newpage

\clearpage

\section{$\eta_c \gamma$ Final Event Selection}

As a final check on the consistancy of the selected events the scatter plot of the $\gamma_1\gamma_2$ invariant mass vs. the center of mass energy of the radiative gamma ($\gamma_3$) is plotted in Figure 5.14. 38 events are left in the final selection, and their properties are given in Tables 5.1 - 5.4. The efficiencies of the various cuts are given in Table 5.5, and their excitation curve is shown in Figure 5.15. The average cross section for the reaction $p\bar p \rightarrow \eta_c\gamma$ is measured to be 16.8 $\pm$ 2.7 pb in the 3523-3529 MeV region. This is compared to the excitation curve from the 3 $\gamma$ data in the $\chi_0$ energy region generated using the same set of cuts, which is shown in Figure 5.16. with an average cross section of 11.3 $\pm$ 3.4 pb. 

\newpage

\clearpage

\begin{figure}[htbp]
\begin{center}
%\vspace*{100pt}
\includegraphics[width=15cm]{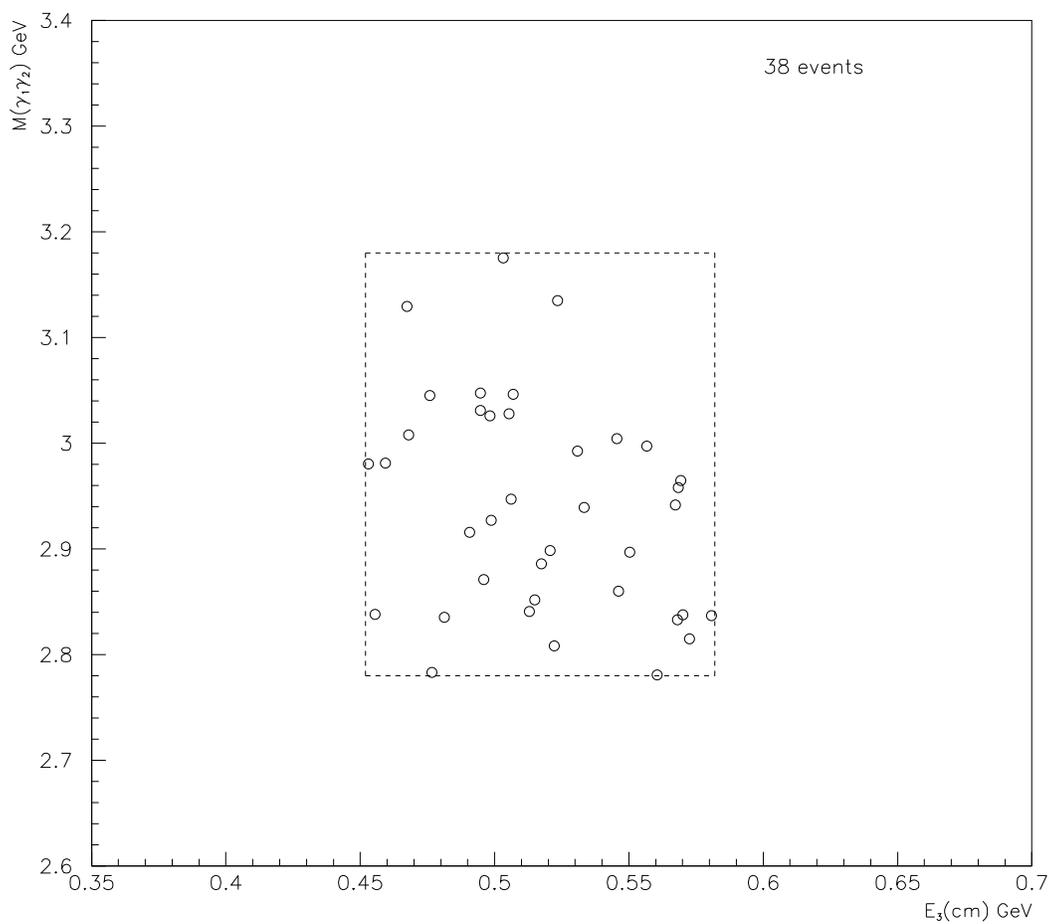}
%\vspace*{-100pt}
\caption{The $M(\gamma_1\gamma_2)$ vs. $E_{cm}(\gamma_3)$ distribution for all 3 $\gamma$ events in the E835p $^1P_1$ data in the final event selection, including cuts on $E_{tot}(lab)$ and $E_{cm}(\gamma_3)$ as shown in Figs 4.38 and 4.39.}
\label{fig:gg1}
\end{center}
\end{figure}

\newpage

\clearpage

\linespread{1.5}
\begin{table}[htbp]
\begin{center}
\begin{tabular}{|c|c|c|c|c|c|c|c|}
\hline
 & & & & & & & \\
Stack & Run & Event & $E_{cm}$ & $M_{12}$ & $E_{cm}(\gamma_3)$ & $M_{13}$ & $M_{23}$ \\
 & & & & & & & \\
 & & & (GeV) & (GeV) & (GeV) & (GeV) & (GeV) \\
 & & & & & & & \\
\hline
 & & & & & & & \\
 15 & 5599 & 4621993 & 3526.0 & 2.976 & 0.455 & 1.288 & 1.183 \\
 & & & & & & & \\
 15 & 5600 & 677914 & 3526.0 & 2.979 & 0.459 & 1.102 & 1.401 \\
 & & & & & & & \\
 15 & 5601 & 758986 & 3526.0 & 3.003 & 0.544 & 1.388 & 1.522 \\
 & & & & & & & \\
 15 & 5610 & 2351648 & 3526.0 & 2.850 & 0.515 & 1.241 & 1.412 \\
 & & & & & & & \\
 16 & 5616 & 1403370 & 3525.7 & 2.808 & 0.573 & 1.475 & 1.336 \\
 & & & & & & & \\
 16 & 5628 & 1823537 & 3525.7 & 2.955 & 0.570 & 1.589 & 1.289 \\
 & & & & & & & \\
 17 & 5648 & 647975 & 3526.5 & 2.841 & 0.512 & 1.439 & 1.240 \\
 & & & & & & & \\
 18 & 5684 & 358509 & 3525.95 & 3.031 & 0.507 & 1.444 & 1.258 \\
 & & & & & & & \\
 19 & 5697 & 3768674 & 3526.8 & 2.834 & 0.569 & 1.626 & 1.107 \\
 & & & & & & & \\
 19 & 5700 & 725256 & 3526.8 & 3.046 & 0.505 & 1.409 & 1.304 \\
 & & & & & & & \\
 19 & 5703 & 158216 & 3526.8 & 2.866 & 0.496 & 1.134 & 1.495 \\
 & & & & & & & \\
 20 & 5716 & 1083628 & 3525.4 & 2.837 & 0.568 & 1.631 & 1.199 \\
 & & & & & & & \\
 20 & 5717 & 3632928 & 3525.4 & 3.132 & 0.468 & 1.350 & 1.271 \\
 & & & & & & & \\
 20 & 5718 & 4506605 & 3525.4 & 2.965 & 0.569 & 1.523 & 1.330 \\
 & & & & & & & \\
 21 & 5736 & 1433245 & 3525.1 & 2.898 & 0.550 & 1.595 & 1.119 \\
 & & & & & & & \\
 21 & 5740 & 917860 & 3525.1 & 2.839 & 0.455 & 1.152 & 1.293 \\
 & & & & & & & \\
 21 & 5741 & 1612904 & 3525.1 & 2.883 & 0.517 & 1.500 & 1.105 \\
 & & & & & & & \\
 21 & 5743 & 1300374 & 3525.1 & 2.927 & 0.496 & 1.107 & 1.524 \\
 & & & & & & & \\
 22 & 5747 & 560676 & 3526.2 & 3.027 & 0.497 & 1.188 & 1.471 \\
 & & & & & & & \\
\hline
 & & & & & & & \\
 Expected & & & & 2.982 & 0.504 & & \\
 & & & & & & & \\
\hline
\end{tabular}
\caption{Properties of the events in the final selection from the E835p $^1P_1 \rightarrow \eta_c \gamma \rightarrow 3 \gamma$ data channel (Part I). The center of mass energy of the third gamma is given as are the three possible invariant masses.}
\end{center}
\end{table}

\newpage

\clearpage

\linespread{1.5}
\begin{table}[htbp]
\begin{center}
\begin{tabular}{|c|c|c|c|c|c|c|c|}
\hline
 & & & & & & & \\
Stack & Run & Event & $E_{cm}$ & $M_{12}$ & $E_{cm}(\gamma_3)$ & $M_{13}$ & $M_{23}$ \\
 & & & & & & & \\
 & & & (GeV) & (GeV) & (GeV) & (GeV) & (GeV) \\
 & & & & & & & \\
\hline
 & & & & & & & \\
 22 & 5751 & 2545491 & 3526.2 & 2.919 & 0.490 & 1.411 & 1.190 \\
 & & & & & & & \\
 23 & 5758 & 500997 & 3523.3 & 3.131 & 0.522 & 1.233 & 1.501 \\
 & & & & & & & \\
 23 & 5760 & 2465962 & 3523.3 & 2.783 & 0.479 & 1.101 & 1.456 \\
 & & & & & & & \\
 23 & 5763 & 461928 & 3523.3 & 3.002 & 0.468 & 1.197 & 1.398 \\
 & & & & & & & \\
 23 & 5763 & 1715630 & 3523.3 & 3.054 & 0.476 & 1.490 & 1.133 \\
 & & & & & & & \\
 24 & 5771 & 594311 & 3526.25 & 2.994 & 0.530 & 1.475 & 1.322 \\
 & & & & & & & \\
 25 & 5779 & 1899147 & 3526.25 & 3.047 & 0.496 & 1.149 & 1.525 \\
 & & & & & & & \\
 27 & 5792 & 3702814 & 3528.9 & 3.026 & 0.496 & 1.498 & 1.215 \\
 & & & & & & & \\
 27 & 5799 & 533891 & 3528.9 & 2.859 & 0.547 & 1.364 & 1.432 \\
 & & & & & & & \\
 27 & 5800 & 584454 & 3528.9 & 2.902 & 0.518 & 1.310 & 1.408 \\
 & & & & & & & \\
 28 & 5811 & 134846 & 3528.5 & 2.835 & 0.580 & 1.440 & 1.412 \\
 & & & & & & & \\
 28 & 5812 & 2006974 & 3528.5 & 2.808 & 0.522 & 1.504 & 1.126 \\
 & & & & & & & \\
 35 & 7005 & 673075 & 3526.2 & 2.782 & 0.563 & 1.255 & 1.523 \\
 & & & & & & & \\
 35 & 7005 & 2551591 & 3526.2 & 3.176 & 0.502 & 1.403 & 1.322 \\
 & & & & & & & \\
 37 & 7021 & 2417964 & 3525.8 & 2.943 & 0.567 & 1.433 & 1.380 \\
 & & & & & & & \\
 38 & 7025 & 3092852 & 3525.8 & 2.839 & 0.482 & 1.183 & 1.313 \\
 & & & & & & & \\
 38 & 7028 & 318008 & 3525.8 & 2.994 & 0.559 & 1.224 & 1.575 \\
 & & & & & & & \\
 39 & 7031 & 2763321 & 3524.7 & 2.940 & 0.532 & 1.392 & 1.363 \\
 & & & & & & & \\
 39 & 7036 & 2967823 & 3524.7 & 2.948 & 0.506 & 1.204 & 1.475 \\
 & & & & & & & \\
\hline
 & & & & & & & \\
 Expected & & & & 2.982 & 0.504 & & \\
 & & & & & & & \\
\hline
\end{tabular}
\caption{Properties of the events in the final selection from the E835p $^1P_1 \rightarrow \eta_c \gamma \rightarrow 3 \gamma$ data channel (Part II). The center of mass energy of the third gamma is given as are the three possible invariant masses.}
\end{center}
\end{table}

\newpage

\clearpage

\linespread{1.5}
\begin{table}[htbp]
\begin{center}
\begin{tabular}{|c|c|c|c|c|c|c|}
\hline
 & & & & & & \\
Stack & Run & Event & $cos(\theta^{*}_1)$ & $cos(\theta^{*}_2)$ & $E_{lab}(tot)$ & $E_{cm}(\gamma_3)$ \\ 
 & & & & & & \\
 & & & & & (GeV) & (GeV) \\
 & & & & & & \\
\hline
 & & & & & & \\
 15 & 5599 & 4621993 & -0.208 & -0.038 & 6.738 & 0.455 \\
 & & & & & & \\
 15 & 5600 & 677914 & -0.263 & 0.289 & 6.421 & 0.459 \\
 & & & & & & \\
 15 & 5601 & 758986 & -0.278 & -0.070 & 6.480 & 0.544 \\
 & & & & & & \\
 15 & 5610 & 2351648 & -0.391 & -0.048 & 6.599 & 0.515 \\
 & & & & & & \\
 16 & 5616 & 1403370 & 0.206 & -0.171 & 6.497 & 0.573 \\
 & & & & & & \\
 16 & 5628 & 1823537 & -0.257 & 0.007 & 6.990 & 0.570 \\
 & & & & & & \\
 17 & 5648 & 647975 & -0.394 & -0.323 & 6.680 & 0.512 \\
 & & & & & & \\
 18 & 5684 & 358509 & -0.338 & 0.360 & 6.732 & 0.507 \\
 & & & & & & \\
 19 & 5697 & 3768674 & -0.088 & 0.363 & 6.618 & 0.569 \\
 & & & & & & \\
 19 & 5700 & 725256 & 0.250 & 0.340 & 6.685 & 0.505 \\
 & & & & & & \\
 19 & 5703 & 158216 & -0.341 & -0.374 & 6.671 & 0.496 \\
 & & & & & & \\
 20 & 5716 & 1083628 & -0.098 & -0.045 & 6.758 & 0.568 \\
 & & & & & & \\
 20 & 5717 & 3632928 & -0.300 & 0.018 & 6.906 & 0.468 \\
 & & & & & & \\
 20 & 5718 & 4506605 & 0.371 & 0.062 & 6.877 & 0.569 \\
 & & & & & & \\
 21 & 5736 & 1433245 & -0.334 & 0.378 & 6.545 & 0.550 \\
 & & & & & & \\
 21 & 5740 & 917860 & 0.084 & 0.028 & 6.393 & 0.455 \\
 & & & & & & \\
 21 & 5741 & 1612904 & 0.265 & 0.400 & 6.728 & 0.517 \\
 & & & & & & \\
 21 & 5743 & 1300374 & -0.304 & 0.043 & 6.588 & 0.496 \\
 & & & & & & \\
 22 & 5747 & 560676 & 0.370 & 0.317 & 6.735 & 0.497 \\
 & & & & & & \\
\hline
 & & & & & & \\
 Expected & & & & & 6.626 & 0.504 \\
 & & & & & & \\
\hline
\end{tabular}
\caption{Properties of the events in the final selection from the E835p $^1P_1 \rightarrow \eta_c \gamma \rightarrow 3 \gamma$ data channel (Part I). The angles of $\gamma_1$ and $\gamma_2$ are given, as well as the total event energy in the lab frame, and the center of mass energy of $\gamma_3$.}
\end{center}
\end{table}

\newpage

\clearpage

\linespread{1.5}
\begin{table}[htbp]
\begin{center}
\begin{tabular}{|c|c|c|c|c|c|c|}
\hline
 & & & & & & \\
Stack & Run & Event & $cos(\theta^{*}_1)$ & $cos(\theta^{*}_2)$ & $E_{lab}(tot)$ & $E_{cm}(\gamma_3)$ \\ 
 & & & & & & \\
 & & & & & (GeV) & (GeV) \\
 & & & & & & \\
\hline
 & & & & & & \\
 22 & 5751 & 2545491 & -0.151 & 0.047 & 6.439 & 0.490 \\
 & & & & & & \\
 23 & 5758 & 500997 & 0.240 & -0.391 & 6.945 & 0.522 \\
 & & & & & & \\
 23 & 5760 & 2465962 & 0.240 & 0.059 & 6.473 & 0.479 \\
 & & & & & & \\
 23 & 5763 & 461928 & 0.373 & 0.366 & 6.539 & 0.468 \\
 & & & & & & \\
 23 & 5763 & 1715630 & 0.038 & 0.080 & 6.699 & 0.476 \\
 & & & & & & \\
 24 & 5771 & 594311 & -0.171 & 0.374 & 6.893 & 0.530 \\
 & & & & & & \\
 25 & 5779 & 1899147 & -0.078 & -0.026 & 6.706 & 0.496 \\
 & & & & & & \\
 27 & 5792 & 3702814 & 0.152 & 0.011 & 6.990 & 0.496 \\
 & & & & & & \\
 27 & 5799 & 533891 & -0.157 & -0.038 & 6.500 & 0.547 \\
 & & & & & & \\
 27 & 5800 & 584454 & 0.219 & 0.041 & 6.668 & 0.518 \\
 & & & & & & \\
 28 & 5811 & 134846 & -0.248 & -0.237 & 6.649 & 0.580 \\
 & & & & & & \\
 28 & 5812 & 2006974 & -0.145 & -0.031 & 6.552 & 0.522 \\
 & & & & & & \\
 35 & 7005 & 673075 & 0.362 & -0.342 & 6.484 & 0.563 \\
 & & & & & & \\
 35 & 7005 & 2551591 & 0.258 & -0.330 & 6.950 & 0.502 \\
 & & & & & & \\
 37 & 7021 & 2417964 & -0.008 & 0.237 & 6.953 & 0.567 \\
 & & & & & & \\
 38 & 7025 & 3092852 & -0.045 & 0.389 & 6.432 & 0.482 \\
 & & & & & & \\
 38 & 7028 & 318008 & 0.267 & 0.302 & 6.741 & 0.559 \\
 & & & & & & \\
 39 & 7031 & 2763321 & 0.277 & -0.363 & 6.665 & 0.532 \\
 & & & & & & \\
 39 & 7036 & 2967823 & 0.282 & 0.067 & 6.616 & 0.506 \\
 & & & & & & \\
\hline
 & & & & & & \\
 Expected & & & & & 6.626 & 0.504 \\
 & & & & & & \\
\hline
\end{tabular}
\caption{Properties of the events in the final selection from the E835p $^1P_1 \rightarrow \eta_c \gamma \rightarrow 3 \gamma$ data channel (Part II). The angles of $\gamma_1$ and $\gamma_2$ are given, as well as the total event energy in the lab frame, and the center of mass energy of $\gamma_3$.}
\end{center}
\end{table}

\newpage

\clearpage

\linespread{1.5}
\begin{table}[htbp]
\begin{center}
\begin{tabular}{|c|c|}
\hline
 & \\
Cut & Efficiency \\
 & \\
\hline
 & \\
 Trigger & $90\% \pm 1\%$ \\
 & \\
 3 Cluster Cut & $94\% \pm 1\%$ \\
 & \\
\hline
 & \\
 Analysis Cuts & \\
 & \\
 2780 MeV $\le M(\gamma_1\gamma_2) \le 3180$ MeV & $76.9\% \pm 0.8\%$ \\
 & \\
 $M(\gamma_1\gamma_3), M(\gamma_2\gamma_3) \le 1100$ MeV & $37.8\% \pm 0.3\%$ \\
 & \\
 $\vert \cos(\theta^{*}_{1,2}) \vert \le$ 0.4 & $51.5\% \pm 0.4\%$ \\
 & \\
6376 MeV $< E_{tot}$(lab)$ <$ 7000 GeV & $91.8\% \pm 4.0\%$ \\
 & \\
454 MeV $< E_{\gamma_3}(cm) <$ 580 MeV & $88.8\% \pm 4.0\%$ \\
 & \\
 Total Analysis Cuts & $(12.2 \pm 0.8) \%$ \\ 
 & \\
\hline
 & \\
 Acceptance & $(48.7 \pm 0.1) \%$ \\
 & \\
\hline
 & \\
 Total & $(5.0 \pm 0.3)\%$ \\
 & \\
\hline
\end{tabular}
\caption{Efficiencies of all cuts used in $\eta_c \gamma$ channel.}
%\label{tab:3}
\end{center}
\end{table}

\newpage

\clearpage

\begin{figure}[htbp]
\begin{center}
\includegraphics[width=15cm]
{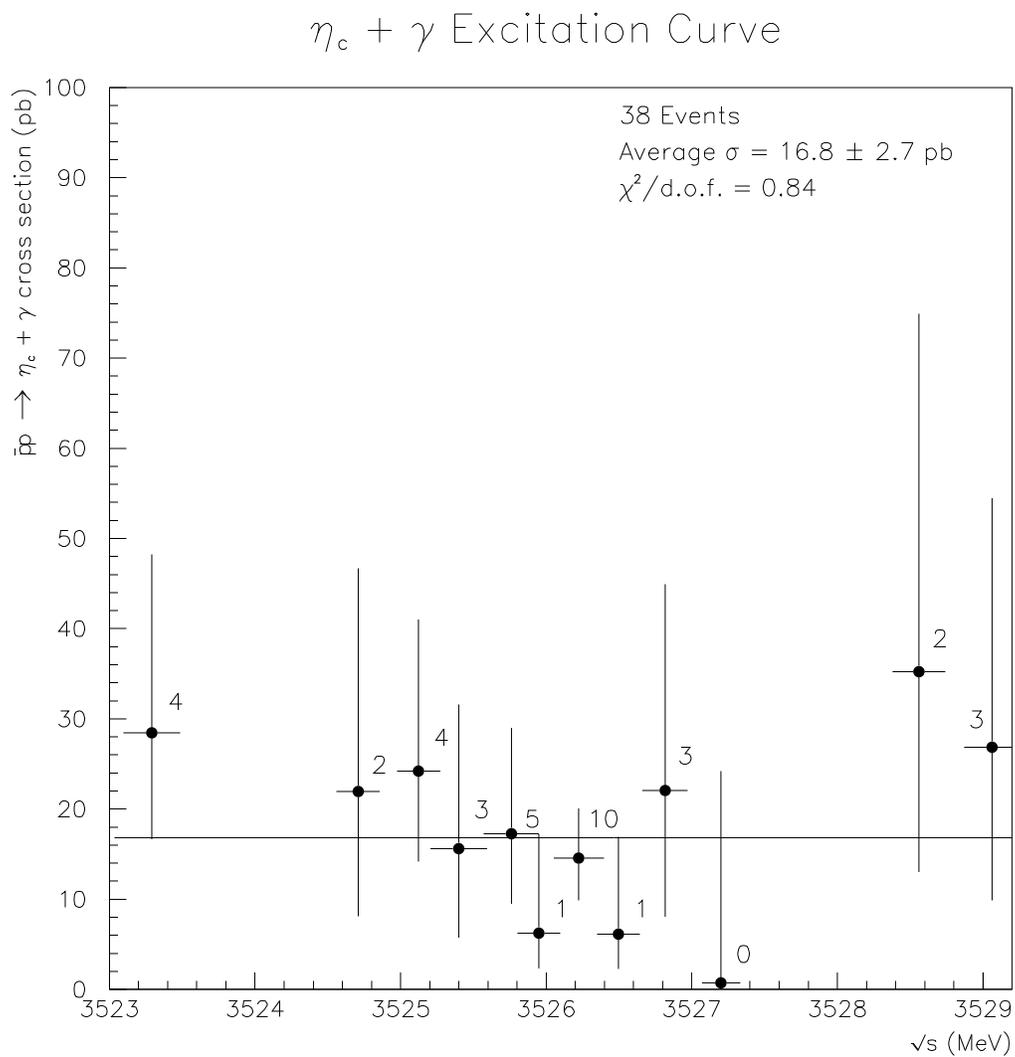}
\caption[Efficiency corrected excitation curve for the $\eta_c + \gamma$ channel.]{Efficiency corrected excitation curve for the $\eta_c + \gamma$ channel. Stacks with energies less than 0.15 MeV apart have been combined. Numbers beside each point denote number of observed events at that energy. The straight line is the fit to a constant cross section.}
\label{fig:9}
\end{center}
\end{figure}

\newpage

\clearpage

\begin{figure}[htbp]
\begin{center}
\includegraphics[width=15cm]
{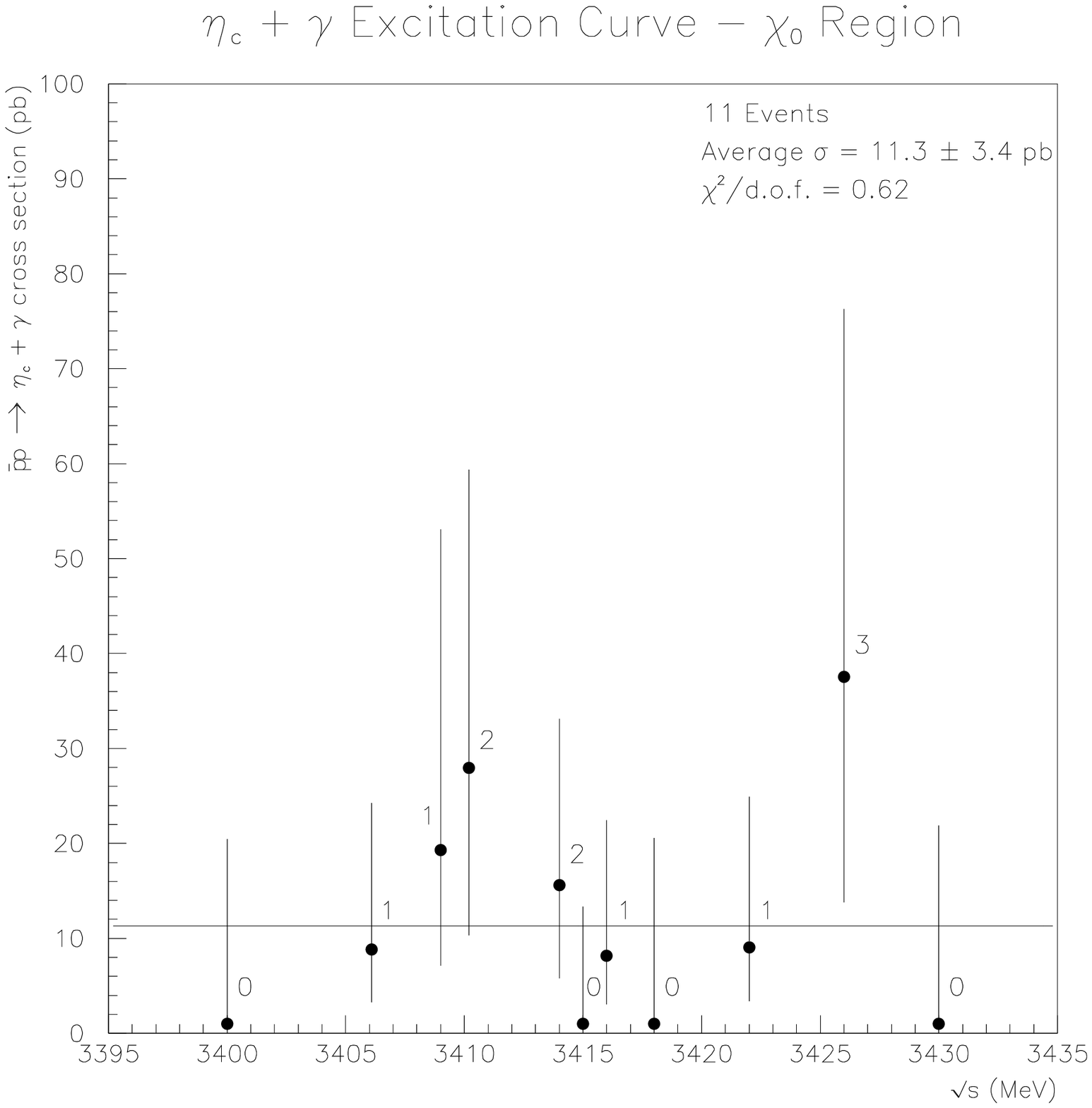}
\caption[Efficiency corrected excitation curve for the $\eta_c + \gamma$ channel in the $\chi_0$ energy region.]{Efficiency corrected excitation curve for the $\eta_c + \gamma$ channel in the $\chi_0$ energy region. The straight line is the fit to a constant cross section.}
\label{fig:9}
\end{center}
\end{figure}

%\include{chapter5_oldneutraltrigger}

%\addcontentsline{toc}{chapter}{Appendices}

\appendix
\baselineskip=24pt
\chapter{$p\bar p \rightarrow ^1P_1 \rightarrow J/\psi + \pi^0 \rightarrow e^{+}e^{-}\gamma\gamma$ Selected Events}

Tables A.1-5 list the properties of all 107 events included in the final event selection for the channel $p\bar p \rightarrow ^1P_1 \rightarrow J/\psi + \pi^0 \rightarrow e^{+}e^{-} \gamma\gamma$, as described in Chapter 4. Properties included in the tables are the stack, run, and event numbers of each event, the two electron weights (EW(1), EW(2)), the invariant mass of the $e^{+}e^{-}$ pair $M(e^{+}e^{-})$, the invariant mass of the $\gamma\gamma$ pair $M(\gamma\gamma)$ which make a $\pi^0$, the $J/\psi + \pi^0$ probability as calculated by a 6 constraint kinematic fit, and the width of the beam for each run.

\linespread{1.5}
\begin{table}[htbp]
\begin{center}
\begin{tabular}{|c|c|c|c|c|c|c|c|c|}
\hline
 & & & & & & & & \\
Stack & Run & Event & EW1 & EW2 & $M(e^{+}e^{-})$ & $M(\gamma\gamma)$ & Prob. & $\sigma_{beam}$ \\
 & & & & & & & & \\
 & & & & & (GeV) & (GeV) & & (keV) \\
 & & & & & & & & \\
\hline
 & & & & & & & & \\
 15 & 5605 & 4479537 & 181.1 & 7.1 & 2.970 & 0.113 & 0.338 & 431.4 \\
 & & & & & & & & \\
 15 & 5609 & 4355890 & 226.7 & 34.7 & 3.101 & 0.105 & 0.546 & 387.5 \\
 & & & & & & & & \\
 15 & 5610 & 5077058 & 33.3 & 358.4 & 3.056 & 0.115 & 0.159 & 380.9 \\
 & & & & & & & & \\
 15 & 5611 & 2957443 & 1431.1 & 91.6 & 3.025 & 0.125 & 0.540 & 368.0 \\
 & & & & & & & & \\
 16 & 5618 & 2875000 & 1382.2 & 0.7 & 3.025 & 0.133 & 0.651 & 392.7 \\
 & & & & & & & & \\
 16 & 5619 & 4926219 & 0.4 & 21.7 & 3.097 & 0.125 & 0.884 & 373.5 \\
 & & & & & & & & \\
 16 & 5629 & 866076 & 271.3 & 35.0 & 3.047 & 0.126 & 0.276 & 368.9 \\
 & & & & & & & & \\
 16 & 5629 & 1920441 & 2373.7 & 1007.9 & 3.195 & 0.125 & 0.809 & 368.9 \\
 & & & & & & & & \\
 16 & 5630 & 555181 & 0.2 & 101.5 & 2.932 & 0.133 & 0.627 & 358.8 \\
 & & & & & & & & \\
 16 & 5630 & 3527765 & 299.1 & 12.3 & 3.118 & 0.117 & 0.344 & 358.8 \\
 & & & & & & & & \\
 16 & 5631 & 1492459 & 1.5 & 2.9 & 2.961 & 0.138 & 0.922 & 348.8 \\
 & & & & & & & & \\
 16 & 5631 & 3251874 & 2541.2 & 1955.1 & 3.095 & 0.128 & 0.709 & 348.8 \\
 & & & & & & & & \\
 17 & 5638 & 321303 & 3120.0 & 1454.7 & 3.198 & 0.119 & 0.138 & 334.4 \\
 & & & & & & & & \\
 17 & 5638 & 718261 & 251.0 & 83.4 & 3.212 & 0.135 & 0.857 & 334.4 \\
 & & & & & & & & \\
 17 & 5638 & 838091 & 7096.2 & 151.8 & 3.081 & 0.133 & 0.618 & 334.4 \\
 & & & & & & & & \\
 17 & 5638 & 3539721 & 108.9 & 482.1 & 3.149 & 0.136 & 0.883 & 334.4 \\
 & & & & & & & & \\
 17 & 5641 & 1675526 & 666.6 & 0.6 & 3.045 & 0.149 & 0.695 & 258.4 \\
 & & & & & & & & \\
 17 & 5647 & 613292 & 1464.1 & 76.5 & 3.188 & 0.144 & 0.844 & 240.8 \\
 & & & & & & & & \\
 17 & 5648 & 3327131 & 1801.8 & 27.7 & 3.044 & 0.132 & 0.639 & 229.4 \\
 & & & & & & & & \\
 18 & 5671 & 69884 & 482.5 & 4147.7 & 3.110 & 0.128 & 0.989 & 330.7 \\
 & & & & & & & & \\
 18 & 5671 & 279531 & 206.4 & 72.2 & 3.254 & 0.160 & 0.874 & 330.7 \\
 & & & & & & & & \\
\hline
\end{tabular}
\caption{Properties of the events in the final selection from the E835p $^1P_1 \rightarrow J/\psi \pi^0 \rightarrow e^{+}e^{-} \gamma\gamma$ data channel (Part 1).}
\end{center}
\end{table}

\newpage

\clearpage

\linespread{1.5}
\begin{table}[htbp]
\begin{center}
\begin{tabular}{|c|c|c|c|c|c|c|c|c|}
\hline
 & & & & & & & & \\
Stack & Run & Event & EW1 & EW2 & $M(e^{+}e^{-})$ & $M(\gamma\gamma)$ & Prob. &
 $\sigma_{beam}$ \\
 & & & & & & & & \\
 & & & & & (GeV) & (GeV) & & (keV) \\
 & & & & & & & & \\
\hline
 & & & & & & & & \\
 18 & 5673 & 1834690 & 83.8 & 431.9 & 3.130 & 0.126 & 0.642 & 297.3 \\
 & & & & & & & & \\
 18 & 5673 & 1959953 & 47.0 & 0.5 & 3.237 & 0.129 & 0.260 & 297.3 \\
 & & & & & & & & \\
 18 & 5677 & 3058269 & 30.7 & 651.2 & 3.048 & 0.121 & 0.272 & 256.1 \\
 & & & & & & & & \\
 18 & 5684 & 2155715 & 19.5 & 45.0 & 2.952 & 0.149 & 0.168 & 233.1 \\
 & & & & & & & & \\
 19 & 5697 & 2169647 & 935.8 & 3.5 & 3.193 & 0.139 & 0.872 & 307.6 \\
 & & & & & & & & \\
 19 & 5700 & 211487 & 742.4 & 2289.8 & 3.134 & 0.116 & 0.763 & 322.9 \\
 & & & & & & & & \\
 19 & 5700 & 667949 & 122.5 & 1330.7 & 3.149 & 0.145 & 0.851 & 322.9 \\
 & & & & & & & & \\
 19 & 5702 & 3013363 & 1464.0 & 0.4 & 3.149 & 0.132 & 0.439 & 308.8 \\
 & & & & & & & & \\
 19 & 5702 & 4119270 & 3.3 & 1398.0 & 3.118 & 0.116 & 0.342 & 308.8 \\
 & & & & & & & & \\
 19 & 5705 & 772006 & 21.2 & 6.8 & 3.057 & 0.120 & 0.238 & 288.6 \\
 & & & & & & & & \\
 20 & 5708 & 1372983 & 164.8 & 5873.4 & 3.005 & 0.152 & 0.125 & 416.4 \\
 & & & & & & & & \\
 20 & 5711 & 2510267 & 25.6 & 939.4 & 3.176 & 0.124 & 0.556 & 337.1 \\
 & & & & & & & & \\
 20 & 5713 & 1177089 & 3077.6 & 0.03 & 3.080 & 0.119 & 0.755 & 336.0 \\
 & & & & & & & & \\
 20 & 5717 & 5072111 & 85.1 & 1289.6 & 3.087 & 0.139 & 0.849 & 280.7 \\
 & & & & & & & & \\
 20 & 5719 & 4043168 & 203.4 & 33.0 & 3.135 & 0.169 & 0.490 & 251.9 \\
 & & & & & & & & \\
 20 & 5719 & 6160608 & 8000.1 & 4.1 & 3.249 & 0.121 & 0.093 & 251.9 \\
 & & & & & & & & \\
 20 & 5720 & 209317 & 322.6 & 28.9 & 2.989 & 0.093 & 0.074 & 242.9 \\
 & & & & & & & & \\
 20 & 5720 & 582624 & 56.8 & 299.5 & 3.113 & 0.141 & 0.946 & 242.9 \\
 & & & & & & & & \\
 20 & 5720 & 2377875 & 46.5 & 65.8 & 3.111 & 0.142 & 0.922 & 242.9 \\
 & & & & & & & & \\
 20 & 5720 & 2696972 & 199.2 & 0.03 & 3.029 & 0.109 & 0.140 & 242.9 \\
 & & & & & & & & \\
 20 & 5721 & 5431093 & 220.7 & 100.1 & 3.032 & 0.137 & 0.625 & 226.9 \\
 & & & & & & & & \\
 21 & 5734 & 372235 & 659.4 & 31.6 & 3.179 & 0.131 & 0.979 & 319.0 \\
 & & & & & & & & \\
\hline
\end{tabular}
\caption{Properties of the events in the final selection from the E835p $^1P_1 \rightarrow J/\psi \pi^0 \rightarrow e^{+}e^{-} \gamma\gamma$ data channel (Part 2).}
\end{center}
\end{table}

\newpage

\clearpage

\linespread{1.5}
\begin{table}[htbp]
\begin{center}
\begin{tabular}{|c|c|c|c|c|c|c|c|c|}
\hline
 & & & & & & & & \\
Stack & Run & Event & EW1 & EW2 & $M(e^{+}e^{-})$ & $M(\gamma\gamma)$ & Prob. & $\sigma_{beam}$ \\
 & & & & & & & & \\
 & & & & & (GeV) & (GeV) & & (keV) \\
 & & & & & & & & \\
\hline
 & & & & & & & & \\
 21 & 5735 & 3284977 & 1.6 & 193.1 & 3.052 & 0.117 & 0.658 & 316.6 \\
 & & & & & & & & \\
 21 & 5738 & 3315627 & 26.6 & 0.9 & 3.289 & 0.124 & 0.763 & 293.5 \\
 & & & & & & & & \\
 21 & 5740 & 1562701 & 516.0 & 2551.7 & 3.134 & 0.131 & 0.901 & 284.0 \\
 & & & & & & & & \\
 21 & 5741 & 380925 & 31.5 & 1266.1 & 3.115 & 0.144 & 0.573 & 293.0 \\
 & & & & & & & & \\
 21 & 5742 & 392712 & 344.5 & 25.4 & 3.087 & 0.131 & 0.050 & 282.0 \\
 & & & & & & & & \\
 22 & 5749 & 748352 & 66.2 & 247.1 & 3.265 & 0.148 & 0.755 & 286.3 \\
 & & & & & & & & \\
 22 & 5750 & 183453 & 0.3 & 7.6 & 3.063 & 0.153 & 0.261 & 270.7 \\
 & & & & & & & & \\
 22 & 5751 & 2158600 & 2.4 & 767.3 & 3.181 & 0.136 & 0.199 & 274.0 \\
 & & & & & & & & \\
 22 & 5752 & 850605 & 5.3 & 521.5 & 2.975 & 0.107 & 0.083 & 268.8 \\
 & & & & & & & & \\
 22 & 5753 & 765130 & 0.7 & 3.5 & 2.968 & 0.157 & 0.071 & 256.0 \\
 & & & & & & & & \\
 22 & 5754 & 1927087 & 215.0 & 601.9 & 3.065 & 0.142 & 0.531 & 254.3 \\
 & & & & & & & & \\
 23 & 5757 & 3611740 & 481.8 & 50.7 & 3.175 & 0.157 & 0.964 & 396.0 \\
 & & & & & & & & \\
 23 & 5759 & 851465 & 55.8 & 1.4 & 3.156 & 0.130 & 0.222 & 493.2 \\
 & & & & & & & & \\
 23 & 5759 & 1560950 & 918.3 & 7.7 & 3.036 & 0.135 & 0.301 & 493.2 \\
 & & & & & & & & \\
 23 & 5761 & 720780 & 2393.0 & 0.05 & 3.255 & 0.121 & 0.511 & 401.1 \\
 & & & & & & & & \\
 23 & 5763 & 1460515 & 3787.0 & 20.1 & 3.064 & 0.134 & 0.786 & 356.2 \\
 & & & & & & & & \\
 23 & 5763 & 2995924 & 14.1 & 2898.2 & 3.123 & 0.123 & 0.465 & 356.2 \\
 & & & & & & & & \\
 24 & 5770 & 802154 & 1090.5 & 469.7 & 3.058 & 0.120 & 0.443 & 365.0 \\
 & & & & & & & & \\
 25 & 5779 & 3513302 & 104.5 & 0.07 & 3.136 & 0.137 & 0.972 & 320.4 \\
 & & & & & & & & \\
 25 & 5780 & 1230432 & 1102.3 & 1928.7 & 3.178 & 0.136 & 0.895 & 288.6 \\
 & & & & & & & & \\
 25 & 5780 & 3397637 & 2606.3 & 76.2 & 3.006 & 0.102 & 0.287 & 288.6 \\
 & & & & & & & & \\
\hline
\end{tabular}
\caption{Properties of the events in the final selection from the E835p $^1P_1 \rightarrow J/\psi \pi^0 \rightarrow e^{+}e^{-} \gamma\gamma$ data channel (Part 3).}
\end{center}
\end{table}

\newpage

\clearpage

\linespread{1.5}
\begin{table}[htbp]
\begin{center}
\begin{tabular}{|c|c|c|c|c|c|c|c|c|}
\hline
 & & & & & & & & \\
Stack & Run & Event & EW1 & EW2 & $M(e^{+}e^{-})$ & $M(\gamma\gamma)$ & Prob. & $\sigma_{beam}$ \\
 & & & & & & & & \\
 & & & & & (GeV) & (GeV) & & (keV) \\
 & & & & & & & & \\
\hline
 & & & & & & & & \\
 26 & 5784 & 445076 & 178.3 & 0.2 & 3.157 & 0.115 & 0.358 & 420.3 \\
 & & & & & & & & \\
 26 & 5786 & 103702 & 411.8 & 234.6 & 3.076 & 0.126 & 0.076 & 316.6 \\
 & & & & & & & & \\
 26 & 5786 & 743320 & 0.4 & 20.8 & 3.046 & 0.128 & 0.172 & 316.6 \\
 & & & & & & & & \\
 26 & 5787 & 1619584 & 14.5 & 0.08 & 3.214 & 0.139 & 0.790 & 307.9 \\
 & & & & & & & & \\
 26 & 5789 & 2537904 & 7.9 & 5.7 & 2.976 & 0.115 & 0.247 & 249.0 \\
 & & & & & & & & \\
 27 & 5792 & 3409678 & 7.7 & 52.8 & 3.109 & 0.111 & 0.211 & 408.4 \\
 & & & & & & & & \\
 27 & 5794 & 350079 & 1038.8 & 563.2 & 3.073 & 0.125 & 0.941 & 395.5 \\
 & & & & & & & & \\
 27 & 5794 & 2560476 & 16.0 & 1514.2 & 3.161 & 0.134 & 0.987 & 395.5 \\
 & & & & & & & & \\
 27 & 5795 & 1613337 & 815.0 & 186.9 & 3.010 & 0.136 & 0.450 & 404.0 \\
 & & & & & & & & \\
 27 & 5799 & 1207572 & 706.0 & 8034.7 & 3.018 & 0.116 & 0.118 & 402.6 \\
 & & & & & & & & \\
 28 & 5811 & 3658020 & 563.9 & 3.3 & 3.077 & 0.125 & 0.867 & 393.4 \\
 & & & & & & & & \\
 28 & 5814 & 124370 & 965.4 & 0.01 & 3.080 & 0.123 & 0.179 & 334.2 \\
 & & & & & & & & \\
 35 & 7002 & 1109495 & 3294.5 & 1252.9 & 3.088 & 0.135 & 0.985 & 331.1 \\
 & & & & & & & & \\
 35 & 7003 & 909378 & 145.7 & 3.1 & 3.033 & 0.157 & 0.162 & 323.6 \\
 & & & & & & & & \\
 35 & 7003 & 1491672 & 25.6 & 134.5 & 3.013 & 0.135 & 0.061 & 323.6 \\
 & & & & & & & & \\
 35 & 7004 & 281849 & 7.8 & 200.0 & 3.095 & 0.126 & 0.789 & 289.9 \\
 & & & & & & & & \\
 35 & 7005 & 2418496 & 286.4 & 372.2 & 3.116 & 0.121 & 0.851 & 244.4 \\
 & & & & & & & & \\
 36 & 7009 & 1181123 & 68.4 & 910.6 & 3.172 & 0.154 & 0.661 & 311.6 \\
 & & & & & & & & \\
 36 & 7010 & 1697124 & 0.1 & 104.9 & 2.982 & 0.124 & 0.442 & 263.0 \\
 & & & & & & & & \\
 36 & 7010 & 2469465 & 83.3 & 749.1 & 3.039 & 0.110 & 0.115 & 263.0 \\
 & & & & & & & & \\
 36 & 7010 & 3059379 & 17.6 & 48.2 & 3.146 & 0.138 & 0.810 & 263.0 \\
 & & & & & & & & \\
\hline
\end{tabular}
\caption{Properties of the events in the final selection from the E835p $^1P_1 \rightarrow J/\psi \pi^0 \rightarrow e^{+}e^{-} \gamma\gamma$ data channel (Part 4).}
\end{center}
\end{table}

\newpage

\clearpage

\linespread{1.5}
\begin{table}[htbp]
\begin{center}
\begin{tabular}{|c|c|c|c|c|c|c|c|c|}
\hline
 & & & & & & & & \\
Stack & Run & Event & EW1 & EW2 & $M(e^{+}e^{-})$ & $M(\gamma\gamma)$ & Prob. & $\sigma_{beam}$ \\
 & & & & & & & & \\
 & & & & & (GeV) & (GeV) & & (keV) \\
 & & & & & & & & \\
\hline
 & & & & & & & & \\
 36 & 7011 & 1428728 & 241.0 & 70.3 & 3.080 & 0.125 & 0.482 & 261.7 \\
 & & & & & & & & \\
 36 & 7015 & 1358608 & 2517.1 & 4.9 & 3.112 & 0.134 & 0.518 & 251.2 \\
 & & & & & & & & \\
 36 & 7015 & 1959109 & 1414.0 & 0.06 & 3.037 & 0.145 & 0.095 & 251.2 \\
 & & & & & & & & \\
 37 & 7020 & 2073871 & 907.5 & 48.1 & 3.046 & 0.134 & 0.699 & 259.8 \\
 & & & & & & & & \\
 37 & 7020 & 2120235 & 84.1 & 57.6 & 3.110 & 0.135 & 0.966 & 259.8 \\
 & & & & & & & & \\
 38 & 7026 & 29150 & 552.2 & 0.05 & 3.071 & 0.143 & 0.948 & 296.1 \\
 & & & & & & & & \\
 38 & 7026 & 1164402 & 446.9 & 155.2 & 3.142 & 0.108 & 0.278 & 296.1 \\
 & & & & & & & & \\
 38 & 7026 & 2053173 & 47.5 & 67.5 & 3.075 & 0.103 & 0.285 & 296.1 \\
 & & & & & & & & \\
 38 & 7027 & 2346296 & 0.03 & 966.5 & 3.156 & 0.136 & 0.764 & 290.6 \\
 & & & & & & & & \\
 38 & 7028 & 2446455 & 227.2 & 939.6 & 3.042 & 0.116 & 0.765 & 285.7 \\
 & & & & & & & & \\
 38 & 7028 & 2661788 & 56.7 & 223.8 & 3.084 & 0.133 & 0.516 & 285.7 \\
 & & & & & & & & \\
 38 & 7029 & 447123 & 7.8 & 202.5 & 3.057 & 0.121 & 0.469 & 280.0 \\
 & & & & & & & & \\
 39 & 7031 & 1426725 & 4.6 & 139.0 & 3.074 & 0.143 & 0.735 & 308.8 \\
 & & & & & & & & \\
 39 & 7032 & 3016075 & 3349.8 & 907.0 & 3.119 & 0.130 & 0.987 & 304.4 \\
 & & & & & & & & \\
 39 & 7036 & 1516166 & 50.9 & 3.4 & 3.093 & 0.122 & 0.926 & 292.4 \\
 & & & & & & & & \\
 40 & 7046 & 914749 & 68.9 & 28.4 & 3.130 & 0.132 & 0.426 & 269.8 \\
 & & & & & & & & \\
 40 & 7047 & 2333095 & 298.1 & 103.1 & 3.281 & 0.135 & 0.185 & 268.1 \\
 & & & & & & & & \\
 41 & 7054 & 491455 & 0.2 & 4510.6 & 3.154 & 0.124 & 0.805 & 279.9 \\
 & & & & & & & & \\
 41 & 7054 & 2455302 & 277.1 & 144.7 & 3.092 & 0.133 & 0.492 & 279.9 \\
 & & & & & & & & \\
 41 & 7058 & 1145967 & 32.4 & 53.6 & 3.170 & 0.134 & 0.998 & 223.3 \\
 & & & & & & & & \\
 41 & 7058 & 2609649 & 3071.3 & 99.8 & 3.106 & 0.106 & 0.097 & 223.3 \\
 & & & & & & & & \\
 41 & 7059 & 93833 & 1306.1 & 2012.6 & 3.162 & 0.115 & 0.461 & 210.1 \\
 & & & & & & & & \\
\hline
\end{tabular}
\caption{Properties of the events in the final selection from the E835p $^1P_1 \rightarrow J/\psi \pi^0 \rightarrow e^{+}e^{-} \gamma\gamma$ data channel (Part 5).}
\end{center}
\end{table}

\newpage

\clearpage

 \baselineskip=24pt
\chapter{Luminosity Monitor Software}

The following is the FORTRAN code used to read data from the E835 luminosity monitor and calculate the luminosities for each data run:

\vspace*{30pt}

\small

\quad     PROGRAM LUMREAD

\noindent C=  READLUM.F

\noindent C=  ADC Channel 1: fixed detector 1 (FBL)

\noindent C=  ADC Channel 2: fixed detector 2 (FBR)

\noindent C=  ADC Channel 3: unused

\noindent C=  ADC Channel 4: movable detector (4 degree)

\noindent C=  Description:

\noindent C=  \quad 1) Open Luminosity Data File, Read the Spectrum From File by DAFT. 

\noindent C=  \quad 2) Fill HBOOK Histograms (for Spectrums)

\noindent C=  \quad 3) Print out the Integrated Luminosity for input run 

\noindent C=  \quad 4) Print out the Instantaneous Luminosity for input run

\noindent C=  \quad 5) Print out the Elapsed Time and Live Time for input run

\noindent C=  \quad  Modified For 

\noindent C=  \quad\quad 1) DAQ Spikes 

\noindent C=  \quad\quad 2) DAQ hangs

\noindent C=  \quad\quad 3) Background calculation

\noindent C=  \quad\quad 4) Cross Section calculation

\noindent C=  \quad\quad 5) Add movable detector for the beam displacement correction

      IMPLICIT NONE

      include 'daft.inc'

      INTEGER NEVT,I,J,K,IEVENT, ii

      CHARACTER*80 FILEN

      CHARACTER*4 RUNN

      INTEGER BUFFERSIZE, STATUS

      PARAMETER (BUFFERSIZE = 32768)

      INTEGER BUFFER(BUFFERSIZE)

      INTEGER FDS(DAFTFDSIZE)

      COMMON/E835DAFT/fds

      INTEGER*2 LENRC

      INTEGER EVENTSIZE, ICYCLE, ISTAT, TIMESIZE

      PARAMETER (EVENTSIZE = 512)

      REAL DATARAW(4,EVENTSIZE),DATACOR(4,EVENTSIZE)

      INTEGER DATALUM(EVENTSIZE)

      REAL RDATALUM(EVENTSIZE),CDATALUM(EVENTSIZE)

      REAL RCOUNTS(4), RCOUNTPS(4), CCOUNTS(4), CCOUNTPS(4)

      REAL RBGCNT(4,42),CBGCNT(4,42), BGCNL(4,42)

      REAL RCOUNTBG(4), CCOUNTBG(4)

      REAL RCOUNTT(4), RCOUNTOT(4)

      DATA RCOUNTS/0.0, 0.0, 0.0, 0.0/

      DATA RCOUNTOT/0.0, 0.0, 0.0, 0.0/

      DATA RCOUNTBG/0.0, 0.0, 0.0, 0.0/

      DATA CCOUNTS/0.0, 0.0, 0.0, 0.0/

      DATA CCOUNTBG/0.0, 0.0, 0.0, 0.0/

      INTEGER STARTTIME(6), STOPTIME(6)

      INTEGER BL(4), BR(4), INDEX

      INTEGER ELTIME, LITIME, COUNTS

      REAL XI,YI,XIYI,XI2,BGSUM,AA,BB

      REAL LDDTEL, LDDTLI, RRIO, RELTIME, RLITIME

      DATA LDDTEL/0.0/

      DATA LDDTLI/0.0/

      LOGICAL DOFIT

      REAL INSLUMFIXL,INSLUMFIXR,INSLUMMOV

      REAL INTLUMFIXL,INTLUMFIXR,INTLUMMOV 

      REAL RINTLUMFIXL,RINTLUMFIXR,RINTLUMMOV 

      REAL INSLUM, INTLUM

\noindent C=    Beam Displacement
      
      REAL BEAMXM(451), ASYMM(451), BEAMXP(451), ASYMP(451)

      REAL LUMDIFF, LUMTRUET, LUMTRUES, X, OFFSET

      REAL CFR, CFL, CFM 

      REAL INTLUMTRUEL,INTLUMTRUER,INTLUMTRUEM

      REAL INSLUMTRUEL,INSLUMTRUER,INSLUMTRUEM

      REAL BMMIN ,BMMAX, BPMIN, BPMAX

      DATA BMMIN/2.0774424E-02/        ! beam displacement -1.5 mm      

      DATA BMMAX/1.016171/             ! beam displacement -5.0 mm   

      DATA BPMIN/-1.7685434E-02/       ! beam displacement  1.5 mm

      DATA BPMAX/-1.017238/            ! beam displacement  5.0 mm      

\noindent C=    Cross section calculation

      REAL PI, MP, ALPHA, HBC2, DETRAD, B, RHO, ST, PLABFILE

      INTEGER RUNNUM, RUNFILE

      DATA PI/3.141592654/

      DATA MP/0.93827231/

      DATA ALPHA/0.00729927/

      DATA HBC2/0.38937966/

      REAL AST, BST, NST, BBA, BBB

      DATA AST/34.48/                ! const for SigmaT calculation

      DATA BST/89.7/

      DATA NST/-0.702/

      DATA BBA/13.643/               ! const for b calculation

      DATA BBB/-0.2053/

      REAL DETANGL, DETANGR, DETANGM, DETANG

      DATA DETANGL/3.496/            ! in degree, left detectors 

      DATA DETANGR/3.511/            ! in degree, left detectors 
   
      DATA DETANGM/1.704/            ! in degree, movable detectors 

      REAL OMEGAL, OMEGAR, OMEGAM, OMEGA

      DATA OMEGAL/0.91469e-4/        ! solid angle, left detectors 

      DATA OMEGAR/0.91377e-4/        ! solid angle, right detectors 

      DATA OMEGAM/1.4572e-4/         ! solid angle, movable detectors 

      REAL LUMCONSTL, LUMCONSTR, LUMCONSTM, LUMCONST, RATIO

      REAL SIGMAC, SIGMAN, SIGMAI, SIGMAT, DTDO

      REAL PLAB, ELAB, BETA, RS, ECM, KA, T, CGT, GT, GT4, CC

      REAL AA1, AA2, AA3, BB1, BB2, BB3

      real vaa(4), vbb(4)
 
\noindent C**   histogram stuff

      CHARACTER*40 HINAME

      INTEGER HINO

      REAL ELTIME2, INSLUM1, INSLUM2, INSLUM4, INSLUMI, LUMEST

      REAL HMEMOR

      COMMON/PAWC/HMEMOR(1500000)

\noindent C=    Program starts here

\noindent C=    Open lookup data file for beam displacement correction
 
      OPEN(unit=2, name='ASYMMETRY.TABLEM', type='old')

      DO I = 1, 451

\quad          READ(2,*) BEAMXM(I), ASYMM(I)

      END DO

      CLOSE(2)

      OPEN(unit=2, name='ASYMMETRY.TABLEP', type='old')

      DO I = 1, 451

      END DO

      CLOSE(2)

\noindent C=    Cross section calculation 

\noindent C==   Input the RUN NUMBER first

      TYPE *, 'Enter The RUN Number: '

      Accept *, RUNNUM

      OPEN(UNIT=2, NAME='../dat/runplab.txt', TYPE='old')

\noindent 1    READ(2, *, END=10) RUNFILE, PLABFILE

      IF(RUNNUM.EQ.RUNFILE) THEN

\quad            PLAB = PLABFILE

\quad            CLOSE(2)

\quad            GO TO 20

      END IF

      GO TO 1

\noindent 10   CLOSE(2)

      TYPE *, 'This RUN can not be found in the runplab.txt file'

      TYPE *, 'Please input the Pbar momentum in GeV: '

      Accept *, PLAB

\noindent 20   DO 60 I = 1, 3
         
        IF (I.EQ.1) THEN 

\quad          DETANG = DETANGL

\quad          OMEGA = OMEGAL

        ELSE IF(I.EQ.2) THEN

\quad          DETANG = DETANGR

\quad          OMEGA = OMEGAR

        ELSE IF(I.EQ.3) THEN 

\quad          DETANG = DETANGM

\quad          OMEGA = OMEGAM

        END IF

        DETRAD = DETANG*PI/180.0           ! Angle in radian 

        B = BBA + BBB*PLAB

        RHO = 0.0

        ST = AST + BST*PLAB**NST

        ELAB = SQRT(PLAB**2 + MP**2)

        BETA = PLAB/ELAB

        RS = (SQRT(PLAB**2+MP**2)+MP)**2 - PLAB**2

        ECM = SQRT(RS)                     ! Ecm calculation

        KA = (ELAB+MP)/(ELAB-MP) 

\noindent C-    t value evaluation

        T = 2.0*MP*DETRAD*DETRAD/KA        ! rough estimation 

        t = 2.0*MP*T

\noindent C-    Calculate dS/dt, first, Sigma-c 

        cgt = (1.0+(abs(t))/0.71)

        gt = 1.0/cgt/cgt

        gt4 = gt*gt*gt*gt

        cc = 4.0*PI*alpha*alpha/beta/beta/t/t

        sigmac = cc*hbc2*gt4

\noindent C-    Calculate dS/dt, second, Sigma-n 

        aa1 = (1.0+rho*rho)*st*st*exp(-b*abs(t))

        aa2 = 16.0*PI*hbc2

        sigman = aa1/aa2

\noindent C-    Calculate dS/dt, third, Sigma-i 

        aa1 = alpha*st*gt*gt*exp(-b*abs(t)/2.0)

        aa2 = beta*abs(t)

        bb1 = alog(b*abs(t)/2.0)

        bb2 = alog(1.0 + 8.0/0.71/b)

        cc  = 4.0*abs(t)/0.71

        bb3 = cc*alog(cc) + cc/2.0

        aa3 = alpha*(0.577+bb1+bb2+bb3)

        sigmai = aa1*(rho*cos(aa3)+sin(aa3))/aa2

        sigmat = sigmac + sigman + sigmai               ! total sigma 

\noindent C-    calculate dt/dOmega

        aa1 = (t + 4.0*mp*mp)

        aa2 = t*aa1*aa1*aa1/ka

        dtdo = sqrt(aa2)/4.0/PI/mp/mp

\noindent C-    get the lumconst

        lumconst = sigmat * dtdo * omega

        IF(I.EQ.1) THEN  

\quad            LUMCONSTL = LUMCONST

        ELSE IF(I.EQ.2) THEN

\quad            LUMCONSTR = LUMCONST

        ELSE IF(I.EQ.3) THEN

\quad            LUMCONSTM = LUMCONST

        END IF

\noindent  60   CONTINUE

\noindent C=    Boundaries for three detectors

      BL(1) = 75        ! left fixed detector

      BR(1) = 165

      BL(2) = 75        ! right fixed detector

      BR(2) = 165

      BL(4) = 200         ! movable detector

      BR(4) = 320

      BL(3) = 150         ! unused 

      BR(3) = 335

\noindent C==   Do some initializations

      DO I = 1, 4

\quad         DO J = 1, 21

\quad\quad           BGCNL(I,J) = FLOAT(BL(I) - 11+J)

\quad         END DO

\quad         DO J = 1, 21

\quad\quad           BGCNL(I,J+21) = FLOAT(BR(I) - 11+J)

\quad         END DO

\quad         DO J = 1, 42

\quad\quad           RBGCNT(I,J) = 0.0

\quad\quad           CBGCNT(I,J) = 0.0

\quad         ENDDO

      ENDDO

      DO I = 1, 512

\quad        RDATALUM(I) = 0.0

\quad        CDATALUM(I) = 0.0

\quad        DATALUM(I) = 0

      END DO

      DO I = 1, 4

\quad        DO J = 1, 512

\quad\quad          DATARAW(I,J) = 0.0

\quad\quad          DATACOR(I,J) = 0.0

\quad        END DO

      END DO

\noindent C==   HBOOK stuff here 

      CALL HLIMIT(1500000)

      call hbook1(101,'left detector  \$',512,0.0,512.0,0.)

      call hbook1(102,'right detector \$',512,0.0,512.0,0.)

      call hbook1(103,'unused \$',512,0.0,512.0,0.)

      call hbook1(104,'movable detector \$',512,

     +            0.0,512.0,0.)

      call hbook1(105,'Ratio = luminosity L / R \$',100,0.0,100.0,0.)

      call hbook1(106,'Ratio = lumonosity L / M \$',100,0.0,100.0,0.)

\noindent C=    Input the name of luminosity data file

       IF(RUNNUM.LE.352) THEN

\quad         WRITE(FILEN, 70) RUNNUM

\noindent  70      FORMAT('/lumin/runs00010352/lum000',I3,'.dat')

\quad         RUNN = FILEN(27:30)

       ELSE IF(RUNNUM.GT.352.AND.RUNNUM.LE.605) THEN

\quad         WRITE(FILEN, 71) RUNNUM

\noindent  71      FORMAT('/lumin/runs03530605/lum000',I3,'.dat')

\quad         RUNN = FILEN(27:30)

       ELSE IF(RUNNUM.GT.605.AND.RUNNUM.LE.786) THEN

\quad         WRITE(FILEN, 72) RUNNUM

\noindent  72      FORMAT('/lumin/runs06060786/lum000',I3,'.dat')

\quad         RUNN = FILEN(27:30)

       ELSE IF(RUNNUM.GT.786.AND.RUNNUM.LE.835) THEN

\quad         WRITE(FILEN, 73) RUNNUM

\noindent  73      FORMAT('/lumin/runs07910835/lum000',I3,'.dat')

\quad         RUNN = FILEN(27:30)

       ELSE IF(RUNNUM.GT.835.AND.RUNNUM.LE.999) THEN

\quad         WRITE(FILEN, 74) RUNNUM

\noindent  74      FORMAT('/lumin/runs08360999/lum000',I3,'.dat')

\quad         RUNN = FILEN(27:30)

       ELSE IF(RUNNUM.GT.999.AND.RUNNUM.LE.1999) THEN

\quad         WRITE(FILEN, 75) RUNNUM

\noindent  75      FORMAT('/lumin/runs10001999/lum00',I4,'.dat')

\quad         RUNN = FILEN(27:30)

       ELSE IF(RUNNUM.GE.2000.AND.RUNNUM.LE.2999) THEN

\quad         WRITE(FILEN, 76) RUNNUM

\noindent  76      FORMAT('/lumin/runs20002999/lum00',I4,'.dat')

\quad         RUNN = FILEN(27:30)

       ELSE IF(RUNNUM.GE.3000.AND.RUNNUM.LE.3999) THEN

\quad         WRITE(FILEN, 77) RUNNUM

\noindent  77      FORMAT('/lumin/runs30003999/lum00',I4,'.dat')

\quad         RUNN = FILEN(27:30)

       ELSE IF(RUNNUM.GE.4000) THEN

\quad         WRITE(FILEN, 78) RUNNUM

\noindent  78      FORMAT('../dat/lum00',I4,'.dat')

\quad         RUNN = FILEN(13:16)

       END IF

\noindent C==   Open file

      STATUS = daftopenf(fds, filen, DAFTRDONLY, buffer, BUFFERSIZE)

      IF(STATUS.NE.0) THEN

\quad         TYPE *, 'Error Opening File, Status = ', STATUS

\quad         STOP

      ENDIF

       NEVT = 1000000000

\noindent C=    Read the start time

      TIMESIZE = 6

      LENRC = daftgeteventf(fds, STARTTIME, TIMESIZE,

     \&        DAFTBORSKIP.or.DAFTEORSKIP.OR.DAFTLABSKIP.or.

     \&        DAFTBOFSKIP.or.DAFTEOFSKIP)    

      IF (STARTTIME(1) .GT. 99) STARTTIME(1) = 0

      IF(LENRC.LT.0)THEN

\quad          TYPE *, 'Error During First Read, Status = ', LENRC

\quad          STOP

      ENDIF

\noindent C=    Read the spectrum 

      DO 200 IEVENT = 1, NEVT

     DO 100 I = 1,4

\quad      LENRC = daftgeteventf(fds, datalum, EVENTSIZE,

\quad     \&        DAFTBORSKIP.or.DAFTEORSKIP.OR.DAFTLABSKIP.or.

\quad     \&        DAFTBOFSKIP.or.DAFTEOFSKIP)    

\quad      IF(LENRC.EQ.6) THEN 

\noindent C=    Read the stop time

\quad\quad          DO J = 1, 6

\quad\quad\quad          STOPTIME(J) = DATALUM(J)

\quad\quad          ENDDO

\quad\quad          IF (STOPTIME(1) .GT. 99) STOPTIME(1) = 0

\quad\quad          TYPE *, 'You have reached the end of this file'

\quad\quad          status = daftclosef(fds)

\quad\quad          GOTO 300

\quad      ENDIF

\quad      IF(LENRC.LT.0)THEN

\quad\quad          TYPE *, 'Error During Reading Spectrum, Status = ', LENRC

\quad\quad          goto 300

\quad      ENDIF

\quad      DO J = 1, EVENTSIZE

\quad\quad         DATARAW(I,J) = float(DATALUM(J))

\quad\quad         DATACOR(I,J) = float(DATALUM(J))

\quad      END DO 

\noindent 100  CONTINUE  

\noindent C=    Read the ELTIME and LITIME

      TIMESIZE = 1

      LENRC = daftgeteventf(fds, ELTIME, TIMESIZE,

     \&        DAFTBORSKIP.or.DAFTEORSKIP.OR.DAFTLABSKIP.or.

     \&        DAFTBOFSKIP.or.DAFTEOFSKIP)    

      LENRC = daftgeteventf(fds, LITIME, TIMESIZE,

     \&        DAFTBORSKIP.or.DAFTEORSKIP.OR.DAFTLABSKIP.or.

     \&        DAFTBOFSKIP.or.DAFTEOFSKIP)    

      IF(LENRC.LT.0)THEN

\quad          TYPE *, 'Error During Time Reading , Status = ', LENRC

\quad          STOP

      ENDIF

\noindent C==   Correction by live time for each spectrum

      DO I = 1, 4

\quad        DO J = 1, EVENTSIZE

\quad\quad            DATACOR(I,J) = DATACOR(I,J)*FLOAT(LITIME)/FLOAT(ELTIME) 

\quad        END DO

      END DO

      DO 150 I = 1, 4                

        DO J = 1, EVENTSIZE

\quad           RDATALUM(J) = DATARAW(I,J)

\quad           CDATALUM(J) = DATACOR(I,J)  

        END DO 
      
\noindent C---  book individual 120 second interval histogram

        WRITE(HINAME, 777) IEVENT

\noindent  777    FORMAT('intervall no. ',I5)

        HINO = (IEVENT*10 + I) + 1000

        CALL HBOOK1(HINO, HINAME, 512, 0., 512., 0.)

\noindent C=    Fill the spectrums  

        DO 105 J = 1, EVENTSIZE

\quad          CALL HF1(100+I,FLOAT(J),RDATALUM(J))

\quad          CALL HFILL(HINO, FLOAT(J), 0., RDATALUM(J))

\noindent  105    CONTINUE 

\noindent C=    Sum the boundaries counts for background calculation.  

\noindent C-    Raw counts

      DO J = 1, 21

\quad          RBGCNT(I,J) = RBGCNT(I,J) + RDATALUM(BL(I)-11+J)

      END DO
     
      DO J = 1, 21

\quad          RBGCNT(I,J+21) = RBGCNT(I,J+21) + RDATALUM(BR(I)-11+J)

      END DO

\noindent C-    Live time corrected counts

      DO J = 1, 21 

\quad          CBGCNT(I,J) = CBGCNT(I,J) + CDATALUM(BL(I)-11+J)

      END DO

      DO J = 1, 21

\quad          CBGCNT(I,21+J) = CBGCNT(I,21+J) + CDATALUM(BR(I)-11+J)

      END DO

\noindent C=    Sum the peak counts 

\noindent C-    Raw counts

      RCOUNTPS(I) = 0.0

      DO J = BL(I), BR(I)

\quad        RCOUNTPS(I) = RCOUNTPS(I) + RDATALUM(J)

      END DO

      RCOUNTS(I) = RCOUNTS(I) + RCOUNTPS(I)

      RCOUNTT(I) = 0.0

      DO J = 1, 512

\quad        RCOUNTT(I) = RCOUNTT(I) + RDATALUM(J)

      END DO

      RCOUNTOT(I) = RCOUNTOT(I) + RCOUNTT(I)

\noindent C-    Live time corrected counts

      CCOUNTPS(I) = 0.0

      DO J = BL(I), BR(I)

\quad        CCOUNTPS(I) = CCOUNTPS(I) + CDATALUM(J)

      END DO

      CCOUNTS(I) = CCOUNTS(I) + CCOUNTPS(I)

\noindent  150  CONTINUE 

      if ((rcountps(2).gt.0) .and. (rcountps(4).gt.0)) then 

\quad         ratio = rcountps(1)/rcountps(2)

\quad         call hf1(105,float(ievent),ratio)

\quad         if (rcountps(4) .gt. 0.) then

\quad\quad            ratio = rcountps(1)*lumconstm/rcountps(4)/lumconstl

\quad\quad            call hf1(106,float(ievent),ratio)

\quad         endif

      endif

      LDDTEL = LDDTEL + 1.0E-6*FLOAT(ELTIME)

      LDDTLI = LDDTLI + 1.0E-6*FLOAT(LITIME)
      
      ELTIME2 = 1.0E-6*FLOAT(ELTIME)

      INSLUM1 = RCOUNTPS(1)*1.0E+27/LUMCONSTL/ELTIME2

      INSLUM2 = RCOUNTPS(2)*1.0E+27/LUMCONSTR/ELTIME2

      INSLUM4 = RCOUNTPS(4)*1.0E+27/LUMCONSTM/ELTIME2

\noindent C --  in case movable detector is 0

      INSLUMI = (INSLUM1+INSLUM2)/2.0

      write(6, *) IEVENT, 1.0E-6*FLOAT(ELTIME), 1.0e-6*float(litime),

     \$            INSLUMI

 \noindent 200  CONTINUE

\noindent C==           Determine the background calculation method

\noindent C==   Raw counts

\noindent  300  DOFIT = .TRUE.

      DO 305 I = 1,4

\quad         IF(I.EQ.3) GOTO 305

\quad         DO J = 1, 42

\quad\quad            IF(RBGCNT(I,J).LT.1.0) DOFIT = .FALSE. 

\quad         ENDDO

\noindent  305  CONTINUE

\noindent C==   Background calculation

      IF(DOFIT) THEN

\quad        DO 306 I = 1,4

\quad           IF(I.EQ.3) GOTO 306

\quad           XI = 0.0

\quad           YI = 0.0

\quad           XIYI = 0.0

\quad           XI2 = 0.0

\quad           DO J = 1, 42

\quad\quad             RBGCNT(I,J) = ALOG(RBGCNT(I,J))

\quad\quad             YI = YI + RBGCNT(I,J)

\quad\quad             XI = XI + BGCNL(I,J)

\quad\quad             XIYI = XIYI + RBGCNT(I,J)*BGCNL(I,J)

\quad\quad             XI2 = XI2 + BGCNL(I,J)*BGCNL(I,J)

\quad           END DO

\quad           AA = (XI*XIYI - YI*XI2)/(XI**2 - 42.0*XI2)

\quad           BB = (42.0*XIYI - YI*XI)/(XI**2 - 42.0*XI2)

\quad           AA = EXP(AA)

\quad           type *, 'aa =', aa, 'bb =', bb

\quad           vaa(i) = aa

\quad           vbb(i) = bb

\quad           DO K = BL(I), BR(I)

\quad           RCOUNTBG(I) = RCOUNTBG(I) + AA*EXP(-BB*FLOAT(K))

           ENDDO

\noindent  306    CONTINUE

      ELSE

\quad        DO 308 I = 1,4

\quad\quad          IF(I.EQ.3) GOTO 308

\quad\quad          BGSUM = 0.0

\quad\quad          DO J = 1, 42

\quad\quad\quad             BGSUM = BGSUM + RBGCNT(I,J)

\quad\quad          ENDDO

\quad\quad          RCOUNTBG(I) = BGSUM*FLOAT(BR(I)-BL(I)+1)/42.0

\noindent  308   CONTINUE

      END IF

\noindent C==   Live time corrected counts

      DOFIT = .TRUE.

      DO 405 I = 1,4

\quad         IF(I.EQ.3) GOTO 405

\quad         DO J = 1, 42

\quad\quad            IF(CBGCNT(I,J).LT.1.0) DOFIT = .FALSE. 

\quad         ENDDO

\noindent  405  CONTINUE

      IF(DOFIT) THEN

\quad        DO 406 I = 1,4

\quad\quad           IF(I.EQ.3) GOTO 406

\quad\quad           XI = 0.0

\quad\quad           YI = 0.0

\quad\quad           XIYI = 0.0

\quad\quad           XI2 = 0.0

\quad\quad           DO J = 1, 42

\quad\quad\quad             CBGCNT(I,J) = ALOG(CBGCNT(I,J))

\quad\quad\quad             YI = YI + CBGCNT(I,J)

\quad\quad\quad             XI = XI + BGCNL(I,J)

\quad\quad\quad             XIYI = XIYI + CBGCNT(I,J)*BGCNL(I,J)

\quad\quad\quad             XI2 = XI2 + BGCNL(I,J)*BGCNL(I,J)

\quad\quad           END DO

\quad\quad           AA = (XI*XIYI - YI*XI2)/(XI**2 - 42.0*XI2)

\quad\quad           BB = (42.0*XIYI - YI*XI)/(XI**2 - 42.0*XI2)

\quad\quad           AA = EXP(AA)

\quad\quad           DO K = BL(I), BR(I)

\quad\quad\quad           CCOUNTBG(I) = CCOUNTBG(I) + AA*EXP(-BB*FLOAT(K))

\quad\quad           ENDDO

\noindent  406    CONTINUE

\quad      ELSE

\quad\quad        DO 408 I = 1,4

\quad\quad\quad          IF(I.EQ.3) GOTO 408

\quad\quad\quad          BGSUM = 0.0

\quad\quad\quad          DO J = 1, 42

\quad\quad\quad             BGSUM = BGSUM + CBGCNT(I,J)

\quad\quad          ENDDO

\quad\quad          CCOUNTBG(I) = BGSUM*FLOAT(BR(I)-BL(I)+1)/42.0

\noindent  408   CONTINUE

      END IF

\noindent C==   ======= Background calculation done ========

\noindent C==   Output the statistics

      TYPE *, '         '

      TYPE *, '         '

      TYPE *, '                      LUMINOSITY MONITOR OUTPUT'

      TYPE *, '                      ========================='

      TYPE *, '                         RUN NUMBER = ', RUNN 

      WRITE(6, 414) ECM

 \noindent 414  FORMAT(26X, 'Ecm        = ', F5.3) 

      TYPE *, '         '     

      WRITE(6,415) (STARTTIME(I), I=1,3) 

\noindent  415  FORMAT(1X,'RUN BEGIN DATE:  ',I2.2,'/',I2.2,'/',I2.2)

      WRITE(6,416) (STARTTIME(I), I=4,6) 

\noindent  416  FORMAT(1X,'RUN BEGIN TIME:  ',I2.2,':',I2.2,':',I2.2)

      WRITE(6,417) (STOPTIME(I), I=1,3) 

\noindent  417  FORMAT(1X,'RUN STOP DATE:   ',I2.2,'/',I2.2,'/',I2.2)

      WRITE(6,418) (STOPTIME(I), I=4,6) 

\noindent  418  FORMAT(1X,'RUN STOP TIME:   ',I2.2,':',I2.2,':',I2.2)

      TYPE *, '         '     

      TYPE *, 'ELAPSED TIME FOR THIS RUN, ET =', LDDTEL, 's'

      IF(LDDTLI.LE.1.0) LDDTLI = LDDTEL

      TYPE *, 'LIVE TIME FOR THIS RUN,    LT =', LDDTLI, 's'

      type *, 

     +'------------------------------------------------------------'

      TYPE *, '         '     

      type *, 

     +'Detector     Counts in       Luminosity    Effective lum.'

      type *, 

     +'            recoil peak        (nb-1)      corr for beam '

      type *, 

     +'                                           offset, LT (nb-1)'

      type *, 

     +'------------------------------------------------------------'

      write(6, *) 

      DO 500 I = 1,4

\quad        IF (I.EQ.3) GOTO 500

\quad        lumest = rcountot(i)*0.60*(lddtli/lddtel)*1.0E-6/lumconstl

\quad        write (6, *) i, rcountot(i), rcounts(i), rcountbg(i),

\quad     \$       rcountbg(i)/rcounts(i)

\quad        write (6, *) i, (rcounts(i)-rcountbg(i))/rcountot(i),

\quad     \$       lddtli/lddtel, lumest

\quad        RCOUNTS(I) = RCOUNTS(I) - INT(RCOUNTBG(I))

\quad        CCOUNTS(I) = CCOUNTS(I) - INT(CCOUNTBG(I))

\noindent  500  CONTINUE

      RINTLUMFIXL = RCOUNTS(1)*1.0E-6/LUMCONSTL

      RINTLUMFIXR = RCOUNTS(2)*1.0E-6/LUMCONSTR

      RINTLUMMOV = RCOUNTS(4)*1.0E-6/LUMCONSTM

      INTLUMFIXL = CCOUNTS(1)*1.0E-6/LUMCONSTL

      INTLUMFIXR = CCOUNTS(2)*1.0E-6/LUMCONSTR

      INTLUMMOV = CCOUNTS(4)*1.0E-6/LUMCONSTM

      INTLUM = (INTLUMFIXL+INTLUMFIXR+INTLUMMOV)/3.0

      INSLUMFIXL = CCOUNTS(1)*1.0E+27/LUMCONSTL/LDDTEL

      INSLUMFIXR = CCOUNTS(2)*1.0E+27/LUMCONSTR/LDDTEL

      INSLUMMOV = CCOUNTS(4)*1.0E+27/LUMCONSTM/LDDTEL

      INSLUM = (INSLUMFIXL+INSLUMFIXR+INSLUMMOV)/3.0

\noindent C==                   Beam Offset Correction

      LUMDIFF = 2.0*(INTLUMFIXL - INTLUMFIXR)

     +           /(INTLUMFIXL + INTLUMFIXR)

      X = LUMDIFF

      IF(LUMDIFF.GT.BPMIN.AND.LUMDIFF.LT.BMMIN) THEN

\quad         INTLUMTRUEL = INTLUMFIXL 

\quad         INTLUMTRUER = INTLUMFIXR

\quad         INTLUMTRUEM = INTLUMMOV

\quad         LUMTRUET = INTLUM

\quad         LUMTRUES = INSLUM        

      ELSE IF(LUMDIFF.LT.BPMAX.OR.LUMDIFF.GT.BMMAX) THEN

\quad        IF(LUMDIFF.LT.0.0) THEN

\quad\quad          CFL = 0.9972-1.1062*X-0.00507*X**2-1.0338*X**3

\quad\quad          CFR = 0.9997+0.00749*X+0.0500*X**2

\quad\quad          CFM = 1.0007-0.08216*X+0.2775*X**2

\quad        ELSE

\quad\quad          CFL = 0.9997-0.0103*X+0.0539*X**2

\quad\quad          CFR = 0.9939+1.1045*X-0.01369*X**2+1.01459*X**3

\quad\quad          CFM = 1.0008+0.0771*X+0.2833*X**2

\quad        END IF

\quad        INTLUMTRUEL = CFL*INTLUMFIXL 

\quad        INTLUMTRUER = CFR*INTLUMFIXR

\quad        INTLUMTRUEM = CFM*INTLUMMOV

\quad        LUMTRUET = (INTLUMTRUEL+INTLUMTRUER+INTLUMTRUEM)/3.0        

\quad        INSLUMTRUEL = CFL*INSLUMFIXL 

\quad        INSLUMTRUER = CFR*INSLUMFIXR

\quad        INSLUMTRUEM = CFM*INSLUMMOV

\quad        LUMTRUES = (INSLUMTRUEL+INSLUMTRUER+INSLUMTRUEM)/3.0        

      ELSE

\quad        IF(LUMDIFF.LT.0.0) THEN

\quad\quad          DO I = 50, 451

\quad\quad\quad            IF(LUMDIFF.GE.ASYMP(I)) THEN

\quad\quad\quad\quad            OFFSET = BEAMXP(I) + (BEAMXP(I-1)-BEAMXP(I))

\quad\quad\quad\quad     +               *(LUMDIFF-ASYMP(I))/(ASYMP(I-1)-ASYMP(I))

\quad\quad\quad\quad            GOTO 585

\quad\quad\quad            END IF

\quad\quad          END DO

\quad        ELSE 

\quad\quad          DO I = 50, 451

\quad\quad\quad            IF(LUMDIFF.LE.ASYMM(I)) THEN

\quad\quad\quad\quad            OFFSET = BEAMXM(I) + (BEAMXM(I-1)-BEAMXM(I))

\quad\quad\quad\quad     +               *(LUMDIFF-ASYMM(I))/(ASYMM(I-1)-ASYMM(I))

\quad\quad\quad\quad            GOTO 585

\quad\quad\quad            END IF

\quad\quad          END DO

\quad        END IF
        
\noindent  585    IF(LUMDIFF.LT.0.0) THEN

\quad\quad          CFL = 0.9972-1.1062*X-0.00507*X**2-1.0338*X**3

\quad\quad          CFR = 0.9997+0.00749*X+0.0500*X**2

\quad\quad          CFM = 1.0007-0.08216*X+0.2775*X**2

\quad        ELSE

\quad\quad          CFL = 0.9997-0.0103*X+0.0539*X**2

\quad\quad          CFR = 0.9939+1.1045*X-0.01369*X**2+1.01459*X**3

\quad\quad          CFM = 1.0008+0.0771*X+0.2833*X**2

\quad        END IF

\quad        INTLUMTRUEL = CFL*INTLUMFIXL 

\quad        INTLUMTRUER = CFR*INTLUMFIXR

\quad        INTLUMTRUEM = CFM*INTLUMMOV

\quad        LUMTRUET = (INTLUMTRUEL+INTLUMTRUER+INTLUMTRUEM)/3.0        

\quad        INSLUMTRUEL = CFL*INSLUMFIXL 

\quad        INSLUMTRUER = CFR*INSLUMFIXR

\quad        INSLUMTRUEM = CFM*INSLUMMOV

\quad        LUMTRUES = (INSLUMTRUEL+INSLUMTRUER+INSLUMTRUEM)/3.0        

      END IF

        TYPE *, '         '     

        COUNTS = INT(RCOUNTS(1))

        WRITE(6, 592) COUNTS, RINTLUMFIXL, INTLUMTRUEL

\noindent  592    FORMAT(1X,'Left ', 6X, I8, 9X, F9.3, 5X, F9.3)

        COUNTS = INT(RCOUNTS(2))

        WRITE(6, 593) COUNTS, RINTLUMFIXR, INTLUMTRUER

\noindent  593    FORMAT(1X,'Right ', 5X, I8, 9X, F9.3, 5X, F9.3)

        COUNTS = INT(RCOUNTS(4))

        WRITE(6, 594) COUNTS, RINTLUMMOV, INTLUMTRUEM

\noindent 594    FORMAT(1X,'Middle ', 4X, I8 ,9X, F9.3, 5X, F9.3)

        TYPE *, '     ' 

        WRITE(6, 600) LUMTRUET

\noindent  600    FORMAT(8X,'Average Integrated    Luminosity = ', F9.3, ' nb-1')

        WRITE(6, 601) LUMTRUES

\noindent 601    FORMAT(8X,'Average Instantaneous Luminosity = ',2X,E9.3)

      type *, 

     +'------------------------------------------------------------'

      TYPE *, '         '

      TYPE *, 'Asymmetry =', LUMDIFF

      IF(LUMDIFF.GT.BPMIN.AND.LUMDIFF.LT.BMMIN) THEN

\quad         TYPE *, 'Beam offset = less than 1.5 mm'

      ELSE IF(LUMDIFF.LT.BPMAX.OR.LUMDIFF.GT.BMMAX) THEN

\quad         TYPE *, 'Beam offset = greater than 5.0 mm'

\quad         TYPE *, 'WARNING: CALL MCR !!!'

      ELSE

\quad        IF(OFFSET.GT.0.0) THEN

\quad\quad           WRITE(6, 590) OFFSET

\noindent 590       FORMAT(1X,'Beam offset = +', F5.3, ' mm')

\quad        ELSE

\quad           WRITE(6, 591) OFFSET

\noindent 591       FORMAT(1X,'Beam offset = ', F6.3, ' mm')

\quad        END IF

      END IF

      TYPE *, '         '

      WRITE(6, 780) LUMCONSTL*1.0E06 

\noindent 780  FORMAT(1X,'Luminosity constant, left detector = ',

     \$     F12.6, ' nb')

      WRITE(6, 781) LUMCONSTR*1.0E06 

\noindent 781  FORMAT(1X,'Luminosity constant, right detector = ',

     \$     F12.6, ' nb')

      WRITE(6, 782) LUMCONSTM*1.0E06 

\noindent 782  FORMAT(1X,'Luminosity constant, movable detector = ',
     \$     F12.6, ' nb')

      CALL HRPUT(0, 'luminst.hbook', 'NT')

      OPEN(10, file='lum.vec', form='formatted', type='unknown')

      vaa(3) = 1

      vbb(3) = 1

      do i = 1, 4

\quad         write(10, *) vaa(i), vbb(i), bl(i), br(i)

      enddo

      close(10)

      STOP 

      END

%Bibliography
%\addcontentsline{toc}{chapter}{Bibliography}
%\include{bibliography}
%\include{thesis_bib}

\end{document}